\documentclass[a4paper,11pt]{report}
\usepackage{jheppub}

\makeatletter
\renewcommand\ps@myplain{%
  \renewcommand\@oddfoot{\hfill-- \thepage\ --\hfill}%
  \renewcommand\@evenfoot{\hfill-- \thepage\ --\hfill}%
  \renewcommand\@oddhead{}%
  \renewcommand\@evenhead{}}
\let\ps@plain\ps@myplain
\makeatother

\usepackage{etoolbox}
\usepackage{url}
\usepackage[capitalize,nameinlink]{cleveref}

\usepackage{minitoc} 

\input{Qcircuit} 

\usepackage[T1]{fontenc}

\newcommand{\tr}{\text{tr}}
\newcommand{\rank}{\text{rank}}

\newcommand{\coQNP}{\mathsf{coQNP}}
\newcommand{\QNP}{\mathsf{QNP}}
\newcommand{\NP}{\mathsf{NP}}

\newcommand{\FR}{\mathsf{FR}}
\newcommand{\cc}{\mathsf{cc}}

\newcommand{\NC}{\mathsf{NC}}
\newcommand{\LogSpace}{\mathsf{L}}
\newcommand{\PolyTime}{\mathsf{P}}
\newcommand{\BQP}{\mathsf{BQP}}
\newcommand{\GH}{\mathsf{GH}}

\newcommand{\Rent}{\mathsf{R}\|^*}
\newcommand{\Rsim}{\mathsf{R}\|}
\newcommand{\Qent}
{\mathsf{Q}\|^*}
\newcommand{\Qsim}{\mathsf{Q}\|}
\newcommand{\Dsim}{\mathsf{D}\|}
\newcommand{\Qone}{\mathsf{Q}1}
\newcommand{\Rone}{\mathsf{R}1}
\newcommand{\Qtwo}{\mathsf{Q}2}
\newcommand{\Rtwo}{\mathsf{R}2}

\newcommand{\SWAP}{\textnormal{SWAP}}

\newcommand{\CDS}{\mathsf{CDS}}
\newcommand{\CDQS}{\mathsf{CDQS}}
\newcommand{\pp}{\textnormal{pp}}
\newcommand{\pc}{\textnormal{pc}}

\newcommand{\ModkL}{\mathsf{Mod}_k\mathsf{L}}

\newcommand{\ketbra}[2]{|#1\rangle\!\langle#2|}
\newcommand{\braket}[2]{\langle#1|#2\rangle}  

\newtheorem{theorem}{Theorem}

\newtheorem{corollary}[theorem]{Corollary}

\newtheorem{definition}[theorem]{Definition}

\newtheorem{lemma}[theorem]{Lemma}

\newtheorem{remark}[theorem]{Remark}

\newenvironment{proof}[1][Proof]{\noindent\textbf{#1.} }{\ \rule{0.5em}{0.5em}}

\usepackage{epigraph}
\usepackage{subcaption}
\usepackage{tikz}
\usepackage{makecell}
\usetikzlibrary{positioning, calc}
\usetikzlibrary{arrows}
\usepackage{tikz-3dplot}
\usepackage{graphicx}
\usepackage{mdwlist}
\usepackage{color}

\usetikzlibrary{decorations.pathreplacing,decorations.markings,snakes}
\tikzset{late arrow/.style={postaction={decorate},
  decoration={markings,mark=at position 0.85 with {\arrow{stealth}}}}}
\usetikzlibrary{decorations.pathmorphing}
\tikzset{snake it/.style={decorate, decoration=snake}}
\usetikzlibrary{decorations.pathreplacing,decorations.markings}

\tikzset{
    >=stealth',
    punkt/.style={
           rectangle,
           rounded corners,
           draw=black, very thick,
           text width=6.5em,
           minimum height=2em,
           text centered},
    pil/.style={
           ->,
           thick,
           shorten <=2pt,
           shorten >=2pt,},
  on each segment/.style={
    decorate,
    decoration={
      show path construction,
      moveto code={},
      lineto code={
        \path [#1]
        (\tikzinputsegmentfirst) -- (\tikzinputsegmentlast);
      },
      curveto code={
        \path [#1] (\tikzinputsegmentfirst)
        .. controls
        (\tikzinputsegmentsupporta) and (\tikzinputsegmentsupportb)
        ..
        (\tikzinputsegmentlast);
      },
      closepath code={
        \path [#1]
        (\tikzinputsegmentfirst) -- (\tikzinputsegmentlast);
      },
    },
  },
  mid arrow/.style={postaction={decorate,decoration={
        markings,
        mark=at position .5 with {\arrow[#1]{stealth'}}
      }}}
}

\title{Entanglement cost in non-local quantum computation}
\author[a]{Alex May}
\affiliation[a]{Perimeter Institute for Theoretical Physics}
\emailAdd{amay@perimeterinstitute.ca}

\abstract{This is a book-length treatment of the subject of non-local quantum computation (NLQC). 
NLQC is a method for implementing quantum operations that interact two systems without directly bringing the systems together. 
Instead, a single round of communication and shared entanglement is used.
NLQC has appeared in the context of quantum cryptography, computational complexity, communication complexity, quantum gravity, and other applications. 
The understanding of entanglement cost in NLQC is closely tied to questions in all of these areas.
We review upper and lower bounds on entanglement cost, as well as some of the applications of NLQC and its connections to other subjects. Video lectures that cover a portion of the material here are available: \url{https://pirsa.org/c26009}.}

\begin{document} 

\dominitoc
\maketitle

\part{Preliminaries}\label{part:prelim}

\chapter{Introduction}\label{chapter:introduction}

\minitoc

\section{The central problem of NLQC}

In a non-local quantum computation (NLQC), a local interaction is replaced with shared entanglement and a single, simultaneous round of quantum communication. 
See figure \ref{fig:non-localandlocal} for the general form, and figure \ref{fig:CNOTexample} for a simple example showing how to implement the CNOT gate as an NLQC.\footnote{We thank Richard Cleve for finding this simple example.}  
In this book, we are interested in understanding this transformation, and especially in understanding how much entanglement is necessary to re-express a given local interaction in this `non-local' form. 
This question of entanglement cost is arguably the central problem in the study of NLQC.

\vspace{0.2cm}
\noindent \textbf{Central problem of NLQC:} \emph{Given a quantum channel $\mathcal{N}_{AB}$, how much entanglement is needed to implement it as a non-local quantum computation?}
\vspace{0.2cm}

\noindent Why do we want to understand this question?
One broad perspective is that quantum information theory is a theory of quantum resources, where we seek to understand what can be done with a given amount of a resource and how different resources can be used to replace one another. 
From this perspective NLQC is the study of how two natural quantum resources relate to one another, in particular entanglement and local interactions.
More concretely, NLQC appears in a long and growing list of applications, including cryptography, quantum gravity, complexity, and others.
In each of these applications, core questions in those areas relate to the central problem of NLQC. 

\begin{figure*}
    \centering
    \begin{subfigure}{0.45\textwidth}
    \centering
    \begin{tikzpicture}[scale=0.6]
    
    \draw[thick] (-1,-1) -- (-1,1) -- (1,1) -- (1,-1) -- (-1,-1);
    
    \draw[thick] (-3.5,-3) to [out=90,in=-90] (-0.5,-1);
    \node[below] at (-3.5,-3) {$A$};
    \draw[thick] (3.5,-3) to [out=90,in=-90] (0.5,-1);
    \node[below] at (3.5,-3) {$B$};
    
    \draw[thick] (0.5,1) to [out=90,in=-90] (3.5,3);
    \node[above] at (3.5,3) {$B$};
    \draw[thick] (-0.5,1) to [out=90,in=-90] (-3.5,3);
    \node[above] at (-3.5,3) {$A$};
    
    \node at (0,0) {$\mathcal{T}$};

    \node at (0,-5) {$ $};
    
    \end{tikzpicture}
    \caption{}
    \label{fig:local}
    \end{subfigure}
    \hfill
    \begin{subfigure}{0.45\textwidth}
    \centering
    \begin{tikzpicture}[scale=0.4]
    
    \draw[thick] (-5,-5) -- (-5,-3) -- (-3,-3) -- (-3,-5) -- (-5,-5);
    \node at (-4,-4) {$\mathcal{V}^L$};
    
    \draw[thick] (5,-5) -- (5,-3) -- (3,-3) -- (3,-5) -- (5,-5);
    \node at (4,-4) {$\mathcal{V}^R$};
    
    \draw[thick] (5,5) -- (5,3) -- (3,3) -- (3,5) -- (5,5);
    \node at (4,4) {$\mathcal{W}^R$};
    
    \draw[thick] (-5,5) -- (-5,3) -- (-3,3) -- (-3,5) -- (-5,5);
    \node at (-4,4) {$\mathcal{W}^L$};
    
    \draw[thick] (-4.5,-3) -- (-4.5,3);
    
    \draw[thick] (4.5,-3) -- (4.5,3);
    
    \draw[thick] (-3.5,-3) to [out=90,in=-90] (3.5,3);
    
    \draw[thick] (3.5,-3) to [out=90,in=-90] (-3.5,3);
    
    \draw[thick] (-3.5,-5) to [out=-90,in=-90] (3.5,-5);
    \draw[black] plot [mark=*, mark size=3] coordinates{(0,-7.05)};
    
    \draw[thick] (-4.5,-6) -- (-4.5,-5);
    \node[below] at (-4.5,-6) {$A$};
    \draw[thick] (4.5,-6) -- (4.5,-5);
    \node[below] at (4.5,-6) {$B$};
    
    \draw[thick] (4.5,5) -- (4.5,6);
    \node[above] at (-4.5,6) {$A$};
    \draw[thick] (-4.5,5) -- (-4.5,6);
    \node[above] at (4.5,6) {$B$};
    
    \end{tikzpicture}
    \caption{}
    \label{fig:non-localcomputation}
    \end{subfigure}
    \caption{Local and non-local computations. a) A channel $\mathcal{T}_{AB\rightarrow AB}$ is implemented by directly interacting the input systems. b) A non-local quantum computation. The goal is for the action of this circuit on the $AB$ systems to approximate the channel $\mathcal{T}_{AB\rightarrow AB}$.}
    \label{fig:non-localandlocal}
\end{figure*}

\begin{figure*}
    \centering
    \begin{subfigure}{0.45\textwidth}
    \centering
    \begin{tikzpicture}[scale=0.6]

    \draw[thick] (-3.5,-3) to [out=90,in=-90] (-0.5,-1);
    \node[below] at (-3.5,-3) {$A$};
    \draw[thick] (3.5,-3) to [out=90,in=-90] (0.5,-1);
    \node[below] at (3.5,-3) {$B$};
    
    \draw[thick] (0.5,1) to [out=90,in=-90] (3.5,3);
    \node[above] at (3.5,3) {$B$};
    \draw[thick] (-0.5,1) to [out=90,in=-90] (-3.5,3);
    \node[above] at (-3.5,3) {$A$};

    \draw[thick] (-0.5,-1) -- (-0.5,1);
    \draw[thick] (0.5,-1) -- (0.5,1);

    \draw[black] plot [mark=*, mark size=3] coordinates{(-0.5,0)};
    \draw[thick] (0.5,0) circle (0.3);
    \draw[thick] (-0.5,0) -- (0.8,0);

    \node at (0,-5) {$ $};
    
    \end{tikzpicture}
    \caption{}
    \end{subfigure}
    \hfill
    \begin{subfigure}{0.45\textwidth}
    \centering
    \begin{tikzpicture}[scale=0.4]
    
    \draw[thick] (-4.5,-5) -- (-4.5,5);
    \draw[black] plot [mark=*, mark size=3] coordinates{(-4.5,-4)};
    \draw[thick] (-3.5,-4) circle (0.3);
    \draw[thick] (-4.5,-4) -- (-3.2,-4);
    
    \draw[black] plot [mark=*, mark size=3] coordinates{(-4.5,4)};
    \draw[thick] (-3.5,4) circle (0.3);
    \draw[thick] (-4.5,4) -- (-3.2,4);
    
    \draw[thick] (4.5,-5) -- (4.5,5);
    \draw[black] plot [mark=*, mark size=3] coordinates{(3.5,-4)};
    \draw[thick] (4.5,-4) circle (0.3);
    \draw[thick] (3.5,-4) -- (4.8,-4);
    
    \draw[black] plot [mark=*, mark size=3] coordinates{(3.5,4)};
    \draw[thick] (4.5,4) circle (0.3);
    \draw[thick] (3.5,4) -- (4.8,4);
    
    \draw[thick] (-3.5,-3) to [out=90,in=-90] (3.5,3) -- (3.5,6);
    
    \draw[thick] (3.5,-3) to [out=90,in=-90] (-3.5,3) -- (-3.5,6);
    
    \draw[thick] (-3.5,-3) -- (-3.5,-5) to [out=-90,in=-90] (3.5,-5) -- (3.5,-3);
    \draw[black] plot [mark=*, mark size=3] coordinates{(0,-7.05)};
    
    \draw[thick] (-4.5,-6) -- (-4.5,-5);
    \node[below] at (-4.5,-6) {$A$};
    \draw[thick] (4.5,-6) -- (4.5,-5);
    \node[below] at (4.5,-6) {$B$};
    
    \draw[thick] (4.5,5) -- (4.5,6);
    \node[above] at (-4.5,6) {$A$};
    \draw[thick] (-4.5,5) -- (-4.5,6);
    \node[above] at (4.5,6) {$B$};
    
    \end{tikzpicture}
    \caption{}
    \end{subfigure}
    \caption{a) The CNOT gate, implemented as a local computation. b) The CNOT gate implemented as an NLQC. The reader can check the basis states $\ket{a,b}_{AB}$ are transformed correctly by the circuit. An interesting additional fact is that the entanglement used is left unchanged at the end of the protocol.}
    \label{fig:CNOTexample}
\end{figure*}

Let's consider what's known about the central problem of NLQC. 
Let $\mathcal{N}_{AB}$ be a quantum channel with $A$ and $B$ both $n$ qubit Hilbert spaces, and denote by $E(\mathcal{N}_{AB})$ the number of qubits of shared entanglement needed to implement $\mathcal{N}_{AB}$ as an NLQC. 
Then, using a technique known as port-teleportation, we will see in chapter \ref{chapter:allchannels} that
\begin{align}\label{eq:PTupperbound}
    \forall \,\mathcal{N}, \,\,\,\,E(\mathcal{N}) \leq 2^{\alpha \,n}.
\end{align}
For certain special families of channels we have better upper bounds.
However, these apply only in limited cases, and even then may not be tight upper bounds. 
Meanwhile, considering lower bounds we know that
\begin{align}
    \exists\, \mathcal{N}: E(\mathcal{N}) \geq \beta \,n.
\end{align}
Further, we know some explicit examples of channels that satisfy this lower bound. 
However, we don't have any lower bounds on NLQC that are stronger than linear, at least without assuming any mathematical conjectures. 
Thus for most channels we have at best a linear lower bound, sometimes no lower bound at all, and an exponential upper bound. 
It's clear we have a lot left to understand about entanglement cost in NLQC. 

Aside from the exponential upper bound that applies for all unitaries, we've also discovered that the \emph{complexity} of the interaction to be implemented can provide an upper bound on the entanglement cost of implementing it. 
One sharp manifestation of this occurs in the context of a special class of NLQCs, known as $f$-routing. 
An instance of $f$-routing is defined by fixing a Boolean function $f:\{0,1\}^{2n}\rightarrow \{0,1\}$. 
The input on the left will be a quantum system consisting of $O(1)$ qubits, along with a classical string $x$ of length $n$. 
The input on the right will be a classical string $y$ of length $n$. 
In this setting, it has been proven that the number of EPR pairs needed to route on a function $f$, $E(f)$, is upper bounded by
\begin{align}
    E(f)\leq 2^{O(\text{Memory}(f))}
\end{align}
where $\text{Memory}(f)$ means the number of bits of memory used by a Turing machine that computes $f$.
This means the functions we know how to do efficiently include the functions in the complexity class $\LogSpace$, called ``logspace'', a well studied object in classical complexity theory. 

Reflecting on this result, it's not clear at all why complexity theory should appear, especially classical complexity theory --- we started with a distributed problem, about redirecting a quantum system based on some split up inputs $(x,y)$. 
Where is the Turing machine in the problem statement? 
In fact it's not even clear why any model of computation, which is about how much of some local resource is needed to run a computation, should be relevant to this non-local, distributed, problem, but nonetheless this emerges naturally. 

Another fascinating appearance of complexity in NLQC is as follows. 
Suppose that we want to implement a unitary $U_{AB}$ as an NLQC. 
Then there is an upper bound on the entanglement cost that relates to how many gates are needed to implement $U$ in a quantum circuit that implements $U$. 
In particular choose a gate set consisting of the Clifford gates plus the $T$ gate. 
We will consider writing $U$ as a layer of Cliffords, then a layer of $T$ gates, then a layer of Cliffords and so on. 
We call the minimal number of layers of $T$ gates the $T$-depth of $U$. 
Letting $E(U)$ be the minimal number of EPR pairs needed to implement $U$ as an NLQC, we have that
\begin{align}
    E(U)\leq O((68n)^{T\text{-depth}(U)}).
\end{align}
So again the entanglement cost is upper bounded by a notion of complexity. 

The relationship between complexity and NLQC turns out to be even more rich than has been outlined so far. 
This relationship provides a number of opportunities: it means for instance that good entanglement lower bounds in NLQC would provide complexity lower bounds.
While complexity lower bounds are notoriously difficult to prove, at least some progress on this difficult problem might be possible using entanglement lower bounds. 

One hint that NLQC may be useful in understanding quantum computation broadly comes from an analogy with the classical subject of communication complexity. 
In communication complexity, Alice holds an input $x$ and Bob holds an input $y$. 
They communicate back and forth to compute a function of their joint inputs, $f(x,y)$. 
Two things turn out to be true about this setting: 1) putting lower bounds on communication is often possible and 2) complexity upper bounds communication cost. 
Combining these two facts leads to interesting lower bounds on complexity in some cases. 
More broadly communication complexity has developed into a useful tool in the study of computational complexity. 

NLQC is to quantum computation something like communication complexity is to classical computation: in NLQC and communication complexity both, the goal is to perform a computation but the focus is on the cost in terms of non-local, distributed, resources.
We can hope that, just like for communication complexity and classical computation, NLQC can be developed into a powerful tool for studying quantum computation. 

\section{Applications of NLQC}

One reason we're motivated to study NLQC is because a broad set of applications relate to this simple setting. 
These settings span physics, complexity theory, and cryptography, and consequently NLQC provides a link among these areas. 
In this book we will understand several of these connections. 

Historically, starting with a patent application in 2006 \cite{kent2006tagging}, NLQC first arose in the context of \emph{quantum position verification} (QPV).
This is a subject in cryptography which studies the possibility of verifying someone's location in space.
Cheating strategies in QPV amount to examples of NLQCs.
This leads to a desire to prove good lower bounds on NLQC, so that we can show cheating in certain QPV schemes is difficult.

\begin{figure*}
\begin{center}
\begin{subfigure}[b]{.8\textwidth}
\begin{center}
\begin{tikzpicture}[scale=0.55]
    
    \draw[fill=gray,opacity=0.4] (-1.5,1) -- (1.5,1) -- (1.5,7) -- (-1.5,7) -- (-1.5,1);
    
    \draw[->] (-7,-1) -- (-7,0);
    \node[above] at (-7,0) {$t$};
    \draw[->] (-7,-1) -- (-6,-1);
    \node[right] at (-6,-1) {$x$};
    
    \draw[->] (-4,-1) -- (-4,-0.1);
    \node[below] at (-4,-1) {$A_0$};
    
    \draw[->] (4,-1) -- (4,-0.1);
    \node[below] at (4,-1) {$A_1$};
    
    \node[left] at (-4,0) {$c_0$};
    \draw[fill=black] (-4,0) circle (0.15);
    
    \node[right] at (4,0) {$c_1$};
    \draw[fill=black] (4,0) circle (0.15);
    
    \draw[->] (4,8) -- (4,9);
    \node[above] at (4,9) {$B_1$};
    \draw[->] (-4,8) -- (-4,9);
    \node[above] at (-4,9) {$B_0$};

    \node[right] at (4,8) {$r_1$};
    \draw[fill=blue] (4,8) circle (0.15);

    \node[left] at (-4,8) {$r_0$};
    \draw[fill=blue] (-4,8) circle (0.15);
    
    \node[below] at (0,-0.53) {$ $};
    
    \end{tikzpicture}
\end{center}
\caption{}
\label{subfig:PVset-up}
\end{subfigure}
\begin{subfigure}[b]{.45\textwidth}
\begin{center}
\begin{tikzpicture}[scale=0.55]
    
    \draw[fill=gray,opacity=0.4] (-1.5,1) -- (1.5,1) -- (1.5,7) -- (-1.5,7) -- (-1.5,1);
    
    \draw[postaction={on each segment={mid arrow}}] (-4,0) -- (0,4);
    \draw[postaction={on each segment={mid arrow}}] (4,0) -- (0,4);
    \draw[postaction={on each segment={mid arrow}}] (0,4) -- (4,8);
    \draw[postaction={on each segment={mid arrow}}] (0,4) -- (-4,8);
    
    \node[below left] at (-4,0) {$c_0$};
    \draw[fill=black] (-4,0) circle (0.15);

    \node[below right] at (4,0) {$c_1$};
    \draw[fill=black] (4,0) circle (0.15);

    \node[above right] at (4,8) {$r_1$};
    \draw[fill=blue] (4,8) circle (0.15);

    \node[above left] at (-4,8) {$r_0$};
    \draw[fill=blue] (-4,8) circle (0.15);
    
    \draw[fill=yellow] (0,4) circle (0.30);
    
    \node[below] at (0,-0.53) {$ $};
    
    \node[above left] at (-3,1) {$A_0$};
    \node[above right] at (3,1) {$A_1$};
    
    \node[below left] at (-3,7) {$B_0$};
    \node[below right] at (3,7) {$B_1$};
    
    \end{tikzpicture}
\end{center}
\caption{}\label{subfig:localschematic}
\end{subfigure}
\hfill
\begin{subfigure}[b]{.45\textwidth}
\begin{center}
\begin{tikzpicture}[scale=0.55]

    \draw[fill=gray,opacity=0.4] (-1.5,1) -- (1.5,1) -- (1.5,7) -- (-1.5,7) -- (-1.5,1);

    \node[below left] at (-4,0) {$c_0$};
    \draw[fill=black] (-4,0) circle (0.15);

    \node[below right] at (4,0) {$c_1$};
    \draw[fill=black] (4,0) circle (0.15);

    \node[above right] at (4,8) {$r_1$};
    \draw[fill=blue] (4,8) circle (0.15);

    \node[above left] at (-4,8) {$r_0$};
    \draw[fill=blue] (-4,8) circle (0.15);
    
    \draw[postaction={on each segment={mid arrow}}] (-4,0) -- (-2,2) -- (-2,6) -- (-4,8);
    \draw[postaction={on each segment={mid arrow}}] (4,0) -- (2,2) -- (2,6) -- (4,8);
    \draw[postaction={on each segment={mid arrow}}] (-2,2) -- (0,4) -- (2,6);
    \draw[postaction={on each segment={mid arrow}}] (2,2) -- (0,4) -- (-2,6);
    
    \draw[dashed] (2,2) -- (0,0) -- (-2,2);
    \node[below] at (0,0) {$\Psi_{LR}$};
    
    \draw[fill=yellow] (-2,2) circle (0.3);
    \draw[fill=yellow] (2,2) circle (0.3);
    \draw[fill=yellow] (-2,6) circle (0.3);
    \draw[fill=yellow] (2,6) circle (0.3);
    
    \node[above left] at (-3,1) {$A_0$};
    \node[above right] at (3,1) {$A_1$};
    
    \node[below left] at (-3,7) {$B_0$};
    \node[below right] at (3,7) {$B_1$};
    
\end{tikzpicture}
\end{center}
\caption{}\label{subfig:nonlocalschematic}
\end{subfigure}
\caption{Time proceeds upwards in all of these diagrams, and the horizontal direction is a spatial dimension. Light rays follow lines with slope $\pm 1$. (a) Set-up for the challenge given by the verifier to check the prover's location in a QPV scheme. Input systems $A_0$ and $A_1$ are received at spacetime locations $c_0$ and $c_1$, respectively, and $B_0$ and $B_1$ should be returned at $r_0$ and $r_1$, respectively. The inputs and outputs should be related by some designated channel $\mathcal{N}_{A_0A_1\rightarrow B_0B_1}$. (b) Completing the task in a local form. The yellow circle represents a channel acting on input systems $A_0$ and $A_1$, and producing output systems $B_0$ and $B_1$. The prover acts within the grey region, corresponding to an honest strategy. (c) A computation happening in the cheating, non-local, form. $A_0$ is interacted with the $L$ system, and $A_1$ with the $R$ system, where $\Psi_{LR}$ is entangled. Then, a round of communication is exchanged, and a second round of operations on each side are performed. All operations happen outside of the spacetime region.}
\label{fig:1dQPV}
\end{center}
\end{figure*}

In more detail, in QPV a prover and verifier interact to establish the spatial location of the prover. 
To do this, the verifier issues a challenge to the prover, which they believe can only be accomplished if the prover applies quantum or classical operations within the spacetime region of interest. 
The challenge involves quantum and classical systems input at one set of spacetime locations and another set of input and output systems returned at a second, later set of spacetime points. 

We illustrate the typical position verification set-up in figure \ref{subfig:PVset-up}. 
At spacetime locations $c_0$ and $c_1$, which are spatially separated but occur at the same time, inputs $A_0$ and $A_1$ are sent towards the grey shaded region. 
The prover is supposed to transform those inputs in some way, and return output systems $B_0$ and $B_1$ to points $r_0$ and $r_1$, respectively. 
The prover can either do this in an honest way, as shown in figure \ref{subfig:localschematic}, or in the dishonest way, as shown in figure \ref{subfig:nonlocalschematic}.
In the honest way, the prover brings $A_0$ and $A_1$ together inside of the grey region, acts on them, and returns the outputs. 
Completing the task in this way requires that the prover can implement quantum operations inside of the region. 
Alternatively, the prover can act in the cheating form, which avoids using any quantum operations inside of the grey region. 
Notice that when translated to a quantum circuit this is exactly an NLQC. 

Ideally, the challenge used in the context of position verification is easy to complete in the honest strategy, and as hard as possible to complete in the NLQC form. 
This is why the central problem of NLQC is important for QPV: we want to find NLQCs which require a lot of entanglement to complete, so that it is hard to cheat in the corresponding QPV scheme. 
Ideally, we would also find a scheme which is easy to implement in the honest, local, form. 
This is one reason that $f$-routing has been considered: to complete it honestly, $f$-routing involves computing a classical function, and then doing $O(1)$ quantum gates on the quantum system involved. 
Meanwhile, it is hoped that for some choices of function $f$ the quantum resources needed to $f$-route will grow with $n$, the classical input size.
Indeed, this has been proven when considering protocols that implement $f$-routing exactly (with zero error) or when considering the number of quantum gates used in the NLQC, but is open when considering the entanglement use in a realistic, noisy, setting.

Before moving on to some other applications, it is useful to note some differing language that you might find in the literature. 
In our viewpoint of QPV, there is a verifier and prover, and the prover may act in the local or non-local forms. 
An alternative viewpoint is to consider an honest player located inside the grey region, and ask if the verifier can be sure they are communicating with that person or with two collaborating attackers sitting on either side of the grey region, who try to pretend to be the player inside. 
This is a different setting and application for QPV, but both scenarios motivate the study of NLQC. 

Since its original appearance in QPV, NLQC has appeared in several other places in quantum information theory and beyond. 

Around 2019, a connection between NLQC and quantum gravity was realized \cite{may2019quantum,may2020holographic}. 
A key idea in gravity is the holographic principle, which asserts that gravity in $d$ dimensions should have an alternative description in terms of a $d-1$ dimensional quantum mechanical theory. 
AdS/CFT is a concrete realization of this idea. 
In this context NLQC turns out to give insight into how interactions in $d$ dimensions can be reproduced in just $d-1$ dimensions. 
The reason for this is fairly simple: already we saw in the QPV context that NLQC lets us re-express things happening inside of a region as something happening instead near the boundary of a region. 
This is also what happens in the context of AdS/CFT and holography, and it turns out that NLQC plays a role in how it realizes this re-expression. 

The realization of the role for NLQC in quantum gravity coincided with an increased interest in developing practical QPV protocols, which was happening around the same time. 
These two developments together led to an increased interest in the subject. 
One consequence of this was the discovery of another connection between NLQC and another unexpected subject, this time within classical information-theoretic cryptography. 
In 2023 \cite{allerstorfer2024relating}, it was understood that two subjects in classical cryptography known as conditional disclosure of secrets (CDS) and private simultaneous messages (PSM) are each closely related to special cases of NLQC. 
In fact, CDS turns out to be a close classical analogue of $f$-routing, and CDS protocols for a given function $f$ actually imply $f$-routing protocols with similar efficiency. 
This classical-quantum connection has a number of surprising consequences, including new $f$-routing upper bounds and new CDS lower bounds.
More broadly, these developments relate the understanding of the cost of privacy in information-theoretic cryptography to the study of entanglement cost in NLQC. 

Recently, in another surprising connection, NLQC was used to develop new protocols in the context of \emph{communication complexity}. 
Communication complexity studies how much communication is needed to compute Boolean functions $f(x,y)$ of two inputs, where $x$ and $y$ are initially separated. 
A long standing goal in communication complexity is to understand when quantum resources provide an advantage over classical resources, and in particular to find the weakest quantum resources which still allow a quantum advantage.  
NLQC techniques allowed \cite{girish2025magic} the development of a new protocol that showed a quantum advantage using even weaker resources than had been achieved before. 

In fact, there are several further applications of NLQC, which are not described here, but include applications to uncloneable secret sharing \cite{ananth2024unclonable} and Hamiltonian complexity \cite{apel2024security}.
As we will highlight in this book, all these applications inform NLQC, which in turn informs the applications.

\section{Comments on this book}

I believe it is an exciting time to study NLQC: the subject is connected closely with many other areas of quantum information science, and these connections have for the most part only emerged in the last 5 years, leaving plenty of ground for exploration.
I find it exciting to work on problems which are motivated simultaneously by the aim to understand the limits of quantum resources in communication complexity, the role of entanglement and complexity in the emergence of spacetime, and the aim to develop secure and practical cryptographic protocols, among many other widely dispersed goals. 
I hope that this text will find use among students and researchers interested in further developing the subject of NLQC.

This book is divided into four parts. 
Part I covers basic material, including simple examples of NLQC, some initial upper and lower bounds on these simple examples, as well as the $f$-routing example of NLQC and its connection to information theoretic cryptography. 
Part II covers protocols for implementing NLQC, including the generic port-teleportation based protocol that shows every channel can be implemented as an NLQC.
Part III covers lower bound techniques.
A very brief part IV studies mappings among examples of NLQCs, a subject which has just begun to be explored.  
The second and third parts each conclude with applications of the techniques covered; an application to communication complexity in the case of upper bounds, and to quantum gravity  and QPV in the case of lower bounds. 
Other applications are mentioned in the core chapters. 
The chapters in this book do not need to be read linearly; dependencies among the chapters are shown in figure \ref{fig:dependencies}.  

The choice of topics here is of course biased by my own research interests. 
One peculiarity is that this book is not focused on QPV, which historically has been the main driving application behind the study of NLQC. 
I do include one chapter, chapter \ref{chapter:QPV}, which discusses QPV in some detail and in particular the challenges that arise in proving the security of (noisy) QPV experiments. 
This chapter captures what is (in my view) the most practical current approach to security and discusses a number of other approaches and variations of the standard QPV setting, but doesn't describe the full breadth of work in this area. 
Outside of QPV, I also haven't covered a number of interesting developments in both applications and theory.
For instance, I have not covered an approach to lower bounding NLQC based on the geometry of Banach spaces \cite{junge2022geometry}, the connection to uncloneable secret sharing \cite{ananth2024unclonable}, or the connection to Hamiltonian complexity \cite{apel2024security}. 

\begin{figure}
    \centering
    \begin{tikzpicture}
    \coordinate (CIntro) at (0,10);
    \coordinate (CNLQC) at (0,8);
    \coordinate (CAll) at (3,8);
    \coordinate (Cf-routing) at (0,6);
    \coordinate (CTdepth) at (3,6);
    \coordinate (Ccomm) at (6,6);
    \coordinate (CUnit) at (-3,6);
    \coordinate (Cf-lower) at (2,4);
    \coordinate (Cmono) at (-3,8);
    \coordinate (Cgrav) at (-6,8);
    \coordinate (CITC) at (-2,4);
    \coordinate (Creduction) at (-5,4);
    \coordinate (CTcount) at (6,8);
    \coordinate (QPV) at (-6,6);
    
    \draw[mid arrow] (CIntro) -- (CNLQC);
    \draw[mid arrow] (CNLQC) -- (CAll);
    \draw[mid arrow] (CNLQC) -- (Cf-routing);
    \draw[mid arrow] (Cf-routing) -- (CTdepth);
    \draw[mid arrow] (CTdepth) -- (Ccomm);
    \draw[mid arrow] (Cf-routing) -- (Cf-lower);
    \draw[mid arrow] (Cf-routing) -- (CITC);
    \draw[mid arrow] (CNLQC) -- (CUnit);
    \draw[mid arrow] (Cmono) -- (Cgrav);
    \draw[mid arrow] (Cf-routing) -- (Creduction);
    \draw[mid arrow] (CNLQC) -- (Cmono);
    \draw[mid arrow] (CAll) -- (CTcount);
    \draw[mid arrow] (CNLQC) -- (QPV);
    
    \filldraw[color=white, fill=white] (CIntro) ellipse (1.1 and 0.5);
    \filldraw[color=white, fill=white] (CNLQC) ellipse (1.3 and 0.75);
    \filldraw[color=white, fill=white] (CAll) ellipse (1.1 and 0.5);
    \filldraw[color=white, fill=white] (Cf-routing) ellipse (1.1 and 0.5);
    \filldraw[color=white, fill=white] (CTdepth) ellipse (1.1 and 0.5);
    \filldraw[color=white, fill=white] (Ccomm) ellipse (1.1 and 0.5);
    \filldraw[color=white, fill=white] (CUnit) ellipse (1.1 and 0.5);
    \filldraw[color=white, fill=white] (Cf-lower) ellipse (1.1 and 0.5);
    \filldraw[color=white, fill=white] (Cmono) ellipse (1.1 and 0.5);
    \filldraw[color=white, fill=white] (Cgrav) ellipse (1.1 and 0.5);
    \filldraw[color=white, fill=white] (CITC) ellipse (1.1 and 0.5);
    \filldraw[color=white, fill=white] (Creduction) ellipse (1.1 and 0.5);
    \filldraw[color=white, fill=white] (CTcount) ellipse (1.1 and 0.5);
    \filldraw[color=white, fill=white] (QPV) ellipse (1.1 and 0.5);
    
    \node at (CIntro) {Chapter \ref{chapter:introduction}};
    \node[below, align=center, font=\tiny, yshift=-0.5ex] at (CIntro) {Introduction};
    
    \node at (CNLQC) {Chapter \ref{chapter:NLQC}};
    \node[below, align=center, font=\tiny, yshift=-0.5ex] at (CNLQC) {Non-local quantum \\ computation};
    
    \node at (CAll) {Chapter \ref{chapter:allchannels}};
    \node[below, align=center, font=\tiny, yshift=-0.5ex] at (CAll) {Any channel as \\ an NLQC};
    
    \node at (Cf-routing) {Chapter \ref{chapter:f-routing}};
    \node[below, align=center, font=\tiny, yshift=-0.5ex] at (Cf-routing) {$f$-routing};
    
    \node at (CTdepth) {Chapter \ref{chapter:Tdepth}};
    \node[below, align=center, font=\tiny, yshift=-0.5ex] at (CTdepth) {$T$-depth and NLQC};
    
    \node at (Ccomm) {Chapter \ref{chapter:communicationcomplexity}};
    \node[below, align=center, font=\tiny, yshift=-0.5ex] at (Ccomm) {Application: Separation of \\ $\Rent$ and $\Rtwo$};
    
    \node at (CUnit) {Chapter \ref{chapter:quantumlowerbound}};
    \node[below, align=center, font=\tiny, yshift=-0.5ex] at (CUnit) {Lower bounds \\ for unitaries};
    
    \node at (Cf-lower) {Chapter \ref{chapter:mostlyclassicallowerbounds}};
    \node[below, align=center, font=\tiny, yshift=-0.5ex] at (Cf-lower) {Lower bounds \\on $f$-routing};
    
    \node at (Cmono) {Chapter \ref{chapter:monogamygames}};
    \node[below, align=center, font=\tiny, yshift=-0.5ex] at (Cmono) {Monogamy games};
    
    \node at (Cgrav) {Chapter \ref{chapter:gravity}};
    \node[below, align=center, font=\tiny, yshift=-0.5ex] at (Cgrav) {Application: \\Quantum gravity};
    
    \node at (CITC) {Chapter \ref{chapter:ITCandNLQC}};
    \node[below, align=center, font=\tiny, yshift=-0.5ex] at (CITC) {NLQC and ITC};
    
    \node at (Creduction) {Chapter \ref{chapter:reductions}};
    \node[below, align=center, font=\tiny, yshift=-0.5ex] at (Creduction) {Reductions};

    \node at (CTcount) {Chapter \ref{chapter:Tcount}};
    \node[below, align=center, font=\tiny, yshift=-0.5ex] at (CTcount) {Upper bound from \\ $T$-count};

    \node at (QPV) {Chapter \ref{chapter:QPV}};
    \node[below, align=center, font=\tiny, yshift=-0.5ex] at (QPV) {Application: Quantum \\ position verification};

\end{tikzpicture}
    \caption{Dependencies among the chapters in this book.}
    \label{fig:dependencies}
\end{figure}

\section{History and further reading}

The roots of NLQC can be traced all the way back to a line of thought by Landau and Peierls \cite{landau1931erweiterung}, who considered a problem they called ``instantaneous measurement''. 
In this problem Alice and Bob, who are spatially separated, are given systems $A$ and $B$ respectively. 
They want to make a measurement, described by a set of projectors $\{\Pi_i\}$. 
They will try to do this ``instantaneously'' in the sense that they will each make a measurement separately on their own systems, and then send the outcomes to a referee, who should then be able to determine the measurement outcome of the projectors $\{\Pi_i\}$.
Instantaneous measurement is similar to NLQC, but with only measurements performed in the first round.

Landau and Peierls concluded that this was not possible for all measurements. 
Later, it was realized \cite{aharonov1981can} that some additional measurements were made possible if entanglement was shared by Alice and Bob, a possibility not considered by Landau and Peierls. 
Finally, Vaidman \cite{vaidman2003instantaneous} showed that with enough entanglement, any measurement could be implemented instantaneously in this sense. 

The earliest appearance of quantum position verification, and along with it NLQC, is in a patent application filed by Kent, Munroe, and Spiller in 2006 \cite{kent2006tagging}. 
Later, QPV appeared in the academic literature in an article due to Malaney \cite{malaney2010location}. 
Kent, Munroe and Spiller then put their ideas into the academic literature as well \cite{kent2011quantum}, and pointed out that at least the simplest proposals for QPV schemes are insecure due to entanglement based attacks. 
Buhrman et al. then introduced a general attack that breaks all QPV schemes in principle, though uses a doubly-exponential amount of entanglement \cite{buhrman2014position}.  
They borrowed from the technique employed by Vaidman for instantaneous measurement to develop their protocol. 

\chapter{Non-local quantum computation}\label{chapter:NLQC}

\minitoc

In this chapter, we consider two simple examples of non-local quantum computation. 
These examples will start to give a sense of the setting and the role of entanglement in NLQC. 

To describe our examples, it is helpful to introduce some language and conventions. 
We consider two parties, Alice and Bob, who we can think of as spatially separated, with Alice on the left and Bob on the right.
Alice and Bob share a joint quantum state $\Psi_{LR}$, with $L$ held by Alice and $R$ held by Bob. 
Alice is given an input system $A$, and Bob is given an input system $B$.
Alice and Bob execute a process of the form shown in figure \ref{fig:non-localandlocal}.  
In words, Alice acts on $AL$ and Bob acts on $RB$. 
They can then exchange a single simultaneous round of quantum communication, act again on the systems they hold after the communication, and then return output systems. 
Their goal is for the map from their inputs and outputs to approximate a desired quantum channel acting jointly on $AB$. 

\section{First examples of NLQC}

Before going further, we first provide some definitions. 
We first define what an NLQC aims to achieve, which is to complete what we call a $2\rightarrow 2$ quantum task. 
\begin{definition}
    A \textbf{$2\rightarrow 2$ quantum task} is defined by a pair of input systems $A$, $B$, a pair of output systems $A'$, $B'$, and a set of input/output state pairs $\mathcal{S}=\{ (\rho_{RAB},\sigma_{RA'B'})\}$. We require that there exists at least one quantum channel $\mathcal{M}_{AB \to A'B'}$ such that $(\mathcal{I}_R \otimes \mathcal{M}_{AB\rightarrow A'B'})(\rho_{RAB}) = \sigma_{RA'B'}$ for all $\rho_{RAB}$.
\end{definition}

Next, we can define an NLQC. 
\begin{definition}
    A \textbf{non-local quantum computation (NLQC)} is a channel in the form
    \begin{align}
        \mathcal{N}_{AB\rightarrow A'B'}(\,\cdot\,) =(\mathcal{W}_{K_aM_a\rightarrow A'}\otimes\mathcal{W}_{K_bM_b\rightarrow B'})\circ (\mathcal{V}_{AL\rightarrow K_{a}M_b}\otimes \mathcal{V}_{RB\rightarrow M_aK_b})(\,\cdot\, \otimes\Psi_{LR}).
    \end{align}
    We refer to $\Psi_{LR}$ as the resource system, the $\mathcal{V}$ channels as the first round operations, and the $\mathcal{W}$ channels as the second round operations.

    We say an NLQC is an $\epsilon$-correct implementation of a $2\rightarrow 2$ task $F_n$ if the channel $\mathcal{N}_{AB\rightarrow A'B'}$ implemented as an NLQC is $\epsilon$-close in diamond norm to at least one channel $\mathcal{M}_{AB\rightarrow A'B'}$ relating the input and output states in the definition of the $2\rightarrow 2$ task.
\end{definition}

The first example we consider is called \emph{routing} \cite{kent2011quantum}.

\vspace{0.2cm}
\noindent \textbf{Routing task:}
\begin{itemize*}
    \item \textbf{Inputs:} On the left, a quantum system $Q$ in unknown state $\ket{\psi}_Q$. On the right, a single classical bit $b\in \{0,1\}$. 
    \item \textbf{Outputs:} Return $\ket{\psi}_Q$ on the left if $b=0$, and on the right if $b=1$. 
\end{itemize*}
Before we consider how to complete this task as an NLQC, it's worth considering what it is that seems to be hard about doing so. 
Consider in particular Alice on the left: She holds $\ket{\psi}_Q$, but $b$ is far away. 
Since $\ket{\psi}_Q$ is an unknown quantum state, she can't copy it and send it to both sides.
Further, she can't wait to find out $b$ and then send it to the appropriate location. 
This is because there is only one round of communication: once she's received information from the right, it's too late to send anything back. 
Apparently, Alice's best strategy is to guess $b$, in which case she succeeds with probability $1/2$.

Using a maximally entangled state $\ket{\Psi^+}_{LR}$ shared between the left and right however, Alice can complete this task with probability $1$. 
To do this, Alice measures $QL$ in the Bell basis, as if she were teleporting $Q$ onto the $R$ system.
Doing so she obtains a measurement outcome $s$. 
This partly solves her problem: since $s$ is classical, she can copy it and send it to both the left and right. 
Additionally, the state on $R$ is now $P^s\ket{\psi}_R$ for some Pauli $P^s$ fixed by $s$. 
Since $R$ is located on the right, Bob can send the quantum part, $P^s\ket{\psi}_R$, to the appropriate location.  
Thus both $s$ and $P^s\ket{\psi}_R$ arrive at the needed location.
Alice or Bob can then undo $P^s$ and hand in $\ket{\psi}$ as needed, completing their task. 
In this example we see that the apparent obstruction presented by the no-cloning theorem is evaded by the use of entanglement. 

\begin{figure}
\centering
\begin{subfigure}[b]{0.45\textwidth}
\begin{tikzpicture}[scale=0.6]

\draw[->] (-0.5,0) -- (-0.5,1);

\node[below] at (-0.5,0) {$\ket{\psi}$};

\draw[black] (-1,1) -- (1,1) -- (1,2) -- (-1,2) -- (-1,1);
\draw (-0.6,1.1) arc (180:0:0.6);
\draw (0,1.1) -- (0.7,1.8);

\draw[double,postaction={on each segment={mid arrow}}] (0,2) -- (4,7);

\draw[postaction={on each segment={mid arrow}}] (5,-0.5) to [out=100,in = -80] (4.5,7);
\node[below] at (5,-0.5) {$R$};

\node[above] at (1.75,4.5) {$s$};

\draw[postaction={on each segment={mid arrow}}] (1,-0.5) to [out=110,in=-90] (0.5,1);
\node[below] at (1,-0.5) {$L$};

\draw[dashed] (1.5,-1) -- (5,-1);
\node[below] at (3.25,-1) {$\ket{\Psi^+}$};

\draw[black] (3.75,7) -- (4.75,7) -- (4.75,8) -- (3.75,8) -- (3.75,7);
\node at (4.25,7.5) {$P_s$};
\draw[black,->] (4.25,8) -- (4.25,9);
\node[above] at (4.25,9) {$\ket{\psi}$};

\end{tikzpicture}
\caption{}
\label{fig:BellTeleportation}
\end{subfigure}
\hfill
\begin{subfigure}[b]{0.45\textwidth}
    \begin{tikzpicture}[scale=0.7]
    
    \draw[postaction={on each segment={mid arrow}}] (-4,0) -- (-2,2) -- (-2,6) -- (-4,8);
    \draw[postaction={on each segment={mid arrow}}] (4,0) -- (2,2) -- (2,6) -- (4,8);
    \draw[postaction={on each segment={mid arrow}}] (-2,2) -- (0,4) -- (2,6);
    \draw[postaction={on each segment={mid arrow}}] (2,2) -- (0,4) -- (-2,6);
    
    \draw[dashed] (2,2) -- (0,0) -- (-2,2);
    \node[below] at (0,0) {$\ket{\Psi^+}_{LR}$};
    
    \draw[fill=yellow] (-2,2) circle (0.3);
    \draw[fill=yellow] (2,2) circle (0.3);
    \draw[fill=yellow] (-2,6) circle (0.3);
    \draw[fill=yellow] (2,6) circle (0.3);
    
    \node[below left] at (-4,0) {$c_1$};
    \draw[fill=black] (-4,0) circle (0.15);

    \node[below right] at (4,0) {$c_2$};
    \draw[fill=black] (4,0) circle (0.15);

    \node[below right] at (4,8) {$r_2$};
    \draw[fill=blue] (4,8) circle (0.15);

    \node[below left] at (-4,8) {$r_1$};
    \draw[fill=blue] (-4,8) circle (0.15);
    
    \node[above left] at (-3,0.9) {\small{$\ket{\psi}_Q$}};
    \node[above right] at (3,0.9) {\small{$b$}};
    
    \node[below left] at (-3,7) {\small{$\ket{\psi}_Q$ if}};
    \node[below left] at (-3,6.4) {\small{$b=0$}};
    
    \node[below right] at (3,7) {\small{$\ket{\psi}_Q$ if}};
    \node[below right] at (3,6.4) {\small{$b=1$}};
    
    \node[left] at (-2,4) {\small{$s$}};
    \node[right] at (2,4) {\small{$R$ if $b=1$}};
    
    \node[right] at (-1.6,2.4) {\small{$s$}};
    \node[left] at (1.4,2.4) {\small{$R$ if}};
    \node[below] at (0.9,2.2) {\small{$b=0$}};
    
    \end{tikzpicture}
    \caption{}
    \label{fig:singlesystemsummoning}
    \end{subfigure}
    \caption{(a) The quantum teleportation protocol. Before the beginning of the protocol, there is a quantum state $\ket{\psi}$ held on the left and an entangled state $\ket{\Psi^+}$ shared between left and right. (b) An implementation of the routing task in the form of a non-local quantum computation. On the left, before the communication, Alice measures $QL$ in the Bell basis. Bob on the right then redirects $R$ to the left or right conditioned on the value of $b$. }
    \label{fig:teleportationinspacetime}
\end{figure}

As a second example, let's look at the following task. 

\vspace{0.2cm}
\noindent \textbf{Measurement task:} 
\begin{itemize*}
    \item \textbf{Input:} On the left a qubit system $Q$ in one of the states $H^{q}\ket{b}_Q$, with $q,b\in\{0,1\}$ and $H$ the Hadamard. On the right, the bit $q$. 
    \item \textbf{Output:} $b$ on both the left and the right. 
\end{itemize*}
Here, Alice on the left faces a different obstacle compared to before: she holds one of the four states $H^q\ket{b}_Q$, where $q,b\in\{0,1\}$. 
She wants to know $b$, but doesn't know $q$ so doesn't know which basis she should measure in to learn $b$. 
If she waits to find out $q$, she'll be too late to get $b$ out on both sides. 

\begin{figure}
    \centering
    \subfloat[\label{fig:nonlocalcircuit}]{
    \begin{tikzpicture}[node distance = 2cm, auto,rotate=90]

    \tikzset{meter/.append style={draw, inner sep=10, rectangle, font=\vphantom{A}, minimum width=30, line width=.8,
 path picture={\draw[black] ([shift={(.1,.3)}]path picture bounding box.south west) to[bend left=50] ([shift={(-.1,.3)}]path picture bounding box.south east);\draw[black,-latex] ([shift={(0,.1)}]path picture bounding box.south) -- ([shift={(.3,-.1)}]path picture bounding box.north);}}}
    
    \draw (0,0) -- (1,2);
    \draw (0,0) -- (1,-2);
    \node[below] at (0,0) {\large{$\ket{\Psi^+}_{LR}$}};
    
    \draw (1,-2) -- (2,-2);
    
    \draw (2,-2.5) -- (2,-1.5) -- (3,-1.5) -- (3,-2.5) -- (2,-2.5); 
    \node at (2.5,-2) {\large{$H^q$}};
    
    \draw (3,-2) -- (5.5,-2);
    
    \node[meter] at (6,-2) {};

    \node[meter] at (6,3.5) {};

    \draw[thick] (4.5,3.5) -- (5.5,3.5);
    
    \draw (1,2) -- (5.5,2);
    \draw (0,3.5) -- (3.5,3.5);
    
    \node[below] at (0,3.5) {$H^q\ket{b}$};
    
    \draw[fill=black] (2.5,3.5) circle (0.1);
    \draw (2.5,3.5) -- (2.5,1.8);
    \draw (2.5,2) circle (0.2);
    
    \draw (3.5,4) -- (4.5,4) -- (4.5,3) -- (3.5,3) -- (3.5,4);
    \node at (4,3.5) {\large{$H$}};
    
    \node[meter] at (6,2) {};
    
    \draw[blue,->] (6.5,2) -- (7,2);
    \node[above] at (7,2) {$s_2$};
    
    \draw[blue,->] (6.5,3.5) -- (7,3.5);
    \node[above] at (7,3.5) {$s_1$};
    
    \draw[blue,->] (6.5,-2) -- (7,-2);
    \node[above] at (7,-2) {$s_3$};
    
    \draw[blue] (0,-2.25) -- (2,-2.25);
    \node[below] at (0,-2.25) {$q$};
    
    \end{tikzpicture}
    }
    \hfill
    \subfloat[\label{fig:nonlocalschematic}]{
    \begin{tikzpicture}[scale=0.7]
    
    \draw[postaction={on each segment={mid arrow}}] (-4,0) -- (-2,2) -- (-2,6) -- (-4,8);
    \draw[postaction={on each segment={mid arrow}}] (4,0) -- (2,2) -- (2,6) -- (4,8);
    \draw[postaction={on each segment={mid arrow}}] (-2,2) -- (0,4) -- (2,6);
    \draw[postaction={on each segment={mid arrow}}] (2,2) -- (0,4) -- (-2,6);
    
    \draw[dashed] (2,2) -- (0,0) -- (-2,2);
    \node[below] at (0,0) {$\ket{\Psi^+}_{LR}$};
    
    \draw[fill=yellow] (-2,2) circle (0.3);
    \draw[fill=yellow] (2,2) circle (0.3);
    \draw[fill=yellow] (-2,6) circle (0.3);
    \draw[fill=yellow] (2,6) circle (0.3);
    
    \node[below left] at (-4,0) {$c_1$};
    \draw[fill=black] (-4,0) circle (0.15);

    \node[below right] at (4,0) {$c_2$};
    \draw[fill=black] (4,0) circle (0.15);

    \node[below right] at (4,8) {$r_2$};
    \draw[fill=blue] (4,8) circle (0.15);

    \node[below left] at (-4,8) {$r_1$};
    \draw[fill=blue] (-4,8) circle (0.15);
    
    \node[above left] at (-3,0.9) {\small{$H^q\ket{b}$}};
    \node[above right] at (3,0.9) {\small{$q$}};
    
    \node[below left] at (-3,7) {\small{$b$}};
    \node[below right] at (3,7) {\small{$b$}};
    
    \node[left] at (-2,4) {\small{$s_1,s_2$}};
    \node[right] at (2,4) {\small{$s_3,q$}};
    
    \node[right] at (-1.6,2.4) {\small{$s_1,s_2$}};
    \node[left] at (1.6,2.4) {\small{$s_3,q$}};
    
    \end{tikzpicture}
    }
    \caption{a) Circuit diagram for the first set of operations used to complete the $BB84$ measurement task. Blue lines indicate classical inputs and outputs. The protocol uses one EPR pair, $\ket{\Psi^+}_{LR}=\frac{1}{\sqrt{2}}(\ket{00}+\ket{11})$ as a resource. The measurements are in the $\{\ket{0},\ket{1}\}$ basis. b) To complete the non-local computation, the classical measurement outcomes should be copied and sent to the left and right. The value $b$ is then computed on both sides from $s_1,s_2,s_3,q$.}
    \label{fig:measurestrategy}
\end{figure}

Again, there is a work around here using entanglement. 
Alice and Bob share an EPR pair $\ket{\Psi^+}_{LR}$ and Alice measures $QL$ in the Bell basis, producing the state $X^{s_1}Z^{s_2}H^q\ket{b}_R$ on $R$. 
Bob also holds $q$, so he applies $H^q$ leaving the state
\begin{align}
    H^q X^{s_1}Z^{s_2}H^q\ket{b}_R.
\end{align}
Bob then measures this in the computational basis. 
This is illustrated in figure \ref{fig:measurestrategy}.
The key observation is that the Pauli correction $X^{s_1}Z^{s_2}$ can flip $\ket{+}$ into $\ket{-}$ or vice versa, or flip $\ket{0}$ into $\ket{1}$ or vice versa, but it can't move you between the computational and Hadamard bases. 
Thus the above is always a state in the computational basis and the measurement has a definite outcome, call it $b'$. 
This measurement outcome along with $q,s_1,s_2$ then determines $b$. 
So Alice and Bob make copies of all the classical measurement outcomes $b',q,s_1,s_2$, send them to both sides, and then on each side compute $b$ from this data. 

Again, we see that entanglement allows us to evade the apparent restriction placed on Alice (not knowing which of two incompatible bases to measure in) and allows us to complete the task. 

\section{Some simple lower bounds}

In the last section we introduced two examples of NLQC, routing and the BB84 measurement, and gave simple protocols for completing them. 
In both cases there is a fairly clear intuition that the NLQC should be impossible if we don't share entanglement.
One of the main concerns in this book will be to prove lower bounds on entanglement in NLQC, so it's worthwhile to dwell on these intuitions a bit further and to try and make them precise. 

\subsection{Lower bound on routing}

Let's start by revisiting the routing task. 
We can show a lower bound on the entanglement cost in terms of the dimension of $Q$, the routed quantum system.
We will assume perfect correctness of the routing protocol in our arguments but a generalization to the robust setting is not too difficult. 

We take the shared resource system to be $\ket{\Psi}_{LR}$.\footnote{For simplicity, we are assuming the resource state is pure. This strategy can be extended to the mixed state case however.}
In a fully general protocol, Alice, on the left, applies a quantum channel $\mathcal{N}_{QL\rightarrow A'B}$.
$A'$ is the system Alice keeps while $B$ is the system she sends to Bob.
Without decreasing their success probability we can take the isometric extension of this channel, call it $V_{QL\rightarrow A'PB}$ and have Alice keep the purifying system $P$. 
We relabel $A'P$ as $A$, so that the isometry applied is $V_{QL\rightarrow AB}$ with $A$ kept by Alice and $B$ sent to Bob.
Bob, who knows $b$, knows where system $Q$ should be brought so can, without decreasing the success probability, simply forward the $R$ system to whichever party should receive $Q$. 
Correctness of the protocol then requires that both $AR$ and $BR$ can recover $Q$. 

Let $Q$ be in the maximally entangled state $\Psi^+_{\bar{Q}Q}$ with reference system $\bar{Q}$. 
Then because $Q$ is maximally entangled with $\bar{Q}$ and can be recovered from both $AR$ and $BR$, we have
\begin{align}\label{eq:recoveryMIs}
    I(\bar{Q}:AR) = 2\log d_{\bar{Q}}, \nonumber \\
    I(\bar{Q}:BR) = 2\log d_{\bar{Q}}.
\end{align}
We claim this implies that $S(R)\geq \log d_{\bar{Q}}$. 
To see why, use the entropic statement
\begin{align}
    I(\bar{Q}:A)+I(\bar{Q}:BR)=2\log d_{\bar{Q}}
\end{align}
which holds for pure states $\ket{\psi}_{\bar{Q}ABR}$ with $\psi_{\bar{Q}}$ maximally mixed, and the second equation from \eqref{eq:recoveryMIs} to conclude that $I(\bar{Q}:A)=0$.  
Then we apply the inequality, 
\begin{align*}
    I(\bar{Q}:AR) \leq I(\bar{Q}:A) + 2S(R),
\end{align*}
which can be proven from strong subadditivity.
Combining this with the first line of \eqref{eq:recoveryMIs} we obtain $S(R)\geq \log d_{\bar{Q}}$ as claimed.  

\subsection{Lower bound on measuring}\label{sec:measuringintro}

Now, let's look at a lower bound on the second NLQC we've introduced, the measuring task. 
Our strategy for the measuring task will be somewhat different than we used for routing. 
For routing, we assumed the protocol worked well and then constrained the resource state. 
Here, we assume the resource state is product, and then show the protocol can't work well.
This immediately implies \emph{some} correlation is necessary to work well, and then with a bit more work we can quantitatively lower bound the needed entanglement. 

To prove an upper bound on their success probability in the product setting we will introduce the notion of a monogamy-of-entanglement (MoE) game, which is a frequently encountered setting in quantum cryptography.
For simplicity, we restrict ourselves to a special case known as the BB84 MoE game, which will be related to the measuring task. 

\begin{definition}\label{def:MoE}
    A BB84 monogamy-of-entanglement game consists of two stages, a state preparation phase and then measurement phase. 
    \begin{itemize}
        \item \textbf{Preparation phase:} Alice and Bob prepare an arbitrary state $\Psi_{\bar{Q}AB}$ with $\bar{Q}$ a single qubit, and give $\bar{Q}$ to the referee. Alice takes system $A$, Bob system $B$, and then the players separate and can no longer communicate. 
        \item \textbf{Measurement phase:} The referee chooses $\theta\in\{0,1\}$ uniformly at random, then measures $\bar{Q}$ in the computational basis if $\theta=0$, or the Hadamard basis if $\theta=1$. She labels her measurement outcome $z$. The referee then reveals $\theta$ to both Alice and Bob. Alice and Bob then respond with values $a,b$ from the set $\{0,1\}$.
    \end{itemize} 
    We say that Alice and Bob win the game if $a=b=z$. We define $C:=\Pr[a=b=z]$. 
\end{definition}
We wouldn't expect to be able to complete the MoE game perfectly.
One intuition for this is in the name of the setting: if system $A$ can be measured to predict both computational and Hadamard basis measurements of $\bar{Q}$, it is telling us that $\bar{Q}:A$ is maximally entangled. 
So then since $\bar{Q}:A$ and $\bar{Q}:B$ cannot both be maximally entangled, we shouldn't be able to win with probability $1$ in the MoE game. 

Next we prove an (optimal) upper bound on the success probability of the BB84 MoE game. 
The strategy here follows \cite{johnston2016extended}. 

\begin{lemma}\label{lemma:BB84MoE}
    Considering the BB84 MoE game, we have $C\leq \cos^2(\pi/8)$.
\end{lemma}

\begin{proof}\,
    We define the measurement operators
    \begin{align}
        \Pi^0 &= \sum_x V_x^0 \otimes A_x^0 \otimes B_x^0, \\
        \Pi^1 &= \sum_x V_x^1 \otimes A_x^1 \otimes B_x^1,
    \end{align}
    where $V_x^\theta := H^\theta \ketbra{x}{x} H^\theta$.
    Similar to \cite{tomamichel2013monogamy} we introduce a `weakened' operator:
    \begin{align}
        \tilde{P} &= \sum_x V_x^0 \otimes A_x^0 \otimes \mathcal{I}_B \\
        \tilde{Q} &= \sum_y V_y^1 \otimes \mathcal{I}_A \otimes B_y^1.
    \end{align}
    Note:
    \begin{align}
        &\Pi^0 \preceq \tilde{P}, &&\Pi^1 \preceq \tilde{Q}, &\frac{\Pi^0 + \Pi^1}{2} \preceq \frac{\tilde P + \tilde Q}{2},
    \end{align}
    since $(\mathcal{I}_A - A^0_x) \succeq 0$, etc. 

    We will use $\sum_x A_x^0 = \mathcal{I}_A, \sum_y B_y^1 = \mathcal{I}_B$, then:
    \begin{align}
        \tilde P &= \sum_x V_x^0 \otimes A_x^0 \otimes \left( \sum_y B_y^1 \right) = \sum_{x,y} V_x^0 \otimes A_x^0 \otimes B_y^1 \\
        \tilde Q &= \sum_y V_y^1 \otimes \left( \sum_x A_x^0 \right) \otimes B_y^1 = \sum_{x,y} V_y^1 \otimes A_x^0 \otimes B_y^1 \\
        \frac{\tilde P + \tilde Q}{2} &= \sum_{x,y} \frac{V_x^0 + V_y^1}{2} \otimes A^0_x \otimes B_y^1.
    \end{align}
    For every pair of $x,y$ the largest eigenvalue of $\frac{V_x^0 + V_y^1}{2}$ is $\cos^2(\pi/8)$, therefore
    \begin{align}
        \frac{V_x^0 + V_y^1}{2} &\preceq \cos^2(\pi/8) \mathcal{I}_Q.
    \end{align}
    Then, summing over $x,y$ gives:
    \begin{align} 
        \frac{\tilde P + \tilde Q}{2} &\preceq \cos^2(\pi/8) \mathcal{I}_Q \otimes \left( \sum_x A_x^0 \right) \otimes \left( \sum_y B_y^1 \right) = \cos^2(\pi/8) \mathcal{I}_{QAB}.
    \end{align}
    Thus,
    \begin{align}
        C &= \tr[ \frac{\Pi^0 + \Pi^1}{2} \rho_{QAB} ] \leq \tr[ \frac{\tilde P + \tilde Q}{2} \rho_{QAB}] \\ &\leq \tr[ \cos^2(\pi/8) \mathcal{I}_{QAB} \, \rho_{QAB}] = \cos^2(\pi/8).
    \end{align}
\end{proof}

To see that this bound is optimal, consider the following strategy. 
Alice and Bob prepare the maximally entangled state $\ket{\Psi^+}_{\bar{Q}Q}$ and give the referee $\bar{Q}$. 
Then, they measure $Q$ in the Breidbart basis, 
\begin{align}\label{eq:breidbartbasis}
    \ket{B_0} &= \cos(\pi/8)\ket{0}+\sin(\pi/8)\ket{1}, \nonumber \\
    \ket{B_1} &= \sin(\pi/8)\ket{0}-\cos(\pi/8)\ket{1}.
\end{align}
Let their measurement outcome be labelled $z'$. 
They each carry a copy of $z'$ with them after separating, and then return $a=b=z'$ as their output (regardless of the value of $\theta$ they later receive). 
A straightforward calculation reveals that this is correct with probability $\cos^2(\pi/8)$. 

Now we're ready to prove the measure task cannot be completed using a product resource state. 
Concretely, we will prove the following theorem. 
\begin{theorem}\label{thm:measuretoBB84MoE}
The success probability for the measuring task when using a product resource state $\Psi_L\otimes \Psi_R$ is bounded according to
\begin{align}
    p_{suc}(\rho_L\otimes \rho_R)\leq \cos^2(\pi/8) \approx 0.85.
\end{align}
\end{theorem}
\begin{proof}
To prove this, we first change our viewpoint on the setting slightly. 
We take the quantum input $Q$ to be maximally entangled with a reference system $\bar{Q}$. 
The referee hands Alice $Q$, and then measures $\bar{Q}$ in either the computational or Hadamard bases, depending on a bit $q$, and obtains outcome $b$. 
After doing so, the post-measurement state on $Q$ is $H^q\ket{b}$, so that this is the same as if Alice had been handed $H^q\ket{b}$ directly as in the original task. 

Next, we notice that if Alice and Bob don't share entanglement, and Bob only gets the classical input $q\in\{0,1\}$, he may as well just copy $q$ and send one copy to Alice. 
An apparently more general thing he could do is to prepare a quantum system $\Psi_{XY}^q$ that depends on $q$, and then send $X$ left and $Y$ right, but actually this can be absorbed into Alice's operations: Alice prepares both $\Psi_{XY}^0$ and $\Psi_{XY}^1$, and then in the second round (after looking at the value of $q$ Bob has sent out) Alice and Bob trace out $\Psi_{XY}^{\neg q}$. 
Thus we can just take Bob to send out $q$, and absorb any more interesting operations into the general channel we allow Alice to apply. 

At this point we have that Alice, Bob, and the referee share a quantum state $\rho_{\bar{Q}AB}=\mathcal{N}_{{Q}\rightarrow AB}(\Psi^+_{Q\bar{Q}})$, and then the referee measures in basis labelled by $q$, and reveals $q$ to both Alice and Bob. 
This is exactly the setting in the BB84 monogamy game, with the added restriction that the shared state has the form $\mathcal{N}_{{Q}\rightarrow AB}(\Psi^+_{Q\bar{Q}})$, thus the success probability in the measuring task is upper bounded by the success probability of the monogamy game. 
Thus by lemma \ref{lemma:BB84MoE} we have
\begin{align}
    p_{suc}(\Psi_L\otimes \Psi_R)\leq \cos^2(\pi/8)
\end{align}
as claimed. 
\end{proof} 

We've now obtained an upper bound on the success probability for the measure task when the resource system is any product state. 
From here, we want to consider any resource state $\rho_{LR}$ which completes the task with high probability and show that this state must be highly correlated. 
One method to do this is to note that the task serves as a method of distinguishing $\rho_{LR}$ from its marginals $\rho_L\otimes \rho_R$, since if we feed $\rho_{LR}$ into the protocol for the task we get a large success probability, while $\rho_L\otimes \rho_R$ gives a small one. 
This leads to a quantitative lower bound on the mutual information, which is also a relative entropy distance $I(L:R)_\rho=D(\rho_{LR}||\rho_L\otimes \rho_R)$. 
Another approach uses the robustness of entanglement; we cover both these techniques in section \ref{sec:robustness}. 

\section{History and further reading}

Initially, there was some confusion over whether every channel could be implemented as an NLQC, and at that time simple examples were studied as a sort of test to understand what features of quantum mechanics might make an NLQC hard (like no-cloning and incompatible bases) and whether entanglement could be used to evade these restrictions. 
The work \cite{buhrman2014position} was the first to settle that all channels could be implemented as an NLQC; we will see a proof of this in chapter \ref{chapter:allchannels}. 

The same work \cite{buhrman2014position} also gave an upper bound on the success probability of the measuring task, which is somewhat weaker than the one we gave here. 
The work \cite{tomamichel2013monogamy} gave the tighter bound of $\cos^2(\pi/8)$ by applying the monogamy game bound.
That work also established \emph{parallel repetition} of the measuring task, meaning that if you repeat the task $n$ times in parallel, with independent inputs, the success probability becomes $\left(\cos^2(\pi/8)\right)^n$. 
We cover the parallel repeated case in chapter \ref{chapter:monogamygames}. 

\chapter{\texorpdfstring{$f$}{TEXT}-routing}\label{chapter:f-routing}

\minitoc

\section{Definition and motivation}

The $f$-routing task is a natural generalization of the routing task we introduced in chapter \ref{chapter:NLQC}. 
An instance of $f$-routing is defined by making a choice of Boolean function $f:\{0,1\}^n\times\{0,1\}^n\rightarrow \{0,1\}$. 
Then, the inputs and outputs required are as follows. 

\vspace{0.2cm}
\noindent \textbf{$f$-routing task:}
\begin{itemize}
    \item \textbf{Inputs:} On the left, Alice receives a quantum system $Q$ in an unknown state, and a classical string $x\in\{0,1\}^n$. This is in state $\ket{\psi}_Q$ which is unknown to Alice. On the right Bob obtains a classical string $y\in \{0,1\}^n$. 
    \item \textbf{Outputs:} Alice should return state $\ket{\psi}_Q$ on the left if $f(x,y)=0$, and on the right if $f(x,y)=1$. 
\end{itemize}

The $f$-routing task was initially defined as a candidate QPV scheme. 
In this context, it is a natural proposal in that the honest player can complete the task by computing $f(x,y)$, and then redirecting the quantum system $Q$, whose size is fixed. 
Thus the honest player only barely needs to manipulate quantum resources --- everything quantum they manipulate or do is $O(1)$ sized. 
We can hope, meanwhile, that the dishonest player does need quantum resources that grow with $n$. 
Indeed, so far all known protocols require entanglement that grows with $n$ to complete $f$-routing (for suitably chosen $f$). 

From this origin however, $f$-routing has come to play a much larger role in the theory of NLQC. 
In chapter \ref{chapter:Tdepth} we will see $f$-routing used as an important sub-routine in NLQC protocols that implement arbitrary unitaries.
As well, we will later see that $f$-routing is equivalent, in a sense we will make precise, to a primitive studied in information-theoretic cryptography known as conditional disclosure of secrets. 
Thus $f$-routing is also an important link connecting NLQC to other topics. 
Finally, $f$-routing is a clean and simple setting where we begin to see the role of complexity theory in NLQC. 
We will begin to bring this out later in this chapter. 

Before moving on to study NLQC, we will record a more formal definition of $f$-routing that allows for small errors in completing the task. 

\begin{definition}\label{def:frouting}
    A \textbf{$f$-routing} task is defined by a choice of Boolean function $f:\{ 0,1\}^{2n}\rightarrow \{0,1\}$, and a $d$ dimensional Hilbert space $\mathcal{H}_Q$.
    Inputs $x\in \{0,1\}^{n}$ and system $Q$ are given to Alice, and input $y\in \{0,1\}^{n}$ is given to Bob.
    Alice and Bob exchange one round of communication, with the combined systems received or kept by Bob labelled $M$ and the systems received or kept by Alice labelled $M'$.
    Label the combined actions of Alice and Bob in the first round as $\mathcal{N}^{x,y}_{Q\rightarrow MM'}$. 
    The $f$-routing task is completed $(\epsilon_0,\epsilon_1)$-correctly if there exist channels $\mathcal{D}^{x,y}_{M\rightarrow Q}$ such that,
    \begin{align}
        \forall (x,y)\in X\times Y \,\,\, s.t. \,\, f(x,y)=1,\,\,\, ||\mathcal{D}^{x,y}_{M\rightarrow Q} \circ\tr_{M'} \circ\mathcal{N}^{x,y}_{Q\rightarrow MM'} -\mathcal{I}_{Q\rightarrow Q}||_\diamond \leq \epsilon_1
    \end{align}
    and there exists a channel $\mathcal{D}^{x,y}_{M'\rightarrow Q}$ such that
    \begin{align}
        \forall (x,y)\in X\times Y \,\,\, s.t. \,\, f(x,y)=0,\,\,\,||\mathcal{D}^{x,y}_{M'\rightarrow Q} \circ\tr_{M} \circ\mathcal{N}^{x,y}_{Q\rightarrow MM'} -\mathcal{I}_{Q\rightarrow Q}||_\diamond \leq \epsilon_0
    \end{align}
    In words, Bob can recover $Q$ if $f(x,y)=1$ and Alice can recover $Q$ if $f(x,y)=0$. 
\end{definition}
If we say an $f$-routing protocol is $\epsilon$-correct, we mean that $\epsilon_0,\epsilon_1\leq \epsilon$. 
We denote the entanglement cost of implementing an $(\epsilon_0,\epsilon_1)$-correct $f$-routing protocol for a function $f$ by $\FR_{\epsilon_0,\epsilon_1}(f)$. 
By the entanglement cost, we will mean in this section the minimal number of shared maximally entangled qubits needed.

The correctness parameters of an $f$-routing protocol can be made arbitrarily small, if they begin below some threshold. 
The reason is that we can take $Q$, encode it into an error correcting code, and then run the $f$-routing protocol on each of the shares of the code. 
The correctness parameters record the noise in each instance of the protocol, so they decrease with use of an appropriate coding strategy. 
Concretely we have the following theorem, proven in \cite{asadi2025conditional}. 

\begin{theorem}\label{thm:fRamplification}
    Let $F_Q$ be an $f$-routing protocol for a function $f$ that supports one qubit input systems with correctness error $\epsilon=0.09$, communication cost $c$, and entanglement cost $E$. Then for every positive integer $k$ there exists an $f$-routing protocol $G_{Q'}$ for $f$ with $k$-qubit secrets, correctness error of $2^{-\Omega(k)}$, communication cost $O(k c)$, and entanglement cost $O(k E)$. 
\end{theorem}

Because of this theorem the correctness parameters are not too important in determining the cost of an $f$-routing protocol, so long as they are at or below the threshold of $0.09$. 
For this reason we sometimes drop them and write $\FR(f):= \FR_{0.09,0.09}(f)$. 

\section{The garden-hose protocol}

Suppose we've chosen a function $f$, and we'd like to try and complete the $f$-routing problem. 
As we commented in the introduction, all NLQCs can be implemented with sufficient entanglement, so we at least know this is always possible. 
We'd like to understand though if we find more efficient protocols that use some of the special structure of $f$-routing.
For instance, perhaps there are efficient protocols when $f$ is a simple function in some appropriate sense. 

\begin{figure*}
    \centering
    \begin{subfigure}{0.45\textwidth}
    \centering
    \begin{tikzpicture}[scale=0.8]
    
    \draw[red, thick] (-1,1)  to [out=-90,in=180] (0,0);
    
    \node[above] at (-1,1) {$Q$};
    \draw[black] plot [mark=*, mark size=2] coordinates{(-1,1)};
    
    \draw[thick] (0,0) -- (5,0);
    \draw[black] plot [mark=*, mark size=2] coordinates{(0,0)};
    \draw[black] plot [mark=*, mark size=2] coordinates{(5,0)};
    
    \draw[thick] (0,-1) -- (5,-1);
    \draw[black] plot [mark=*, mark size=2] coordinates{(0,-1)};
    \draw[black] plot [mark=*, mark size=2] coordinates{(5,-1)};

    \node[right] at (-0.5,0.5) {$x=1$}; 
    
    \draw[blue, thick] (5,0)  to [out=0,in=0] (5.1,-1);
    \node[right] at (5.3,-0.8) {$y=0$};
    
    \end{tikzpicture}
    \caption{}
    \label{fig:disconnectedsurfacesintro}
    \end{subfigure}
    \hfill
    \begin{subfigure}{0.45\textwidth}
    \centering
    \begin{tikzpicture}[scale=0.8]
    
    \draw[blue, thick] (-1,1)  to [out=-90,in=180] (0,0);
    \draw[red, thick] (-1,1)  to [out=-90,in=180] (0,-2);
    
    \node at (-2,-1) {$x=1$};
    
    \node[above] at (-1,1) {$Q$};
    \draw[black] plot [mark=*, mark size=2] coordinates{(-1,1)};
    
    \draw[thick] (0,0) -- (5,0);
    \draw[black] plot [mark=*, mark size=2] coordinates{(0,0)};
    \draw[black] plot [mark=*, mark size=2] coordinates{(5,0)};
    
    \draw[thick] (0,-1) -- (5,-1);
    \draw[black] plot [mark=*, mark size=2] coordinates{(0,-1)};
    \draw[black] plot [mark=*, mark size=2] coordinates{(5,-1)};

    \node[right] at (-0.5,0.5) {$x=0$}; 
    
    \draw[blue, thick] (5,0)  to [out=0,in=0] (5.1,-1);
    \node[right] at (5.3,-0.8) {$y=0$};
    
    \draw[thick] (0,-2) -- (5,-2);
    \draw[black] plot [mark=*, mark size=2] coordinates{(0,-2)};
    \draw[black] plot [mark=*, mark size=2] coordinates{(5,-2)};
    
    \end{tikzpicture}
    \caption{}
    \label{fig:connectedsurfacesintro}
    \end{subfigure}
    \caption{Some simple garden-hose protocols. Blue lines indicate Bell basis measurements. Black lines indicate shared EPR pairs, with the left side of the pairs held by Alice and right side held by Bob. a) Garden-hose protocol for computing $AND(x,y)$. Alice measures $Q$ and the first EPR pair in the Bell basis iff $x=1$. Bob measures the two EPR pairs iff $y=0$. b) Garden-hose protocol for $OR(x,y)$, which uses similar conditional measurements.}
    \label{fig:GHexamples}
\end{figure*}

To get started with a simple example, let's suppose $x$ and $y$ are each single bits, and $f(x,y)=x\wedge y=AND(x,y)$. 
How can Alice and Bob accomplish $f$-routing in this case?
Consider the following protocol, illustrated in figure \ref{fig:GHexamples} and executed upon receiving the inputs. 
\begin{itemize*}
    \item If $x=0$, Alice keeps $Q$. If $x=1$, Alice measures $Q$ along with one end of the maximally entangled state $\Psi^+_{L_1R_1}$ in the Bell basis, receiving measurement outcome $k_1$. Alice broadcasts $k_1$ in the communication round. 
    \item If $y=0$, Bob measures $R_1$ along with $R_2$ (from a second EPR pair $\Psi^+_{L_2R_2}$) in the Bell basis. He then broadcasts the measurement outcome, call it $k_2$. If $y=1$, Bob makes no measurements. 
\end{itemize*}
Let's analyze what happens here, case by case. 
\begin{itemize*}
    \item $x=0$: Then $f(x,y)=x\wedge y=0$. Also, Alice keeps $Q$, so $Q$ is on the left at the end. This is correct. 
    \item $x=1,y=0$: Then $f(x,y)=x\wedge y=0$. Also, the state ends up on $L_2$, but encrypted by the Pauli corrections coming from one or more Bell measurements. Alice corrects these Pauli errors based on the measurement outcomes in the second round. 
    \item $x=y=1$: Then $f(x,y)=1$. Also, Alice measures $QL_1$, producing the input state on $R_1$ up to a Pauli correction. Bob holds $R_1$ and $k_1$ at the end, and can recover $Q$. This is correct. 
\end{itemize*}
We see that in every possible case the protocol works, so the protocol is correct. 
In fact, it is perfectly correct, corresponding to $\epsilon=0$ in the formal definition given above.

Let's start trying to generalize this protocol. 
The first thing to focus on is the role of Bell basis measurements. 
After Alice's initial measurement, her input state $\ket{\psi}_Q$ is moved onto the $R_1$ system, producing $P^{k_1}\ket{\psi}_{R_1}$ with $k_1$ given by the measurement outcome. 
Whenever Alice or Bob makes a measurement, we can always have the measurement outcome broadcast to both sides. 
This means that wherever $R_1$ ends up will be wherever the input $\ket{\psi}$ ends up as well. 
We will give a name to this situation: we say that Alice has teleported$^*$ $Q$ to Bob, with the $*$ indicating that the Pauli corrections are still held only by the sender, but will be broadcast and made available on both sides later. 

We can actually use the teleport$^*$ protocol to execute the $f$-routing task for any choice of function $f$. 
To see this, it is helpful to introduce an analogy, which will make it somewhat easier to think about what is happening in such protocols.\footnote{A short video describing this analogy is here \url{https://youtu.be/KpZFy1xw-L0?si=uybjSidNd-uzV9tx&t=22}.} 
We imagine Alice and Bob are neighbours and share a fence. 
On Alice's side of the fence she has a tap, which she can turn on to produce water. 
Alice and Bob have $E$ pipes running between their two yards. 
Alice can connect the tap to one pipe with a hose, and connect the ends of various pipes together on her end with further hoses. 
She can choose how to do this in a way that depends on her input $x$. 
Meanwhile, Bob connects the ends of some of his pipes together using hoses. 
After the connections are made, Alice turns on the tap. 
The water should spill on Alice's side if $f(x,y)=0$, and on Bob's side if $f(x,y)=1$. 

Two things here turn out to be true: 1) Every function $f(x,y)$ can be computed in this way, and 2) For every garden-hose protocol using $E$ pipes to compute $f$, there is a corresponding $f$-routing protocol using $E$ EPR pairs to route on the same function $f$. 

Let's first of all look at why every function can be computed in the garden-hose model. 
One simple way to do this is as follows. 
Alice and Bob have a set of $2^{n+1}$ pipes, which we think of as $2^n$ pairs of pipes, with each pair labelled with the strings $x$. 
Upon receiving $x$, Alice connects her tap to the first pipe in the $x$th pair. 
Then, for each pair, Bob connects the two pipes in the pair if $f(x,y)=0$ and leaves the pair unconnected if $f(x,y)=1$. 
We can see that this works, since the water will go to Bob's side on the first pipe in the $x$th pair, and then will stay there if $f(x,y)=1$ and will return to Alice's side if $f(x,y)=0$. 

Now let's see why every garden-hose protocol can be turned into an $f$-routing protocol. 
In the analogy, the tap will become the initial quantum state, and each pipe will become an EPR pair. 
Connecting the tap to the first pipe will be performing teleport$^*$, using the input state and the associated EPR pair. 
Subsequent pipe connections become Bell measurements on pairs of ends of EPR pairs. 
The input state, up to Pauli corrections, ends up on whichever side the water has flowed to in the garden-hose analogy. 
In the communication round the measurement outcomes are broadcast to both sides, so that the corrected quantum state is available wherever the water ends up. 

Note that since we can compute every function in the garden-hose model, we can also $f$-route on any function using the associated protocol. 

Given a function $f$, we can ask about its \emph{garden-hose complexity}, denoted $\GH(f)$, and defined to be the minimal number of pipes needed to compute $f$ in the garden-hose model. 
This is also the minimal number of EPR pairs needed to compute $f$ in the simple model where we teleport$^*$ the input state back and forth. 
It's interesting to ask when $\GH(f)$ is polynomial in $n$, the input size to $f$. 
For the garden-hose, it turns out that the functions we can efficiently implement are exactly the functions computable on a Turing machine using a logarithmic size memory \cite{buhrman2013garden}. 
This is also referred to as the class $\LogSpace$. 
We won't prove this fact here, but in the next section we will develop an easier to prove but still non-trivial connection between $f$-routing and classical complexity theory.

\section{\texorpdfstring{$f$}{TEXT}-routing on formulas}

In this section we introduce a different approach to implementing $f$-routing, which we call \emph{code-routing}. 
This strategy was introduced in \cite{cree2023code}. 
The basic idea is that by using error-correcting codes, we can split quantum systems into several parts, and then perform garden-hose-like operations on the parts.
This adds some flexibility, and potentially increases the functions we can implement efficiently. 

\begin{figure*}
    \centering
    \begin{subfigure}{0.45\textwidth}
    \centering
    \begin{tikzpicture}[scale=0.4]
    
    \draw[thick] (0,0) -- (0,3) -- (3,3) -- (3,0) -- (0,0);
    \node at (1.5,1.5) {\large{$x$}};
    
    \draw[thick] (1.5,3) -- (1.5,4.5);
    \draw[black] plot [mark=*, mark size=3] coordinates{(1.5,4.5)};
    
    \draw[gray,<->] (4.5,1.5) -- (7.5,1.5);
    
    \draw[black] plot [mark=*, mark size=3] coordinates{(9,1.5)};
    \draw[blue,->] (9,1.5) -> (11,3.5);
    \node[above right, align=center] at (11,3.5) {if $x=1$, \\ send};
    
    \node[above] at (1.5,4.5) {$v$};
    \node[below] at (9,1.5) {$v$};
    
    \end{tikzpicture}
    \caption{}
    \label{fig:unitroutinga}
    \end{subfigure}
    \hfill
    \begin{subfigure}{0.45\textwidth}
    \centering
    \begin{tikzpicture}[scale=0.4]
    
    \draw[thick] (0,0) -- (0,3) -- (3,3) -- (3,0) -- (0,0);
    \node at (1.5,1.5) {\large{$y$}};
    
    \draw[thick] (1.5,3) -- (1.5,4.5);
    \draw[black] plot [mark=*, mark size=3] coordinates{(1.5,4.5)};
    
    \draw[gray,<->] (4.5,1.5) -- (7.5,1.5);
    
    \draw[black] plot [mark=*, mark size=3] coordinates{(9,4.5)};
    \draw[thick] (9,4.5) to [out=-90,in=180] (11.9,1.5);
    \draw[black] plot [mark=*, mark size=3] coordinates{(12,1.5)};
    \draw[black,thick] (12,1.5) -- (16,1.5);
    \draw[black] plot [mark=*, mark size=3] coordinates{(16,1.5)};
    
    \draw[blue,->] (16,1.5) -- (14,3.5);
    \node[above right, align=center] at (14,3.5) {if $y=0$,\\ send};
    
    \node[above] at (1.5,4.5) {$v$};
    \node[above] at (9,4.5) {$v$};
    
    \end{tikzpicture}
    \caption{}
    \label{fig:unitroutingb}
    \end{subfigure}
    \caption{Illustration of the \emph{unit-routing} protocol. The effect of the protocol is to bring the share $v$ to the side labelled by the input bit. a) For an input bit $x_i$ held by Alice, who holds share $v$, the share is simply sent to the left if $x=0$ and right if $x=1$. b) With $v$ starting on the left but input bit $y$ held on the right, the share is first measured in the Bell basis with one end of an EPR pair that has been shared between left and right. Then, the end of the EPR pair on the right is sent left or right based on the value of $y$.}
    \label{fig:unitrouting}
\end{figure*}

To get a sense of how this can work, let's start with a very simple choice of error-correcting code: a code that takes a single quantum system and stores it into three, in such a way that the input can be recovered from any two out of the three systems. 
In other words, this is a code that corrects one erasure error. 
Protocols using this code to compute an AND function and an OR function are shown in figure \ref{fig:CRexamples}. 
Let's walk through the case of implementing the AND function. 

\vspace{0.2cm}
\noindent \textbf{AND protocol:}
\begin{itemize*}
    \item Alice encodes $Q$ into the code, producing shares $S_1,S_2,S_3$. 
    \item Alice always keeps $S_1$, so it always ends up on the left.  
    \item Alice and Bob perform \emph{unit routing} to bring $S_2$ to the side labelled by $x$.
    \item Alice and Bob perform unit routing to bring $S_3$ to the side labelled by $y$. 
\end{itemize*}
We describe unit-routing below, but assuming it works for a moment you can confirm that this protocol makes the system $Q$ available on the left if $x\wedge y=0$, and on the right if $x\wedge y=1$.  
Unit routing is illustrated in figure \ref{fig:unitrouting}; unit-routing is just a very simple case of a garden-hose protocol, where the input is a single bit. 

\begin{figure*}
    \centering
    \begin{subfigure}{0.3\textwidth}
    \centering
    \begin{tikzpicture}[scale=1.1]
    
    \draw[black,thick] (-3,3) -- (-1,3) -- (-1,2) -- (-3,2) -- (-3,3);
    \node at (-2,2.5) {$\mathcal{E}$};
    \draw[black,thick] (-2,4) -- (-2,3);
    \draw[black] plot [mark=*, mark size=2] coordinates{(-2,4)};
    
    \draw[black,thick] (-1.25,2) -- (-1.25,1.5);
    \draw[black,thick] (-2,2) -- (-2,1.5);
    \draw[black,thick] (-2.75,2) -- (-2.75,1.5);
    
    \draw[black] plot [mark=*, mark size=2] coordinates{(-1.25,1.5)};
    \draw[black] plot [mark=*, mark size=2]
    coordinates{(-2,1.5)};
    \draw[black] plot [mark=*, mark size=2] coordinates{(-2.75,1.5)};
    \node[below] at (-2.75,1.5) {keep};
    
    \node[above] at (-2,4) {$Q$};
    
    \draw[thick] (-1.5,1) -- (-1,1) -- (-1,0.5) -- (-1.5,0.5) -- (-1.5,1);
    \draw[thick] (-1.25,1.5) -- (-1.25,1);
    \node at (-1.25,0.75) {$y$};
    
    \draw[thick] (-2.25,1) -- (-1.75,1) -- (-1.75,0.5) -- (-2.25,0.5) -- (-2.25,1);
    \draw[thick] (-2,1.5) -- (-2,1);
    \node at (-2,0.75) {$x$};
    
    \end{tikzpicture}
    \caption{}
    \label{fig:AND}
    \end{subfigure}
    \hfill
    \begin{subfigure}{0.3\textwidth}
    \centering
    \begin{tikzpicture}[scale=1.1]
    
    \draw[black,thick] (-3,3) -- (-1,3) -- (-1,2) -- (-3,2) -- (-3,3);
    \node at (-2,2.5) {$\mathcal{E}$};
    \draw[black,thick] (-2,4) -- (-2,3);
    \draw[black] plot [mark=*, mark size=2] coordinates{(-2,4)};
    \node[above] at (-2,4) {$Q$};
    
    \draw[black,thick] (-1.25,2) -- (-1.25,1.5);
    \draw[black,thick] (-2,2) -- (-2,1.5);
    \draw[black,thick] (-2.75,2) -- (-2.75,1.5);
    
    \draw[black] plot [mark=*, mark size=2] coordinates{(-1.25,1.5)};
    \draw[black] plot [mark=*, mark size=2]
    coordinates{(-2,1.5)};
    \draw[black] plot [mark=*, mark size=2] coordinates{(-2.75,1.5)};
    
    \node[below] at (-2.75,1.5) {send};
    
    \draw[thick] (-1.5,1) -- (-1,1) -- (-1,0.5) -- (-1.5,0.5) -- (-1.5,1);
    \draw[thick] (-1.25,1.5) -- (-1.25,1);
    \node at (-1.25,0.75) {$y$};
    
    \draw[thick] (-2.25,1) -- (-1.75,1) -- (-1.75,0.5) -- (-2.25,0.5) -- (-2.25,1);
    \draw[thick] (-2,1.5) -- (-2,1);
    \node at (-2,0.75) {$x$};
    
    \end{tikzpicture}
    \caption{}
    \end{subfigure}
    \caption{Some simple code-routing protocols. The map $\mathcal{E}$ takes in the $Q$ system and records it into a 3 share secret sharing scheme where any 2 shares recover the secret. In the protocol, Alice, who initially holds $Q$, performs the encoding map $\mathcal{E}$. The lower boxes indicate the unit-routing protocol should be implemented on the attached shares. a) Code-routing protocol for computing $AND(x,y)$. b) Code-routing protocol for computing $OR(x,y)$.}
    \label{fig:CRexamples}
\end{figure*}

A natural possibility is to concatenate code-routing protocols. 
For instance, we could consider the encoding into the one-erasure error code described above, but now take the share $S_3$ and encode it into a further error-correcting code. 
A protocol like this is shown in figure \ref{fig:CRconcatenatedexample}. 
What functions can we perform $f$-routing for in this way? 
And which ones can we implement efficiently in this way?
To understand these questions we need to make another diversion into complexity theory. 

One basic model of computation studied in classical complexity theory is the \emph{formula}.
The basic ingredients that make up a formula are simple gates, which we will choose to be AND, OR, and NOT. 
We will denote the AND operation on two bits $x,y$ by $x\wedge y$, the OR operation by $x\vee y$, and the negation by $\neg x$. 
An example of a formula is 
\begin{align}
    (x_0 \vee y_0) \wedge (x_0 \vee (\neg y_1))
\end{align}
The formula evaluates to either $0$ or $1$. 
To evaluate it, we insert the given values of the variables $x_0,y_0,y_1$, etc, and evaluate the terms in brackets first.

\begin{figure}
    \centering
    \begin{subfigure}{0.3\textwidth}
    \centering
    \begin{tikzpicture}[scale=1.1]
    
    \draw[black,thick] (-3,3) -- (-1,3) -- (-1,2) -- (-3,2) -- (-3,3);
    \node at (-2,2.5) {$\mathcal{E}$};
    \draw[black,thick] (-2,4) -- (-2,3);
    \draw[black] plot [mark=*, mark size=2] coordinates{(-2,4)};
    \node[above] at (-2,4) {$Q$};
    
    \draw[black,thick] (-1.25,2) -- (-1.25,1.5);
    \draw[black,thick] (-2,2) -- (-2,1.5);
    \draw[black,thick] (-2.75,2) -- (-2.75,1.5);
    
    \draw[black] plot [mark=*, mark size=2] coordinates{(-1.25,1.5)};
    \draw[black] plot [mark=*, mark size=2]
    coordinates{(-2,1.5)};
    \draw[black] plot [mark=*, mark size=2] coordinates{(-2.75,1.5)};
    
    \node[below] at (-2.75,1.5) {keep};
    
    \draw[thick] (-1.25,1.5) -- (-1.25,0);
    
    \draw[thick] (-2.25,1) -- (-1.75,1) -- (-1.75,0.5) -- (-2.25,0.5) -- (-2.25,1);
    \draw[thick] (-2,1.5) -- (-2,1);
    \node at (-2,0.75) {$\neg x$};
    
    \draw[thick] (-2,0) -- (-0.5,0) -- (-0.5,-0.75) -- (-2,-0.75) -- (-2,0);
    \node at (-1.25,-0.375) {$\mathcal{E}$};
    
    \draw[thick] (-1.25,-0.75) -- (-1.25,-1.25);
    \draw[black] plot [mark=*, mark size=2] coordinates{(-1.25,-1.25)};
    
    \draw[thick] (-0.65,-0.75) -- (-0.65,-1.25);
    \draw[black] plot [mark=*, mark size=2] coordinates{(-0.65,-1.25)};
    
    \draw[thick] (-1.85,-0.75) -- (-1.85,-1.25);
    \draw[black] plot [mark=*, mark size=2] coordinates{(-1.85,-1.25)};
    \node[below] at (-1.85,-1.25) {send};
    
    \draw[thick] (-1.25,-1.25) -- (-1.25,-1.75);
    \draw[thick] (-0.65,-1.25) -- (-0.65,-1.75);
    
    \draw[thick] (-0.4,-1.75) -- (-0.9,-1.75) -- (-0.9,-2.25) -- (-0.4,-2.25) -- (-0.4,-1.75);
    \node at (-0.65,-2) {$y$};
    
    \draw[thick] (-1.5,-1.75) -- (-1,-1.75) -- (-1,-2.25) -- (-1.5,-2.25) -- (-1.5,-1.75);
    \node at (-1.25,-2) {$x$};
    
    \end{tikzpicture}
    \caption{}
    \end{subfigure}
    \hfill
    \begin{subfigure}{0.3\textwidth}
    \centering
    \begin{tikzpicture}[scale=1.1]
    
    \draw[black,thick] (-3,3) -- (-1,3) -- (-1,2) -- (-3,2) -- (-3,3);
    \node at (-2,2.5) {$AND$};
    \draw[black,thick] (-2,4) -- (-2,3);
    \node[above] at (-2,4) {$f(x,y)$};
    
    \draw[black,thick] (-1.25,2) -- (-1.25,1.5);
    \draw[black,thick] (-2,2) -- (-2,1.5);
    
    
    \draw[thick] (-1.25,1.5) -- (-1.25,0);
    \draw[thick] (-2.25,1) -- (-1.75,1) -- (-1.75,0.5) -- (-2.25,0.5) -- (-2.25,1);
    \draw[thick] (-2,1.5) -- (-2,1);
    \node at (-2,0.75) {$\neg x$};
    
    \draw[thick] (-2,0) -- (-0.5,0) -- (-0.5,-0.75) -- (-2,-0.75) -- (-2,0);
    \node at (-1.25,-0.375) {$OR$};
    
    \draw[thick] (-1.25,-0.75) -- (-1.25,-1.25);
    
    \draw[thick] (-0.65,-0.75) -- (-0.65,-1.25);
    
    
    \draw[thick] (-1.25,-1.25) -- (-1.25,-1.75);
    \draw[thick] (-0.65,-1.25) -- (-0.65,-1.75);
    
    \draw[thick] (-0.4,-1.75) -- (-0.9,-1.75) -- (-0.9,-2.25) -- (-0.4,-2.25) -- (-0.4,-1.75);
    \node at (-0.65,-2) {$y$};
    
    \draw[thick] (-1.5,-1.75) -- (-1,-1.75) -- (-1,-2.25) -- (-1.5,-2.25) -- (-1.5,-1.75);
    \node at (-1.25,-2) {$x$};
    
    \end{tikzpicture}
    \caption{}
    \label{fig:CRformula}
    \end{subfigure}
    \caption{a) Code-routing protocol for computing $f(x,y)=AND(NOT(x),OR(x,y))$. Note that the encoding map takes in the $Q$ wire and outputs the bottom-most wires, which are then unit-routed. b) Boolean formula for computing the same function. The code-routing protocol can be designed directly by studying the formula: AND gates become encodings where the third share is always kept, while OR gates become encodings where the third share is always sent. Note that the formula should be evaluated by starting with the leaves and simplifying the formula upwards.}
    \label{fig:CRconcatenatedexample}
\end{figure}

We can also represent a formula more diagrammatically, as illustrated in figure \ref{fig:CRconcatenatedexample}. 
In such a diagram, the formula is represented by a graph that takes the form of a tree. 
The leaves of the tree are individual variables, which may be negated. 
The nodes are either AND or OR gates.
The tree is evaluated from the bottom to the top: we set the values of the variables, and then work our way up the tree, evaluating each node. 
Notice that every node in the tree can have only a single output. 
This distinguishes formulas from circuits, which are similar but allow each node to have multiple outputs. 
How hard it is to compute a function using the formula model is measured by the \emph{size} of the formula. 
The size is defined to be the number of leaves in the formula. 

Notice that in our definition of a formula, we only allowed negations on the leaves, but not anywhere else in the expression. 
This seems a bit restrictive, but in fact whenever there are negations elsewhere in the expression we can move them to the leaves using De Morgan's laws, 
\begin{align}
    \neg (x\wedge y) &= (\neg x) \vee (\neg y), \nonumber \\
    \neg (x\vee y) &= (\neg x) \wedge (\neg y).
\end{align}
These are basic rules from Boolean logic. 

Now we're ready to return to the code-routing idea and see how it relates to formulas. 
It's perhaps easiest to see this first with an example. 
Consider the code-routing protocol shown in figure \ref{fig:CRconcatenatedexample}. 
We claim this performs $f$-routing on the function
\begin{align}
    f(x,y) = AND(\neg x, OR(x,y)).
\end{align}
More generally, taking the pattern of concatenation in a code-routing protocol, we can directly read off the formula it evaluates by the following identifications:
\begin{itemize*}
    \item Encodings where one share is always kept $\rightarrow$ AND gates
    \item Encodings where one share is always sent $\rightarrow$ OR gates
    \item Unit-routings on a variable $\rightarrow$ Leaves of the formula
\end{itemize*}

In the code-routing protocol, each unit-routing costs a single EPR pair (at most) to evaluate. 
Since each unit-routing corresponds to a leaf in the corresponding formula, this means that the size of the formula (the number of leaves) sets the number of EPR pairs needed to implement the function in this scheme,
\begin{align}
    E(f) \leq \text{FormulaSize}(f).
\end{align}
The derivation of the upper bound on $f$-routing from formula size is simple, at least in so far as we just needed to concatenate some very simple codes together to achieve the bound. 
We've presented this upper bound as a way of illustrating, in a simple setting, the connection between NLQC protocols and complexity theory. 
The more general lesson, seen in several further examples, is that \emph{models of computation emerge naturally from NLQC protocols}. 
This is a somewhat mysterious observation, but it obtains an at least partial explanation when we relate NLQC to the holographic principle and quantum gravity. 
We revisit that idea in chapter \ref{chapter:gravity}. 

\section{History and further reading}

The garden-hose protocol was first given in \cite{buhrman2013garden}. 
Some nice properties of the garden-hose complexity were proven in \cite{klauck2014new}. 
These properties were used, along with many new ideas, to build the $T$-depth based protocols for NLQC which we cover in chapter \ref{chapter:Tdepth}. 

The formula size upper bound for $f$-routing isn't the strongest known upper bound on $f$-routing (it's just the simplest to explain). 
The most powerful known upper bound on $f$-routing is also based on the code-routing idea, but uses more elaborate code constructions based on a computational model known as a span program.  
In particular, we can achieve a complexity class known as $\ModkL$ using this technique. 
This improves on the garden-hose strategy, which achieves the class $\LogSpace$, where $\LogSpace \subseteq \ModkL$, and it is strongly believed that $\LogSpace \subsetneq \ModkL$. 

There have been some other surprises coming from the study of the efficiency of $f$-routing.
One of these is that actually every $f$-routing protocol can be implemented with entanglement cost $2^{O(\sqrt{n \log n})}$, beating the simple exponential protocol we gave using the garden-hose, or the port-teleportation protocol when applied to this context. 
This sub-exponential upper bound was proven by relating $f$-routing to another subject in cryptography called conditional disclosure of secrets; we will see that connection in detail in chapter \ref{chapter:ITCandNLQC}. 
Another surprising $f$-routing protocol coming from the same connection to CDS gives a polynomially efficient protocol for a function believed to be far outside of $\ModkL$, and in fact outside of $\PolyTime$ (but inside of $\BQP$). 
These topics were explored in \cite{allerstorfer2024relating}. 

\chapter{NLQC and information-theoretic classical cryptography}\label{chapter:ITCandNLQC}

\minitoc

For many subjects in quantum information theory we have a classical analogue, which provides guidance and inspiration. 
For instance, classical error-correction informs quantum error correction, and classical computation informs quantum computation. 
In contrast, in NLQC the setting seems to trivialize if we take a naive classical analogue: supposing Alice and Bob's inputs were classical strings $x,y\in\{0,1\}^n$, they can always implement any computation by simply copying $x,y$ and sending copies to both sides, then separately computing the outputs needed on each side. 
Nonetheless, we might ask if there is some more interesting classical analogue setting that can help inform the study of NLQC.  

It came as something of a surprise that the answer here is yes; there is a tightly related classical analogue to NLQC, at least for certain NLQC examples. 
Further, recent work is finding that these examples are less restrictive than first believed, and in fact these classical analogues capture many NLQC settings. 
To develop the classical analogues, we will study aspects of classical information-theoretic cryptography. 
There, the key question is to understand the \emph{cost of privacy} in certain cryptographic settings. 
The connection to NLQC reveals that this question is closely related to understanding entanglement cost in NLQC. 

\section{The cost of privacy}

Information-theoretic cryptography deals with information processing scenarios involving untrusted or partially trusted parties. 
The classic example is of communicating over a public channel. 
Suppose that Alice wishes to send a message to Bob, but their communication channel can be accessed by a third party, Eve, as well. 
We can imagine for instance that Alice will broadcast her message on the radio, so both Bob and Eve can listen in.
Despite this limitation, Alice wants to send a message that only Bob can read. 

Shannon \cite{shannon1949communication} asked about the minimal resources to turn the public channel into a private one; in other words he asked about the cost of privacy in this setting. 
Shannon showed that it is necessary and sufficient for Alice and Bob to share a single bit of randomness per bit of message they want to send. 
With this, to send message bit $m_i$ Alice can compute $m'_i=m_i\oplus r_i$ and send $m'_i$ over the public channel.
Then Bob can compute $m'_i\oplus r_i=m_i$ to recover the message. 
In this context we call the message $m'_i$ the ciphertext. 
Since Bob can recover the message, we say the protocol is \emph{correct}. 
One can check that to a referee who doesn't know $r_i$ there is no correlation between the message $m_i$ and the bit Eve sees, $m_i'$. 
This shows that the protocol is also \emph{secure}. 
This protocol is known as the \emph{one-time pad}. 

A slightly harder problem is to show that one bit of shared randomness is necessary to achieve a correct and secure protocol. 
To show this, we can use entropic arguments. 
In particular security requires that
\begin{align}\label{eq:OTPsecurity}
    I(M:M')=0
\end{align}
since we need the ciphertext not to reveal anything about the message. 
Meanwhile correctness requires that
\begin{align}\label{eq:OTPcorrectness}
    I(M:M'|R)=S(M)
\end{align}
where $R$ is the variable shared by Alice and Bob. 
Combining these, we can show that $S(R)\geq S(M)$. 
To do this, we first show that $S(M)=I(M:R|M')$:
\begin{align}
    I(M:R|M') &= S(MM')+S(RM')-S(M')-S(MM'R) \qquad  \qquad \text{(by definition)} \nonumber \\
              &= S(MM')+S(RM')-S(M') - \nonumber \\
              &\qquad \left(S(MR)+S(M'R)-S(M)-S(R) \right) \qquad \qquad \,\,\, \quad \text{(by \eqref{eq:OTPcorrectness})} \nonumber \\
              &= 2S(M) - S(MR) + S(R) \qquad \qquad \qquad \qquad \qquad \quad \,\,\,\,\text{(by \eqref{eq:OTPsecurity})} \nonumber \\
              &= S(M)
\end{align}
where the last line used that $S(MR)=S(M)+S(R)$, which follows because the message $M$ is independent of the randomness used, $R$. 
Next, use that  $I(A:B|C)\leq S(B)$, so that
\begin{align}
    S(M)=I(M:R|M') &\leq S(R)
\end{align}
as claimed. 

The cost of sending a private message over a public channel, then, is exactly one bit of randomness, so we have a complete understanding of the cost of privacy in this setting. 
However, there are many more settings in information-theoretic cryptography where the cost of privacy is open. 
Two examples, which will be of interest to us here are \emph{conditional disclosure of secrets} (CDS) and \emph{private simultaneous message} (PSM) settings. 

Let's look at CDS in a bit more detail, and return to PSM later (in section \ref{sec:PSM}). 
The CDS scenario is illustrated in figure \ref{fig:CDSandCDQS}. 
The setting involves three parties, Alice, Bob and the referee. 
Alice receives input $x\in X=\{0,1\}^n$, Bob receives input $y\in Y=\{0,1\}^n$, and the referee knows both $x$ and $y$. 
Alice additionally holds a secret $s\in S$. 
An instance of CDS is specified by a choice of Boolean function $f:X\times Y\rightarrow \{0,1\}$. 
Alice and Bob can share randomness. 
From their inputs and shared correlation, Alice and Bob each produce a message which they send simultaneously to the referee. 
Their goal is for the referee to be able to recover $s$ when $f(x,y)=1$ (which is correctness in this setting), but not learn anything about $s$ when $f(x,y)=0$ (which is privacy). 

Without privacy, this setting is trivial: Alice would just send the single bit $s$ to the referee, so there is $O(1)$ communication cost and no randomness cost. 
With privacy, completing this task seems much harder.
In fact, it's not even clear when first considering the problem that this is possible at all for non-trivial functions. 
To get some intuition, it's useful to consider a simple example. 
For instance, suppose that we want to complete CDS for the function $f(x,y)=x\wedge y = AND(x,y)$.
We focus on the case where $x,y$ are single bits. 
To complete the corresponding CDS, Alice and Bob can use the following protocol involving a single bit of shared randomness $r$,

\vspace{0.2cm}
\noindent \textbf{Protocol:} (CDS for $f(x,y)=x\wedge y$)
\begin{itemize*}
    \item If $x=0$, Alice doesn't send anything. If $x=1$, Alice computes $m_A=s\oplus r$ and sends this to the referee.
    \item If $y=0$, Bob sends nothing. If $y=1$, Bob sends $r$. 
\end{itemize*}
You can check that this is correct: if $f=1$ so that $x=y=1$, the referee gets $m_A=s\oplus r$ and $r$, so they can compute $s$. 
To see it is secure, notice that the referee gets only one of $m_A$ and $r$ if either input is $0$, so only one of these whenever $f=0$. 

The basic step used by Alice and Bob in the above protocol is a sort of classical analogue of teleportation.
To see why, notice that after Alice takes the XOR $m_A=s\oplus r$, we can view the bit $s$ as stored in Bob's lab in the `encrypted' form $s\oplus m_A$. 
Taking the XOR is the analogue of performing the Bell measurement in teleportation, where $m_A$ plays the role of the measurement outcome. 
The encrypted bit $s\oplus m_A$ is the analogue of the qubit $X^aZ^b\ket{\psi}$ that appears in Bob's lab after the measurement is made. 

Inspired by this analogy, we can give a protocol for completing CDS for any function that mimics the garden-hose protocol we had for $f$-routing. 
The role of the hoses is now played by shared random bits, and connecting pipes is now played by taking the XOR and sending the outcome to the referee. 
Any unused random bits on Alice's side should be discarded, while unused random bits on Bob's side should be sent to the referee. 
As an example, the reader may want to try to adapt the garden-hose protocol for the OR function (figure \ref{fig:GHexamples}) to the CDS setting.

Relating CDS to the garden-hose model gives us some basic insight into the communication cost in CDS. 
For instance, we know that CDS can be completed for any choice of function $f$. 
However, characterizing the cost of privacy in the CDS setting is much harder than in the case of the one-time pad: the best upper bounds on randomness cost for a generic function are $2^{O(\sqrt{n\log n})}$ \cite{liu2017conditional}, and the best lower bounds are linear.\footnote{The linear lower bounds only hold if we assume perfect correctness or perfect security. Allowing small errors, the best bounds are logarithmic.}

Part of the difficulty in characterizing the cost of privacy in CDS comes from a connection to complexity theory. 
For instance, letting $CDS(f)$ denote the randomness cost of completing CDS for the function $f$, we have that
\begin{align}\label{eq:CDSandformulasize}
    CDS(f)\leq O(\text{FormulaSize}(f)). 
\end{align}
This upper bound can be proven using an upper bound strategy very similar to the quantum upper bound given in chapter \ref{chapter:f-routing}. 
Note that this upper bound means we are unlikely to obtain a complete characterization of the cost of privacy in the CDS context: doing so would, at minimum, fully characterize the formula size of Boolean functions, a problem we expect is far out of the current reach of complexity theory. 

Even if we can't hope to achieve a complete characterization, there is still a lot to be gained from the study of the cost of privacy in CDS and related settings. 
For one thing, CDS is useful as a basic building block for other desired goals in cryptography \cite{gertner1998protecting,gay2015communication,applebaum2020power}, so there are practical reasons to want to find efficient protocols or understand when we can't find one. 
As well, CDS is tied closely to communication complexity \cite{applebaum2021placing}, and in fact the open problem of obtaining linear lower bounds on CDS in the noisy setting has been identified as a ``easier-but-similar'' problem along the way to solving certain long standing problems in communication complexity. 
Finally, we can see from the formula size upper bound that understanding the cost of privacy is closely related to understanding the complexity of Boolean functions.

\section{Conditional disclosure of secrets and \texorpdfstring{$f$}{TEXT}-routing}

\subsection{Classical and quantum CDS}

\begin{figure*}
    \centering
    \begin{subfigure}{0.45\textwidth}
    \centering
    \begin{tikzpicture}[scale=0.4]
    
    \draw[thick] (-5,-5) -- (-5,-3) -- (-3,-3) -- (-3,-5) -- (-5,-5);
    
    \draw[thick] (5,-5) -- (5,-3) -- (3,-3) -- (3,-5) -- (5,-5);
    
    \draw[thick] (5,5) -- (5,3) -- (3,3) -- (3,5) -- (5,5);
    
    \draw[thick, mid arrow] (4,-3) -- (4.5,3);
    
    \draw[thick, mid arrow] (-4,-3) to [out=90,in=-90] (3.5,3);
    
    \draw[thick,dashed] (-3.5,-5.5) -- (3.5,-5.5);
    \draw[black] plot [mark=*, mark size=3] coordinates{(-3.5,-5.5)};
    \draw[black] plot [mark=*, mark size=3] coordinates{(3.5,-5.5)};
    \node[below] at (0,-5.5) {$r$};
    
    \draw[thick] (-4.5,-6) -- (-4.5,-5);
    \node[below] at (-4.5,-6) {$x,s$};
    
    \draw[thick] (4.5,-6) -- (4.5,-5);
    \node[below] at (4.5,-6) {$y$};

    \node[left] at (0,1) {$m_A$};
    \node[right] at (4.5,0) {$m_B$};
    
    \draw[thick] (4,5) -- (4,6);
    \node[above] at (4,6) {$s$ iff $f(x,y)=1$};
    
    \end{tikzpicture}
    \caption{}
    \label{fig:CDS}
    \end{subfigure}
    \hfill
    \begin{subfigure}{0.45\textwidth}
    \centering
    \begin{tikzpicture}[scale=0.4]
    
    \draw[thick] (-5,-5) -- (-5,-3) -- (-3,-3) -- (-3,-5) -- (-5,-5);
    
    \draw[thick] (5,-5) -- (5,-3) -- (3,-3) -- (3,-5) -- (5,-5);
    
    \draw[thick] (5,5) -- (5,3) -- (3,3) -- (3,5) -- (5,5);
    
    \draw[thick, mid arrow] (4,-3) -- (4.5,3);
    
    \draw[thick, mid arrow] (-4,-3) to [out=90,in=-90] (3.5,3);
    
    \draw[thick] (-3.5,-5) to [out=-90,in=-90] (3.5,-5);
    \draw[black] plot [mark=*, mark size=3] coordinates{(0,-7.05)};

    \node[left] at (0,1) {$M_A$};
    \node[right] at (4.5,0) {$M_B$};
    
    \draw[thick] (-4.5,-6) -- (-4.5,-5);
    \node[below] at (-4.5,-6) {$x, Q$};
    
    \draw[thick] (4.5,-6) -- (4.5,-5);
    \node[below] at (4.5,-6) {$y$};
    
    \draw[thick] (4,5) -- (4,6);
    \node[above] at (4,6) {Q iff $f(x,y)=1$};
    
    \end{tikzpicture}
    \caption{}
    \label{fig:CDQS}
    \end{subfigure}
    \caption{(a) A classical CDS protocol. Alice, on the lower left, holds input $x\in \{0,1\}^n$ and a secret $s$ from alphabet $S$. Bob, on the lower right, holds input $y\in \{0,1\}^n$. Alice and Bob can share a random string $r$. The referee, top right, holds $x$ and $y$. Alice sends a message $m_A(x,s,r)$ to the referee; Bob sends a message $m_B(y,r)$. The referee should learn $s$ iff $f(x,y)=1$ for some agreed on choice of Boolean function $f$. (b) A quantum CDS protocol. The secret can be a quantum system $Q$ or classical string $s$ (the two cases are equivalent, as noted in \cite{allerstorfer2024relating}). Alice and Bob can share an entangled quantum state, and send quantum messages to the referee. The referee should be able to recover the secret iff $f(x,y)=1$.} 
    \label{fig:CDSandCDQS}
\end{figure*}

To start our study of CDS and its relationship to NLQC, we begin by defining the classical CDS setting more carefully. 
To do this, we need to make precise what we mean by the correctness and security conditions. 
Correctness is easy to formalize: we want the referee to recover the secret $s$ with high probability. 
This should be with high probability over Alice and Bob's shared randomness, and hold for every choice of the secret $s$. 

Security is a bit trickier to formalize.
The way this is done is to consider a \emph{simulator} notion of security: if the referee should learn very little about $s$, then the distribution on the messages they see should be close to one that doesn't depend on $s$. 
This distribution is called the simulator distribution. 
We now give the following definition of a classical CDS scheme. 

\begin{definition}\label{def:CDS}
    A \textbf{conditional disclosure of secrets (CDS)} task with classical resources is defined by a choice of function $f:\{0,1\}^{2n}\rightarrow \{0,1\}$.
    The scheme involves input $x\in \{0,1\}^{n}$ given to Alice and input $y\in \{0,1\}^{n}$ given to Bob.
    Alice and Bob share a random string $r\in R$.
    Additionally, Alice holds a string $s$ drawn from a distribution $S$, which we call the secret. 
    Alice sends message $m_A(x,s,r)\in M_A$ to the referee, and Bob sends message $m_B(y,r)\in M_B$.  
    We require the following two conditions on a CDS protocol. 
    \begin{itemize}
        \item $\epsilon$\textbf{-correct:} There exists a decoding function $D(m_A,x,m_B,y)$ such that 
        \begin{align}
            \forall s\in S,\,\forall \,(x,y) \in X\times Y \,\,s.t.\,\,f(x,y)=1,\,\,\, \underset{r\leftarrow R}{\mathrm{Pr}}[D(m_A,x,m_B,y)=s] \geq 1-\epsilon \: .
        \end{align}
        \item $\delta$\textbf{-secure:} There exists a simulator producing a distribution $Sim$ taking on values in $M=M_AM_B$ such that
        \begin{align}
            \forall s\in S,\,\forall \,(x,y) \in X\times Y \,\,s.t.\,\, f(x,y)=0,\,\,\, \left\Vert  \text{Sim}_{M|xy} - P_{M|xys} \right\Vert _1\leq \delta \: .
        \end{align}
    \end{itemize}
\end{definition}

We define the randomness cost of a CDS protocol to be the number of bits in the random string $r$. 
The minimal randomness cost for a function $f$ that achieves $\epsilon$-correctness and $\delta$-security we denote by $\text{CDS}_{\epsilon, \delta}(f)$.
By default, we will assume single-bit secrets ($\mathrm{supp}(S)=\{0,1\}$) when discussing CDS complexity. 

The specific correctness and security parameters for CDS are not too important, so long as they are constants (independent of $n$). 
The reason for this is that CDS protocols can be amplified, in the sense that they can be repeated to obtain smaller correctness and security errors. 
This is stated in the next theorem, which is proven in \cite{applebaum2017conditional}.

\begin{theorem}\label{thm:CDSamplification}
    Let F be a $\CDS$ protocol for a function $f$ that supports one bit secrets with correctness error $\epsilon=1/3$ and privacy error $\delta=1/3$. Then for every integer $k$ there exists a $\CDS$ protocol G for $f$ with $k$-bit secrets and privacy and correctness errors of $2^{-\Omega(k)}$. The communication and randomness complexity of $G$ is larger than that of F by a factor of $k$. 
\end{theorem}
The proof of this theorem involves encoding the secret into a classical secret sharing scheme, and then running the CDS protocol on each of the shares of the secret sharing scheme. 
Because of this theorem, we will often drop the subscripts when discussing CDS cost and write $\text{CDS}(f)=\text{CDS}_{0.1, 0.1}(f)$, where the choice of $0.1$ is arbitrary (though it needed to be below 1/3). 

To get familiar with how to work with the simulator notion of security used in the context of CDS, we prove the following simple fact. 
\begin{lemma}
    Suppose that we have a CDS protocol for the function $f(x,y)$, which hides a single bit secret. 
    Then, using twice the communication and shared randomness of the original protocol, it is possible to construct a $2\epsilon$-correct, $2\delta$-secure protocol which hides a 2 bit secret. 
\end{lemma}

\begin{proof}
To understand correctness of the new protocol, notice that on $1$ instances the probability of the referee guessing $s_i$ correctly is at least $1-\epsilon$, so their probability of guessing both $s_i$ correctly is at least $(1-\epsilon)^2 \geq (1-2\epsilon)$. 

To understand security, we define a simulator for the composed protocol by taking the product of the distributions for a single instance of the protocol, 
\begin{align}
    \text{Sim}_{M_1M_2|xy} \equiv \text{Sim}_{M_1|xy}\text{Sim}_{M_2|xy}.
\end{align}
We also note that, using fresh randomness for each instance of the CDS, we have that the distribution of the messages from the two protocols satisfies
\begin{align}
    P_{M_1M_2|xys} = P_{M_1|xys_1}P_{M_2|xys_2}.
\end{align}
Then by repeated application of the triangle inequality, and using security of each instance of the CDS, we have that on $0$ instances
\begin{align}
    ||\text{Sim}_{M_1M_2|xy} - P_{M_1M_2|xys}||_1  &= ||\text{Sim}_{M_1|xy}\text{Sim}_{M_2|xy} - P_{M_1|xys_1}P_{M_2|xys_2}||_1 \nonumber \\
    &= ||\text{Sim}_{M_1|xy}\text{Sim}_{M_2|xy} - P_{M_1|xys_1}\text{Sim}_{M_2|xy}\\ &\qquad +P_{M_1|xys_1}\text{Sim}_{M_2|xy} - P_{M_1|xys_1}P_{M_2|xys_2}||_1 \nonumber \\
    &\leq ||\text{Sim}_{M_1|xy} - P_{M_1|xys_1}||_1+||\text{Sim}_{M_2|xy} - P_{M_2|xys_2}||_1 \nonumber \\
    &\leq 2\delta\nonumber
\end{align}
as claimed. 
\end{proof} 

Our interest in CDS is due to its relationship with $f$-routing.
This relationship goes through a quantum variant of CDS.
To define quantum CDS, we first of all want to understand what the functionality of the protocol should be. 
In particular, we can choose to have the protocol hide classical secrets, or hide quantum secrets. 
It turns out that these settings are essentially equivalent: Given a protocol hiding a quantum secret we can choose the secret to be in a basis state to hide a classical secret. 
Conversely, given a protocol hiding a classical secret we can act on a quantum system with a random Pauli, then send the Pauli encrypted qubit to the referee and run the classical CDS protocol with the choice of Pauli as the secret. 
This hides the quantum system from the referee unless they learn the classical part of the secret.\footnote{We use this strategy again below to show that classical CDS protocols imply quantum CDS protocols. Note that the setting here is a bit different, we are arguing that two variants of quantum CDS (which both involve use of quantum resources) are equivalent.} 
Given this, we are free to define quantum CDS to use a classical or quantum secret; we choose a quantum secret because this will connect more directly to $f$-routing. 

To give a formal definition for quantum CDS, we need to revisit the correctness and security conditions from the classical case and understand how they should be adapted to the quantum context. 
The correctness condition is again fairly clear: letting the secret input to the CDQS protocol be $Q$, whenever $f=1$ there should be a decoder the referee can apply such that the combined action of Alice and Bob's operations and the referee's decoding operation should be close to the identity. 
For security we now need a quantum notion of a simulator. 
This is defined as a state preparation channel $\mathcal{S}_{\varnothing\rightarrow M}$ whose output, whenever $f=0$, is close to the message the referee receives. 
We give the formal definition next. 

\begin{definition}\label{def:CDQS}
    A \textbf{conditional disclosure of secrets task with quantum resources (CDQS)} is defined by a choice of function $f:\{0,1\}^{2n}\rightarrow \{0,1\}$, and a $d_Q$-dimensional Hilbert space $\mathcal{H}_Q$ which holds the secret.
     The task involves inputs $x\in \{0,1\}^{n}$ and system $Q$ given to Alice, and input $y\in \{0,1\}^{n}$ given to Bob.
    Alice sends message system $M_A$ to the referee, and Bob sends message system $M_B$. 
    Alice and Bob share a resource state $\Psi_{LR}$ with $L$ held by Alice and $R$ held by Bob. 
    Label the combined message systems as $M=M_AM_B$.
    Label the quantum channel defined by Alice and Bob's combined actions $\mathcal{N}_{Q\rightarrow M}^{x,y}$.
    We put the following two conditions on a CDQS protocol. 
    \begin{itemize}
        \item $\epsilon$\textbf{-correct:} There exists a channel $\mathcal{D}^{x,y}_{M\rightarrow Q}$, called the decoder, such that
        \begin{align}
            \forall (x,y)\in X\times Y \,\,\, s.t. \,\, f(x,y)=1,\,\,\, \left\Vert\mathcal{D}^{x,y}_{M\rightarrow Q}\circ \mathcal{N}^{x,y}_{Q\rightarrow M} - \mathcal{I}_{Q\rightarrow Q}\right\Vert_\diamond \leq \epsilon \: .
        \end{align}
        \item $\delta$\textbf{-secure:} There exists a quantum channel $\mathcal{S}_{\varnothing \rightarrow M}^{x,y}$, called the simulator, such that
        \begin{align}
            \forall (x,y)\in X\times Y \,\,\, s.t. \,\, f(x,y)=0,\,\,\, \left\Vert \mathcal{S}_{\varnothing \rightarrow M}^{x,y} \circ \tr_Q - \mathcal{N}_{Q\rightarrow M}^{x,y}\right\Vert_\diamond \leq \delta \: .
        \end{align}
    \end{itemize}
\end{definition}
We will take the Hilbert space $Q$ to be 2 dimensional throughout this work. 
The communication pattern of a CDQS protocol is shown in figure \ref{fig:CDQS}. 
We define the entanglement cost of a CDQS protocol to be the log dimension of the $L$ or $R$ Hilbert space, whichever is smaller. 
The minimal entanglement cost for a function $f$ that achieves $\epsilon$-correctness and $\delta$-security we denote by ${\text{CDQS}}_{\epsilon, \delta}(f)$. 
We will also use ${\text{CDQS}}(f) = {\text{CDQS}}_{0.09, 0.09}(f)$.
Quantum CDS protocols can also be amplified \cite{asadi2025conditional} to reduce correctness and security errors, so the constant $0.09$ is somewhat arbitrary, though it must be below a threshold set by the amplification theorem.  

An important fact that we will make use of is that classical CDS protocols imply quantum CDS protocols. 
In more detail we have the following theorem. 
\begin{theorem}\label{thm:CDStoCDQS}
    An $\epsilon$-correct and $\delta$-secure CDS protocol hiding $2n$ bits and using $n_M$ bits of message and $n_E$ bits of randomness gives a CDQS protocol which hides $n$ qubits, is $2\sqrt{\epsilon}$ correct and $\delta$ secure using $n_M$ classical bits of message plus $n$ qubits of message, and $n_E$ classical bits of randomness.
\end{theorem}
The basic reason why this theorem is true is easy to understand: to implement a quantum CDS protocol, we first encrypt $Q$ using the one-time pad, with a key labelled $k$, 
\begin{align}
    \mathcal{P}_{Q\rightarrow QS}(\rho)=\frac{1}{4}\sum_k P^k_Q \rho_Q P^k_Q\otimes \ketbra{k}{k}_S
\end{align}
$P^k$ is a choice of one of the four Pauli operators labelled by $k$. 
The one-time pad has the property that if you don't know the key $k$ ($S$ is traced out) the state on $Q$ looks maximally mixed, and hence independent of the input $Q$.
After encrypting $Q$ then, Alice simply forwards $Q$ to the referee, and inserts the key $k$ into a classical CDS protocol, which she runs along with Bob. 
If $f=1$ the referee gets $k$ and can decrypt $Q$; if $f=0$ then $k$ is hidden from the referee, essentially tracing it out and leaving $Q$ maximally mixed. 
This theorem is proven in \cite{allerstorfer2024relating} as theorem 22.

Notice that as a consequence of this theorem, any upper bounds on randomness complexity in classical CDS become upper bounds on quantum CDS, and lower bounds on quantum CDS become lower bounds on classical CDS. 
More specifically, using the above theorem and the fact that both CDS and CDQS can be amplified, we obtain
\begin{align}\label{eq:CDSandCDQS}
    \CDS(f)= \Omega(\CDQS(f))
\end{align}
Recalling that $\CDS(f)=\CDS_{0.1,0.1}(f)$ and $\CDQS(f)=\CDQS_{0.09,0.09}(f)$, this is expressing that we can amplify the CDS parameters to be small enough, then use the last theorem to obtain a CDQS protocol with parameters $\epsilon=\delta\leq 0.09$. 

\subsection{\texorpdfstring{$f$}{TEXT}-routing is quantum CDS}

In this section we prove the main result of this chapter, which is that quantum CDS and $f$-routing are equivalent in a certain sense. 
In particular, starting from an $f$-routing protocol using resource state $\Psi_{LR}$, one can always construct a CDQS protocol for the same function using the same resource state. 
Conversely, starting from a CDQS protocol using resource state $\Psi_{LR}$, we can always construct an $f$-routing protocol which uses resource state $\ket{\Psi}_{LRR'}$, where system $R'$ purifies the state $\Psi_{LR}$. 

The key intuition as to why CDQS and $f$-routing should be related comes from the \emph{decoupling theorem}. 
Decoupling is an expression of the fact that in quantum mechanics information is not created or destroyed.
To capture this more precisely, first recall the notion of a \emph{complementary channel}. 

\begin{definition}
    Given a quantum channel $\mathcal{N}_{A\rightarrow B}$, a channel $\mathcal{N}^c_{A\rightarrow E}$ is said to be complementary to $\mathcal{N}_{A\rightarrow B}$ if there exists an isometry $V_{A\rightarrow BE}$ such that
    \begin{align}
        \mathcal{N}_{A\rightarrow B}=\tr_E\circ (V_{A\rightarrow BE}(\cdot) V^\dagger_{A\rightarrow BE}), \nonumber \\
        \mathcal{N}_{A\rightarrow E}^c=\tr_B\circ (V_{A\rightarrow BE}(\cdot) V^\dagger_{A\rightarrow BE}).
    \end{align}
\end{definition}
In words, the output of the complementary channel is whatever the direct channel dumps into the environment. 

The following decoupling theorem expresses that if information is not revealed to the environment, then it must be preserved in the system. 
Conversely, if no information about the input to the channel is stored in its output, then all of the information must have been sent to the environment. 

\begin{theorem}\label{thm:newdecoupling}
    Let $\mathcal{N}_{A\rightarrow B}:\mathcal{L}(\mathcal{H}_A)\rightarrow \mathcal{L}(\mathcal{H}_B)$ be a quantum channel, and let $\mathcal{N}^c_{A\rightarrow E}$ be a channel which is complementary to $\mathcal{N}_{A\rightarrow B}$. Let $\mathcal{S}_{A \rightarrow E}$ be a completely depolarizing channel, which traces out the input and replaces it with a fixed state $\sigma_E$. 
    Then we have that if the complementary channel is close to depolarizing, then the direct channel almost preserves information,
    \begin{align}
        \frac{1}{4}\inf_{\mathcal{D}_{B\rightarrow A}} || \mathcal{D}_{B\rightarrow A} \circ \mathcal{N}_{A\rightarrow B}-\mathcal{I}_{A\rightarrow A}||_{\diamond}^2 \leq ||\mathcal{N}^c_{A\rightarrow E} - \mathcal{S}_{A \rightarrow E} ||_\diamond.
    \end{align}
    Conversely if the channel preserves information then its complementary channel is close to totally depolarizing,
    \begin{align}
        ||\mathcal{N}^c_{A\rightarrow E} - \mathcal{S}_{A \rightarrow E} ||_\diamond \leq 2 \inf_{\mathcal{D}_{B\rightarrow A}} ||\mathcal{D}_{B\rightarrow A} \circ \mathcal{N}_{A\rightarrow B} - \mathcal{I}_{A\rightarrow A} ||_\diamond^{1/2}.
    \end{align}
\end{theorem}

With decoupling in mind, we can understand why $f$-routing and quantum CDS should be related. 
The basic idea of the transformation is shown in figure \ref{fig:NLQCandCDQS}. 
Starting with the CDQS protocol, to build an $f$-routing protocol Alice and Bob first purify their operations. 
Then, they take the output of their original channels and send them right in the $f$-routing protocol, and take the purifying systems and send them left. 
Thus the map from $Q$ to the systems on the right and the map from $Q$ to the systems on the left are complementary channels. 
To see that the $f$-routing protocol is correct, let's separately consider $f=0$ and $f=1$ instances. 
In a $f=1$ instance, we know the referee recovers the secret $Q$. 
The systems the referee gets now go to the right, so in the $f$-routing protocol $Q$ is available on the right. 
In a $f=0$ instance, we have from security of the CDQS protocol that no information is revealed on the right, so that the map to the right is a completely depolarizing channel. 
But then the map to the left is the complementary channel, so by decoupling $Q$ is available on the left. 

The inverse transformation from $f$-routing to CDQS is similar: now, we just have Alice and Bob take whatever systems they would have sent left and trace them out. 
Correctness of the CDQS protocol is immediate again, and security follows from the decoupling theorem: now we know that $Q$ is available on the left, so the complementary channel (the map to the right) must not reveal anything about $Q$, so the CDQS protocol is secure.  

\begin{figure*}
    \centering
    \begin{subfigure}{0.45\textwidth}
    \centering
    \begin{tikzpicture}[scale=0.4]
    
    \draw[thick] (-5,-5) -- (-5,-3) -- (-3,-3) -- (-3,-5) -- (-5,-5);
    
    \draw[thick] (5,-5) -- (5,-3) -- (3,-3) -- (3,-5) -- (5,-5);
    
    \draw[thick] (5,5) -- (5,3) -- (3,3) -- (3,5) -- (5,5);
    
    \draw[thick] (4.5,-3) -- (4.5,3);
    
    \draw[thick] (-3.5,-3) to [out=90,in=-90] (3.5,3);
    
    \draw[thick] (-3.5,-5) to [out=-90,in=-90] (3.5,-5);
    \draw[black] plot [mark=*, mark size=3] coordinates{(0,-7.05)};
    
    \draw[thick] (-4.5,-6) -- (-4.5,-5);
    \node[below] at (-4.5,-6) {$x,Q$};
    
    \draw[thick] (4.5,-6) -- (4.5,-5);
    \node[below] at (4.5,-6) {$y$};

    \node[left] at (-1,-2) {$M_0$};
    \node[right] at (4.5,0) {$M_1$};
    
    \draw[thick] (4,5) -- (4,6);
    \node[above] at (4,6) {$Q$};

    \node at (-4,-4) {$\mathbf{\mathcal{N}}^L$};
    \node at (4,-4) {$\mathbf{\mathcal{N}}^R$};
    \node at (4,4) {$\mathbf{W}^R$};
    
    \end{tikzpicture}
    \caption{}
    \label{fig:CDQSagain}
    \end{subfigure}
    \hfill
    \begin{subfigure}{0.45\textwidth}
    \centering
    \begin{tikzpicture}[scale=0.4]
    
    \draw[thick] (-5,-5) -- (-5,-3) -- (-3,-3) -- (-3,-5) -- (-5,-5);
    \node at (-4,-4) {$\mathbf{V}^L$};
    
    \draw[thick] (5,-5) -- (5,-3) -- (3,-3) -- (3,-5) -- (5,-5);
    \node at (4,-4) {$\mathbf{V}^R$};
    
    \draw[thick] (5,5) -- (5,3) -- (3,3) -- (3,5) -- (5,5);
    \node at (4,4) {$\mathbf{W}^R$};
    
    \draw[thick] (-5,5) -- (-5,3) -- (-3,3) -- (-3,5) -- (-5,5);
    \node at (-4,4) {$\mathbf{W}^L$};
    
    \draw[thick] (-4.5,-3) -- (-4.5,3);
    
    \draw[thick] (4.5,-3) -- (4.5,3);
    
    \draw[thick] (-3.5,-3) to [out=90,in=-90] (3.5,3);
    
    \draw[thick] (3.5,-3) to [out=90,in=-90] (-3.5,3);
    
    \draw[thick] (-3.5,-5) to [out=-90,in=-90] (3.5,-5);
    \draw[black] plot [mark=*, mark size=3] coordinates{(0,-7.05)};
    
    \draw[thick] (-4.5,-6) -- (-4.5,-5);
    \draw[thick] (4.5,-6) -- (4.5,-5);
    
    \draw[thick] (4.5,5) -- (4.5,6);
    \draw[thick] (-4.5,5) -- (-4.5,6);
    
    \draw[thick] (3.5,5) -- (3.5,6);
    \draw[thick] (-3.5,5) -- (-3.5,6);

    \node[left] at (-1,-2) {$M_0$};
    \node[right] at (4.5,0) {$M_1$};

    \node[right] at (1,-2) {$M_1'$};
    \node[left] at (-4.5,0) {$M_0'$};
    
    \end{tikzpicture}
    \caption{}
    \label{fig:f-routing}
    \end{subfigure}
    \caption{Corresponding CDQS (left) and $f$-routing (right) protocols. To define the CDQS protocol from the $f$-routing protocol, we have Alice and Bob trace out systems $M_0'$ and $M_1'$. Systems $M_0$ and $M_1$ are sent to the referee rather than to Bob. To define the $f$-routing protocol from the CDQS, purify the local channels $\mathbf{\mathcal{N}}^L$ and $\mathbf{\mathcal{N}}^R$ to isometries $\mathbf{V}^L$ and $\mathbf{V}^R$. Send the original outputs of the channel to Bob on the right, and the purifying systems to Alice on the left. We adopt the notation $M=M_0M_1$ and $M'=M_0'M_1'$.}
    \label{fig:NLQCandCDQS}
\end{figure*}

This reasoning is made precise in the next theorem.
\begin{theorem}\label{thm:CDQSandfRouting}
    An $(\epsilon_0,\epsilon_1)$-correct $f$-routing protocol that routes $n$ qubits implies the existence of an $\epsilon_1$-correct and $\delta=2\sqrt{\epsilon_0}$-secure CDQS protocol that hides $n$ qubits using the same entangled resource state and the same message size. 
    An $\epsilon$-correct and $\delta$-secure CDQS protocol hiding secret $Q$ using at most $n_E$ qubit resource state and $n_M$ qubit messages implies the existence of a $(2\sqrt{\delta},\epsilon)$-correct $f$-routing protocol that routes system $Q$ using $2n_E$ qubits of resource state and $4(n_M+n_E)$ qubits of message. 
\end{theorem}
\begin{proof} \,Begin by considering an $f$-routing protocol. 
Figure \ref{fig:NLQCandCDQS} establishes the subsystem labels we will use here. 
We first show that an $f$-routing protocol is easily modified to construct a CDQS protocol. 
To do so, we send systems $M_0$ and $M_1$ that Bob would receive in the second round of the $f$-routing protocol to the referee of the CDQS protocol. 
Then, if $f(x,y)=1$, $\epsilon_1$-correctness in $f(x,y)=1$ instances of the $f$-routing scheme immediately gives $\epsilon_1$-correctness of the CDQS.  

To show secrecy of the CDQS protocol, we first establish some notation. 
We label the channel realized by the first round operations of Alice and Bob $\mathcal{N}_{Q\rightarrow MM'}$, and let $\mathbf{V}_{Q\rightarrow MM'E}$ be an isometric extension of this channel. 
By correctness in $0$ instances of the $f$-routing scheme, we have that there exists a channel $\mathcal{D}^{xy}_{M'\rightarrow Q}$ such that
\begin{align}
    ||\mathcal{D}^{xy}_{M'\rightarrow Q}\circ [\tr_M \circ \mathcal{N}_{Q\rightarrow M'M}^{x,y}] - \mathcal{I}_Q ||_\diamond = ||\mathcal{D}^{xy}_{M'\rightarrow Q}\circ [\tr_{ME} (\mathbf{V}^{xy}_{Q\rightarrow MM'E} \cdot (\mathbf{V}^{xy}_{Q\rightarrow MM'E})^\dagger)] - \mathcal{I}_Q ||_\diamond \leq \epsilon_0 \nonumber 
\end{align}
Then the decoupling theorem \ref{thm:newdecoupling} tells us that there exists a completely depolarizing channel $\mathcal{S}_{Q\rightarrow ME}$ such that
\begin{align}
    ||\tr_{M'} (\mathbf{V}^{xy}_{Q\rightarrow MM'E} \cdot (\mathbf{V}^{xy}_{Q\rightarrow MM'E})^\dagger) - \mathcal{S}^{xy}_{Q\rightarrow ME}||_\diamond \leq 2\sqrt{\epsilon_0}
\end{align}
Adding a trace over part of the outputs of channels can only make the channels less distinguishable, and hence the diamond norm smaller, so that
\begin{align}
    ||\tr_{M'E} (\mathbf{V}^{xy}_{Q\rightarrow MM'E} \cdot (\mathbf{V}^{xy}_{Q\rightarrow MM'E})^\dagger) - \mathcal{S}^{xy}_{Q\rightarrow M}||_\diamond \leq 2\sqrt{\epsilon_0}
\end{align}
but this is just
\begin{align}
    ||\mathcal{N}^{xy}_{Q\rightarrow M} - \mathcal{S}^{xy}_{Q\rightarrow M}||_\diamond \leq 2\sqrt{\epsilon_0}
\end{align}
which is exactly $2\sqrt{\epsilon_0}$-security of the CDQS. 
Note that the CDQS protocol defined by the $f$-routing protocol uses the same entangled resource state and no more communication. 

Now suppose we have a CDQS protocol which is $\epsilon$-correct and $\delta$-secure. 
Then to build the $f$-routing protocol, purify the channels Alice and Bob perform to isometries, replace the resource state $\Psi_{LR}$ with a purification $\ket{\Psi}_{LRR'}$,\footnote{If $LR$ consist of at most $n_E$ qubits, then $R'$ need not be larger than $n_E$ qubits, so the total resource system size is at most $2n_E$ qubits.} and send the original message systems of the CDQS to Bob and their purifications to Alice. 
Then by $\epsilon$-correctness of the CDQS protocol, we immediately have $\epsilon$-correctness of the $f$-routing protocol when $f(x,y)=1$.

Next consider the case where $f(x,y)=0$. 
Then security of the CDQS implies that there exists a simulator channel $\mathcal{S}_{\varnothing \rightarrow M}^{xy}$ such that
\begin{align}
    ||\mathcal{S}_{\varnothing \rightarrow M}^{xy} \circ \tr_Q - \mathcal{N}^{xy}_{Q\rightarrow M} ||_\diamond \leq \delta.
\end{align}
We will again apply the decoupling theorem. 
Notice that now, because of how we have defined the $f$-routing protocol, the map from $Q$ to $MM'$ is isometric, so $(\mathcal{N}^{xy}_{Q\rightarrow M})^c = (\mathcal{N}^{xy})_{Q\rightarrow M'}$.
Then the decoupling theorem implies the existence of a decoding channel $\mathcal{D}_{M'\rightarrow Q}^{xy}$ such that
\begin{align}
    ||\mathcal{D}^{xy}_{M'\rightarrow Q} \circ (\mathcal{N}^{xy})^c_{Q\rightarrow M'} -\mathcal{I}_Q||_\diamond \leq \sqrt{4 || \mathcal{S}_{\varnothing \rightarrow M}^{xy} \circ \tr_Q - \mathcal{N}^{xy}_{Q\rightarrow M}||} \leq 2\sqrt{\delta}
\end{align}
which gives $2\sqrt{\delta}$ correctness on $0$ instances. 

To see how the communication in the resulting $f$-routing protocol is related to the communication in the original CDQS protocol, we can use that a channel $\mathcal{N}_{A\rightarrow B}$ can always be purified by an isometry $\mathbf{V}_{A\rightarrow B C}$ where $d_C\leq d_Ad_B$. 
Let CDQS have messages that each consist of at most $n_{M}$ qubits, and use an $n_{E}$ qubit resource system on systems $LR$.
Then the most general possible protocol is defined by families of channels 
\begin{align}
\{\mathcal{N}^x_{L\rightarrow M_{0}}\},\,\,\,\, \{\mathcal{N}^y_{R\rightarrow M_{1}}\}
\end{align}
applied on the left and right respectively. 
We define purifications of these, 
\begin{align}
\{\mathbf{V}^x_{L\rightarrow M_{0}M_{0}'}\},\,\,\,\, \{\mathbf{V}^y_{R\rightarrow M_{1}M_1'}\}
\end{align}
We see that the message sizes are now at most $n_M + n_E$
qubits, so the total size of the communication is at most $4(n_M+n_E)$.
The entangled resource system used in the $f$-routing protocol is any purification of the one used in the CDQS protocol.
\end{proof}

As a consequence of this theorem and the fact that $f$-routing and CDQS can both be amplified, we obtain that
\begin{align}\label{eq:FR=CDQS}
    \FR(f) = \Theta(\CDQS(f)).
\end{align}
Note however that in writing this we should understand the `entanglement cost' of the $f$-routing protocol to be the log dimension of the resource state. 

\subsection{Implications for \texorpdfstring{$f$}{TEXT}-routing}\label{sec:CDSimplications}

The connection between classical CDS and $f$-routing opens up many avenues for fruitful interaction between NLQC and classical information-theoretic cryptography. 
We will continue to deepen this connection as we move through the chapters in this book. 
For instance, in chapter \ref{chapter:mostlyclassicallowerbounds}, we will develop a technique for lower bounding the entanglement cost of $f$-routing. 
Given the connections we've proven above, this also gives a lower bound on randomness cost in classical CDS. 
This bound turns out to be new, and in fact there is no known classical proof of this bound. 
As well, in chapter \ref{chapter:reductions} we will see that $f$-routing is equivalent to many other NLQC examples. 
Thus CDS is not just connected to one special isolated example of NLQC, but actually connected to a large and important class of examples. 

More immediately, the CDS and $f$-routing connection reveals a surprising new upper bound for $f$-routing. 
In \cite{liu2017conditional}, the authors prove that for all functions $f$, 
\begin{align}
    \CDS(f) \leq 2^{O(\sqrt{n\log n})}.
\end{align}
That is, all functions can be completed in CDS with sub-exponential randomness. 
The protocol that achieves this is based on a mathematical object known as a matching vector family. 
From equation \eqref{eq:FR=CDQS}, this also means that $f$-routing can be completed for all functions using sub-exponential entanglement. 
This comes as a surprise, since the garden-hose, formula-size, and span-program-based upper bounds require exponential entanglement in the worst case. 

Another surprising result concerns the highest complexity functions which have efficient $f$-routing schemes. 
From the techniques we've developed so far for $f$-routing, the hardest functions with efficient schemes are those in the complexity class $\ModkL$. 
This is a small class --- it is contained for example inside of $\NC^2$, the set of problems computable in circuits of depth $(\log n)^2$. 
This is believed to be well inside of $\mathsf{P}$. 
However, starting from the connection to CDS it is possible to construct efficient schemes for problems believed to be outside of $\mathsf{P}$. 
These are based on a connection between CDS and classical secret sharing schemes, and use the construction of non-linear secret sharing schemes. 
The smallest class these functions are known to be inside of is $\mathsf{BQP}$. 

These observations leave the status of efficient $f$-routing schemes wide open. 
We don't know for which functions there are efficient schemes, or even if all functions could have efficient schemes, and we don't seem to even have a good conjecture.

\section{Private simultaneous messages and coherent function evaluation}\label{sec:PSM}

A second well studied scenario in the information-theoretic cryptography literature is the \emph{private simultaneous message} setting, illustrated in figure \ref{fig:quantumandclassicalPSM}. 
We won't give as many details around PSM as we did for CDS; this section just briefly summarizes what is known about the relationship between PSM and NLQC.\footnote{We omit this partly because less is known about this connection, and partly because the organization of this book emphasizes $f$-routing and we wanted to avoid too significant of a detour.}

The PSM setting involves three players: Alice, Bob, and the referee. 
Alice receives input $x\in\{0,1\}^n$, Bob receives input $y\in\{0,1\}^n$.
All parties agree in advance on a choice of Boolean function $f:\{0,1\}^n\times\{0,1\}^n\rightarrow \{0,1\}$.
Classically, Alice and Bob share randomness and send classical messages to the referee.
Quantumly, Alice and Bob share entanglement and send quantum messages.
The goal is for the referee to compute $f(x,y)$ without learning anything further about the value of $(x,y)$.
This setting finds many applications in classical cryptography, both as a basic primitive and as a toy model for secure multi-party computation \cite{ishai1997private}. 

\begin{figure*}
    \centering
    \begin{subfigure}{0.45\textwidth}
    \centering
    \begin{tikzpicture}[scale=0.4]
    
    \draw[thick] (-5,-5) -- (-5,-3) -- (-3,-3) -- (-3,-5) -- (-5,-5);
    
    \draw[thick] (5,-5) -- (5,-3) -- (3,-3) -- (3,-5) -- (5,-5);
    
    \draw[thick] (5,5) -- (5,3) -- (3,3) -- (3,5) -- (5,5);
    
    \draw[thick] (4,-3) -- (4.5,3);
    
    \draw[thick] (-4,-3) to [out=90,in=-90] (3.5,3);
    
    \draw[thick,dashed] (-3.5,-5) to [out=-90,in=-90] (3.5,-5);
    \node[below] at (-3.5,-6) {$r$};
    \draw[black] plot [mark=*, mark size=3] coordinates{(0,-7.05)};
    \node[below] at (3.5,-6) {$r$};

    \node[left] at (0,1) {$m_0(x,r)$};
    \node[right] at (4.5,0) {$m_1(y,r)$};
    
    \draw[thick] (-4.5,-6) -- (-4.5,-5);
    \node[below] at (-4.5,-6) {$x$};
    
    \draw[thick] (4.5,-6) -- (4.5,-5);
    \node[below] at (4.5,-6) {$y$};
    
    \draw[thick] (4,5) -- (4,6);
    \node[above] at (4,6) {$f(x,y)$};
    
    \end{tikzpicture}
    \caption{}
    \label{fig:PSM}
    \end{subfigure}
    \hfill
    \begin{subfigure}{0.45\textwidth}
    \centering
    \begin{tikzpicture}[scale=0.4]
    
    \draw[thick] (-5,-5) -- (-5,-3) -- (-3,-3) -- (-3,-5) -- (-5,-5);
    
    \draw[thick] (5,-5) -- (5,-3) -- (3,-3) -- (3,-5) -- (5,-5);
    
    \draw[thick] (5,5) -- (5,3) -- (3,3) -- (3,5) -- (5,5);
    
    \draw[thick] (4,-3) -- (4.5,3);
    
    \draw[thick] (-4,-3) to [out=90,in=-90] (3.5,3);
    
    \draw[thick] (-3.5,-5) to [out=-90,in=-90] (3.5,-5);
    \node[below] at (0,-7.25) {$\Psi_{LR}$};
    \draw[black] plot [mark=*, mark size=3] coordinates{(0,-7.05)};

    \node[left] at (0,1) {$M_0$};
    \node[right] at (4.5,0) {$M_1$};
    
    \draw[thick] (-4.5,-6) -- (-4.5,-5);
    \node[below] at (-4.5,-6) {$x$};
    
    \draw[thick] (4.5,-6) -- (4.5,-5);
    \node[below] at (4.5,-6) {$y$};
    
    \draw[thick] (4,5) -- (4,6);
    \node[above] at (4,6) {$f(x,y)$};
    
    \end{tikzpicture}
    \caption{}
    \label{fig:PSQM}
    \end{subfigure}
    \caption{Private simultaneous message protocols (PSM). Again Alice and Bob do not communicate. They hold inputs $x$ and $y$ respectively. The referee should be able to learn $f(x,y)$ but nothing else about $(x,y)$. a) The classical setting, where Alice and Bob share randomness and send classical messages. b) The quantum setting, where Alice and Bob share entanglement and can send quantum messages.}
    \label{fig:quantumandclassicalPSM}
\end{figure*}

The key distinction between CDS and PSM is that in CDS the referee is given the inputs $(x,y)$, while in PSM the inputs must be hidden from the referee. 
This suggests that PSM is a harder task.
Indeed we can show that a PSM protocol for a function $f$ can be modified into a CDS protocol for the same function, and the resulting protocol uses almost the same resources. 
This means the PSM cost is an upper bound on the CDS cost. 
This is true both in the quantum and classical cases --- quantum PSM implies quantum CDS, and classical PSM implies classical CDS. 
We expect that PSM is strictly harder than CDS, in that there are functions for which there is a more efficient CDS protocol than there is PSM protocol. 
For instance, there is no known sub-exponential upper bound on PSM for random functions, but there is for CDS. 

PSM turns out to also be related to a special class of NLQC protocols, which we call coherent function evaluation (CFE). 
These are evaluations of NLQCs of the form
\begin{align}
    \mathbf{V}_f = \sum_{xy} \ket{xy}_{Z'} \ket{f(x,y)}_{Z} \bra{x}_X\bra{y}_Y
\end{align}
where $f$ is a choice of Boolean function. 
Here $X$ is the input on the left, $Y$ is the input on the right, $Z$ is the output on the right, and $Z'$ is the output on the left. 
In words, this coherently computes $f(x,y)$, and sends the input left and the output right. 
This is naturally related to PSM via a similar transformation to the one we saw in the case of CDS, wherein the system sent to the right in the NLQC protocol become the systems sent to the referee in the PSM protocol. 
Because the inputs $\ket{x}\ket{y}$ are treated coherently and sent left, this ensures that the inputs are not also revealed on the right, so are hidden from the referee. 
Meanwhile, because $f(x,y)$ is output on the right in the NLQC, $f(x,y)$ is revealed to the referee in the PSM protocol. 

In the case of CDS the transformation could be reversed: a quantum CDS protocol implies the existence of a similarly efficient $f$-routing protocol. 
This reversal doesn't quite work in the context of PSM. 
Instead, a good PSM protocol implies the existence of a CFE protocol that works with a fidelity of $1/2$. 
Achieving this fidelity is non-trivial, so there is a partial result here, but this leaves the connection between PSM and CFE somewhat loose. 

This leaves an interesting open problem: what is the right NLQC analogue of PSM?
CFE is strong enough to imply PSM, but there may be a weaker NLQC which both implies PSM and is implied by it. 
Aside from completing analogies in a satisfying way, this question addresses something deep about the connection between the cost of privacy in classical cryptography and entanglement cost in NLQC. 
In the context of CDS, the notion of privacy was that the secret should be hidden in 0 instances, and we found a translation of this into a setting with no explicit notion of privacy: on zero instances the secret should be made available in the purifying system. 
In other words, the quantum context allows us to translate a privacy condition into a second correctness condition. 
The fact that we don't know which NLQC PSM implies means that we don't have the right way to do this for the relevant notion of privacy in this case, so we don't yet know how to study this type of privacy in the language of additional (quantum) correctness conditions. 

\section{History and further reading}

Conditional disclosure of secrets was first studied in the context of private information retrieval \cite{gertner1998protecting}, in 1998. 
Meanwhile, $f$-routing was introduced in \cite{kent2011quantum, buhrman2013garden} in 2011. 
These were first observed to be related in \cite{allerstorfer2024relating}. 
The initial observation that these settings must be related was made using the quantum gravity perspective on non-local computation, which we will study in chapter \ref{chapter:gravity}. 
The introduction of quantum CDS made in this setting led to several follow up works studying quantum CDS in its own right. 
Basic properties and lower bounds were established in \cite{asadi2025conditional}, and separations between classical and quantum CDS were found in \cite{girish2025}. 

Quantum PSM was first studied in \cite{kawachi2021communication}, and then reappeared in relation to NLQC in \cite{allerstorfer2024relating}. 
The NLQC perspective has also motivated further consideration of quantum PSM. 
For instance, in \cite{girish2025} new PSM upper bounds are given using NLQC techniques. 
Further in \cite{girish2025magic} PSM lower bounds are related to $T$-depth lower bounds. 
Finally, also in \cite{girish2025magic} quantum PSM is separated from two way classical communication complexity. 

\part{Upper bounds}\label{part:upperbounds}

\chapter{Any channel can be implemented as an NLQC}\label{chapter:allchannels}

\minitoc

In this chapter we will study the most general possible NLQC and give a protocol for completing it.
The protocol uses a subroutine known as port-teleportation, which is of independent interest.

\section{Port-teleportation} \label{sec:portteleportation}

In this section we describe the port-teleportation protocol, first introduced in \cite{ishizaka2008asymptotic}.
Before doing so, it'll be helpful to step back and consider the idea of teleportation more broadly.
The basic steps of any teleportation are as follows. 
Initially, Alice and Bob share an entangled resource state $\ket{\Psi}_{AB}$.
Further, Alice holds the state $\ket{\psi}_{A'}$ she would like to send to Bob. 
Then, 
\begin{enumerate*}
    \item Alice performs a POVM measurement $\mathcal{M}=\{F_x\}_x$ on the $AA'$ system. 
    \item Alice sends Bob the classical measurement outcome $x$. 
    \item Bob applies channel $\mathcal{C}_x$ to $B$.
\end{enumerate*}
The teleportation is successful when the final state on $B$ is $\ket{\psi}_B$. 

The most familiar example of a teleportation procedure occurs when $A'$ is a qubit, $\ket{\Psi}_{AB}=\ket{\Psi^+}_{AB}$ is the maximally entangled state, and the measurement is in the Bell-basis, 
\begin{align}
    \{\ket{\Psi^+}_{A'A}, X_A \ket{\Psi^+}_{A'A},Z_A\ket{\Psi^+}_{A'A},X_AZ_A\ket{\Psi^+}_{A'A}\}
\end{align}
We will call this \emph{Bell-basis teleportation}, or just teleportation when it is clear from context that we mean this procedure specifically. 
An important fact about Bell-basis teleportation is that after Alice's measurement the $B$ system is in one of the states
\begin{align}
    Z_B^{x_1}X_B^{x_2}\ket{\psi}_B
\end{align}
Bob's correction operation is to apply $Z_B^{x_1}X_B^{x_2}$, which he can do once he receives $x=(x_1,x_2)$ from Alice. 

\begin{figure}
    \centering
    \begin{tikzpicture}
    
    \draw[fill=lightgray] (0,0) -- (0,2) -- (2,2) -- (2,0) -- (0,0);
    \node at (1,1) {\Large{$\Pi_i$}};
    
    \draw (0.25,-0.5) -- (0.25,0);
    \node[below] at (0.25,-0.5) {$\ket{\psi}_{A'}$};
    
    \draw[blue,->] (1,2) -- (1,2.5);
    \node[above right] at (0.8,2.5) {$k \in \{1,...,N\}$};
    
    \draw (1.75,0) -- (3.75,-1) -- (5.75,0) -- (5.75,0.25);
    \draw (1.5,0) -- (3.75,-1.15) -- (6,0) -- (6,0.25);
    \draw (1.25,0) -- (3.75,-1.3) -- (6.25,0) -- (6.25,0.25);
    \draw (1,0) -- (3.75,-1.45) -- (6.5,0) -- (6.5,0.25);
    
    \draw (4.5,-0.5) -- (5,-1);
    \node[below] at (5,-1) {$N$};
    
    \node[below] at (3.75,-1.45) {$\bigotimes_{i=1}^N \ket{\Psi^+}_{A_iB_i}$};
    
    \draw[gray,->] (7,0.75) to [out=180,in=90] (6.25,0.25);
    \node at (7,0.75) [gray,align=left,right]{\small{With high fidelity,} $\ket{\psi}$ \\ \small{appears on the $B_{k}$ system.}};
    
    \end{tikzpicture}
    \caption{The port-teleportation protocol. State $\ket{\psi}$ is held in system $A'$, along with $N$ entangled systems $\ket{\Psi^+}_{A_iB_i}$. We denote $A=A_1...A_N$. A POVM $\{\Lambda^k\}$ is performed on the $AA'$ system producing output $k \in \{1,...,N\}$. The state $\ket{\psi}$ then appears on the $B_k$ system with a fidelity controlled by $1/N$.}
    \label{fig:port_teleport}
\end{figure}

Now consider a different teleportation procedure, known as \emph{port-teleportation} and illustrated in figure \ref{fig:port_teleport}. In port-teleportation the entangled state is
\begin{align}
    \ket{\Psi}_{AB} = \bigotimes_{i=1}^N \ket{\Psi^+}_{A_iB_i}.
\end{align}
The measurement will produce an outcome $x\in \{1,...,N\}$, and the correction operation will be to trace out all but the $x$th subsystem $B_i$. 
We will discuss below how the measurement can be chosen to achieve this.
The key distinctions between port-teleportation and Bell-basis teleportation are that 1) $N$ may be quite large, so that the dimensionality of the resource system is much larger than the input system $A'$ and 2) The correction operation is the trace, which has the interesting feature of commuting with unitaries $\bigotimes_{i=1}^N U_{B_i}$ acting on each output `port' $B_i$. 
This will turn out to be the key feature that makes port-teleportation useful in the context of NLQC.

Next we will understand how to choose the measurement step to achieve the desired functionality of port-teleportation. 
Begin by writing the entire teleportation procedure as a quantum channel,
\begin{align}
    \mathcal{T}_{A'\rightarrow B'}(\sigma^{in}_{A'}) &= \sum_k^{N}\tr_{AA'\bar{B_k}}\left( \Lambda^k_{AA'} \left[\left(\bigotimes_{i=1}^N \ketbra{\Psi^+}{\Psi^+}_{A_iB_i}\right)\otimes \sigma^{in}_{A'} \right] \right)_{B_k\rightarrow B'}. \nonumber
\end{align}
The final subscript indicates that we relabel the $B_k$ system as $B'$ after taking the trace. System $A$ refers to the collection $A_1...A_N$ and $\bar{B}_k$ refers to $B_1...B_N\setminus B_k$. 
We can take the trace over $\bar{B}_k$ explicitly, leading to
\begin{align}
    \mathcal{T}_{A'\rightarrow B'}(\sigma^{in}_{A'}) &=\sum_k \tr_{AA'}\left( \Lambda^k_{AA'} \left[ \sigma_{AB'}^k \otimes \sigma^{in}_{A'} \right] \right),
\end{align}
where the states $\sigma_{AB'}^k$ are defined by
\begin{align}
    \sigma_{AB'}^k &= \tr_{\bar{B}_k} \left(\bigotimes_{i=1}^N \ketbra{\Psi^+}{\Psi^+}_{A_iB_i}\right)_{B_k\rightarrow B'}, \nonumber \\
    &= \ketbra{\Psi^+}{\Psi^+}_{A_kB'}\otimes \frac{{I}_{\bar{A}_k}}{d^{N-1}} . \label{eq:explicitsigmak}
\end{align}
Note that we've done the relabelling to $B'$ in the last line.

We will use the \emph{entanglement fidelity} to quantify how close this channel is to the identity. 
The entanglement fidelity is defined from two channels, $\mathcal{N}_Y$ and $\mathcal{M}_Y$, 
\begin{align}
    F_e(\mathcal{N}_Y,\mathcal{M}_Y) = F(\mathcal{I}\otimes \mathcal{N}_Y(\Psi^+_{XY}),\mathcal{I}\otimes \mathcal{M}_Y(\Psi^+_{XY})).
\end{align}
In words, the entanglement fidelity is measuring how similarly the two input channels act on one end of the maximally entangled state.\footnote{You can also (after some work) understand the entanglement fidelity as expressing how similar the two channels are acting on average over states in the Hilbert space.} 

Returning to our problem, consider the entanglement fidelity between $\mathcal{T}$ and the identity, so that we quantify how well our teleportation procedure works. 
This is
\begin{align}
    F_e(\mathcal{T},I) = \tr\left( \Psi^+_{CB'} ({I}_C \otimes \mathcal{T}_{A'\rightarrow B'})(\Psi^{+}_{CA'}) \right).
\end{align}
We will try to make this as large as possible by a careful choice of measurement $\{\Lambda_k\}_k$. 
Inserting the form of our channel into the expression above, we obtain
\begin{align}
    F_e(\mathcal{T},I)= \sum_k \tr (\Psi^+_{CB'} \Lambda_{AA'}^{k} (\sigma_{AB'}^k\otimes \Psi^{+}_{CA'})).
\end{align}
To simplify this further, observe that the $C$ system of the two EPR pairs appearing $\Psi_{CB'}$ and $\Psi_{CA'}$ are contracted, which produces 
\begin{align}\label{eq:SWAPidentity}
    \tr_C(\Psi^+_{CB'}\Psi^+_{CA'}) = \frac{1}{d^2} SWAP_{A'B'}
\end{align}
This is easiest to check using diagrammatic notation, illustrated in figure \ref{fig:SWAPdiagram}. 

\begin{figure*}
    \centering
    \begin{tikzpicture}[scale=0.5]
    
    \draw[thick] (-2,2) to [out=-90,in=-90] (2,2);
    \draw[black] plot [mark=*, mark size=3] coordinates{(0,0.85)};

    \node[above] at (-2,2) {$B'$};
    \node[below] at (-2,-2) {$B'$};

    \draw[thick] (-2,-2) to [out=90,in=90] (2,-2);
    \draw[black] plot [mark=*, mark size=3] coordinates{(0,-0.85)};

    \draw[thick] (2,-2) to [out=-90,in=-90] (6,-2);
    \draw[black] plot [mark=*, mark size=3] coordinates{(4,0.85-4)};

    \draw[thick] (2,-6) to [out=90,in=90] (6,-6);
    \draw[black] plot [mark=*, mark size=3] coordinates{(4,-0.85-4)};

    \node[above] at (6,-2) {$A'$};
    \node[below] at (6,-6) {$A'$};

    \node[left] at (2,-2) {$C$};
    \node[left] at (2,-6) {$C$};

    \node[right] at (2,2) {$C$};

    \draw[thick] (2,-6) to [out=-90,in=-90] (5,-6) -- (5,2) to [out=90,in=90] (2,2);

    \node at (8,-1) {$=$};

    \node at (9,-1) {$\frac{1}{d^2}$};

    \draw[thick] (10,-3) to [out=90,in=-90] (14,1);
    \draw[thick] (14,-3) to [out=90,in=-90] (10,1);

    \node[below] at (10,-3) {$B'$};
    \node[below] at (14,-3) {$A'$};

    \node[above] at (10,1) {$B'$};
    \node[above] at (14,1) {$A'$};
    
    \end{tikzpicture}
    \caption{Diagrammatic proof of equation \eqref{eq:SWAPidentity}.}\label{fig:SWAPdiagram}
\end{figure*}

\begin{figure*}
    \centering
    \begin{tikzpicture}[scale=0.5]
    
    \draw[thick] (-4,-4) -- (0,-4) -- (0,-2) -- (-4,-2) -- (-4,-4);
    \node at (-2,-3) {$\sigma$};

    \draw[thick] (-1,-2) -- (-1,0);

    \draw[thick] (-2,0) -- (2,0) -- (2,2) -- (-2,2) -- (-2,0);
    \node at (0,1) {$\Lambda$};

    \node[left] at (-1,-1) {$A$};
    \node[left] at (1,-1) {$A'$};
    \node[left] at (-3,-1) {$B'$};

    \draw[thick] (-3,-2) -- (-3,3);
    \draw[thick] (-1,2) -- (-1,3);
    \draw[thick] (1,2) -- (1,3);

    \draw[thick] (1,-5) -- (1,0);
    \draw[thick] (-3,-5) -- (-3,-4);
    \draw[thick] (-1,-4) to [out=-90,in=-90] (5,-4);

    \draw[thick] (-3,-5) to [out=-90,in=90] (1,-8);
    \draw[thick] (1,-5) to [out=-90,in=90] (-3,-8);

    \draw[thick] (5,-4) -- (5,3);
    \draw[thick] (5,3) to [out=90,in=90] (-1,3);

    \draw[thick] (1,-8) to [out=-90,in=-90] (3,-8) -- (3,3) to [out=90,in=90] (1,3);
    \draw[thick] (-3,-8) to [out=-90,in=-90] (-5,-8) -- (-5,3) to [out=90,in=90] (-3,3);

    \draw[dashed] (-3.5,-8) -- (1.5,-8) -- (1.5,-5) -- (-3.5,-5) -- (-3.5,-8);

    \node at (6.5,-1) {$=$};

    \draw[thick] (10,0) -- (14,0) -- (14,2) -- (10,2) -- (10,0);
    \node at (12,1) {$\Lambda^k$};
    
    \draw[thick] (10,-4) -- (14,-4) -- (14,-2) -- (10,-2) -- (10,-4);
    \node at (12,-3) {$\sigma$};

    \draw[thick] (11,0) -- (11,-2);
    \draw[thick] (13,0) -- (13,-2);

    \draw[thick] (11,-4) to [out=-90,in=-90] (8,-4) -- (8,2) to [out=90,in=90] (11,2);
    \draw[thick] (13,-4) to [out=-90,in=-90] (16,-4) -- (16,2) to [out=90,in=90] (13,2);
    
    \end{tikzpicture}
    \caption{Diagrammatic proof of expression \eqref{eq:FE}. The part of the diagram within the dashed box is the SWAP found in figure \ref{fig:SWAPdiagram}.}\label{fig:FEdiagram}
\end{figure*}

Using this, one can then check that
\begin{align}\label{eq:FE}
    F_e(\mathcal{T},I) &= \frac{N}{d^2} \left(\frac{1}{N}\sum_{k=1}^N \tr(\Lambda_{AB'}^{k} \sigma_{AB'}^k)\right).
\end{align}
This is shown diagrammatically in figure \ref{fig:FEdiagram}. 
Notice that the expression inside the brackets is a guessing probability: it's the probability of guessing a randomly chosen $\sigma^k$ correctly using the POVM $\{\Lambda^k\}_k$.

From this expression, we can already get a sense of how many EPR pairs we will need to make port-teleportation work: we know the guessing probability is bounded above by $1$, so we'd better have $N\gtrsim d^2$ if we want a fidelity close to $1$. 
In fact this turns out to suffice. 
To show this, it suffices to use a general method of designing measurements to distinguish among a set of density matrices $\{\sigma^k\}$ called the \emph{pretty good measurement}. 
The pretty good measurement takes the POVM elements
\begin{align}
    \Lambda^k = \sigma^{-1/2}\sigma^k\sigma^{-1/2},\,\qquad \sigma=\sum_k\sigma^k.
\end{align}
This is known as the 'pretty good' measurement because the probability of successfully guessing $k$ given a randomly selected $\sigma^k$ turns out to never be worse than $(p_{opt})^2$, where $p_{opt}$ is the best any measurement can do. 
Thus, without having to think too hard about which measurement to make, we know we do `pretty good'. 

Using this choice of measurement the entanglement fidelity is found to be 
\begin{align}
    F_e(\mathcal{T},I) \geq \left(1 - \frac{d^2-1}{2N}\right)^2\geq 1- \frac{d^2-1}{N}. 
\end{align}
We perform a more general calculation in appendix \ref{sec:pauliadaptedPT}, which simplifies to the above bound in our setting. 
From the above, we see that if we want to achieve constant error $\epsilon_e$ as measured by the entanglement fidelity, we need to set 
\begin{align}
    \boxed{N_e=\frac{d^2-1}{\epsilon_e}}
\end{align}
We can also ask about the entanglement needed to achieve constant diamond norm error. 
The entanglement fidelity is related to the diamond norm distance by equation \ref{eq:diamondandFe}, so that
\begin{align}\label{eq:PTdiamondbound}
    ||\mathcal{T}-\mathcal{I} ||_\diamond \leq 2d\sqrt{\frac{d^2-1}{2{N}} } \leq d^2\sqrt{\frac{2}{N}}
\end{align}
Because the diamond distance measures how distinct the two input channels are maximized over input states, this bound shows that, for large enough $N$, the teleportation channel works well for all input states.
To achieve constant diamond norm error $\epsilon_\diamond$ then it suffices to take\footnote{In fact, this turns out to be overkill: the diamond norm error turns out to be related to the entanglement fidelity without dimensional factors \cite{christandl2021asymptotic}, so we can lower the cost by a factor of $d^2$. We don't justify this here, so we stick to quoting the weaker result.}
\begin{align}
    \boxed{N_\diamond = \frac{2d^4}{\epsilon_\diamond^2}.}
\end{align}

Returning to the pretty good measurement, in our setting the POVM elements are
\begin{align}
    \Lambda^k_{AA'} = \sigma^{-1/2} \left(\ketbra{\Psi^+}{\Psi^+}_{A_kA'}\otimes \mathcal{I}_{\bar{A}_k} \right)  \sigma^{-1/2}.
\end{align}
Ignoring for a moment the normalizing $\sigma^{-1/2}$ factors, these are intuitive: they are projecting in an EPR pair between $A'$ and the $A_i$ system, which just means mapping the $A'$ to the $B_i$ system identically, which is exactly what we want.
The addition of the $\sigma^{-1/2}$ factors smears these projectors such that this forms a complete measurement basis. 

\section{Port-teleportation based protocol}

We're now ready to give an NLQC protocol for an arbitrary channel $\mathcal{N}_{A_LA_R\rightarrow B_LB_R}$.
We assume each of $A_L$, $A_R$ consist of $n$ qubits, though the generalization to $A_L$ and $A_R$ being of different dimensions is easy.
Label the input to the channel as $\ket{\psi}_{A_LA_R}$.
The protocol is illustrated in figure \ref{fig:PTprotocol}. 

\begin{figure}
    \centering
    \begin{tikzpicture}

    \draw[thick] (-4.5,0) to [out=-90,in=180] (0,-1.5) to [out=0,in=-90] (6.5,0);
    \draw[black] plot [mark=*, mark size=3] coordinates{(0,-1.5)};

    \draw[thick] (-6,0) -- (-4,0) -- (-4,1) -- (-6,1) -- (-6,0);
    \draw[thick] (-5.5,-1) -- (-5.5,0);
    \node[below] at (-5.5,-1) {$A$};
    
    \draw[thick] (-5.5,0.25) to [out=90,in=90] (-4.5,0.25);
    \draw[thick] (-5,0.25) -- (-4.5,0.75);
    \draw[thick,->] (-5,1) -- (-5,1.5);
    \node[above] at (-5,1.5) {$i$};

    \draw[thick] (4,0) -- (7,0) -- (7,1.5) -- (4,1.5) -- (4,0);

    \draw[very thick] (5.5,0) to [out=-90,in=0] (0,-1) to [out=180,in=-90] (-1,0) -- (-1,1.5);
    \node at (5,-0.25) {$...$};
    \draw[thick] (-3.5,1.5) -- (-2.5,1.5) -- (-2.5,2.5) -- (-3.5,2.5) -- (-3.5,1.5);
    \node at (-3,2) {$P^i$};

    \draw[thick] (4.5,0.25) to [out=90,in=90] (6.5,0.25);
    \draw[thick] (5.5,0.25) -- (6,1);

    \draw[thick,->] (5.5,1.5) -- (5.5,2);
    \node[above] at (5.5,2) {$j$};

    \draw[thick] (6.5,-1.5) -- (6.5,0);
    \node[below] at (6.5,-1.5) {$B$};

    \draw (5.5,2.3) circle [radius=0.25];

    \draw[thick, dashed] (5.5,2.55) to [out=90,in=-90] (7.5,8);
    \draw[thick, dashed] (5.5,2.55) to [out=90,in=-90] (-0.75,8);

    \draw[very thick] (4.5,0) to [out=-90,in=0] (0,-0.75) to [out=180,in=-90] (-3,0) -- (-3,1.5);
    \draw[thick] (-1.5,1.5) -- (-0.5,1.5) -- (-0.5,2.5) -- (-1.5,2.5) -- (-1.5,1.5);
    \node at (-1,2) {$P^i$};

    \draw[very thick] (-1,2.5) -- (-1,3);
    \draw[very thick] (-3,2.5) -- (-3,3);

    \node at (-2,2) {$...$};
    \node at (-2,3.5) {$...$};

    \draw[thick] (-1.5,3) -- (-0.5,3) -- (-0.5,4) -- (-1.5,4) -- (-1.5,3);
    \draw[thick] (-1.25,4) -- (-1.25,8);
    \draw[thick] (-0.75,4) to [out=90,in=-90] (7,8);
    \node at (-1,3.5) {$\mathcal{N}$};

    \draw[thick] (-3.5,3) -- (-2.5,3) -- (-2.5,4) -- (-3.5,4) -- (-3.5,3);
    \draw[thick] (-3.25,4) -- (-3.25,8);
    \draw[thick] (-2.75,4) to [out=90,in=-90] (5,8);
    \node at (-3,3.5) {$\mathcal{N}$};

    \draw[thick] (4.5,8) -- (7.5,8) -- (7.5,9.5) -- (4.5,9.5) -- (4.5,8);
    \node at (6,8.75) {$\tr_{j^c}$};

    \draw[thick] (-0.75,8) -- (-3.75,8) -- (-3.75,9.5) -- (-0.75,9.5) -- (-0.75,8);
    \node at (-2.25,8.75) {$\tr_{j^c}$};

    \draw[thick] (-2.25,9.5) -- (-2.25,10);
    \node[above] at (-2.25,10) {$A$};

    \draw[thick] (6,9.5) -- (6,10);
    \node[above] at (6,10) {$B$};
    
    \end{tikzpicture}
    \caption{The port-teleportation based protocol for implementing any channel as an NLQC. On the left, Alice performs a Bell basis measurement on her system and one end of a maximally entangled state, obtaining measurement outcome $i$. Bob then collects the other end of that maximally entangled state and his input $B$, and executes the measurement from the port-teleportation protocol, obtaining outcome $j$. Thus the full input, up to a Pauli correction determined by $i$, is now available on Alice's side, albeit it is stored in a port determined by $j$. Note that each of the ports (labelled with thick wires) has dimension $d_Ad_B$. Alice undoes the Pauli and applies the needed channel to every port. Then, she splits the two subsystems of each port corresponding to the two outputs of the channel and sends the first subsystem left, and the second subsystem right. Bob broadcasts his measurement outcome $j$. After the communication round Alice and Bob trace out all but the $j$th port.}
    \label{fig:PTprotocol}
\end{figure}

\vspace{0.2cm}
\noindent \textbf{NLQC protocol for arbitrary channels:}\\
\vspace{0.2cm}
Preparation phase:
\begin{enumerate*}
    \item Distribute a maximally entangled system $\ket{\Psi^+}_{F_{L}F_{R}}$ consisting of $n$ EPR pairs, with $F_L$ sent to Alice and $F_R$ to Bob.
    \item Distribute a set of $N$ maximally entangled systems $\otimes_{k=1}^N \ket{\Psi^+}_{E_{L,k}E_{R,k}}$ with each $\ket{\Psi^+}_{E_{L,k}E_{R,k}}$ consisting of $2n$ EPR pairs, with all $E_{L,k}$ sent to Alice and all $E_{R,k}$ to Bob.
\end{enumerate*}
Execution phase:
\begin{enumerate*}
    \item Alice measures $A_L F_{L}$ in the Bell basis, obtaining outcome $i$. Then Bob holds the state
    \begin{align}
    (P^i_{F_{R}}\otimes I_{A_R})\ket{\psi}_{F_{R}A_R}
    \end{align}
    and the index $i$ is held on the left by Alice. 
    \item Perform the ``pretty-good'' measurement described in section \ref{sec:portteleportation}, as if port-teleporting $F_{R}A_R$ to Alice using the $N$ maximally entangled pairs $\ket{\Psi^+}_{E_{L,k}E_{R,k}}$. Call the measurement outcome $j$. Then $j$ is held on the right by Bob and the state
    \begin{align}
        \Psi \approx (P^i\otimes I)\ket{\psi}_{E_{L,j}} \otimes  \rho_{E_{L}\setminus E_{L,j}}
    \end{align} 
    is held on the left by Alice. The $\approx$ symbol indicates that this is correct up to the error induced by the port-teleportation. 
    \item On the left, Alice applies $\mathcal{N}\circ (P^i\otimes I)$ to every subsystem $E_{L,k}$. Then Alice holds
    \begin{align}
        \Psi \approx \mathcal{N}(\ketbra{\psi}{\psi}_{E_{L,j}})\otimes  \mathcal{N}^{\otimes (N-1)}(\rho_{E_{L}\setminus E_{L,j}})
    \end{align}
    with all $E_L$ systems on the left, and $j$ on the right.
    \item Relabel the $E_{L,k}$ qubits as $B_{L,k}B_{R,k}$, and send all of the $B_{L,k}$ systems to Alice on the left and all of the $B_{R,k}$ systems to Bob on the right. Send $j$ from the right to both the right and left. 
    \item Bob traces out all but the $B_{R,j}$ system, and returns $B_{R,j}$ as his output. Similarly Alice on the left traces out all but the $B_{L,j}$ system, and returns $B_{L,j}$.
\end{enumerate*}

This completes the arbitrary channel non-locally, although the use of port-teleportation means this performs the intended channel only approximately. 
The total entanglement cost is as follows. 
From the initial Bell teleportation, we use $n$ EPR pairs. 
From the port-teleportation, we teleport $N$ ports, each consisting of $2n$ qubits, so the cost is $2nN$. 
The needed value of $N$ depends on if we want to achieve constant diamond norm error or constant entanglement fidelity. 
Fixing the entanglement fidelity to be $1-\epsilon_e$, the total cost is
\begin{align}
    E=n+2nN=n+2n\frac{2^{4n}-1}{\epsilon_e}
\end{align}
EPR pairs. 
Fixing the diamond norm error to be $\epsilon_\diamond$, the total cost is
\begin{align}
    E=n+2n\frac{2^{8n+1}}{\epsilon_\diamond^2}.
\end{align}

In the port-teleportation based NLQC protocol, notice that we use a Bell-basis teleportation to first collect Alice and Bob's input systems together, then a port-teleportation to undo the resulting Pauli correction. 
This leaves the full input system together, uncorrupted by any Pauli operations, but in an unknown port. 
It appears to be necessary to use this combination of Bell and port teleportation, in that there is no known protocol that uses only port-teleportation. 

\section{History and further reading}

The singly exponential upper bound described here was given by Beigi and K\"onig \cite{beigi2011simplified}. 
Port-teleportation was introduced earlier in \cite{ishizaka2008asymptotic}.
The initial motivation for introducing port-teleportation had to do with an object known as a \emph{universal quantum programmable processor}.
A programmable processor is a channel $P$ which takes in two systems: an input state $\ket{\psi}_A$ and a program system, $\ket{\phi_U}$. 
The channel should apply a unitary which is specified by the program state, $U\ket{\psi}_A$. 
That is, 
\begin{align}
    \mathcal{P}_{AB}(\ketbra{\psi}{\psi}_A\otimes\ketbra{\phi_U}{\phi_U}) = U_A\ketbra{\psi}{\psi}U_A^\dagger.
\end{align}
This is similar to how a classical computer works, in that a computer can take two inputs, a data input $x$ and a second input describing a program $P$, and then output $P(x)$. 

Nielsen and Chuang \cite{nielsen1997programmable} showed that for every distinct program you want to be able to apply, you need the program space to pick up one extra dimension. 
Thus universal programmable processors are impossible in the sense that there are an infinite number of unitaries (since it's a continuous space), so no program space is large enough to allow universality. 
In the approximate setting, where the output just needs to be close to $U\ket{\psi}$, we can have finite dimensional program spaces. 
The port-teleportation scheme provides one construction of an approximate quantum processor since we can take $\otimes_{i=1}^NU_{B_i}\ket{\Psi^+}_{A_iB_i}$ as the program state, and the channel that applies the measurement $\{\Lambda^k\}$ and then traces out the unused ports as the action of the programmable processor. 

We have a good understanding of how large of a program space an approximate universal processor needs, see e.g. \cite{kubicki2019resource}. 
In particular, the port-teleportation protocol is optimal if we want to be able to apply any unitary up to some fixed accuracy $\epsilon$. 
On the other hand, we have no such understanding for the case of non-local quantum computation. 
We don't know, for instance, if it is necessary that a general NLQC protocol also define a universal processor (as the port-teleportation protocol does). 

In \cite{junge2022geometry}, an attempt is made to adapt the techniques used to characterize universal programmable processors to the NLQC setting. 
They have some success in doing this, but find that their final bound relies on some mathematical conjectures which so far have not been proven. 
Assuming those conjectures though, they show an exponential lower bound on the entanglement needed to implement an arbitrary unitary as an NLQC. 
Without these conjectures though there is so far no exponential lower bound on NLQC, so it's unknown if the port-teleportation protocol is optimal. 
On the other hand, there are some surprising examples of NLQCs that were once expected to have exponential cost but are now known to be sub-exponential, see \cite{allerstorfer2024relating}, so proving exponential lower bounds or showing a generic sub-exponential upper bound remains an important open problem. 

The port-teleportation protocol described here is not quite the optimal one, at least in regard to some parameter scalings. 
In \cite{christandl2021asymptotic}, it is shown that a modified port-teleportation protocol achieves
\begin{align}
    F_e=1-\frac{d^{5}}{4\sqrt{2}N^2}+O(1/N^3).
\end{align}
Thus the entanglement fidelity approaches $1$ with $1/N^2$, rather than $1/N$, as in our result. 
Regarding the entanglement needed to achieve fixed error however, this means we need $N=d^{5/2}/\sqrt{\epsilon_e}$ for constant entanglement fidelity, so the scaling with $d$ is worse than our quoted result.

\chapter{\texorpdfstring{$T$}{TEXT}-depth upper bound}\label{chapter:Tdepth}

\minitoc

In this chapter we give an upper bound on NLQCs that implement a general unitary $U_{AB}$. 
The upper bound is in terms of the number of layers of $T$-gates that appear when we write the unitary in a circuit decomposition. 
This relates the entanglement cost in NLQC to a quantum notion of complexity; this builds on the connection between NLQC and classical complexity we saw in chapter \ref{chapter:f-routing}.

\section{Upper bound for Clifford unitaries}\label{sec:cliffordNLQC}

The Clifford unitaries are easy to implement in the form of an NLQC.
To see why, recall that a Clifford unitary is any unitary such that
\begin{align}
    \forall \,P \in \mathcal{P}_n, \,\exists \,P'\in \mathcal{P}_n \,\,\,\,s.t. \,\,\,\,CPC^\dagger = P'
\end{align}
where $C$ acts on $n$ qubits, and $\mathcal{P}_n$ is the $n$ qubit Pauli group. That is, the Cliffords conjugate Pauli operators to Pauli operators. 
This fact means Cliffords interact in a tidy way with Bell basis teleportation, which lets us do Cliffords efficiently as NLQCs. 
We show this explicitly with the next protocol, which implements a Clifford $C_{AB}$ with $A$ and $B$ each $n$ qubits using $n$ EPR pairs. 
The protocol assumes Alice and Bob begin with $n$ shared EPR pairs.

\vspace{0.2cm}
\noindent \textbf{NLQC for Cliffords:}
\begin{itemize*}
    \item Bob teleports$^*$ his input $B$ to Alice, and broadcasts the measurement outcome $k$ to both sides. 
    \item Alice applies $C_{AB}$ to her input and her end of the EPR pairs. 
    \item Alice keeps the output system $A$, and sends $B$ to Bob. 
    \item Alice and Bob compute $C_{AB} P^k C_{AB}^\dagger=(P')^k$ to determine $P'$ (which is a function of $k$), and then each apply the needed Pauli operators locally to produce $C_{AB}\ket{\psi}$.
\end{itemize*}
Notice what happens in this protocol. 
After the teleportation, Alice holds
\begin{align}
    C_{AB}P_B^k\ket{\psi}_{AB} = (P_{AB}')^k C_{AB}\ket{\psi}_{AB}.
\end{align}
After moving the Pauli through the Clifford, it may now act on both subsystems $A$ and $B$, but importantly (since it is Pauli) it is tensor product across the split between $A$ and $B$. 
This means that the Pauli correction $P'$ can be fixed after the communication round, separately by Alice and Bob. 

\section{Upper bound from \texorpdfstring{$T$}{TEXT}-depth}

Of course, not all unitaries are Clifford. 
Recall that a universal gate set for a unitary is $H$, $CNOT$, $S$, and $T$. 
The first three operations --- $H$, $CNOT$ and $S$ --- generate the Cliffords. 
Adding in the $T$ gate then lets this set generate arbitrary unitaries. 
If we have a unitary which only involves using a few T-gates, can we then implement it efficiently?
It turns out the answer is yes, in a sense we make precise. 

To see this, we will develop a protocol for implementing circuits of low $T$-depth.
The $T$-depth is the number of layers of $T$ gates needed to implement the unitary, where we allow arbitrary Clifford circuits before and after each layer of $T$'s.  
The efficiency of this protocol will be roughly $\sim n^d$ for $d$ the $T$-depth. 
The protocol relies in a crucial way on the garden-hose protocol, which we saw earlier in the context of implementing $f$-routing in chapter \ref{chapter:f-routing}. 
We will need to develop some facts about the garden-hose before proceeding. 

\subsection{More on the garden-hose}

The first lemma regarding the garden-hose model will let us put garden-hose protocols into a standard form amenable to being composed with other operations. 
\begin{lemma}
Suppose there is a garden-hose protocol that computes $f(x,y)$ using $m$ pipes in the sense that water spills on Alice's side if $f(x,y)=0$, and on Bob's side if $f(x,y)=1$. 
Then there is also a garden-hose protocol that computes $f(x,y)$ in the sense of water spilling on Alice's side from one of two designated pipes that uses at most $3m$ pipes. 
\end{lemma}
\begin{proof}
The proof is given by the diagram in figure \ref{fig:standardformGH}. 
There, the input $Q$ is input to a first instance of the garden-hose protocol for $f$. 
Then, all of the spilling pipes on the right are connected to corresponding pipes in a second copy of the protocol, and all of the spilling pipes on the left are connected to the corresponding pipes in a third copy. 
If the water spills on the right of the original protocol, it will emerge from the input pipe of the second protocol. 
Meanwhile, if the water spills from the left of the original protocol, it will emerge from the input pipe of the third protocol. 
These two pipes become the two designated output pipes on Alice's side. 
\end{proof}
\begin{figure}
    \centering
    \begin{tikzpicture}[scale=0.8]

    \draw[thick] (0,0) -- (0,2.5) -- (5,2.5) -- (5,0) -- (0,0);
    \draw[thick,blue] (-1,3) to [out=-90,in=180] (0,2);
    \node[above] at (-1,3) {in};
    \draw[thick] (0,1) -- (-1,1);
    \node at (2.5,1.25) {GH protocol for $f$};

    \begin{scope}[shift={(0,-5)}]
    \draw[thick] (0,0) -- (0,2.5) -- (5,2.5) -- (5,0) -- (0,0);
    \draw[thick,blue] (-1,-0.5) to [out=90,in=180] (0,2);
    \node[below] at (-1,-0.5) {$0$ out};
    \draw[thick] (0,1) -- (-1,1);
    \node at (2.5,1.25) {GH protocol for $f$};
    \end{scope}

    \begin{scope}[shift={(9,-5)}]
    \draw[thick] (0,0) -- (0,2.5) -- (5,2.5) -- (5,0) -- (0,0);
    \draw[thick,blue] (-1,-0.5) to [out=90,in=180] (0,2);
    \node[below] at (-1,-0.5) {$1$ out};
    \draw[thick] (0,1) -- (-1,1);
    \node at (2.5,1.25) {GH protocol for $f$};
    \end{scope}

    \draw[thick] (-1,1) to [out=180,in=180] (-1,-4);
    \draw[thick] (5,-4) -- (6,-4);

    \draw[thick] (5,1) -- (15,1) to [out=0,in=0] (15,-4) -- (14,-4);
        
    \end{tikzpicture}
    \caption{Transformation of a garden-hose protocol into `standard form', which means having two possible spilling pipes on Alice's side, one where water emerges if $f=0$ and another where water emerges if $f=1$. The black wires indicate multiple pipes; black wires on the right indicate all of the open wires on the right, and similarly for black wires on the left.}
    \label{fig:standardformGH}
\end{figure}

Finally, we need the next somewhat more involved statement which tells us about how fast the garden-hose complexity can grow when we compose functions by taking XOR's. 
\begin{theorem}\label{thm:XORlemma}
The garden-hose complexity satisfies:
    \begin{align}
        \GH\left(\bigoplus_i f_i\right) \leq 4 \sum_i \GH(f_i)
    \end{align}
\end{theorem}

\begin{proof}
     Consider garden-hose protocols for each $f_i$, which we label $P_i$. 
     We give a garden-hose protocol for $\oplus_i f_i$ by wiring copies of the $P_i$ together in an appropriate way. 
     Concretely, we take 4 copies of $P_i$, and connect them as shown in figure \ref{fig:XORgadget}. 
     The gadget has four open hoses, which we wire together with further gadgets: we wire the $0$ output of the $f_i$ gadget to the $0$ output of the $f_{i+1}$ gadget, and the $1$ output of the $f_i$ gadget to the $1$ output of the $f_{i+1}$ protocol. 
     By inspection, one can check that the gadget flips the parity of the input if $f_i=1$, and leaves the input unchanged if $f_i=0$.
     To compute $\oplus_i f_i$ then, we connect the tap to the $0$ input of the $f_1$ gadget, and label the $0$ and $1$ outputs of the final $f_i$ as the $0$ and $1$ labelled output hoses of the protocol. 
     After the water flows through gadgets for each $f_i$, we've computed $\oplus_i f_i$. 
\end{proof}

\begin{figure}
    \centering
    \begin{tikzpicture}[scale=0.8]

    \draw[thick] (0,0) -- (0,2.5) -- (5,2.5) -- (5,0) -- (0,0);
    \draw[thick,blue] (-1,3) to [out=-90,in=180] (0,2);
    \node[above] at (-1,3) {$0$ in};
    \draw[thick] (0,1) -- (-1,1);
    \draw[thick] (0,0.5) -- (-1,0.5);
    \node at (2.5,1.25) {GH protocol for $f_i$};

    \begin{scope}[shift={(9,0)}]
    \draw[thick] (0,0) -- (0,2.5) -- (5,2.5) -- (5,0) -- (0,0);
    \draw[thick,blue] (-1,3) to [out=-90,in=180] (0,2);
    \node[above] at (-1,3) {$1$ in};
    \draw[thick] (0,1) -- (-1,1);
    \draw[thick] (0,0.5) -- (-1,0.5);
    \node at (2.5,1.25) {GH protocol for $f_i$};
    \end{scope}

    \begin{scope}[shift={(0,-5)}]
    \draw[thick] (0,0) -- (0,2.5) -- (5,2.5) -- (5,0) -- (0,0);
    \draw[thick,blue] (-1,-0.5) to [out=90,in=180] (0,2);
    \node[below] at (-1,-0.5) {$0$ out};
    \draw[thick] (0,1) -- (-1,1);
    \draw[thick] (0,0.5) -- (-1,0.5);
    \node at (2.5,1.25) {GH protocol for $f_i$};
    \end{scope}

    \begin{scope}[shift={(9,-5)}]
    \draw[thick] (0,0) -- (0,2.5) -- (5,2.5) -- (5,0) -- (0,0);
    \draw[thick,blue] (-1,-0.5) to [out=90,in=180] (0,2);
    \node[below] at (-1,-0.5) {$1$ out};
    \draw[thick] (0,1) -- (-1,1);
    \draw[thick] (0,0.5) -- (-1,0.5);
    \node at (2.5,1.25) {GH protocol for $f_i$};
    \end{scope}

    \draw[thick] (-1,1) to [out=180,in=180] (-1,-4);
    \draw[thick] (8,1) to [out=180,in=180] (8,-4);

    \draw[thick] (-1,0.5) to [out=180,in=180] (-1,-1) -- (5,-1) to [out=0,in=180] (7,-4.5) -- (8,-4.5);

    \draw[thick] (8,0.5)  to [out=180,in=0] (7,-2) -- (-1,-2) to [out=180,in=180] (-1,-4.5);
        
    \end{tikzpicture}
    \caption{XOR gadget for computing $\oplus_i f_i$ in the garden-hose model. It can be checked directly that if water enters on the top left, it exits on the left if $f_i=0$ and on the right if $f_i=1$. Meanwhile if water enters from the top right, it exits from the bottom right if $f_i=0$ and from the bottom left if $f_i=1$. By wiring such gadgets together for each $f_i$ then, we can compute $\oplus_i f_i$.}
    \label{fig:XORgadget}
\end{figure}

\subsection{Undoing \texorpdfstring{$S$}{TEXT} gates with the garden-hose}\label{sec:undoingS}

The protocol that gives the $T$-depth based upper bound is somewhat involved, so it'll be helpful to start with a high level description of the protocol.
This high level description will suffice to motivate a subroutine making use of the garden-hose protocol, which we will describe in detail later in this section. 

To describe the $T$-depth protocol at a high level, we consider a decomposition of the unitary $U_{AB}$ of interest of the form
\begin{align}
    U_{AB} = C_{d+1} \bar{T}_d C_{d} \bar{T}_{d-1}...\bar{T}_1C_1.
\end{align}
Here $\bar{T}_i$ indicates the ith layer of $T$ gates, which note need not include a $T$ gate on every qubit (but could, and this will turn out to be the hardest case).
The first step in the protocol is for Bob to teleport$^*$ all of his input to Alice, who now holds the full input state up to Pauli corrections. 
Then, Alice applies the first layer of the unitary, $\bar{T}_1C_1$. 
The effect of the Pauli corrections is such that Alice now holds the state
\begin{align}
    \bar{T}_1C_1P[m_1]\ket{\psi}_{AB'} = \bar{T}_1P'[m_1]C_1\ket{\psi}_{AB'}
\end{align}
where $m_1$ is the set of measurement outcomes Bob obtained, $P[m_1]$ the resulting Pauli string of corrections, and $P'[m_1]=C_1P[m_1]C_1^\dagger$. 
Next, we use the identities, 
\begin{align}
    ZT &=TZ, \nonumber \\
    TX &=SXT.
\end{align}
This shows that we can actually move the $P'$ correction through the first layer of $T$ gates, but in doing so we will pick up $S$ operators on whichever qubits have both a $T$ acting in the circuit and an $X$ correction appearing. 
We'd like to take care of these $S$ gates before we communicate, and get to a point where Alice holds
\begin{align}
    P''[m_1]\bar{T}_1C_1\ket{\psi}_{AB'}.
\end{align}
We will in fact show how to do this for each layer, always getting ourselves back to this point where the full circuit up to the current layer is done up to Pauli corrections, and the entire quantum state is on Alice's side. 

So how do we take care of the $S$ gates? 
Label the measurement outcomes appearing after layer $i$ has been executed as $M_i$. 
The idea is to take the measurement outcomes appearing so far to be the inputs to functions $f_i^k$ which determine if the $k$th qubit has a phase $S$ correction at this stage in the protocol. 
Then, we run a garden-hose protocol on those inputs, which puts the qubit of interest onto one of two wires, the two output wires of our (standard form) garden-hose protocol. 
We then do $S^\dagger$ to the $f=1$ wire, and nothing to the $f=0$ wire. 
Finally, we run a second copy of the garden-hose protocol to reverse the process that moved the state onto the two possible wires, and get the relevant qubit back to a fixed location. 

This idea lets us undo the $S$ gate, but at the expense of introducing new Pauli corrections, with the measurement outcomes describing those Pauli gates split up between Alice and Bob's side. 
When we want to apply later layers of the circuit, these Pauli corrections also need to be accounted for when doing subsequent $S$ corrections. 
We need to use the garden-hose again to undo those, so we need to know how complex those corrections are. 
This is captured in the next lemma. 

\begin{lemma}\label{lemma:GHgrowth}
Assume Alice holds a single qubit state $S^{f(x,y)}\ket{\psi}$, where Alice knows $x$ and Bob knows $y$. 
Then the following two statements hold:
\begin{enumerate}
    \item There exists an instantaneous protocol (no communication) which uses $2\GH(f)$ EPR pairs after which Alice holds $X^{g(\hat{x})}Z^{h(\hat{x})}\ket{\psi}$, where $\hat{x}$ depends on $x$ and $2GH(f)$ bits that describe Alice and Bob's measurement outcomes. 
    \item The garden-hose complexities of $g$ and $h$ are at most linear in the complexity of $f$, 
    \begin{align}
        \GH(g) &\leq 4\GH(f), \nonumber \\
        \GH(h) &\leq 11\GH(f).
    \end{align}
    \end{enumerate}
\end{lemma}

\begin{proof}
    The first part we described briefly above: we run a garden-hose protocol for $f(x,y)$, act on the $f=1$ output with $S^\dagger$, then undo the garden-hose protocol to put the qubit back onto a single, fixed wire. 

    Note that in the garden-hose protocol used to apply the conditional $S^\dagger$ there is a sequence of teleportation measurements made, which create possible $X$ and $Z$ corrections. 
    Call the bits determining if there is an $X$ correction $b_x^{i,j}$, where $i,j$ label the two EPR pairs involved in the measurement.
    Similarly, there are corrections $b_z^{i,j}$. 
    Note that not all measurements contribute to these corrections, only those that occur in the unbroken chain of EPR pairs connected to the input state.
    Rather than obtain the input state with a $(S^\dagger)^{f(x,y)}$ applied, we actually end up applying, up to a global phase, the operator
    \begin{align}
        X^{\sum_{(i,j)\in A} b_x^{i,j}}Z^{\sum_{(i,j)\in A} b_z^{i,j}} (S^\dagger)^{f(x,y)} X^{\sum_{(i,j)\in B} b_x^{i,j}} Z^{\sum_{(i,j)\in B} b_z^{i,j}}
    \end{align}
    where the pairs of indices $(i,j)\in B$ correspond to measurements in the chain that occur before $(S^{-1})^{f(x,y)}$, while pairs $(i,j)\in A$ occur after. 
    Using that (again up to a global phase)
    \begin{align}
        XZS^{\dagger} &= S^{\dagger}X \nonumber \\
        S^{\dagger}Z &= ZS^{\dagger}
    \end{align}
    the above becomes
    \begin{align}
        X^{\sum_{(i,j)\in A\cup B} b_x^{i,j}}Z^{\sum_{(i,j)\in A\cup B} b_z^{i,j}+\sum_{(i,j)\in B}f(x,y) b_x^{i,j}} (S^{\dagger})^{f(x,y)} 
    \end{align}
    so that
    \begin{align}
        g(\hat{x}) &= \sum_{(i,j)\in A\cup B} b_x^{i,j} \nonumber \\
        h(\hat{x}) &= \sum_{(i,j)\in A\cup B} b_z^{i,j} + f(x,y) \sum_{(i,j)\in B} b_x^{i,j}
    \end{align}
    Thus to compute $g$, we just need to compute the parity of all of the $b_x^{i,j}$ that occur in the chain. 
    Note that which measurements are actually a part of the chain depends on $x$, so this is a function of the original input $x$ as well as the measurement outcomes $b_x^{(i,j)}$.
    The function $h$ is somewhat more involved, in particular there is an additional correction based on the $b_{x}^{(i,j)}$ for measurements that occur before the conditional $S^{\dagger}$. 

    Let's begin with designing a garden-hose protocol to compute $g(\hat{x})$. 
    To do this, we create a ``rail'', consisting of two EPR pairs, one for each EPR pair in the initial protocol. 
    Then, we connect subsequent rails in the ordering defined by the sequence of EPR pairs used in the original protocol. 
    We connect the rails end to end if $b_x^{i,j}=0$, and we connect them crosswise if $b_x^{i,j}=1$. 
    Thus after running over all $(i,j)$, the input is crossed if the parity of the $b_x^{i,j}$ is odd, and left unchanged if the parity is even. 
    This protocol uses twice the EPR pairs used in the protocol for applying $(S^{-1})^{f(x,y)}$, which itself was $2\GH(f)$, so the cost is $4\GH(f)$. 

    Now we consider the function $h(\hat{x})$. 
    We need a somewhat more involved protocol that treats EPR pairs before and after the conditional $S^{\dagger}$ differently, and accounts for the value of $f(x,y)$. 
    To do this, we first run a garden-hose protocol to compute $f(x,y)$, then feed the two output hoses into two different subsequent garden-hose protocols.
    The $f=0$ pipe is input to a protocol computing the parity of just the $Z$ corrections. 
    We do this using the ``rail'' construction, just as in computing $g(\hat{x})$. 
    The $f=1$ pipe is input to a similar rail protocol, which flips the rails if $b_{x}^{(i,j)}\oplus b_z^{(i,j)}=1$ for pipes occurring before the $S^{\dagger}$, and flips the pipes after the $S^{\dagger}$ if $b_z^{(i,j)}=1$.  
    The garden-hose complexity of this protocol is composed of:
    \begin{itemize}
        \item The complexity of computing $f(x,y)$, in a way that uses just two spilling pipes, which is $3\GH(f)$.
        \item The complexity of computing the $Z$ corrections only, in the sub-protocol that is used when $f(x,y)=0$. This is $4\GH(f)$, where the 4 comes from using the rail construction to double the pipes in the initial protocol, which itself was the protocol that involved computing $f$, applying $S^{\dagger}$, then running the protocol for $f$ in reverse.
        \item The complexity of computing the parity of the $b_{x}^{(i,j)}\oplus b_z^{(i,j)}$ for the first part of the protocol (before $S^{\dagger}$) along with the parity of the $b_z^{(i,j)}$ in the later part of the protocol. This is $4\GH(f)$ again.  
    \end{itemize}
    In total then the garden-hose complexity of $h(\hat{x})$ is $11\GH(f)$. 
\end{proof}

\subsection{\texorpdfstring{$T$}{TEXT}-depth protocol}

We are ready to prove our upper bound on the entanglement cost of implementing a unitary based on the $T$-depth. 
The protocol we use was already described at a high level in section \ref{sec:undoingS}, and in fact since we now know how to undo the unwanted $S$ gates, we have all the ingredients to implement the protocol. 
The remaining issue is to understand the entanglement cost. 

Before delving into the detailed proof, we give a heuristic understanding of where the dominant scaling of the entanglement cost comes from. 
The protocol involves applying Clifford circuits to states with uncorrected Pauli operators acting on it. 
We conjugate these Pauli operators through the Clifford, leading to new corrections and in particular to phase gates after we apply the layer of $T$'s.
A Pauli on any input wire can conjugate through to Pauli operators on an arbitrary subset of the output wires. 
To capture this more precisely, define $g_{i,j}=1$ if there is an $X$ correction on the $j$th input wire and $0$ otherwise, along with $g_{i+1,j}=1$ if an $X$ correction appears on the $j$th output wire after conjugation. 
Similarly, we define $h_{i,j}$ and $h_{i+1,j}$ to be 1 to indicate a $Z$ correction on the input or output $j$th wire, respectively. 
Then, we can see that the output wire functions are related to the input wire functions by
\begin{align}
    g_{i+1,k}&=\bigoplus_{j\in S_{g,k}} g_{i,j} \oplus \bigoplus_{j\in S_{g,k}'}h_{i,j},\nonumber \\
    h_{i+1,k}&=\bigoplus_{j\in S_{h,k}} g_{i,j} \oplus \bigoplus_{j\in S_{h,k}'}h_{i,j}.
\end{align}
The subsets $S_{g,k}^{(\prime)}$ and $S_{h,k}^{(\prime)}$ depend on the choice of Clifford and the wire $k$ being considered, and have size at most $n$. 
From lemma \ref{thm:XORlemma}, we know how the garden-hose complexity of the XOR of many functions behaves, and in particular we can bound it by something of order $n$ times the worst-case garden-hose complexity of the $g_{i,j}$ and $h_{i,j}$. 
The garden-hose complexity of the worst single qubit correction at layer $i+1$, call it $t_{i+1}$, then is related to the complexity at the previous layer by $t_{i+1}\lesssim n t_i$.
At the $d$th layer, which occurs for a circuit of $T$-depth $d$, we get that $t_d\lesssim (Kn)^d$ for a constant $K$. 
An added complication is that these Pauli corrections move through the $T$ gates at this layer to give $S$ gates, and then to correct the $S$ gates we apply lemma \ref{lemma:GHgrowth}. 
This blows up the garden-hose complexity of the Pauli corrections on that wire, but only by a constant factor that contributes to the value of $K$. 

\begin{theorem}\label{thm:Tdepth}
    Given a unitary $U_{AB}$ that can be implemented in a Clifford+T decomposition using a circuit of $T$-depth $d$, we have that
    \begin{align}
        E(U_{AB})\leq O((Kn)^d)
    \end{align}
    where $U_{AB}$ acts on $n$ qubits, $E(U)$ denotes the number of EPR pairs used to implement $U$, and $K$ is a constant. 
\end{theorem}
\begin{proof} We first have Bob teleport$^*$ his system $B$ to Alice, who then holds
    \begin{align}
        X^{\vec{g}_{0}(y)}Z^{\vec{h}_0(y)}\ket{\psi}_{AB}
    \end{align}
    where, less succinctly, we mean 
    \begin{align}
        X^{\vec{g}_0(y)}&=X_1^{g_{0,1}(y)}...X_n^{g_{0,n}(y)}\nonumber \\
        Z^{\vec{h}_0(y)}&=Z_1^{h_{0,1}(y)}...Z_n^{h_{0,n}(y)}
    \end{align}
    where $X_i$ and $Z_i$ act on the $i$th qubit. 
    Note that the entries of both $\vec{h}$ and $\vec{g}$ all have constant garden-hose complexity, since they are functions only of Bob's inputs. 
    
    This will serve as our base case in an inductive argument. 
    We induct on the level $i$, and assume Alice holds the state
    \begin{align}        X^{\vec{g}_i(x,y)}Z^{\vec{h}_i(x,y)}\bar{T}_iC_i...\bar{T}_1C_1\ket{\psi}_{AB}.
    \end{align}
    where Alice holds $x$ and Bob holds $y$, and the entries of $\vec{g}_i$ and $\vec{h}_i$ have known garden-hose complexities. 
    Define
    \begin{align}
        t_i=\max\{\max_j\{\GH(g_{i,j})\},\max_j\{ \GH(h_{i,j})\}\}.
    \end{align}
    In words $t_i$ is the worst-case garden-hose complexity of any single $X$ or $Z$ correction in the $i$th layer. 
    We have from above that $t_0=2$. 
    
    To induct, have Alice apply $\bar{T}_{i+1}C_{i+1}$, obtaining 
    \begin{align}        \bar{T}_{i+1}C_{i+1}X^{\vec{g}_i(x,y)}Z^{\vec{h}_i(x,y)}\bar{T}_iC_i...\bar{T}_1C_1\ket{\psi}_{AB} = \bar{S}^{\vec{f}_i(x,y)}X^{\vec{g}_i'(x,y)}Z^{\vec{h}_i'(x,y)}\bar{T}_{i+1}C_{i+1}...\bar{T}_1C_1\ket{\psi}_{AB} \nonumber
    \end{align}
    Then, we use the procedure of lemma \ref{lemma:GHgrowth} to undo the phase gates, obtaining
    \begin{align}    X^{\vec{g}_{i}''(x,y)\oplus\vec{g}_i'(x,y)}Z^{\vec{h}_i''(x,y)\oplus\vec{h}_i'(x,y)}\bar{T}_{i+1}C_{i+1}...\bar{T}_1C_1\ket{\psi}_{AB}. 
    \end{align}
    The functions $\vec{g}'_{i}, \vec{h}'_{i}$ arise from commuting the $X, Z$ operators through $\bar{T}_{i+1}C_{i+1}$, while the $\vec{g}''_{i}, \vec{h}''_{i}$ operators appear when correcting the phase gates.
    The singly-primed operators are of the form
    \begin{align}\label{eq:parityform}
        g'_{i,j}&=\bigoplus_{l\in S_{g,j}}{g}_{i,l}\oplus\bigoplus_{k\in S'_{g,j}}{h}_{i,k} \nonumber \\
        h'_{i,j}&=\bigoplus_{l\in S_{h,j}}{g}_{i,l}\oplus\bigoplus_{k\in S'_{h,j}}{h}_{i,k}
    \end{align}
    where the subsets $S_{g/h,j}^{(\prime)}$ depend on the Clifford. 
    
    The functions $\vec{g}''_{i}, \vec{h}''_{i}$ appear when undoing the $S^{\vec{f}(x,y)}$ operator, which we do using the procedure in lemma \ref{lemma:GHgrowth}. 
    We are also provided with upper bounds on the garden-hose complexity of these functions from that lemma. 
    We want to determine the garden-hose complexity of $g_{i+1,j}:=g_{i,j}''\oplus g_{i,j}'$ and $h_{i+1,j}:=h_{i,j}''\oplus h_{i,j}'$. 
    Starting with $g_{i+1,j}$, we have
    \begin{align}
        \GH(g_{i+1,j})&=\GH\left(\bigoplus_{l\in S_{g,j}}{g}_{i,l}\oplus\bigoplus_{k\in S'_{g,j}}{h}_{i,k}\oplus g_{i,j}'' \right)\nonumber \\
        &\leq 4\left(\sum_{l\in S_{g,j}}\GH\left(g_{i,l} \right)+\sum_{k\in S'_{g,j}} \GH(h_{i,k}) + \GH(g''_{i,j})\right) \nonumber \\
        &\leq  4\left(\sum_{l\in S_{g,j}}GH\left(g_{i,l} \right)+\sum_{k\in S'_{g,j}} \GH(h_{i,k}) + 4\GH(f_{i,j})\right) \nonumber \\
        &\leq  4\left(nt_i+nt_i + 4\GH(f_{i,j})\right)
    \end{align}
    where the first inequality uses lemma \ref{thm:XORlemma} (the XOR lemma), the second line uses that $\GH(g''_{i,j})\leq 4\GH(f_{i,j})$ which comes from lemma \ref{lemma:GHgrowth}, and the last uses the definition of $t_i$. 
    
    It remains to bound the garden-hose complexity of $f_{i,j}$. 
    Notice that since $f_{i,j}$ is itself a parity function of the $g_{i,j}$ and $h_{i,j}$ (it is of the form \eqref{eq:parityform}), so again its garden-hose complexity is at most $4nt_i$ by a use of the XOR lemma.
    Overall then this gives
    \begin{align}\label{eq:gupper}
        \GH(g_{i+1,j})&\leq 4(nt_i+nt_i + 4\cdot 4nt_i) = 72nt_i
    \end{align}
    An upper bound can be determined for $\GH(h_{i+1,j})$ in a similar way. 
    The only difference is that where before we used $\GH(g''_{i,j})\leq 4\GH(f_{i,j})$, we now need $\GH(h''_{i,j})\leq 11\GH(f_{i,j})$, which is given in lemma \ref{lemma:GHgrowth}.
    This changes the constant but gives a similar upper bound, 
    \begin{align}\label{eq:hupper}
        \GH(h_{i+1,j})&\leq 184nt_i
    \end{align}
    Using equation \eqref{eq:gupper} and equation \eqref{eq:hupper}, we get that
    \begin{align}
        t_{i+1}\leq 184nt_i
    \end{align}
    In fact, our numerical constant here is not optimal --- it can be reduced to 68 \cite{speelman2015instantaneous} --- but we ignore this for now and focus on the scaling with $n$, and just write $t_{i+1}=Knt_i$. 
    This relation and our earlier computation showing that $t_0=2$ is solved by
    \begin{align}
        t_k\leq (Kn)^k.
    \end{align}
    Then, the total cost of all of the garden-hose corrections used in the protocol is
    \begin{align}
        E = n\sum_{i}t_i=n\sum_{i=1}^d(Kn)^i =n\frac{(Kn)^{d+1}-1}{Kn-1} = O((Kn)^{d})
    \end{align}
    as claimed. 
\end{proof}

\section{History and further reading}

Prior to the development of the $T$-depth based NLQC protocol given here \cite{speelman2015instantaneous}, Broadbent developed a protocol to implement arbitrary unitaries using linear entanglement and access to PR box correlations \cite{broadbent2016popescu}.
This result was circulated privately and inspired Speelman to develop his protocol. 

In the same paper \cite{speelman2015instantaneous}, Speelman gives a similar technique for achieving a scaling like $n2^{T\text{-count}(f)}$. 
We explore $T$-count based upper bounds using other techniques in the next chapter. 

Another interesting upper bound, which can be related to $T$-count, is proven in \cite{gonzales2019bounds}. 
Applying the $T$-depth protocol here to a two qubit gate, one finds that to implement the gate with error $\epsilon$ costs $O(1/\epsilon)$ entanglement (to implement the gate with high precision may require many $T$ gates). 
In \cite{gonzales2019bounds}, they find a new protocol with $O(\log(1/\epsilon))$ entanglement cost. 
It is not known if similar improvements can be made for operations with larger numbers of qubits. 

\chapter{\texorpdfstring{$T$}{TEXT}-count upper bounds}\label{chapter:Tcount}

\minitoc

In chapter \ref{chapter:allchannels} we gave a proof that every channel can be implemented as an NLQC. 
Our proof used port-teleportation, which leads to an exponential entanglement cost. 
In chapter \ref{chapter:Tdepth}, we saw that at low T-depth more efficient protocols are possible, finding a cost like $O((Kn)^d)$ for $d$ the T-depth of the unitary being implemented. 
Here, we give another upper bound approach, which is based on the $T$-count rather than the $T$-depth. 
We find that we can implement a unitary with $T$-count $t$ using $\sim 2^t$ EPR pairs, with the exact scaling given below. 

One strategy to achieve this uses techniques similar to those used in the last chapter; this was developed in \cite{speelman2015instantaneous}. 
We give two different strategies which involve a revisiting of the port-teleportation protocol, and careful exploitation of the Clifford-like structure that remains when only a few $T$-gates are present in a unitary.

\section{Clifford structure and \texorpdfstring{$T$}{TEXT}-gates}

Clifford unitaries are those unitaries that conjugate all Pauli strings to other Pauli strings. 
It turns out that this structure is robust, in the sense that if a $T$-gate is added to a Clifford circuit it is still the case that many of the Pauli operators are conjugated to other Pauli operators. 
In fact, letting $G(U)$ denote the group of Pauli operators that are mapped to Pauli operators by the unitary $U$, we will show that
\begin{align}
    |G(U)|\geq 2^{2n-t}
\end{align}
where $n$ is the number of qubits the unitary acts on and $t$ is the number of $T$ gates in a circuit decomposition of $U$. 
This Clifford-like aspect of unitaries with low $T$-count will turn out to make them easier than a generic unitary to implement as an NLQC. 
The rest of this section develops a proof that the Pauli conjugation properties behave in the way we've described. 

We begin with the following definition.

\begin{definition} \textbf{Stabilizer subgroup:} Consider a pure state $\ket{\psi}$ in an $n$ qubit Hilbert space. We define the stabilizer subgroup of $\ket{\psi}$, call it $G(\psi)$, as 
\begin{align}
    G(\psi)=\{P\in \mathcal{P}_n:P\ket{\psi}=\ket{\psi}\}.
\end{align}
\end{definition}
Recall that $\mathcal{P}_n$ is the $n$ qubit Pauli group. 
Notice that since this is a subgroup of $\mathcal{P}_n$, it satisfies $|G(\psi)|=2^k$ for a non-negative integer $k$. 
We call $k$ the \textbf{stabilizer dimension of} $\ket{\psi}$. 

\begin{definition}
    \textbf{Unitary stabilizer subgroup:} Consider an $n$ qubit unitary $U$. The unitary stabilizer subgroup $G(U)$ is defined by
    \begin{align}
        G(U)=\{P\in \mathcal{P}_n: UPU^\dagger \in \mathcal{P}_n\}.
    \end{align}
\end{definition}
Again, since this is a subgroup of $\mathcal{P}_n$, it has $|G(U)|=2^k$ for a non-negative integer $k$. 
We call $2^k$ the \textbf{stabilizer dimension of} $U$. 

We can relate the state and unitary notions of the stabilizer dimension as follows. 
\begin{lemma}\label{lemma:choistabdim}
    Given a unitary $V$, the stabilizer dimension of $V$ equals the stabilizer dimension of the Choi state of $V$, $\ket{\psi_V}_{AB}=\mathcal{I}_A\otimes V_B\ket{\Psi^+}_{AB}$. 
\end{lemma}
\begin{proof}
Suppose that the Choi state of $V$, $\ket{\psi_V}$, is stabilized by $2^{m}$ Paulis. Then there are $2^{m}$ Pauli strings $P\otimes Q$ such that
\begin{align}
    (P\otimes Q)\ket{\psi_V} = \ket{\psi_V}.
\end{align}
Using the definition of the Choi state, this becomes
\begin{align}
    (P\otimes QV)\ket{\Psi^+}_{LR} = (\mathcal{I}\otimes V)\ket{\Psi^+}_{LR}.
\end{align}
Now we use the transpose trick, $\mathcal{O}_A\ket{\Psi^+}_{AB}=\mathcal{O}_B^T\ket{\Psi^+}_{AB}$
to move $P$ to the right, obtaining
\begin{align}
    \mathcal{I}\otimes QVP^T\ket{\Psi^+} = \mathcal{I}\otimes V\ket{\Psi^+}
\end{align}
Finally, we note that the Choi state of an operator uniquely fixes the operator, so that we can conclude
\begin{align}
    QVP^T=V \Rightarrow P^T = V^\dagger QV.
\end{align}
So we get that for each stabilizer $P\otimes Q$ of $\ket{\psi_V}$, $V$ maps $Q$ to $P$ under conjugation and hence $P\otimes Q\in G(V)$.  
Using the same relations, we can show that if $Q$ maps to $P$ under conjugation by $V$, then $P\otimes Q$ is a stabilizer of $\ket{\psi_V}$. 

To show there are exactly $2^m$ Pauli operators that are mapped to Pauli operators by $V$, it remains to show that for any two distinct stabilizers $P_1\otimes Q_1$, $P_2\otimes Q_2$ of $\ket{\psi_V}$, $Q_1$ and $Q_2$ must be distinct. 
To see this, suppose by way of contradiction that $Q_1=Q_2=Q$, so then
\begin{align}
    P_1\otimes Q\ket{\psi_V}=P_2\otimes Q\ket{\psi_V}
\end{align}
which, using the definition of $\ket{\psi_V}$ and the transpose trick leads to
\begin{align}
    \mathcal{I}\otimes QVP_1^T\ket{\Psi^+}=\mathcal{I}\otimes QVP_2^T\ket{\Psi^+}
\end{align}
so then also $QVP_1^T=QVP_2^T$ and then $P_1=P_2$, a contradiction with $P_1\otimes Q$ and $P_2\otimes Q$ being distinct.
\end{proof}

Our strategy for understanding the size of $G(U)$ when $U$ contains a given number of $t$ gates will involve first understanding the stabilizer dimension of the Choi state of $U$, then using the last lemma to relate this to the stabilizer dimension of $U$. 
We bound the stabilizer dimension of a state prepared with a small number of $t$ gates next. 

\begin{lemma}\label{lemma:Tcountandstabdim}
    A state on $\ell$ qubits $\ket{\psi}$ prepared by a Clifford+$T$ circuit with $T$-count at most $t$ has stabilizer dimension at least $2^{\ell-t}$. 
\end{lemma}
\begin{proof}
    Suppose we have a state $\ket{\psi}$ with at least $k$ stabilizers, meaning the stabilizer group
    \begin{align}
        S(\psi)=\{P\in P_\ell: P\ket{\psi}=\ket{\psi}\}
    \end{align}
    has at least $k$ elements, $|S(\psi)|\geq k$. 

    Now we apply a single $T$ gate to the $i$th qubit and would like to understand what happens to the size of the stabilizer group. 
    Call this new state $\psi'$. 
    We consider the map $\varphi:S(\psi)\rightarrow \{-1,+1\}$ defined by sending $P\in S(\psi)$ to $-1$ if it anti-commutes with $Z_i$, and to $+1$ if it commutes. 
    It is straightforward to see this is a homomorphism, so that $|S(\psi)|=|\ker \varphi|\cdot |\text{Im} \varphi|$. 
    Since the image is of dimension at most 2, we get 
    \begin{align}
        |S(\psi)|/2\leq |\ker \varphi|.
    \end{align}
    But now notice that $\ker{\varphi}$ contains only Pauli strings with a $Z$ or $I$ on the $i$th qubit, so that $\ker{\varphi}\subseteq S(\psi')$, and then $|\ker{\varphi}|\leq |S(\psi')|$. 
    Combining this with the above expression, we get
    \begin{align}
        |S(\psi)|/2\leq |S(\psi')|
    \end{align}
    so that the $T$-gate, at worst, reduces the size of the stabilizer group by $1/2$. 

    Meanwhile, the Clifford layers will not change the size of the stabilizer group. 
    Thus if we start with the all $\ket{0}$ state, which has $2^{\ell}$ stabilizers, a Clifford+$T$ circuit with $T$-count $t$ will prepare a state with $2^{\ell-t}$ stabilizers, as claimed. 
\end{proof}

\begin{corollary}
    A $n$ qubit unitary $U$ constructed from $t$ $T$-gates has $|G(U)|\geq 2^{2n-t}$. 
\end{corollary}
\begin{proof}
    Consider the Choi state of $U$, $\ket{\psi_U}$, which is a state on $2n$ qubits. The circuit decomposition of $U$ gives a circuit preparing $\ket{\psi_U}$ using $t$ $T$ gates, so then $|G(\ket{\psi_U})|\geq 2^{2n-t}$ by lemma \ref{lemma:Tcountandstabdim}. But then by lemma \ref{lemma:choistabdim} we have $|G(U)|=|G(\ket{\psi_U})|$, so also $|G(U)| \geq 2^{2n-t}$, as claimed. 
\end{proof}

\section{Upper bound from Pauli adapted teleportation}

\subsection{Pauli adapted teleportation}

Next, we will describe a new form of teleportation, which is intermediate in some sense between Bell teleportation and port-teleportation.  
This idea was explored in \cite{eum2024asymptotic}.

Recall that in Bell-basis teleportation we measure in the basis
\begin{align}
    \{\Psi^s_{A_0A_1}\}_{i} 
\end{align}
where $\Psi^s$ ranges over the maximally entangled states.
Systems $A_1B_1$ are prepared in the maximally entangled state.
In this case the teleportation corrections are Pauli operators applied to $B_1$. 
In port teleportation, we consider a set of POVM elements given by 
\begin{align}
    \{G^{-1/2}(\Psi_{A_0A_i} \otimes \mathcal{I}_{\bar{A}_i})G^{-1/2}\}
\end{align}
where $i$ ranges from $1$ to $N$, and $G=\sum_i\Psi^0_{A_0A_i} \otimes \mathcal{I}_{\bar{A}_i}/d_{\bar{A}_i}$. 
Systems $A_iB_i$ for $i\in[N]$ are prepared in the maximally entangled state.
In this setting, the teleportation correction is to trace out all but one of the $B_i$. 

To interpolate naturally between these two settings, we can allow both a port-correction (trace out all but one of the $B_i$) and a limited set of Pauli corrections. 
We will follow \cite{eum2024asymptotic} and call this \emph{port-based quantum correction} (PBQC) teleportation.
To specify the measurement we should make in this case, define states
\begin{align}
    g^{i,s} = \frac{1}{d_{\bar{A}_i}} \Psi^s_{A_0A_i}\otimes \mathcal{I}_{\bar{A}_i}
\end{align}
where $\Psi^s_{A_0A_i}$ is the maximally entangled state acted on by Pauli $P[s]$.
We will allow $s\in S$, for $S$ a subset of the full Pauli group. 
To define the POVM elements of the measurement, first define
\begin{align}\label{eq:PauliAdapatedPOVM}
    G &=\sum_{i,s} g^{i,s} \nonumber \\
    \Sigma^{i,s} &= G^{-1/2} g^{i,s}G^{-1/2} \nonumber \\
    \Delta &= \frac{1}{m}\left(I - \sum_{i,s} \Sigma^{i,s}\right)
\end{align}
Here $m=N\cdot |S|$, i.e. $m$ is the total number of distinct states $g^{i,s}$. 
If $G$ is not full rank, then we interpret $G^{-1/2}$ to be defined by inverting $G$ on its support and leaving it zero on its kernel. 
This means that $G^{-1/2}GG^{-1/2}=I_{\text{span}(\{g^{i,s}\})}\oplus 0_{\ker(G)}$.
Then we define the POVM elements 
\begin{align}
    \Pi^{i,s}=\Sigma^{i,s} + \Delta.
\end{align}
Notice that this is a complete measurement, 
\begin{align}
    \sum_{i,s} \Pi^{i,s} &= \sum_{i,s} \Sigma^{i,s} + \sum_{i,s} \Delta \nonumber \\
    &= G^{-1/2}\left(\sum_{i,s} g^{i,s} \right)G^{-1/2} + \Delta m \nonumber \\
    &= G^{-1/2}GG^{-1/2} + I - G^{-1/2}GG^{-1/2} \nonumber \\
    &= I.
\end{align}

We are interested in the channel defined as follows: we prepare the maximally entangled state $\ket{\Psi}_{A_iB_i}$ for $i\in[N]$, then measure $A_0A$ using the POVM $\{\Pi^{i,s}\}$. 
After obtaining outcome $(i,s)$, we trace out all systems except $B_i$, do $P^s_{B_i}$, then finally relabel $B_i$ as $A_0$.  
We label this channel as $\Lambda^S$. 
Expressing the above steps in notation, the channel is
\begin{align}
    \Lambda^S(\rho_{A_0}) = \sum_{i,s} \mathcal{P}^s\left(\tr_{A_0A}\tr_{\bar{B}_{i}}\left(\sqrt{\Pi^{i,s}_{A_0A}}\left(\rho_{A_0} \otimes \Psi_{AB} \right) \sqrt{\Pi^{i,s}_{A_0A}}\right)\right)_{B_i\rightarrow A_0}.
\end{align}
Here $\mathcal{P}_{B_i}^s$ is the channel that applies the $P^s$ Pauli to $B_i$, and the subscript $B_i\rightarrow A_0$ indicates the $B_i$ system is relabelled as $A_0$. 
In appendix \ref{sec:pauliadaptedPT}, we compute the entanglement fidelity of this channel, finding:
\begin{align}\label{eq:mixedteleportation}
    \boxed{F_e(\Lambda^S) \geq \left(1-\frac{d^2-|S|}{2N |S|} \right)^2.}
\end{align}
We defer this computation to the appendix and continue below to understand how this can be used in an NLQC protocol. 
Before doing so however, let's briefly pause to understand the behaviour of the above fidelity. 
Notice that if $|S|=d^2$, so that $S$ is the full Pauli group, then $F_e(\Lambda^S)=1$ even with $N=1$. 
This recovers the behaviour of Bell basis teleportation. 

To teleport with entanglement fidelity $1-\epsilon_e$, we need a number of EPR pairs $N_e$ given by
\begin{align}\label{eq:Nepauliadapted}
    N_e = \frac{d^2-|S|}{\epsilon_e|S|}.
\end{align}
To teleport with diamond norm error $\epsilon$, we recall from equation \ref{eq:diamondandFe} that a channel $\mathcal{N}$ achieving entanglement fidelity $F_e(\mathcal{N})$ has
\begin{align}
    \Vert \mathcal{N}-\mathcal{I} \Vert_\diamond \leq d \sqrt{1-F_e(\mathcal{N})}.
\end{align}
Thus we get that
\begin{align}
    \epsilon_\diamond=\Vert \Lambda^S-\mathcal{I} \Vert_\diamond \leq d \sqrt{\frac{d^2-|S|}{N|S|}}
\end{align}
so we should set
\begin{align}
    N_\diamond= \frac{d^2(d^2-|S|)}{\epsilon_\diamond^2|S|}
\end{align}

\subsection{\texorpdfstring{$T$}{TEXT}-count upper bound from adapted teleportation}

Consider a unitary $U_{AB}$ with stabilizer group $S$. 
We will give a method of implementing $U_{AB}$ as a non-local quantum computation.
The protocol is similar to the port-teleportation based protocol for generic channels, but replaces the port-teleportation step with a Pauli-adapted teleportation, with the choice of allowed Pauli operators chosen to be the stabilizer subgroup $S$ associated with $U_{AB}$. 
This allows for a smaller value of $N$ when $|S|$ is large, reducing the entanglement cost. 
We give the protocol next (see also figure \ref{fig:PTprotocolwithPauli}). 

\vspace{0.2cm}
\noindent \textbf{NLQC protocol for unitary $U_{AB}$:}\\
Preparation phase:
\begin{enumerate*}
    \item Distribute a maximally entangled system $\ket{\Psi^+}_{F_{L}F_{R}}$ consisting of $n$ EPR pairs, with $F_L$ sent to Alice and $F_R$ to Bob.
    \item Distribute a set of $N$ maximally entangled systems $\otimes_{k=1}^N \ket{\Psi^+}_{E_{L,k}E_{R,k}}$ with each $\ket{\Psi^+}_{E_{L,k}E_{R,k}}$ consisting of $2n$ EPR pairs, with all $E_{L,k}$ sent to Alice and all $E_{R,k}$ to Bob.
\end{enumerate*}
Execution phase:
\begin{enumerate*}
    \item Alice measures $A_LF_{L}$ in the Bell basis, obtaining outcome $s$. Then Bob holds the state
    \begin{align}
    (P^s_{F_{R}}\otimes I_{A_R})\ket{\psi}_{F_{R}A_R}
    \end{align}
    and the index $s$ is held on the left by Alice. 
    \item Bob performs the measurement described in equation \ref{eq:PauliAdapatedPOVM}, as if port-teleporting $F_{R}A_R$ to Alice using the $N$ maximally entangled pairs $\ket{\Psi^+}_{E_{L,k}E_{R,k}}$. Call the measurement outcome $(j,t)$, where $j$ labels a port and $t$ a Pauli correction. Then $(j,t)$ is held on the right by Bob and the state
    \begin{align}
        \Psi \approx (P^tP^s\otimes I)\ket{\psi}_{E_{L,j}} \otimes  \rho_{E_{L}\setminus E_{L,j}}
    \end{align} 
    is held on the left by Alice. The $\approx$ symbol indicates that this is correct up to the error induced by the Pauli-adapted teleportation. 
    \item On the left, Alice applies $\mathcal{U}\circ (P^s\otimes I)$ to every subsystem $E_{L,k}$, where $\mathcal{U}(\cdot)=U(\cdot)U^\dagger$. Then Alice holds
    \begin{align}
        \Psi \approx \mathcal{U}(P^t\ketbra{\psi}{\psi}_{E_{L,j}}P^t)\otimes  \sigma_{E_{L}\setminus E_{L,j}}=\tilde{P}^t\mathcal{U}(\ketbra{\psi}{\psi}_{E_{L,j}})\tilde{P}^t\otimes \sigma_{E_L\setminus E_{L,j}}
    \end{align}
    with all $E_L$ systems on the left, and $(j,t)$ on the right. Note that because we have $P^t\in S$, there is a choice of Pauli $\tilde{P}^t$ such that the last equality holds. 
    \item Relabel the $E_{L,k}$ qubits as $B_{L,k}B_{R,k}$, and send all of the $B_{L,k}$ systems to Alice on the left and all of the $B_{R,k}$ systems to Bob on the right. Send $(j,t)$ from the right to both the right and left. 
    \item Bob traces out all but the $B_{R,j}$ system, relabels this as the $B$ system applies $\tilde{P}^t_B$, then returns $B_{R,j}$ as his output. Similarly Alice on the left traces out all but the $B_{L,j}$ system, relabels this as the $A$ system, applies $\tilde{P}_A^t$, and returns $B_{L,j}$.
\end{enumerate*}

\begin{figure}
    \centering
    \begin{tikzpicture}

    \draw[thick] (-4.5,0) to [out=-90,in=180] (0,-1.5) to [out=0,in=-90] (6.5,0);
    \draw[black] plot [mark=*, mark size=3] coordinates{(0,-1.5)};

    \draw[thick] (-6,0) -- (-4,0) -- (-4,1) -- (-6,1) -- (-6,0);
    \draw[thick] (-5.5,-1) -- (-5.5,0);
    \node[below] at (-5.5,-1) {$A$};
    
    \draw[thick] (-5.5,0.25) to [out=90,in=90] (-4.5,0.25);
    \draw[thick] (-5,0.25) -- (-4.5,0.75);
    \draw[thick,->] (-5,1) -- (-5,1.5);
    \node[above] at (-5,1.5) {$s$};

    \draw[thick] (4,0) -- (7,0) -- (7,1.5) -- (4,1.5) -- (4,0);

    \draw[very thick] (5.5,0) to [out=-90,in=0] (0,-1) to [out=180,in=-90] (-1,0) -- (-1,1.5);
    \node at (5,-0.25) {$...$};
    \draw[thick] (-3.5,1.5) -- (-2.5,1.5) -- (-2.5,2.5) -- (-3.5,2.5) -- (-3.5,1.5);
    \node at (-3,2) {$P^s$};

    \draw[thick] (4.5,0.25) to [out=90,in=90] (6.5,0.25);
    \draw[thick] (5.5,0.25) -- (6,1);

    \draw[thick,->] (5.5,1.5) -- (5.5,2);
    \node[above] at (5.5,2) {$(j,t)$};

    \draw[thick] (6.5,-1.5) -- (6.5,0);
    \node[below] at (6.5,-1.5) {$B$};

    \draw[thick, dashed] (5.5,2.55) to [out=90,in=-90] (7.5,8);
    \draw[thick, dashed] (5.5,2.55) to [out=90,in=-90] (-0.75,8);

    \draw[very thick] (4.5,0) to [out=-90,in=0] (0,-0.75) to [out=180,in=-90] (-3,0) -- (-3,1.5);
    \draw[thick] (-1.5,1.5) -- (-0.5,1.5) -- (-0.5,2.5) -- (-1.5,2.5) -- (-1.5,1.5);
    \node at (-1,2) {$P^s$};

    \draw[very thick] (-1,2.5) -- (-1,3);
    \draw[very thick] (-3,2.5) -- (-3,3);

    \node at (-2,2) {$...$};
    \node at (-2,3.5) {$...$};

    \draw[thick] (-1.5,3) -- (-0.5,3) -- (-0.5,4) -- (-1.5,4) -- (-1.5,3);
    \draw[thick] (-1.25,4) -- (-1.25,8);
    \draw[thick] (-0.75,4) to [out=90,in=-90] (7,8);
    \node at (-1,3.5) {$\mathcal{U}$};

    \draw[thick] (-3.5,3) -- (-2.5,3) -- (-2.5,4) -- (-3.5,4) -- (-3.5,3);
    \draw[thick] (-3.25,4) -- (-3.25,8);
    \draw[thick] (-2.75,4) to [out=90,in=-90] (5,8);
    \node at (-3,3.5) {$\mathcal{U}$};

    \draw[thick] (4.5,8) -- (7.5,8) -- (7.5,9.5) -- (4.5,9.5) -- (4.5,8);
    \node at (6,8.75) {$\tr_{j^c}$};

    \draw[thick] (-0.75,8) -- (-3.75,8) -- (-3.75,9.5) -- (-0.75,9.5) -- (-0.75,8);
    \node at (-2.25,8.75) {$\tr_{j^c}$};

    \draw[thick] (-2.25,9.5) -- (-2.25,10);
    
    \draw[thick] (-2.75,10) -- (-1.75,10) -- (-1.75,11) -- (-2.75,11) -- (-2.75,10);

    \draw[thick] (-2.25,11) -- (-2.25,11.5);
    \node at (-2.25,10.5) {$\tilde{P}^t_A$};

    \draw[thick] (6,9.5) -- (6,10);
    \draw[thick] (5.5,10) -- (6.5,10) -- (6.5,11) -- (5.5,11) -- (5.5,10);
    \draw[thick] (6,11) -- (6,11.5);
    
    \node at (6,10.5) {$\tilde{P}^t_B$};
    
    \end{tikzpicture}
    \caption{The NLQC protocol for implementing unitaries $U$ using Pauli adapted teleportation. On the left, Alice performs a Bell basis measurement on her system and one end of a maximally entangled state, obtaining measurement outcome $s$. Bob then collects the other end of that maximally entangled state and his input $B$, and executes the measurement from the Pauli adapted teleportation protocol, obtaining outcome $(j,t)$, where $j$ labels a port and $t$ a Pauli. Thus the full input, up to a Pauli correction $P^tP^s$, is now available on Alice's side, stored in a port determined by $j$. Note that each of the ports (labelled with thick wires) has dimension $d_Ad_B$. Alice undoes the Pauli $P^s$ (she doesn't know $t$) and applies the needed unitary $U$ to every port. Then, she splits the two subsystems of each port corresponding to the two outputs of the channel and sends the first subsystem left, and the second subsystem right. Bob broadcasts his measurement outcome $j$. After the communication round Alice and Bob trace out all but the $j$th port, then undo the Pauli $\tilde{P}_A^t\otimes \tilde{P}_B^t=UP^t U^\dagger$.}
    \label{fig:PTprotocolwithPauli}
\end{figure}

Next let's understand the entanglement cost of this protocol. 
The initial Bell teleportation uses $n$ EPR pairs. 
We then do the Pauli adapted teleportation, which when achieving entanglement fidelity $1-\epsilon_e$ uses $2nN_e$ (equation \ref{eq:Nepauliadapted}) EPR pairs. 
The total cost is
\begin{align}
    E_e=n+2n\left(\frac{2^{4n}-|S|}{|S|\epsilon_e}\right).
\end{align}
Now, we suppose that the $2n$ qubit unitary $U_{AB}$ has $T$-count $t$. 
Then, $|S|\geq 2^{4n-t}$, which substituted into the above leads to
\begin{align}
    E_e \leq n + 2n\frac{2^{t}-1}{\epsilon_e}. 
\end{align}
Note that if we had required the protocol be correct to constant diamond norm error, applying the generic bound \ref{eq:diamondandFe} we would have additional factors of $d=2^{2n}$ above, and the overall scaling would not be dominated by $2^t$. 
In the next section we discuss another approach which does achieve a $2^t$ dominated scaling, even when fixing the diamond norm error.\footnote{In standard port-teleportation, one can actually exploit symmetries of the port-teleportation measurement to observe that the entanglement fidelity bounds the diamond norm error, without introducing dimensional factors \cite{christandl2021asymptotic} (Proposition 1.7). It is not known if this remains true for the Pauli adapted teleportation used here. It is possible then that this protocol also achieves the $2^t$ scaling in diamond norm.}

\section{Upper bound from the magic compression theorem}

In this section, we give a second upper bound technique based on the $T$-count. 
Compared to the technique based on Pauli adapted teleportation, we will see that this technique achieves a cost of $E\sim n + t2^{\alpha t}$ even when requiring a fixed diamond norm error. 

The key result we make use of in designing our protocol is the following theorem. 

\begin{theorem} \label{thm:Tcompression} \text{(T-compression \cite{leone2024learning})} Let $U_t$ be a $n$-qubit unitary which maps at least $2^{2n-t}$ Paulis to other Paulis.
Then there exists an integer $s \geq n-t$, a subset $K$ of $s$ qubits and Clifford operations $C_1, C_2$ that allows the following decomposition for $U_t$,
\begin{align}
    U_t = C_2(\mathcal{I}_K\otimes u_{\bar{K}})C_1
\end{align}
for some unitary operator $u$ acting on subsystem $\bar{K}$ consisting of at most $(n-s)\leq t$ qubits.
\end{theorem}

With this theorem in mind, the protocol is not too hard to devise -- the basic idea is to use port-teleportation to handle the non-Clifford part, $u_{\bar{K}}$, and Bell teleportation to handle the Clifford parts. 
We skip a formal protocol description of the form used above, and instead show the protocol in figure \ref{fig:compressionNLQCprotocol}.

\begin{figure}
    \centering
    \begin{tikzpicture}

    \draw[thick] (-4.5,0) to [out=-90,in=180] (0,-1.5) to [out=0,in=-90] (6,-1) -- (6,0);
    \draw[black] plot [mark=*, mark size=2] coordinates{(0,-1.5)};

    \draw[thick] (-6,0) -- (-4,0) -- (-4,1) -- (-6,1) -- (-6,0);
    \draw[thick] (-5.5,-1) -- (-5.5,0);
    \node[below] at (-5.5,-1) {$A$};
    
    \draw[thick] (-5.5,0.25) to [out=90,in=90] (-4.5,0.25);
    \draw[thick] (-5,0.25) -- (-4.5,0.75);
    \draw[thick,->] (-5,1) -- (-5,1.5);
    \node[above] at (-5,1.5) {$s$};
    \draw[dashed] (-5,1.7) circle [radius=0.3];

    \draw[dashed] (-5,1.7) to [out=90,in=-100] (5.75,10);
    \draw[dashed] (-5,1.7) to [out=90,in=-90] (-2.5,10);

    \draw[thick] (4,0) -- (6.5,0) -- (6.5,1.25) -- (4,1.25) -- (4,0);
    \draw[thick] (4.5,0.25) to [out=90,in=90] (6,0.25);
    \draw[thick] (5.25,0.25) -- (5.8,0.75);
    
    \node at (5,-0.25) {$...$};
    \draw[thick] (-3.5,1.5) -- (-2.5,1.5) -- (-2.5,2.5) -- (-3.5,2.5) -- (-3.5,1.5);
    \node at (-3,2) {$u_{\bar{K}}$};

    \draw[thick,->] (5.25,1.25) -- (5.25,2);
    \node[left] at (5.25,1.55) {$j$};

    \draw[thick] (4,2) -- (6.5,2) -- (6.5,3.25) -- (4,3.25) -- (4,2);
    \node at (5.25,2.5) {tr$_{j^c}$};

    \draw[thick,blue] (6,3.25) -- (6,4);
    \draw[thick,blue] (6,-1) -- (6,0);

    \draw[thick] (7,-1.5) -- (7,-0.5);
    \node[below] at (7,-1.5) {$B$};

    \draw[fill=white,thick] (5.75,-1) -- (7.25,-1) -- (7.25,-0.5) -- (5.75,-0.5) -- (5.75,-1);
    \node at (6.5,-0.75) {$C_1$};
    
    \draw[thick,green] (7,-0.5) -- (7,4);

    \draw[fill=white,thick] (5.75,4) -- (7.25,4) -- (7.25,4.5) -- (5.75,4.5) -- (5.75,4);
    \node at (6.5,-0.75+5) {$C_2$};

    \draw[thick] (6,4.5) to [out=90,in=-90] (-2.25,9.5);
    \draw[thick] (7,4.5) to [out=90,in=-90] (6,9.5);

    \draw[blue,thick] (4.5,0) to [out=-90,in=0] (0,-0.75) to [out=180,in=-90] (-3,0) -- (-3,1.5);
    \draw[blue] plot [mark=*, mark size=2] coordinates{(0,-0.75)};
    
    \draw[blue,thick] (5.5,0) to [out=-90,in=0] (0,-1) to [out=180,in=-90] (-1,0) -- (-1,1.5);
    \draw[blue] plot [mark=*, mark size=2] coordinates{(0,-1)};
    
    \draw[blue,thick] (-1,2.5) -- (-1,3);
    \draw[blue,thick] (-3,2.5) -- (-3,3);

    \draw[thick, fill=white] (-3.5,0) -- (-2.5,0) -- (-2.5,1) -- (-3.5,1) -- (-3.5,0);
    \node at (-3,0.5) {$\tilde{P}_{\bar{K}}^s$};

    \draw[thick, fill=white] (-1.5,0) -- (-0.5,0) -- (-0.5,1) -- (-1.5,1) -- (-1.5,0);
    \node at (-1,0.5) {$\tilde{P}_{\bar{K}}^s$};
    
    \draw[thick] (-1.5,1.5) -- (-0.5,1.5) -- (-0.5,2.5) -- (-1.5,2.5) -- (-1.5,1.5);
    \node at (-1,2) {$u_K$};

    \node at (-2,2) {$...$};

    \draw[thick] (-1.5,3) -- (0,3) -- (0,3.75) -- (-1.5,3.75) -- (-1.5,3);
    \draw[thick] (-1.25,3.25) to [out=90,in=90] (-0.25,3.25);
    \draw[thick] (-0.75,3.25) -- (-0.4,3.6);
    \draw[->,thick] (-0.75,3.75) -- (-0.75,4.25);
    \node[above] at (-0.75,4.25) {$t_N$};
    \draw[dashed] (-0.75,4.5) circle [radius=0.3];
    \draw[dashed] (-0.75,4.5) to [out=90,in=-90] (5.75,10);
    \draw[dashed] (-0.75,4.5) to [out=90,in=-90] (-2.5,10);

    \draw[thick] (-3.5,3) -- (-2,3) -- (-2,3.75) -- (-3.5,3.75) -- (-3.5,3);
    \draw[thick] (-3.25,3.25) to [out=90,in=90] (-2.25,3.25);
    \draw[thick] (-2.75,3.25) -- (-2.4,3.6);
    \draw[->,thick] (-2.75,3.75) -- (-2.75,4.25);
    \node[above] at (-2.75,4.25) {$t_1$};
    \draw[dashed] (-2.75,4.5) circle [radius=0.3];
    \draw[dashed] (-2.75,4.5) to [out=90,in=-90] (5.75,10);
    \draw[dashed] (-2.75,4.5) to [out=90,in=-90] (-2.5,10);

    \draw[thick,blue] (-0.25,3) to [out=-90,in=-180] (1,2.5) -- (4,2.5);
    \draw[blue] plot [mark=*, mark size=2] coordinates{(1,2.5)};
    
    \draw[thick,blue] (-2.25,3) to [out=-90,in=-180] (-1,2.75) -- (4,2.75);
    \draw[blue] plot [mark=*, mark size=2] coordinates{(1,2.75)};

    \draw[thick] (-2.25,9.5) -- (-2.25,10);
    
    \draw[thick] (-2.75,10) -- (-1.75,10) -- (-1.75,11) -- (-2.75,11) -- (-2.75,10);

    \draw[thick] (-2.25,11) -- (-2.25,11.5);
    \node at (-2.25,10.5) {$\tilde{P}^t_A$};

    \draw[thick] (6,9.5) -- (6,10);
    \draw[thick] (5.5,10) -- (6.5,10) -- (6.5,11) -- (5.5,11) -- (5.5,10);
    \draw[thick] (6,11) -- (6,11.5);
    
    \node at (6,10.5) {$\tilde{P}^t_B$};

    \end{tikzpicture}
    \caption{Black wires consist of $n$ qubits, blue wires consist of $t$ qubits, and green wires consist of $2n-t$ qubits. The protocol implements a unitary of the form $U_{AB}=C_2(\mathcal{I}_K\otimes u_{\bar{K}})C_1$. A description of the protocol is given in the main text.}
    \label{fig:compressionNLQCprotocol}
\end{figure}

Informally, the protocol is as follows: Alice first Bell teleports $A$ to Bob, who then implements the first Clifford $C_1$, obtaining outcome $s$.  
Bob then divides the output into $K$ and $\bar{K}$. 
The $\bar{K}$ system will have Pauli operators acting on it, and Bob doesn't know which Pauli operators. 
Bob then port-teleports $\bar{K}$ back to Alice, obtaining measurement outcome $j$. 
Alice undoes the Pauli $P^s_{\bar{K}}$ on every port, then acts on every port with $u_{\bar{K}}$. 
Alice now Bell teleports every port back to Bob, who traces out all but the correct port $j$. 
Finally, Bob implements $C_2$. 
At this point the full unitary has been implemented, up to Pauli corrections coming from the uncorrected Pauli operators on the $K$ system, and from the Bell teleportation back to Bob.
Bob distributes the $A$ system back to Alice in the communication round, Alice distributes all her Bell measurement outcomes, and in the final stage both players correct all the remaining Pauli operators. 

The entanglement cost of this protocol comes from: 1) $n$ from Alice's initial Bell teleportation of the $A$ system 2) $tN_\diamond$ from Bob's port-teleportation (of the $\bar{K}$ system, consisting of $t$ qubits) and 3) $tN$ from Alice Bell teleporting each port back to Bob. 
The total cost then is $n+2tN$. 
To achieve constant diamond norm distance $\epsilon_\diamond$, we should take $N=2^{4t+1}/\epsilon_\diamond^2$, so the final cost is
\begin{align}
    E_\diamond = n+\frac{t2^{4t+2}}{\epsilon_\diamond^2}
\end{align}
as claimed. 

\section{History and further reading}

The protocol in this chapter is original to this book. 
The Pauli-adapted port-teleportation protocol was introduced in \cite{eum2024asymptotic}. 
They do not perform the calculation to determine the entanglement fidelity as a function of the input dimension and stabilizer dimension, though they deal with some special cases. 
We give the more general calculation in appendix \ref{sec:pauliadaptedPT}.

\chapter{Application: Separation of \texorpdfstring{$\Rent$}{TEXT} and \texorpdfstring{$\Rtwo$}{TEXT}}\label{chapter:communicationcomplexity}

\minitoc

\section{Communication complexity}

Communication complexity deals with settings where separated parties, usually called Alice and Bob, communicate to compute a function of their combined inputs. 
For instance, suppose Alice holds $x\in\{0,1\}^n$, Bob holds $y\in\{0,1\}^n$, and they wish to compute $f(x,y)$. 
We are interested in how much communication Alice and Bob need to do this. 
We can consider many different scenarios distinguished by the resources Alice and Bob share and the pattern of communication they are allowed to use. 
For instance, perhaps the most common model allows Alice and Bob to communicate back and forth over many rounds; this is called the \emph{two-way communication model}. 
Alternatives are to allow only Bob to send a message to Alice (who should output $f(x,y)$) but not vice versa (the \emph{one-way communication model}), or have both Alice and Bob send messages to a third party (who initially knows neither $x$ nor $y$) who should compute $f(x,y)$. 
These three models are illustrated in figure \ref{fig:CCmodels}. 

\begin{figure*}
    \centering
    \begin{subfigure}{0.3\textwidth}
    \centering
    \begin{tikzpicture}[scale=0.4]
    
    \draw[thick] (-5,-5) -- (-5,-2) -- (-2,-2) -- (-2,-5) -- (-5,-5);
    
    \draw[thick] (5,-5) -- (5,-2) -- (2,-2) -- (2,-5) -- (5,-5);
    
    \draw[thick, mid arrow] (-2,-4.5) -- (2,-4.5);
    \draw[thick, mid arrow] (2,-2.5) -- (-2,-2.5);
    
    \draw[thick] (-3.5,-6) -- (-3.5,-5);
    \node[below] at (-3.5,-6) {$x$};
    
    \draw[thick] (3.5,-6) -- (3.5,-5);
    \node[below] at (3.5,-6) {$y$};
    
    \draw[thick] (-3.5,-2) -- (-3.5,-1);
    \node[above] at (-3.5,-1) {$f(x,y)$};

    \node at (0,-3.25) {\vdots};
    
    \end{tikzpicture}
    \caption{}\label{fig:twoway}
    \end{subfigure}
    \hfill
    \centering
    \begin{subfigure}{0.3\textwidth}
    \centering
    \begin{tikzpicture}[scale=0.4]
    
    \draw[thick] (-5,-5) -- (-5,-2) -- (-2,-2) -- (-2,-5) -- (-5,-5);
    
    \draw[thick] (5,-5) -- (5,-2) -- (2,-2) -- (2,-5) -- (5,-5);
    
    \draw[thick, mid arrow] (2,-3.5) -- (-2,-3.5);
    
    \draw[thick] (-3.5,-6) -- (-3.5,-5);
    \node[below] at (-3.5,-6) {$x$};
    
    \draw[thick] (3.5,-6) -- (3.5,-5);
    \node[below] at (3.5,-6) {$y$};
    
    \draw[thick] (-3.5,-2) -- (-3.5,-1);
    \node[above] at (-3.5,-1) {$f(x,y)$};
    
    \end{tikzpicture}
    \caption{}\label{fig:oneway}
    \end{subfigure}
    \hfill
    \begin{subfigure}{0.3\textwidth}
    \centering
    \begin{tikzpicture}[scale=0.4]
    
    \draw[thick] (-5,-5) -- (-5,-3) -- (-3,-3) -- (-3,-5) -- (-5,-5);
    
    \draw[thick] (5,-5) -- (5,-3) -- (3,-3) -- (3,-5) -- (5,-5);
    
    \draw[thick] (5,5) -- (5,3) -- (3,3) -- (3,5) -- (5,5);
    
    \draw[thick, mid arrow] (4,-3) -- (4.5,3);
    
    \draw[thick, mid arrow] (-4,-3) to [out=90,in=-90] (3.5,3);
    
    
    \draw[thick] (-4.5,-6) -- (-4.5,-5);
    \node[below] at (-4.5,-6) {$x$};
    
    \draw[thick] (4.5,-6) -- (4.5,-5);
    \node[below] at (4.5,-6) {$y$};
    
    \draw[thick] (4,5) -- (4,6);
    \node[above] at (4,6) {$f(x,y)$};
    
    \end{tikzpicture}
    \caption{}
    \label{fig:simultaneous}
    \end{subfigure}
    \caption{a) The two way quantum communication model. Alice receives $x\in \{0,1\}^n$, Bob receives $y\in \{0,1\}^n$. Alice and Bob communicate back and forth, until eventually Alice outputs $f(x,y)$. b) The one-way communication model. Bob sends a message to Alice, who should output $f(x,y)$. c) The simultaneous message model. Alice and Bob both send a message to the referee, who should output $f(x,y)$.} 
    \label{fig:CCmodels}
\end{figure*}

Communication complexity has many applications. 
The most direct is considering network applications where multiple computers need to compute a function of their joint inputs, and they wish to minimize the communication resources needed to do so. 
Less obviously, communication complexity is also relevant to the design of chip layouts, where there is a need to minimize the number of bits sent across the device. 
From a more theoretical perspective, communication complexity is related to computational complexity in that lower bounds on communication complexity provide computational complexity lower bounds. 
Indeed, it is often possible to prove lower bounds on explicit functions in the communication complexity setting, which give us some of the few examples of explicit lower bounds in computational complexity, albeit weak ones. 

Communication complexity has also been extensively studied in the quantum context. 
A key area of study there is to understand when quantum resources provide advantages over classical settings.
In fact, quantum advantages in the communication context are closely related to quantum advantages in the computational setting, though we won't explore that in detail here. 

To organize our understanding of quantum and classical communication complexity, it is helpful to define \emph{communication complexity classes}. 
These are the analogues of computational complexity classes, now in the communication setting. 
To define a class, we pick a communication model and consider all those families of functions that can be computed in that model efficiently. 
For instance consider the two-way communication model. 
The class of function families which can be computed in that model using $O(\text{poly}(\log(n)))$ communication is referred to as the class $\Rtwo$. 
Notice that `efficient' here means $O(\text{poly}(\log(n)))$; this is because all functions can be computed using $n$ bits, so efficient is taken to mean a value similar to the log of the maximum.
Some other communication complexity classes are defined in table \ref{table:communicationclasses}.

\begin{table}
\centering
\begin{tabular}{|l|l|l|l|l|}
\hline
Models & Communication pattern & Error  & Correlation               & Messages \\ \hline
$\Dsim$ & Simultaneous & 0 & None & Classical \\ \hline
$\Rsim$ & Simultaneous & 1/3 & Classical randomness & Classical \\ \hline
$\Qsim$ & Simultaneous & 1/3 & Classical randomness & Quantum \\ \hline
$\Rent$  & Simultaneous & 1/3 & Entanglement & Classical \\ \hline
$\Qent$ & Simultaneous & 1/3 & Entanglement & Quantum \\  \hline
$\Rone$ & One-way & 1/3 & None & Classical \\ \hline 
$\Qone$ & One-way & 1/3 & None & Quantum \\ \hline 
$\Rtwo$ & Two-way & 1/3 & None & Classical \\ \hline 
$\Qtwo$ & Two-way & 1/3 & None & Quantum \\ \hline 
\end{tabular}
\caption{Various models of communication complexity. The three possible communication patterns --- simultaneous, one-way, and two way --- are shown in figure \ref{fig:CCmodels}.}
\label{table:communicationclasses}
\end{table} 

A simple question we can ask is: what are the minimal quantum resources needed to provide an advantage, for some function, over the standard classical class $\Rtwo$?
If another class can be shown to contain a function not in $\Rtwo$, we say it has been separated from $\Rtwo$. 
In fact, proving such functional separations is difficult. 
Instead, what we usually aim for is to find a \emph{partial function} in the other class but not in $\Rtwo$. 
A partial function is similar to a function, but its value is only defined for some of the inputs. 
Another way to phrase this is to say that we provide a promise on the inputs, so that the inputs where the partial function isn't defined never appear.  

The question of finding the weakest quantum class separated (by a partial function) from $\Rtwo$ has been intensely studied, with a series of successive results gradually reducing the quantum resources needed to find an advantage. 
These results are summarized in figure \ref{fig:2}. 
The weakest quantum class separated from $\Rtwo$ is $\Rent$, which allows shared entanglement and simultaneous classical communication. 
We will prove this separation here; it turns out that non-local quantum computation techniques are key to the proof.
This is pretty strong: the communication model is the weakest one (simultaneous), and the communication is only classical. 
The only quantum resource we need is initial shared entanglement. 

\tikzset{every picture/.style={line width=0.75pt}} 
\begin{figure}
    \centering

\tikzset{every picture/.style={line width=0.75pt}} 

\begin{tikzpicture}[x=0.75pt,y=0.75pt,yscale=-1,xscale=1]

\draw   (251,55) .. controls (251,43.95) and (260.5,35) .. (272.22,35) .. controls (283.94,35) and (293.44,43.95) .. (293.44,55) .. controls (293.44,66.05) and (283.94,75) .. (272.22,75) .. controls (260.5,75) and (251,66.05) .. (251,55) -- cycle ;

\draw   (123,153) .. controls (123,141.95) and (132.5,133) .. (144.22,133) .. controls (155.94,133) and (165.44,141.95) .. (165.44,153) .. controls (165.44,164.05) and (155.94,173) .. (144.22,173) .. controls (132.5,173) and (123,164.05) .. (123,153) -- cycle ;

\draw   (121,54) .. controls (121,42.95) and (130.5,34) .. (142.22,34) .. controls (153.94,34) and (163.44,42.95) .. (163.44,54) .. controls (163.44,65.05) and (153.94,74) .. (142.22,74) .. controls (130.5,74) and (121,65.05) .. (121,54) -- cycle ;

\draw   (397,53) .. controls (397,41.95) and (406.5,33) .. (418.22,33) .. controls (429.94,33) and (439.44,41.95) .. (439.44,53) .. controls (439.44,64.05) and (429.94,73) .. (418.22,73) .. controls (406.5,73) and (397,64.05) .. (397,53) -- cycle ;

\draw   (510,55) .. controls (510,43.95) and (519.5,35) .. (531.22,35) .. controls (542.94,35) and (552.44,43.95) .. (552.44,55) .. controls (552.44,66.05) and (542.94,75) .. (531.22,75) .. controls (519.5,75) and (510,66.05) .. (510,55) -- cycle ;

\draw   (261,231) .. controls (261,219.95) and (270.5,211) .. (282.22,211) .. controls (293.94,211) and (303.44,219.95) .. (303.44,231) .. controls (303.44,242.05) and (293.94,251) .. (282.22,251) .. controls (270.5,251) and (261,242.05) .. (261,231) -- cycle ;

\draw   (409,236) .. controls (409,224.95) and (418.5,216) .. (430.22,216) .. controls (441.94,216) and (451.44,224.95) .. (451.44,236) .. controls (451.44,247.05) and (441.94,256) .. (430.22,256) .. controls (418.5,256) and (409,247.05) .. (409,236) -- cycle ;

\draw    (526.44,76.52) -- (451.86,215.88) ;
\draw [shift={(450.44,218.52)}, rotate = 298.16] [fill={rgb, 255:red, 0; green, 0; blue, 0 }  ][line width=0.08]  [draw opacity=0] (10.72,-5.15) -- (0,0) -- (10.72,5.15) -- (7.12,0) -- cycle    ;
\draw    (426.22,75) -- (428.39,210.52) ;
\draw [shift={(428.44,213.52)}, rotate = 269.08] [fill={rgb, 255:red, 0; green, 0; blue, 0 }  ][line width=0.08]  [draw opacity=0] (10.72,-5.15) -- (0,0) -- (10.72,5.15) -- (7.12,0) -- cycle    ;

\draw [color={rgb, 255:red, 0; green, 0; blue, 0 }  ,draw opacity=1 ]   (400.44,65.52) -- (299.15,212.05) ;

\draw [shift={(297.44,214.52)}, rotate = 304.66] [fill={rgb, 255:red, 0; green, 0; blue, 0 }  ,fill opacity=1 ][line width=0.08]  [draw opacity=0] (10.72,-5.15) -- (0,0) -- (10.72,5.15) -- (7.12,0) -- cycle ;

\draw [color={rgb, 255:red, 0; green, 0; blue, 0 }  ,draw opacity=1 ]   (282.44,74.52) -- (416.37,215.35) ;
\draw [shift={(418.44,217.52)}, rotate = 226.44] [fill={rgb, 255:red, 0; green, 0; blue, 0 }  ,fill opacity=1 ][line width=0.08]  [draw opacity=0] (10.72,-5.15) -- (0,0) -- (10.72,5.15) -- (7.12,0) -- cycle ;

\draw   (123,232) .. controls (123,220.95) and (132.5,212) .. (144.22,212) .. controls (155.94,212) and (165.44,220.95) .. (165.44,232) .. controls (165.44,243.05) and (155.94,252) .. (144.22,252) .. controls (132.5,252) and (123,243.05) .. (123,232) -- cycle ;
\draw    (144.22,173) -- (144.22,209) ;
\draw [shift={(144.22,212)}, rotate = 270] [fill={rgb, 255:red, 0; green, 0; blue, 0 }  ][line width=0.08]  [draw opacity=0] (10.72,-5.15) -- (0,0) -- (10.72,5.15) -- (7.12,0) -- cycle    ;
\draw    (142.22,74) -- (144.12,130) ;
\draw [shift={(144.22,133)}, rotate = 268.06] [fill={rgb, 255:red, 0; green, 0; blue, 0 }  ][line width=0.08]  [draw opacity=0] (10.72,-5.15) -- (0,0) -- (10.72,5.15) -- (7.12,0) -- cycle    ;
\draw [color={rgb, 255:red, 255; green, 0; blue, 0 }  ,draw opacity=1 ][line width=1.5]    (157.44,68.52) -- (398.04,217.42) ;

\draw [shift={(401.44,219.52)}, rotate = 211.75] [fill={rgb, 255:red, 255; green, 0; blue, 0 }  ,fill opacity=1 ][line width=0.08]  [draw opacity=0] (13.4,-6.43) -- (0,0) -- (13.4,6.44) -- (8.9,0) -- cycle;

\draw (128.66,44.97) node [anchor=north west][inner sep=0.75pt]  [font=\large] [align=left] {$\Rent$};
\draw (130.49,143.39) node [anchor=north west][inner sep=0.75pt]  [font=\large] [align=left] {$\Qsim$};
\draw (257.99,44.35) node [anchor=north west][inner sep=0.75pt]  [font=\large] [align=left] {$\Qent$};
\draw (405.94,43.3) node [anchor=north west][inner sep=0.75pt]  [font=\large] [align=left] {$\Qone$};
\draw (271.47,220.93) node [anchor=north west][inner sep=0.75pt]  [font=\large] [align=left] {$\Rone$};
\draw (416.4,226.64) node [anchor=north west][inner sep=0.75pt]  [font=\large] [align=left] {$\Rtwo$};
\draw (516.81,45.82) node [anchor=north west][inner sep=0.75pt]  [font=\large] [align=left] {$\Qtwo$};
\draw (135.47,221.93) node [anchor=north west][inner sep=0.75pt]  [font=\large] [align=left] {$\Rsim$};
\draw (500,142) node [anchor=north west][inner sep=0.75pt]  [font=\small] [align=left] {~\cite{raz1999exponential}};
\draw (434,131) node [anchor=north west][inner sep=0.75pt]  [font=\small] [align=left] {~\cite{klartagregev}};

\draw (339,100) node [anchor=north west][inner sep=0.75pt]  [font=\small] [align=left] {~\cite{gavinskyetal}};

\draw (300,70) node [anchor=north west][inner sep=0.75pt]  [font=\small] [align=left] {~\cite{gavinsky2016entangled,girish2022quantum,arunachalam2023one}};

\draw (104,182) node [anchor=north west][inner sep=0.75pt]  [font=\small] [align=left] {~\cite{gavinsky2019}};
\draw (105,94) node [anchor=north west][inner sep=0.75pt]  [font=\small] [align=left] {~\cite{tradeoffs23}};
\draw (201.09,67.95) node [anchor=north west][inner sep=0.75pt]  [font=\small,rotate=-33.54] [align=left] {\textbf{\textcolor[rgb]{1,0,0}{\cite{girish2025magic}}}};

\end{tikzpicture}
    \caption{Quantum versus Classical Communication. Here, an arrow from $A$ to $B$ denotes that $A$ exponentially outperforms $B$ for some task. We use $2$ to denote interactive protocols, $1$ to denote one-way protocols and $\|$ to denote simultaneous protocols. The red arrow uses NLQC techniques and is the focus of this section.}
    \label{fig:2}
\end{figure}

The proof of this separation has two components. 
After choosing a partial function, we need to 1) prove that function is not in $\Rtwo$ and 2) give a communication protocol showing it can be done efficiently in $\Rent$. 
The function we will use is called \emph{forrelation}, which we describe later. 
Part 1), putting a classical lower bound on the communication cost of forrelation, is quite difficult and not discussed here, but see \cite{girish2022quantum}. 
Our main interest is in establishing the second part, which connects to non-local computation. 
In fact we will find that the non-local quantum computation technique of chapter \ref{chapter:Tdepth} provides the needed tool to design a protocol for forrelation in $\Rent$.

\section{Transforming \texorpdfstring{$\Qent$}{TEXT} protocols into \texorpdfstring{$\Rent$}{TEXT} protocols}

Consider an arbitrary $\Qent$ protocol. 
This has the form shown in figure \ref{fig:simultaneous}, with the messages allowed to be quantum. 
Although not shown there, we also allow Alice and Bob to share entanglement. 
We can always view the referee's actions as first applying a unitary $U$ and then measuring the first qubit to determine $f(x,y)$. 
In this section, we show a technique to convert such protocols into $\Rent$ protocols, that is, to make the communication classical.  
When $U$ has low $T$-depth, this transformation is efficient.

The basic idea behind our transformation is as follows. 
We start with a $\Qent$ protocol, and then apply the $T$-depth based NLQC protocol to the operation applied by the referee. 
The crucial observation is that before Alice and Bob communicate in the $T$-depth NLQC protocol, Alice already holds $P^{(k)}U\ket{\psi}$ with $\ket{\psi}$ the input and $U$ the unitary being applied. 
Because of this, we can have Alice measure the first qubit and obtain a measurement outcome $m'$, which will be related in a simple way to the measurement outcome the referee would have obtained, call it $m$. 
In particular if $b$ is a bit fixing if there is an $X$ correction on the first qubit in $P^{(k)}U\ket{\psi}$, then we get that $m=m'\oplus b$. 
Then, we just need to look at Speelman's protocol and determine how many bits Alice and Bob need to send to determine $b$. 
This is worked out in the next theorem. 

\begin{theorem}\label{thm:Q||toR||}
    Consider a $\Qent$ protocol for function $f$, which uses $k$ qubits of message. 
    Suppose that this protocol involves the referee applying a $T$-depth-$d$ unitary to the messages received from Alice and Bob, along with at most $a$ qubits of ancilla and then measuring the first qubit to return the output. 
    Then there is a $\Rent$ protocol for $f$ which uses $O((68(k+a))^d)$ classical bits of communication.
\end{theorem}

\begin{proof}
To begin, suppose Alice and Bob have already executed their own actions in the $\Qent$ protocol, and now hold message system $M_A$ and $M_B$. 
Then, considering the unitary $U_{M_AM_BE}$ the referee would apply to their systems ($E$ represent any ancilla the referee uses), Alice and Bob execute the first round operations in Speelman's NLQC protocol for $U_{M_AM_BE}$.
After doing so, Alice holds the state
\begin{align}
    X^{\vec{g}_d(x,y)}Z^{\vec{h}_d(x,y)}U_{M_AM_BE}\ket{\psi}.
\end{align}
Note that we've adopted the notation from chapter \ref{chapter:Tdepth} to describe the Pauli corrections. 
Alice now measures the first qubit, obtaining outcome
\begin{align}
    m'=m\oplus g_{d,1}
\end{align}
where $m$ will be distributed as if a measurement was made on $U_{M_AM_BE}\ket{\psi}$. 
Alice will send the referee $m'$, and then we need to understand the (simultaneous, classical) communication complexity of Alice and Bob computing $g_{d,1}$. 
One way to do this is for Alice and Bob to send the referee all of their measurement outcomes from all teleportations used in the instantiation of Speelman's protocol. 
From theorem \ref{thm:Tdepth}, the number of such teleportations is $O((K(k+a))^d)$, where $k$ is the number of message qubits and $a$ is the number of ancilla qubits. 
Using the analysis from \cite{speelman2015instantaneous}, we can take $K=68$. 

Using the value of $m'$ and $g_{d,1}$, the referee can compute $m$, which recall is distributed exactly like the referee's measurement outcome in the original protocol. 
Thus if the $\Qent$ protocol was correct with probability $p$, then the $\Rent$ protocol will be correct with probability $p$ as well.
\end{proof}

\section{Separating \texorpdfstring{$\Rent$}{TEXT} and \texorpdfstring{$\Rtwo$}{TEXT}}

The transformation given in the last section can be used to give a partial function separation of $\Rent$ and $\Rtwo$. 
Our starting point will be earlier results that show $\Qent$ contains partial functions not in $\Rtwo$. 
In particular, we use the partial function constructed in \cite{girish2022quantum}, which is called \emph{forrelation}.
Another function constructed in \cite{arunachalam2023one} would also work equally well, but we focus on one example for simplicity. 

To describe the forrelation problem, let $n$ be a power of 2. 
Define the forrelation of a string $x\in\{-1,1\}^{n}$ as
\begin{align}
    \text{forr}(x) := \frac{1}{n}\bra{x_1}H^{\otimes \log_2(n/2)}\ket{x_2}
\end{align}
where $x_1$ is the first half of $x$ and $x_2$ is the second half of $x$. 
Define a communication problem as follows.
\begin{definition}
Alice gets $x\in\{-1,1\}^n$ and Bob gets $y\in \{-1,1\}^n$, where $n$ is a power of 2. The goal of the players is to output $f(x,y)$ defined by
\begin{align}\label{def:forrelation}
    f(x,y) = \begin{cases}
        -1 \quad \text{if}\,\, \text{forr}(x\cdot y)\geq \alpha \\
        +1 \quad \text{if}\,\, \text{forr}(x\cdot y)\leq \alpha/2
    \end{cases}
\end{align}
where $\alpha>0$ is a constant. Here, $x\cdot y$ denotes the point-wise product of $x$ and $y$.
\end{definition}
This is a partial function because of the promise we give on the forrelation of the inputs. 

It was shown in \cite{girish2022quantum} that forrelation is not in $\Rtwo$, and further they gave a $\Qent$ protocol which uses $O(\text{poly}(\log(n)))$ communication and has constant $T$-depth. 
Further, the protocol doesn't involve the referee using any ancilla. 
Consequently, the transformation in theorem \ref{thm:Q||toR||} immediately gives a $O(\text{poly}(\log(n)))$ cost $\Rent$ protocol, so we are done. 

We won't review the $\Qent$ protocol we start with here, but see \cite{girish2022quantum} for details. 

\part{Lower bounds}\label{part:lower}

\chapter{Lower bounds for unitaries}\label{chapter:quantumlowerbound}

\minitoc

In part \ref{part:upperbounds} of this book we discussed upper bound techniques for NLQC. 
In this part we begin the study of entanglement lower bounds for NLQC. 

We can see already, given our upper bound techniques, that we shouldn't expect to be able to prove very strong lower bounds. 
To see why, recall that in the context of $f$-routing we showed, for instance,
\begin{align}
    E(f)\leq \text{FormulaSize}(f)
\end{align}
so that entanglement lower bounds on $f$-routing also imply lower bounds on formula size. 
This also holds for other complexity measures, including span program size, a measure we mentioned briefly in chapter \ref{chapter:f-routing}. 
In classical complexity theory very little is known about lower bounds on these measures, so if we do too well at proving entanglement lower bounds, we would make complexity theory breakthroughs. 
As well, $T$-depth lower bounds seem to be hard to prove in quantum complexity theory as well --- the best such bound known is of $T$-depth 2 --- so entanglement lower bounds on any unitaries would also imply impressive quantum complexity statements. 

In this chapter we will focus on two simple lower bound techniques which apply to unitaries. 
These lower bound techniques prove, at best, linear lower bounds, so don't run into any complexity barriers.
Nonetheless, linear lower bounds already have useful applications in quantum position verification and quantum gravity, among other areas. 
We will explore quantum position verification in chapter \ref{chapter:QPV}, and the gravity connection in chapter \ref{chapter:gravity}.

\section{From entropy to entanglement}

This chapter focuses on lower bounding the entanglement cost of implementing \emph{unitaries} as NLQCs. 
The same techniques do not seem to apply easily to more general quantum channels. 
One of the properties that makes unitaries easier to lower bound is that they are extreme points in the convex space of quantum channels. 
This is a key ingredient in showing that lower bounds against pure state resources can be upgraded to lower bounds against general (mixed) resource states. 
In other words, classical correlation doesn't help to implement extreme points in the space of quantum channels, so in some sense we can reduce to the simpler setting of studying entanglement in pure states. 

In this section we will first develop the general argument showing that, for unitary channels, a lower bound on entropy in pure state resources gives a lower bound on entanglement in any mixed state resource. 
We start by recalling some tools for quantifying correlation and entanglement in quantum systems. 

\subsection{Correlation and entanglement measures}

This section can be skipped and returned to when lemmas recorded here are referenced later. 

\vspace{0.2cm}
\noindent \textbf{The mutual information:} The mutual information quantifies correlations in quantum states. 
It is defined as follows. 
\begin{definition}
    Given a density matrix $\rho_{AB}$, the mutual information $I(A:B)_\rho$ is defined by 
    \begin{align}
    I(A:B)_\rho = S(A)_\rho+S(B)_\rho - S(AB)_\rho.
\end{align}
\end{definition}
The mutual information is continuous in the choice of quantum state \cite{winter2016tight}, as expressed in the next lemma. 
\begin{lemma}\label{lemma:MIcontinuity} Suppose that $\Vert\rho-\sigma\Vert_1=\epsilon$. 
Then
\begin{align}
    |I(A:B)_\rho-I(A:B)_\sigma| \leq 4n_A\epsilon  + (1+2\epsilon)h\left( \frac{2\epsilon}{1+2\epsilon}\right).
\end{align}
where $h(x)=-x\log x -(1-x)\log(1-x)$ is the binary entropy function.
\end{lemma}
The mutual information also can't grow too much when adding subsystems, as expressed in the next lemma. 
\begin{lemma} The mutual information satisfies 
    \begin{align}
        I(A:BC)_\rho\leq I(A:B)_\rho+2S(C)_\rho.
    \end{align}
\end{lemma}
This follows from subadditivity and the Araki-Lieb inequality. 

We also make use of the conditional quantum mutual information, 
\begin{align}
    I(A:B|C)_\rho = S(AC)_\rho + S(BC)_\rho - S(C)_\rho - S(ABC)_\rho.
\end{align}
The mutual and conditional mutual informations are related by the chain rule, 
\begin{align}
    I(A:BC)_\rho = I(A:B|C)_\rho+ I(A:C)_\rho.
\end{align}
We have the following statement about the conditional mutual information.
\begin{lemma}\label{lemma:dataprocessing} The quantum conditional mutual information satisfies the data processing inequality, 
    \begin{align}
    I(A:B|C)_\rho \geq I(A:B|C)_{\mathcal{N}_{B}(\rho)}.
\end{align}
\end{lemma}
\begin{proof}
This is more commonly stated for the mutual information (corresponding to $C=\varnothing$) but the statement for the conditional mutual information follows immediately, 
\begin{align}
    I(A:B|C)_\rho &= I(A:BC)_\rho - I(A:C)_\rho \nonumber \\
    &\geq I(A:BC)_{\mathcal{N}_B(\rho)} - I(A:C)_\rho \nonumber \\
    &=I(A:BC)_{\mathcal{N}_B(\rho)} - I(A:C)_{\mathcal{N}_B(\rho)} \nonumber \\
    &=I(A:B|C)_{\mathcal{N}_B(\rho)}.
\end{align}
We use the chain rule in the first line, data processing for the mutual information in the second line, the fact that the channel doesn't act on $C$ in the third line, and the chain rule again in the last line.
\end{proof}

\vspace{0.2cm}
\noindent \textbf{The entanglement of formation:} The entanglement of formation \cite{hill1997entanglement,wootters1998entanglement} is a tool for quantifying entanglement, which we will make use of in this chapter.  
\begin{definition}
    The \textbf{entanglement of formation} is defined as
\begin{align}
    E_f(A:B)_\rho = \min_{\{p_i,\ket{\psi_i}\}} \sum_i p_i S(B)_{\psi_i},
\end{align}
where the minimization is over ensembles $\{p_i,\ket{\psi_i}\}$ such that $\rho=\sum_i p_i\ketbra{\psi_i}{\psi_i}$. 
\end{definition}
Note that an equivalent definition would replace $S(B)_{\psi_i}$ with $S(A)_{\psi_i}$ above. 
The entanglement of formation is a faithful measure of entanglement, meaning that it is zero if and only if $\rho$ is separable. 
This is easy to see from its definition: if it is separable so that 
\begin{align}
    \rho_{AB} = \sum_i p_i \rho_A^i\otimes \rho^i_B
\end{align}
then we introduce decompositions $\rho_A^i=\sum_k \lambda_k^i \ketbra{\phi_k^i}{\phi_k^i}_A$ and $\rho_B^i=\sum_k \mu_k^i \ketbra{\varphi_k^i}{\varphi_k^i}_B$ and we see that $E_f(A:B)_\rho=0$. 
Conversely, if $E_f(A:B)_\rho=0$ then there exists a decomposition into states $\psi_i$ such that $S(B)_{\psi_i}=0$ for all $i$, which means all $\psi_i$ are tensor product, and hence $\rho$ is separable. 

The entanglement of formation satisfies the following property, which shows that it can't grow too much as you add subsystems.
\begin{lemma} The entanglement of formation satisfies
    \begin{align}\label{eq:SRcontinuity}
    E_f(A:BC)_\rho\leq E_f(A:B)_\rho + \log d_C.
\end{align}
\end{lemma}
This statement follows from the definition of the entanglement of formation, subadditivity of the von Neumann entropy, and the statement $S(X)\leq \log d_X$. 

We also have a data processing inequality for the entanglement of formation \cite{bennett1996mixed}.
\begin{lemma}\label{lemma:Efdataprocessing} The entanglement of formation is non-increasing under the action of a local quantum channel,\footnote{In fact the entanglement of formation is also decreasing under LOCC operations, although we will not need that stronger property here.} 
\begin{align}
    E_f(A:B)_{\rho_{AB}} \geq E_f(A:B')_{\mathcal{I}_{A}\otimes \mathcal{N}_{B\rightarrow B'}(\rho_{AB})}.
\end{align}
\end{lemma}
We will use subadditivity of the entanglement of formation across tensor products. 
\begin{lemma}\label{lemma:Efsubadditive}
    The entanglement of formation is subadditive across tensor products, 
    \begin{align}
        E_f(AA':BB')_{\sigma_{AB}\otimes \rho_{A'B'}}\leq E_f(A:B)_{\sigma_{AB}}+ E_f(A':B')_{\rho_{A'B'}}
    \end{align}
\end{lemma}
\begin{proof}
    This is immediate from the definition, because ensemble decompositions of $\sigma$ and $\rho$ together define an ensemble decomposition of $\sigma\otimes \rho$ which achieves the value on the right hand side of the inequality. 
\end{proof}

We recall a Fannes type continuity bound \cite{nielsen2000continuity, winter2016tight} for the entanglement of formation. 
\begin{lemma}\label{lemma:Efcontinuity}
Consider two states $\rho_{AB}$, $\sigma_{AB}$ with $\Vert\sigma_{AB}-\rho_{AB}\Vert_1 \leq \epsilon$, define $\eta_\epsilon=2\sqrt{\epsilon(1-\epsilon)}$ and $d=\min\{d_A,d_B\}$.
Then, the entanglement of formation of $\rho_{AB}$ and $\sigma_{AB}$ cannot be too different:
\begin{align}
    |E_f(A:B)_\rho - E_f(A:B)_\sigma| &\leq \eta_\epsilon \log d + H(\eta_\epsilon)
\end{align}
where $H(x) = (1+x)h\left(\frac{x}{1+x}\right)$ and $h(x)=-x\log x -(1-x)\log (1-x)$. 
\end{lemma}

We will also use a regularized version of the entanglement of formation. 
Considering a density matrix $\sigma_{A'B'}$, we consider the following quantity.
\begin{definition}
Given a bipartite density matrix $\sigma_{A'B'}$, we define the \textbf{entanglement cost} of $\sigma_{A'B'}$ to be
\begin{align}
    E_C(A':B')_\sigma:=\lim_{m\rightarrow \infty}\frac{E_f(A:B)_{\sigma^{\otimes m}}}{m}.
\end{align}
\end{definition}
Here $A$ consists of $m$ copies of the $A'$ Hilbert space, and $B$ consists of $m$ copies of the $B'$ Hilbert space. 
This quantity will appear in the context of the controllable entanglement lower bound on NLQC (section \ref{sec:controllableentanglement}).
This quantity was proven in \cite{hayden2001asymptotic} to be equal to the asymptotic rate of EPR pairs needed to prepare copies of $\sigma_{A'B'}$ using LOCC, which is why it is given the name of entanglement cost. 
Further, it is known that the entanglement cost is strictly positive whenever $\sigma_{A'B'}$ is entangled (see Theorem 2 of  \cite{yang2005irreversibility}), and of course is zero when $\sigma$ is separable. 

A useful mathematical lemma in the context of the entanglement cost is Fekete's lemma, which states that for sequences $\{a_n\}_n$ which are subadditive, $a_{k+\ell}\leq a_k+a_\ell$, we have that
\begin{align}
    \lim_{n\rightarrow \infty} \frac{a_n}{n} = \inf_{n\geq 1} \frac{a_n}{n}.
\end{align}
Using that $a_m=E_f(A:B)_{\sigma^{\otimes m}}$ is subadditive (by lemma \ref{lemma:Efsubadditive}) along with Fekete's lemma, we obtain the following result.
\begin{lemma}\label{lemma:ECandfinite_m}
    For any integer $m\geq 1$, we have that
    \begin{align}
        \frac{E_f(A:B)_{\sigma^{\otimes m}}}{m} \geq E_C(A':B')_\sigma.
    \end{align}
\end{lemma}

\subsection{Entanglement lower bounds from entropy lower bounds}

We claim that once we've lower bounded the entropy of any resource state needed to implement a unitary as an NLQC, then we've automatically lower bounded the entanglement, as quantified by the entanglement of formation. 
To prove this, we will need a lemma stating that if a convex mixture of quantum channels approximates a unitary, then, on average, each term in the convex mixture must be close to the unitary. 

To prove this lemma, we define an average case notion of the fidelity, which captures how well a quantum channel preserves information on average.
\begin{definition}
The average case fidelity of a quantum channel $\mathcal{N}_A$ is defined as
\begin{align}
    \bar{F}(\mathcal{N}_A) = \int d\psi \, F(\psi_A, \mathcal{N}_A(\psi))
\end{align}
where the integral is over the Haar measure. 
\end{definition}

Then, we need the following statement. 
\begin{lemma}Given a quantum channel $\mathcal{N}$ acting on a $d$ dimensional space, we have
\begin{align}
    \frac{d+1}{d}\left(1-\bar{F}(\mathcal{N})\right)\leq \frac{1}{2}\Vert \mathcal{N}-\mathcal{I}\Vert_\diamond \leq \sqrt{d(d+1)}\sqrt{1-\bar{F}(\mathcal{N})}.
\end{align}
\end{lemma}
This is proposition 9 in \cite{wallman2014randomized}. 

Now we are ready to prove the following lemma regarding unitary channels. 
\begin{lemma}\label{lemma:Uextreme} 
    Suppose that $\{\mathcal{N}^i\}_i$ are quantum channels, let $U$ be a unitary, and let $\mathcal{U}(\cdot)=U(\cdot)U^\dagger$ be a unitary channel. 
    The $\mathcal{N}_i$ and $U$ act on a $d$ dimensional Hilbert space.
    Then if $\Vert \sum_i p_i \mathcal{N}^i-\mathcal{U}\Vert_\diamond \leq \epsilon$, then we have that $\sum_i p_i \Vert \mathcal{N}^i-\mathcal{U}\Vert_\diamond \leq d\sqrt{2\epsilon}$.
\end{lemma}
\begin{proof}
    We have by assumption that
    \begin{align}
        \epsilon \geq \left\Vert\sum_i p_i\mathcal{N}^i - \mathcal{U}\right\Vert_\diamond.
    \end{align}
    We would like to bound a similar quantity but with the sum moved outside the diamond norm. 
    To do this, we first relate the above to the average case fidelity, 
    \begin{align}\label{eq:epsilonFbar}
        \epsilon \geq \left\Vert\sum_i p_i\mathcal{N}^i - \mathcal{U}\right\Vert_\diamond = \left\Vert\sum_i p_i\mathcal{U}^\dagger \circ \mathcal{N}^i - \mathcal{I}\right\Vert_\diamond \geq 2\frac{d+1}{d}\left(1-\bar{F}\left(\sum_ip_i\mathcal{U}^\dagger\circ \mathcal{N}^i\right)\right).
    \end{align}
    Now we use that the average case fidelity is linear, in the sense that
    \begin{align}
        \bar{F}\left(\sum_ip_i\mathcal{U}^\dagger\circ \mathcal{N}^i\right) &= \int d\psi \, F\left( \sum_ip_i\mathcal{U}^\dagger\circ \mathcal{N}^i(\psi), \psi\right) \nonumber \\
        &=\sum_i p_i \int d\psi F\left( \mathcal{U}^\dagger\circ \mathcal{N}^i(\psi), \psi\right) \nonumber \\
        &= \sum_i p_i\bar{F}(\mathcal{U}^\dagger \circ \mathcal{N}^i)
    \end{align}
    The first equality used that the integral is over pure states. 
    Returning to equation \eqref{eq:epsilonFbar}, we have now
    \begin{align}
        \epsilon \geq 2\frac{d+1}{d} \left(1-\sum_ip_i\bar{F}\left(\mathcal{U}^\dagger\circ \mathcal{N}^i\right)\right).
    \end{align}
    or equivalently,
    \begin{align}
        \sum_ip_i\bar{F}\left(\mathcal{U}^\dagger\circ \mathcal{N}^i\right) \geq 1-\frac{1}{2}\frac{d}{d+1} \epsilon.
    \end{align}
    Now we consider the quantity we want to upper bound, which is $\sum_i p_i\Vert \mathcal{U}^\dagger\circ \mathcal{N}^i-\mathcal{I}\Vert_\diamond$, 
    \begin{align}
        \sum_i p_i\Vert \mathcal{U}^\dagger\circ \mathcal{N}^i-\mathcal{I}\Vert_\diamond \leq 2\sqrt{d(d+1)}\sum_i p_i\sqrt{1-\bar{F}(\mathcal{U}^\dagger\circ \mathcal{N}^i)}.
    \end{align}
    Now, use that $f(x)=\sqrt{1-x}$ is concave to move the sum inside the square root, to obtain
    \begin{align}
        \sum_i p_i\Vert \mathcal{U}^\dagger\circ \mathcal{N}^i-\mathcal{I}\Vert_\diamond \leq 2\sqrt{d(d+1)}\sqrt{1-\sum_i p_i\bar{F}(\mathcal{U}^\dagger\circ \mathcal{N}^i)} \leq d\sqrt{2\epsilon}
    \end{align}
    as claimed. 
\end{proof}

Now, we're ready to prove the following theorem. 
\begin{theorem}\label{thm:entropytoentanglement}
    Suppose that any $\epsilon$-correct (in diamond norm) implementation of unitary $U_{AB}$ as an NLQC using a pure resource state $\ket{\Psi}_{LR}$ must have $S(R)_\Psi\geq \lambda-F(\epsilon)$, with $F(x)$ a concave, monotone increasing, function. Then, any protocol using a mixed state resource $\Psi_{LR}$ must satisfy $E_f(L:R)_{\Psi}\geq \lambda - F(d_{AB}\sqrt{2\epsilon})$.
\end{theorem}
\begin{proof}
    By assumption, we have that the channel implemented by the NLQC, call it $\mathcal{N}$, is $\epsilon$-close to the target unitary channel,
    \begin{align}
        \Vert \mathcal{N}-\mathcal{U}\Vert_\diamond\leq \epsilon.
    \end{align}
    Let the resource system used in the NLQC be denoted $\Psi_{LR}$. 
    Let
    \begin{align}
        \Psi_{LR}=\sum_i p_i \ketbra{\Psi^i}{\Psi^i}_{LR}
    \end{align}
    be any ensemble decomposition of $\Psi_{LR}$ into pure states. 
    Then, the channel $\mathcal{N}$ is given by $\mathcal{N}=\sum_i p_i \mathcal{N}^i$ where $\mathcal{N}^i$ is the channel implemented by the NLQC protocol when given pure state $\ket{\Psi^i}$ as a resource state. 
    We have then that
    \begin{align}
        \left\Vert \sum_ip_i\mathcal{N}^i-\mathcal{U}\right\Vert_\diamond\leq \epsilon.
    \end{align}
    Now from lemma \ref{lemma:Uextreme}, we get that we can pull the sum out of the diamond norm at the expense of a relaxed error, 
    \begin{align}
        \sum_i p_i\Vert \mathcal{N}^i-\mathcal{U} \Vert_\diamond \leq d_{AB}\sqrt{2\epsilon}.
    \end{align}
    Label $\Vert \mathcal{N}^i-\mathcal{U} \Vert_\diamond=\epsilon_i$, so that $\sum_i p_i\epsilon_i\leq d_{AB}\sqrt{2\epsilon}$.
    Now consider the entanglement of formation. 
    \begin{align}
        E_f(L:R)_\Psi &= \min_{p_i,\Psi^i} \sum_i p_i S(R)_{\Psi^i} \nonumber \\
        &\geq\min_{p_i,\Psi^i} \sum_i p_i\left(\lambda-F(\epsilon_i) \right) \nonumber \\
        &=\lambda - \max_{p_i,\Psi^i} \sum_i p_i F(\epsilon_i) \nonumber \\
        &\geq \lambda -F(d_{AB}\sqrt{2\epsilon})
    \end{align}
    where in the first line we used that the channel implemented when using resource state $\Psi^i$ is $\epsilon_i$ close to $\mathcal{U}$, and in the last line we used that $F$ is concave and monotone increasing.
\end{proof}

The requirement that $F$ is concave and monotone increasing may seem somewhat specific and unlikely to be satisfied in practice. 
However, the monotone increasing property just expresses that the lower bound should weaken as the error increases, so is expected. 
The concavity property holds in practice in the two settings where we apply this below. 

\section{Lower bound from controllable correlation}

Given a unitary $U_{AB}$, what properties of the unitary could lower bound the entanglement cost? 
For inspiration, return to the simple example of routing that we studied in chapter \ref{chapter:NLQC}. 
There, we have a quantum input $A$ on the left and a classical bit $b$ on the right.
The bit $b$ should control where $A$ ends up: $A$ should go left if $b=0$, and go right if $b=1$. 

To lower bound this, we considered placing $A$ in the maximally entangled state with a reference, which we call $\bar{Q}$. 
We labelled the quantum systems produced on the right in the first round by $R$. 
Without lowering how well the protocol works, we can always send all of $R$ to the side labelled by $b$ since there is no output required on the other side. 
Meanwhile the system kept on the left we label $A$, and the system sent from left to right we label $B$.
Then we noticed that completing the routing task implies that
\begin{align}
    I(\bar{Q}:AR)_{b=0}=2\log d_{\bar{Q}}, \nonumber \\
    I(\bar{Q}:BR)_{b=1}=2\log d_{\bar{Q}}.
\end{align}
From this starting point, we used entropic inequalities to lower bound $S(R)$, which also lower bounds the entropy of the portion of the resource system held on the right. 

\begin{figure*}
    \centering
    \begin{tikzpicture}[scale=0.4]
    
    \draw[thick] (-5,-5) -- (-5,-3) -- (-3,-3) -- (-3,-5) -- (-5,-5);
    \node at (-4,-4) {$\mathcal{V}^L$};
    
    \draw[thick] (5,-5) -- (5,-3) -- (3,-3) -- (3,-5) -- (5,-5);
    \node at (4,-4) {$\mathcal{V}^R$};
    
    \draw[thick] (5,5) -- (5,3) -- (3,3) -- (3,5) -- (5,5);
    \node at (4,4) {$\mathcal{W}^R$};
    
    \draw[thick] (-5,5) -- (-5,3) -- (-3,3) -- (-3,5) -- (-5,5);
    \node at (-4,4) {$\mathcal{W}^L$};
    
    \draw[thick] (-4.5,-3) -- (-4.5,3);
    
    \draw[thick] (4.5,-3) -- (4.5,3);
    
    \draw[thick] (-3.5,-3) to [out=90,in=-90] (3.5,3);
    
    \draw[thick] (3.5,-3) to [out=90,in=-90] (-3.5,3);
    
    \draw[thick] (-3.5,-5) to [out=-90,in=-90] (3.5,-5);
    \draw[black] plot [mark=*, mark size=3] coordinates{(0,-7.05)};
    \node[below right] at (-3.2,-5) {$L$};
    \node[below left] at (3.2,-5) {$R$};
    
    \draw[thick] (-4.5,-6) -- (-4.5,-5);
    \node[left] at (-4.5,-6) {$A$};
    \draw[thick] (4.5,-6) -- (4.5,-5);
    \node[below] at (4.5,-6) {$B$};
    
    \draw[thick] (4.5,5) -- (4.5,6);
    \draw[thick] (-4.5,5) -- (-4.5,6);
    \node[above] at (-4.5,6) {$A$};
    \node[above] at (4.5,6) {$B$};

    \node[left] at (-4.5,0) {$M_1$};
    \node[right] at (-2.5,1.75) {$M_2$};
    \node[left] at (2.5,1.75) {$M_3$};
    \node[right] at (4.5,0) {$M_4$};

    \draw[thick] (-8.5,-6) -- (-8.5,5);
    \draw[thick] (-4.5,-6) to [out=-90,in=-90] (-8.5,-6);
    \node[below left] at (-8.5,-6) {$\bar{Q}$};
    \draw[black] plot [mark=*, mark size=3] coordinates{(-6.5,-7.15)};
    
    \end{tikzpicture}
    \caption{A non-local quantum computation implementing a unitary $U_{AB}$. To prove lower bounds on the entanglement cost, we consider placing the input system $A$ in the maximally entangled state with $\bar{Q}$. We find that if adjusting the input on $B$ changes the amount of correlation between $A$ and $\bar{Q}$ in the final state then $L:R$ must be entangled.}
    \label{fig:NLQC_correlation}
\end{figure*}

The key property of the routing task we make use of is that \emph{the input on the right controls where the correlation with $\bar{Q}$ goes.}
This occurs even though $A$, which begins correlated with $\bar{Q}$, is on the left and causally separated from $B$. 
To address the entanglement cost in a more general setting, we will take this as our starting point.
We want to capture the extent to which a quantum channel can redirect the correlation in $A$ shared with $\bar{Q}$, in a way controlled by the input on $B$. 
The property this defines we will call the \emph{controllable correlation}, which we then prove lower bounds the entanglement in the resource system. 

In fact, we will modify this idea a little bit, and look at when the correlation between $\bar{Q}$ and $A$ can be made either small or large by adjusting the input on $B$. 
This is equivalent when we have pure state inputs and a unitary channel because in that case the output is pure which means,
\begin{align}
    I(\bar{Q}:A)+I(\bar{Q}:B)=2S(\bar{Q})
\end{align}
So $I(\bar{Q}:A)$ small is equivalent to $I(\bar{Q}:B)$ large. 
This formulation turns out to be more convenient, and to be more similar to the next lower bound we present. 

For convenience, we will switch our system labels to those shown in figure \ref{fig:NLQC_correlation}. 
We then make the following definition. 
\begin{definition}
    Considering a unitary $U_{AB}$, define the states
    \begin{align}
        \rho^1_{\bar{Q}AB}&=U_{AB}(\Psi^+_{\bar{Q}A}\otimes \phi^1_B)U_{AB}^\dagger, \nonumber \\
        \rho^2_{\bar{Q}AB}&=U_{AB}(\Psi^+_{\bar{Q}A}\otimes \phi^2_B )U_{AB}^\dagger.
    \end{align}
    We say that $U_{AB}$ has $(\lambda_1,\lambda_2)$-controllable correlation if there exist states $\phi^1_B, \phi^2_B$ such that
    \begin{align}
        \lambda_1=I(\bar{Q}:A)_{\rho^1}, \qquad\lambda_2=I(\bar{Q}:A)_{\rho^2}
    \end{align}
    with $\lambda_1>\lambda_2$. If there is no choice of states $\phi^1_B,\phi^2_B$ such that $\lambda_1>\lambda_2$ we say $U_{AB}$ is not controllably correlated. 
\end{definition}
As a simple example, the CNOT$_{A\rightarrow B}$ gate is controllably correlated: taking the input on $B$ to be $\ket{+}_B$, CNOT$_{A\rightarrow B}$ acts identically on $A$, leaving $\bar{Q}A$ in the maximally entangled state, so $\lambda_1=2$. 
On the other hand choosing the input on $B$ to be $\ket{0}_B$ creates the GHZ state on $\bar{Q}AB$, so $\lambda_2=1$.
Thus CNOT has controllable correlation with $\lambda_1=2, \lambda_2=1$. 
In contrast, the SWAP$_{AB}$ gate is not controllably correlated --- regardless of the input on $B$, the final state on $\bar{Q}A$ will be product, so we always have $\lambda_1=\lambda_2=0$.
Similarly, the identity is not controllably correlated since the state on $\bar{Q}A$ will not be influenced by the input on $B$, so we always have $\lambda_1=\lambda_2$.

With this notion of controllable correlation in hand, we can prove the following theorem, which lower bounds the entropy of one end of the resource state. 
We can then employ theorem \ref{thm:entropytoentanglement} to upgrade this to a lower bound on the entanglement of formation in the resource state.
\begin{theorem}\label{thm:controllablecorrelation}
    Suppose that unitary $U_{AB}$ has $(\lambda_1,\lambda_2)$-controllable correlation.
    If an NLQC protocol using a pure resource state $\ket{\Psi}_{LR}$ gives an $\epsilon$-correct (in diamond norm distance with $\epsilon<1/4$) implementation of $U_{AB}$, then
    \begin{align}
        S(R)_\Psi \geq \frac{\lambda_1-\lambda_2}{2}-\Delta(2\sqrt{\epsilon},n_A)
    \end{align}
    where 
    \begin{align}
        \Delta(x,n) = 4nx+(1+2x)h\left(\frac{2x}{1+2x} \right).
    \end{align}
\end{theorem}

\begin{proof}
    Recall that we defined the states
    \begin{align}
        \rho^1_{\bar{Q}AB}&=U_{AB}(\Psi^+_{\bar{Q}A}\otimes \phi^1_B)U_{AB}^\dagger, \nonumber \\
        \rho^2_{\bar{Q}AB}&=U_{AB}(\Psi^+_{\bar{Q}A}\otimes \phi^2_B )U_{AB}^\dagger.
    \end{align}
    These are the states resulting from the exact implementation of the unitary $U_{AB}$. 
    When $U_{AB}$ is replaced by the $\epsilon$-close implementation, we label the resulting states as $\sigma^1$ and $\sigma^2$, and note that we have
    \begin{align}
        \Vert\rho^1_{\bar{Q}AB}-\sigma^1_{\bar{Q}AB}\Vert_1\leq \epsilon, \nonumber \\
        \Vert\rho^2_{\bar{Q}AB}-\sigma^2_{\bar{Q}AB}\Vert_1\leq \epsilon,
    \end{align}
    which follows from the definition of the diamond norm distance.
    Note further that we write $\sigma_{\bar{Q}M_1M_2}^{1,2}$ for the states produced mid-way through the NLQC protocol (see figure \ref{fig:NLQC_correlation}) upon giving input $\phi^{1,2}_B$. 
    As well, we will drop the state label when considering the entropy of $\bar{Q}$, since this is unaffected by the state on $B$. 
    Thus $S(\bar{Q})=S(\bar{Q})_{\sigma^1}=S(\bar{Q})_{\sigma^2}$. 

    We wish to understand how systems $M_1$ and $M_2$ are related to system $\bar{Q}$. 
    First observe that by the causal structure of the circuit,
    \begin{align}
        \sigma_{M_2\bar{Q}}^{i} = \sigma_{M_2}^i\otimes \rho_{\bar{Q}}.
    \end{align}
    This holds regardless of the input on $B$, so for both $\sigma^1$ and $\sigma^2$.
    We will use this below in the form
    \begin{align}\label{eq:M2}
\boxed{S(M_2\bar{Q})_{\sigma^1}=S(M_2)_{\sigma^1}+S(\bar{Q})_{\sigma^1}.}
    \end{align}
    Next, consider $M_1$.
    We have by assumption that
    \begin{align}
        I(\bar{Q}:A)_{\rho^2}=\lambda_2.
    \end{align}
    We need to undo the last step of the NLQC circuit so as to construct the state on $M_1M_2$ from that on $A$.
    To do this, we consider taking a dilation of the channel $\mathcal{W}^L$ applied on Alice's side in the second round, and label the resulting unitary by $W^L_{M_1M_2\rightarrow AE}$ where $E$ is the ancillary system produced by the unitary. 
    This produces a density matrix $\sigma^2_{\bar{Q}AE}$.
    We claim this is close to $\rho^2_{\bar{Q}A}\otimes \kappa_E$ for some choice of density matrix $\kappa_E$.  
    Recall that
    \begin{align}
        \Vert \sigma^2_{\bar{Q}AB}-\rho^2_{\bar{Q}AB} \Vert_1\leq \epsilon.
    \end{align}
    We consider a purification of $\phi^2_B$, to a state $\phi^{2}_{B\bar{B}}$, and define $\rho^2_{\bar{Q}AB\bar{B}}=U_{AB}(\Psi^+_{\bar{Q}A}\otimes \phi^2_{B\bar{B}})U_{AB}^\dagger$.
    Considering the extension of $\sigma^{2}_{\bar{Q}AB}$ to $\sigma^{2}_{\bar{Q}AB\bar{B}E}$, we apply lemma \ref{lemma:traceUhlmann} to find that there exists an extension of $\rho_{\bar{Q}AB\bar{B}}$ such that
    \begin{align}
        \Vert\rho^2_{\bar{Q}AB\bar{B}E} - \sigma^2_{\bar{Q}AB\bar{B}E}\Vert_1\leq 2\sqrt{\epsilon}
    \end{align}
    But, then notice that $\rho^2_{\bar{Q}AB\bar{B}}$ is a pure state, since $\Psi^+_{\bar{Q}A}$, $\phi^2_{B\bar{B}}$ are pure and $\rho^2_{\bar{Q}AB\bar{B}}=U_{AB}(\Psi^+_{\bar{Q}A}\otimes \phi^2_{B\bar{B}})U_{AB}^\dagger$. 
    This means every extension must be of the form $\rho^2_{\bar{Q}AB\bar{B}}\otimes \kappa_E$ for some density matrix $\kappa_E$. 
    Then we trace out $B\bar{B}$ to give
    \begin{align}
        \Vert\rho_{\bar{Q}A}^2\otimes \kappa_E - \sigma^2_{\bar{Q}AE}\Vert_1 \leq 2\sqrt{\epsilon}.
    \end{align}
    The state on $\bar{Q}M_1M_2$ then satisfies
    \begin{align}\label{eq:rooteps}
        \Vert(W^L_{M_1M_2\rightarrow AE})^\dagger( \rho_{\bar{Q}A}^2\otimes \kappa_E)W^L_{M_1M_2\rightarrow AE} - \sigma^2_{\bar{Q}M_1M_2}\Vert_1\leq 2\sqrt{\epsilon}.
    \end{align}
    Consider the state as above produced on giving input $\phi^2_B$, and consider the mutual information $I(\bar{Q}:M_1)_{\sigma^2}$,
    \begin{align}
        I(\bar{Q}:M_1)_{\sigma^2} &\leq I(\bar{Q}:M_1M_2)_{\sigma^2} \nonumber \\ 
        &\leq I(\bar{Q}:M_1M_2)_{\rho^2} + \Delta(2\sqrt{\epsilon},n_{\bar{Q}}) \nonumber \\
        &=I(\bar{Q}:A)_{\rho^2} + \Delta(2\sqrt{\epsilon},n_{\bar{Q}}) \nonumber \\
        &=\lambda_2+\Delta(2\sqrt{\epsilon},n_{\bar{Q}}).
    \end{align}
    The first inequality uses data processing, the second inequality uses equation \eqref{eq:rooteps} and the continuity statement lemma \ref{lemma:MIcontinuity} and the third uses that $\rho_{\bar{Q}M_1M_2}=(W^L_{M_1M_2\rightarrow AE})^\dagger(\rho_{\bar{Q}A}^2\otimes \kappa_E)W^L_{M_1M_2\rightarrow AE}$.
    
    Finally, notice that the state on $M_1\bar{Q}$ cannot depend on the input on $B$, so that
    \begin{align}
        I(\bar{Q}:M_1)_{\sigma^1} = I(\bar{Q}:M_1)_{\sigma^2} \leq \lambda_2 +\Delta(2\sqrt{\epsilon}, n_{\bar{Q}}).
    \end{align}
    This statement is key to our proof and worth commenting on.
    This is telling us that even when inputting state $\phi^1_B$, the correlation across $\bar{Q}:M_1$ must be small, and in particular similar to its value when inputting $\phi^2_B$.
    But, when the input state is $\phi^1_B$, a lot of correlation has to end up in $A$. 
    This means $M_1M_2$ is highly correlated with $\bar{Q}$ even while $M_1$ is not. 
    We will use this below in the form
    \begin{align}\label{eq:M1}
        \boxed{S(M_1\bar{Q})_{\sigma^1}\geq S(M_1)_{\sigma^1}+S(\bar{Q})_{\sigma^1}-\lambda_2 - \Delta(2\sqrt{\epsilon},n_{\bar{Q}})}.
    \end{align}
    Continuing, we make use of the statement 
    \begin{align}
        I(\bar{Q}:A)_{\rho^1}=\lambda_1.
    \end{align}
    To do so, we first use continuity of the mutual information to turn this into a statement about $\sigma^1$, 
    \begin{align}
        \lambda_1 - \Delta(\epsilon,n_{\bar{Q}})\leq I(\bar{Q}:A)_{\sigma^1} .
    \end{align}
    Then observe that from data processing,
    \begin{align}
        I(\bar{Q}:A)_{\sigma^1} \leq I(\bar{Q}:M_1M_2)_{\sigma^1}
    \end{align}
    so then
    \begin{align}
        \lambda_1-\Delta(\epsilon,n_{\bar{Q}}) \leq S(M_1M_2)_{\sigma^1}+S(\bar{Q})_{\sigma^1}-S(M_1M_2\bar{Q})_{\sigma^1}
    \end{align}
    or, rearranging,
    \begin{align}\label{eq:fixed}
        \boxed{S(M_1M_2\bar{Q})_{\sigma^1} \leq S(M_1M_2)_{\sigma^1}+S(\bar{Q}) - \lambda_1 + \Delta(\epsilon,n_{\bar{Q}}).}
    \end{align}
    which we will use below.

    Now, consider the conditional mutual information $I(M_1:M_2|\bar{Q})_{\sigma^1}$. 
    This is
    \begin{align}
        I(M_1:M_2|\bar{Q})_{\sigma^1} &= S(M_1\bar{Q})_{\sigma^1}+S(M_2\bar{Q})_{\sigma^1}-S(\bar{Q})_{\sigma^1}-S(M_1M_2\bar{Q})_{\sigma^1} \qquad \,\,\,\qquad\qquad \qquad \text{definition of CMI} \nonumber \\
        &= S(M_1\bar{Q})_{\sigma^1}+S(M_2)_{\sigma^1}-S(M_1M_2\bar{Q})_{\sigma^1} \,\qquad \qquad\qquad\qquad \qquad \quad\,\,\,\,\,\,\,\,\, \qquad \text{eq. \eqref{eq:M2}} \nonumber \\
        &\geq S(M_1)_{\sigma^1} + S(\bar{Q}) + S(M_2)_{\sigma^1} -\lambda_2 - S(M_1M_2\bar{Q})_{\sigma^1} -\Delta(2\sqrt{\epsilon},n_{\bar{Q}}) \quad \quad \,\,\,\,\,\,\,\,\,\text{eq. \eqref{eq:M1}} \nonumber \\
        &\geq S(M_1)_{\sigma^1} + S(M_2)_{\sigma^1} - S(M_1M_2)_{\sigma^1} +\lambda_1 - \lambda_2 -\Delta(\epsilon,n_{\bar{Q}})-\Delta(2\sqrt{\epsilon},n_{\bar{Q}})\quad\,\,\,\,\text{eq. \eqref{eq:fixed}}\nonumber \\
        &\geq\lambda_1-\lambda_2 -2\Delta(2\sqrt{\epsilon},n_{\bar{Q}})\,\,\,\,\quad  \qquad\qquad\qquad\qquad \,\,\,\qquad \qquad \quad\qquad\,\qquad\qquad \text{subadditivity} \nonumber 
    \end{align}
    so that the conditional mutual information is bounded below. In the last line we used that $\Delta(2\sqrt{\epsilon},n_{\bar{Q}}) \geq \Delta(\epsilon,n_{\bar{Q}})$ to simplify the error term. 

    Next, we would like to translate this to a bound on the mutual information of the resource state.
    Using that
    \begin{align}
        \sigma_{\bar{Q}M_1M_2}^1=\mathcal{V}^L_{AL\rightarrow M_1}\otimes \mathcal{V}^R_{RB\rightarrow M_2}(\sigma^{1}_{\bar{Q}ALRB})
    \end{align}
    and data processing for the CMI (lemma \ref{lemma:dataprocessing}) we have that 
    \begin{align}
        I(M_1:M_2|\bar{Q})_{\sigma^1_{M_1M_2\bar{Q}}}\leq I(AL:RB|\bar{Q})_{\sigma_{\bar{Q}ALRB}^1} = I(L:R)_{\Psi}
    \end{align}
    where in the equality we used that
    \begin{align}
        \sigma^{1}_{\bar{Q}ALRB} = \Psi^+_{\bar{Q}A}\otimes \Psi_{LR}\otimes \phi^1_B.
    \end{align}
    Combined with our lower bound on the CMI, we have then
    \begin{align}
        \lambda_1-\lambda_2-2\Delta(2\sqrt{\epsilon},n_{\bar{Q}})\leq I(R:L)_{\Psi}.
    \end{align}
    Considering in particular a pure state resource, this leads to
    \begin{align}\label{eq:SRlowerboundCC}
        S(R)_\Psi \geq \frac{\lambda_1-\lambda_2}{2}-\Delta(2\sqrt{\epsilon},n_{\bar{Q}})
    \end{align}
    as claimed.
\end{proof}

Using this lower bound on the entropy along with theorem \ref{thm:entropytoentanglement} that lets us upgrade lower bounds on entropy to lower bounds on entanglement (when considering unitaries), we obtain the following. 
\begin{corollary}
    Suppose that unitary $U_{AB}$ has $(\lambda_1,\lambda_2)$-controllable correlation. Then, any NLQC protocol which implements $U_{AB}$ $\epsilon$-correctly must use a resource state with 
    \begin{align}
        E_f(L:R)_\Psi \geq \frac{\lambda_1-\lambda_2}{2}-\Delta(2\sqrt{d_{AB}\sqrt{2\epsilon}},n_A).
    \end{align}
\end{corollary}
\begin{proof}
    This follows immediately from theorem \ref{thm:controllablecorrelation}, theorem \ref{thm:entropytoentanglement}, and the fact that $\Delta(x,n)$ is concave in its first argument. 
\end{proof}

\vspace{0.2cm}
\noindent \textbf{Example applications:} Now that we have our first general purpose technique for lower bounding entanglement in NLQC, we can start trying it out on some commonly encountered gates. 

For the $CNOT$ gate, we saw that choosing $\phi^1_B=\ketbra{+}{+}_B$, and $\phi^2_B=\ketbra{0}{0}_B$ gives $\lambda_1=2$, $\lambda_2=1$, so a lower bound of $1/2$ on the entanglement of formation in the resource state.
More generally, we can choose any unitary $U$ and then numerically optimize the choice of inputs $\phi^1_B, \phi^2_B$ to obtain as strong of a lower bound as we can. 
The results of doing such an optimization are shown in table \ref{table:lowerbounds}.

\begin{figure}
\centering
\begin{tabular}{| p{3.59cm} | p{2.75cm} |}
\hline
\textbf{Gate} & \textbf{Lower bound from CC}\\ 
\hline
CNOT & 0.500 \\  
\hline
DCNOT & 0.500 \\
\hline
Berkeley B &  0.500  \\
\hline
$\exp(-i \frac{\pi}{4}X\otimes X)$ & 0.500 \\
\hline
iSWAP & 0.500 \\
\hline
$\sqrt{\SWAP}$ & 0.300 \\
\hline
Sycamore & 0.476 \\
\hline
Magic & 0.500 \\
\hline
Dagwood Bumstead & 0.446 \\
\hline
CS & 0.300 \\
\hline
CT & 0.277  \\
\hline
ECR & 0.500 \\
\hline
CSX & 0.300 \\
\hline
Random unitary & $\langle (\lambda_1-\lambda_2)/2\rangle \approx 0.400$ \\
\hline
\end{tabular}
\caption{Results of a numerical optimization computing the controllable correlation for some simple gates; these values are lower bounds on the entanglement of formation in any resource state that suffices to complete the corresponding gate as an NLQC. Matrix expressions for the listed gates appear in appendix \ref{sec:gates}. Code to compute these values is available here: \url{https://colab.research.google.com/drive/1GNDzOggTM1_vvFiJjYS9Ip3EYUlDy4zJ?usp=sharing}.}\label{table:lowerbounds}
\end{figure}

\begin{figure}
    \centering
    \includegraphics[width=0.75\linewidth]{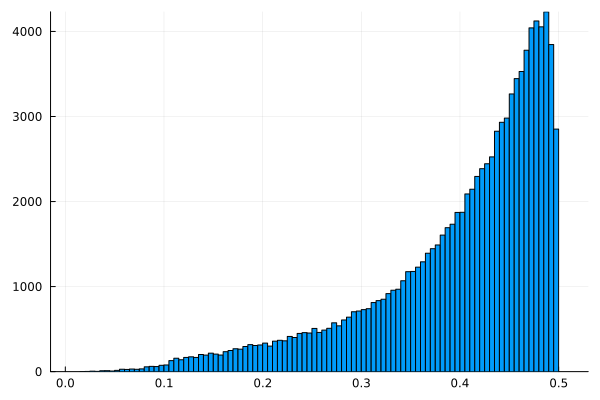}
    \caption{Histogram showing the value of the controllable correlation lower bound computed for $100,000$ two qubit unitaries drawn from the Haar distribution. The average value of the lower bound is $\approx 0.400$. Code to compute this histogram is available here: \url{https://colab.research.google.com/drive/1GNDzOggTM1_vvFiJjYS9Ip3EYUlDy4zJ?usp=sharing}.}
    \label{fig:histogram}
\end{figure}

To build up some further intuition for when this technique works well, we can try running our optimization procedure on random choices of unitary. 
The result of doing so for 100,000 random two qubit unitaries is shown in figure \ref{fig:histogram}. 
The most striking feature of the resulting data is that we never encounter a unitary where this technique fails to return a positive lower bound, and in fact it appears that the distribution tapers off to zero probability near zero entanglement cost.  
We know examples with zero cost exist --- for instance the SWAP or identity unitaries --- but they seem to be rare. 

These observations let us start to get intuition for which unitaries require entanglement and which don't. 
Apparently most do, and our only known examples that do not are the SWAP and identity. 
An interesting open problem is to characterize if these are the only examples that don't require entanglement (up to equivalence under local unitaries). 

\subsection*{Parallel repetition}

Suppose that we have a lower bound on the entanglement needed to implement a channel $\mathcal{N}_{AB}$ as an NLQC, say of $E_f\geq \lambda$.
Does this mean that the channel $\mathcal{N}_{AB}^{\otimes n}$ requires entanglement at least $n\lambda$?
This question is known as the \emph{parallel repetition} question for an NLQC; similar questions appear in the context of non-local games and other settings. 
We do not have a general understanding of when parallel repetition holds for an NLQC. 
We can notice however that the lower bounds proven from the controllable correlation satisfy parallel repetition.  

Concretely we obtain the following.
\begin{corollary}
    Consider a unitary $G$ with $(\lambda_1,\lambda_2)$-controllable correlation.
    Then an (exact) implementation of $G^{\otimes n}$ as an NLQC requires entanglement of formation in the resource state $\Psi$ lower bounded by
    \begin{align}
        E_f(L:R)_\Psi\geq n\left( \frac{\lambda_1-\lambda_2}{2}\right)
    \end{align}
\end{corollary}
\begin{proof}
    Suppose that $\phi^1, \phi^2$ can be used to show $G$ has $(\lambda_1, \lambda_2)$-controllable correlation. 
    Then considering $G^{\otimes n}$, use inputs $(\phi^1)^{\otimes n}$, $(\phi^2)^{\otimes n}$ to obtain $\lambda_1'=n\lambda_1$, $\lambda_2'=n\lambda_2$, which lead to the stated lower bound.
\end{proof}

There is an open problem around parallel repetition for the controllable correlation lower bound in the noisy context. 
In particular, we do not have a good lower bound on implementations of $(G')^{\otimes n}$ where $G'$ is a noisy version of $G$.

\section{Lower bound from controllable entanglement}\label{sec:controllableentanglement}

In the last section we proved a lower bound of $1/2$ on the entanglement of formation needed to implement a $CNOT$ gate as an NLQC. 
Our best upper bound for implementing $CNOT$ comes from the teleportation based protocol seen in section \ref{sec:cliffordNLQC}, which uses 1 EPR pair, which has $E_f=1$. 
Our upper and lower bounds don't quite match, which is a bit unsatisfying, especially considering that the $CNOT$ is probably the simplest possible (non-trivial) unitary we could consider. 
Can we improve our bound, and get a lower bound of $E_f\geq 1$?

In this section we give such an improved lower bound. 
The technique is a little bit harder than the controllable correlation lower bound. 
In particular we will need to invoke some results on Schumacher compression, which are reasonably involved. 

To motivate our second lower bound technique, let's revisit the $CNOT$ gate. 
Consider the set-up we've shown in figure \ref{fig:NLQC_correlation}, where we take $CNOT_{A\rightarrow B}$ to act on $\ket{\Psi^+}_{\bar{Q}A}$. 
Choosing the input on $B$ to be $\ket{+}_B$, we find that $\bar{Q}\!:\!A$ is left maximally entangled, so $\lambda_1=I(\bar{Q}\!:\!A)_{\rho^1}=2$. 
Choosing the input on $B$ to be $\ket{0}_B$, we find that $\bar{Q}:A:B$ is placed in a GHZ state, and $\bar{Q}\!:\!A$ becomes a maximally classically correlated state $\frac{1}{2}(\ketbra{00}{00}_{\bar{Q}A}+\ketbra{11}{11}_{\bar{Q}A})$. 
This gives $\lambda_2=1$. 
This leads to the lower bound of $E_f\geq (2-1)/2= 1/2$ for the $CNOT$ using the controllable correlation technique.

To improve this, a natural idea is to try and use a measure of \emph{entanglement}, rather than correlation, between $\bar{Q}$ and $A$: while the correlation as measured by the mutual information drops to half its maximal value, if we measure entanglement the correlation measure would drop from maximal to zero, since $\bar{Q}:A$ is only classically correlated in the GHZ state. 
Perhaps this behaviour would lead to a tighter bound. 

This motivates the following definition. 
\begin{definition}\label{def:controllableentanglement} Consider a unitary $U_{AB}$. We say that $U_{AB}$ has $\lambda$-\textbf{controllable entanglement} if there exist states $\phi^1_B$, $\phi^2_B$ such that
\begin{align}
    \rho^1_{\bar{Q}AB}=U_{AB}(\Psi^+_{\bar{Q}A}\otimes \phi^1_B)U^\dagger_{AB},\qquad
    \rho^2_{\bar{Q}AB}=U_{AB}(\Psi^+_{\bar{Q}A}\otimes \phi^2_B)U^\dagger_{AB}
\end{align}
with 
\begin{align}
    \lambda:=E_C(\bar{Q}:A)_{\rho^1}>0,\qquad 0=E_C(\bar{Q}:A)_{\rho^2}.
\end{align} 
Recall that $E_C(\bar{Q}:A)_{\rho^1}$ denotes the entanglement cost of $\rho^1$.
If there are no such choices of input state, we say that $U_{AB}$ is not controllably entangled. 
\end{definition}
Recall that the entanglement cost is zero if and only if the density matrix is separable \cite{yang2005irreversibility}, so we can equivalently understand the above conditions as stating there is an input on $B$ such that $\bar{Q}:A$ is entangled, and another input on $B$ such that it is separable. 

This notion of the controllable entanglement leads to the following lower bound. 
\begin{theorem}\label{thm:mainlowerbound}
Suppose that an NLQC protocol using resource state $\Psi_{LR}$ implements a unitary $U_{AB}$ to within diamond norm distance $\gamma$, and where $U_{AB}$ has $\lambda$-controllable entanglement.
Then for small enough $\gamma$, we have that
\begin{align}
    E_f(L:R)_\Psi \geq \lambda-P(d_{AB}\sqrt{2\gamma})
\end{align}
where $P(x)=2\sqrt{2}n_{\bar{Q}}x^{1/8}$. 
\end{theorem} 
As an example, the $CNOT$ gate is controllably entangled with $\lambda=1$, leading to a tight lower bound of $E_f(L:R)\geq 1$ for the CNOT gate. 
To see this, we can use $\phi^1_B=\ketbra{+}{+}_B$, $\phi^2=\ketbra{0}{0}$.
Then $(\rho^1)_{\bar{Q}A}:=\tr_B (U_{AB}(\Psi^+_{\bar{Q}A}\otimes \phi_B^1)U_{AB}^\dagger)=\Psi^+_{\bar{Q}A}$ is maximally entangled, which has an entanglement cost of 1.
Meanwhile, when inputting $\ket{0}$, we find $\rho^2_{\bar{Q}A}=\frac{1}{2}(\ketbra{00}{00}+\ketbra{11}{11})$, which is separable, so has an entanglement cost of 0.
A similar argument shows that for $CNOT^{\otimes n}$, the controllable entanglement technique gives a lower bound of $E_f(L:R)\geq n$. 

We will prove the above result in several steps. 
First, (lemma \ref{lemma:dimensionlowerbound}) we lower bound the dimension of the resource system.
Then we use a compression argument to translate this to a lower bound on the entropy of one side of any pure state resource. 
Finally, we apply theorem \ref{thm:entropytoentanglement} to translate this to a lower bound on the entanglement of formation. 

\subsection*{Dimension lower bound}\label{sec:dimensionbound}

As our first step towards a proof of theorem \ref{thm:mainlowerbound}, we prove the following lemma which lower bounds the dimension of the resource system.  
\begin{lemma}\label{lemma:dimensionlowerbound}
    Suppose $U_{AB}$ has $\lambda$-controllable entanglement. 
    A non-local quantum computation which implements $U_{AB}^{\otimes m}$ to within diamond norm $\epsilon$ and uses a resource system $\Psi_{LR}$ must have 
    \begin{align}
        n_R \geq E_f(\bar{Q}:A)_{(\rho^1)^{\otimes m}} -m\,g_1(\epsilon) -  g_2(\epsilon)
    \end{align}
    where $(\rho^1)_{\bar{Q}AB}=U_{AB}(\Psi^+_{\bar{Q}A}\otimes \phi^1_B)U^\dagger_{AB}$, and $g_1(\epsilon),g_2(\epsilon)$ are positive functions that go to $0$ as $\epsilon\rightarrow 0$. More specifically, they are given by
    \begin{align}
        g_1(\epsilon) &= n_{\bar{Q}}(\eta_{\epsilon} + \eta_{2\sqrt{\epsilon}}), \nonumber \\
        g_2(\epsilon) &= H(\eta_\epsilon) + H(\eta_{2\sqrt{\epsilon}}),
    \end{align}
    where $\eta_x=2\sqrt{x(1-x)}$, and $H(y)=(1+y)h\left(\frac{y}{1+y}\right)$.
\end{lemma}

\begin{proof} Suppose we have an NLQC protocol that implements $\tilde{U}$, with $\Vert\tilde{U}-U^{\otimes m}\Vert_\diamond\leq \epsilon$ using a resource state $\Psi_{LR}$. 
We take as input on the left one end of the maximally entangled state $\ket{\Psi^+}^{\otimes m}_{\bar{Q}A}$, and label the reference system by $\bar{Q}$. 
The set up is shown in figure \ref{fig:NLQC_correlation}; we will use the operation and system labels shown there. 

We consider two scenarios. 
First, consider inputting the $(\phi^1)^{\otimes m}$ state into the remaining input, labelled $B$. 
We label the state resulting from a perfect implementation of $U^{\otimes m}$ on this input as $\rho^{1,m}$, and from the imperfect implementation as $\sigma^{1,m}$. 
In this case, a perfect implementation of $U^{\otimes m}$ would give, 
\begin{align}\label{eq:TP}
    E_f(\bar{Q}:A)_{\rho^{1,m}} = E_f(\bar{Q}:A)_{(\rho^1)^{\otimes m}}.
\end{align} 
Since the protocol implements $\tilde{U}$ which is close to $U^{\otimes m}$, we need to use the continuity bound of lemma \ref{lemma:Efcontinuity} and we obtain
\begin{align}
    E_f(\bar{Q}:A)_{\rho^{1,m}} - E_f(\bar{Q}:A)_{\sigma^{1,m}} \leq m \,n_{\bar{Q}}\,\eta_\epsilon + H(\eta_\epsilon).
\end{align}
Inserting equation \eqref{eq:TP}, we have
\begin{align}
    E_f(\bar{Q}:A)_{\sigma^{1,m}} \geq E_f(\bar{Q}:A)_{(\rho^1)^{\otimes m}}-m\,n_{\bar{Q}}\eta_\epsilon-H(\eta_\epsilon).
\end{align}
We can also observe that, 
\begin{align}\label{eq:correct}
    E_f(\bar{Q}:M_1M_2)_{\sigma^{1,m}}\geq E_f(\bar{Q}:A)_{(\rho^1)^{\otimes m}}-m\,n_{\bar{Q}}\eta_\epsilon - H(\eta_\epsilon)
\end{align}
which follows from the previous line and the data processing inequality (lemma \ref{lemma:Efdataprocessing}) for the entanglement of formation. 

Second, we input $(\phi^2)^{\otimes m}$ into $B$. 
We label the state created in this case by $\rho^{2,m}=(\rho^2)^{\otimes m}$ in the perfect case, and as $\sigma^{2,m}$ in the imperfect case.
Recall that $\rho^2_{\bar{Q}A}$ is separable.
Consider purifying the operation $\mathcal{W}^L_{M_1M_2\rightarrow A}$ to a unitary $W^L_{M_1M_2\rightarrow AE}$.
This produces a state $\sigma^{2,m}_{\bar{Q}AE}$. 
By lemma \ref{lemma:traceUhlmann}, we can extend the $\sigma^{2,m},\rho^{2,m}$ states to the $\bar{Q}ABEX$ Hilbert space ($X$ is an additional purifying system) and obtain
\begin{align}
    \Vert \sigma^{2,m}_{\bar{Q}ABEX} - \rho^{2,m}_{\bar{Q}ABEX} \Vert_1 \leq 2\sqrt{\epsilon}.
\end{align}
Since $\rho^{2,m}_{\bar{Q}ABX}$ is pure, the extension of $\rho^{2,m}$ must be product across $\bar{Q}AB:EX$, so
\begin{align}
    \Vert \sigma^{2,m}_{\bar{Q}ABEX} - \rho^{2,m}_{\bar{Q}ABX}\otimes \rho_{E} \Vert_1 \leq 2\sqrt{\epsilon}.
\end{align}
Next trace out $BX$, 
\begin{align}
    \Vert \sigma_{\bar{Q}AE}^{2,m} - \rho_{\bar{Q}A}^{2,m}\otimes \rho_{E} \Vert_1 \leq 2\sqrt{\epsilon}.
\end{align}
Now apply $(W^L)^\dagger$ to both states, which won't change the trace distance, and choose an explicit decomposition of $\rho^{2,m}$ into a convex sum over product states $\rho^{2,m}(i)$ (recall that by assumption it is separable),  
\begin{align}
    2\sqrt{\epsilon} &\geq \left\Vert\sum_{i}p^i\rho^{2,m}_{\bar{Q}}(i)\otimes (W^L)^\dagger_{M_1M_2\rightarrow AE}(\rho^{2,m}_{A}(i)\otimes \rho_E) W^L_{M_1M_2\rightarrow AE} - (W^L)^\dagger_{M_1M_2\rightarrow AE}(\sigma^{2,m}_{\bar{Q}AE})W^L_{M_1M_2\rightarrow AE}\right\Vert_1 \nonumber \\
    &=\left\Vert\sum_{i}p^i\rho^{2,m}_{\bar{Q}}(i)\otimes \rho^{2,m}_{M_1M_2}(i) - \sigma^{2,m}_{\bar{Q}M_1M_2}\right\Vert_1.
\end{align}
From this we also obtain that the state on $\sigma^{2,m}_{\bar{Q}M_1}$ is close to separable, which using continuity of $E_f$ gives, 
\begin{align}
    E_f(\bar{Q}:M_1)_{\sigma^{2,m}} \leq m n_{\bar{Q}}\eta_{2\sqrt{\epsilon}} + H(\eta_{2\sqrt{\epsilon}}).
\end{align}
Finally, notice that by causality the state on $\bar{Q}M_1$ must be the same regardless of the input at $B$, so that
\begin{align}\label{eq:smallonQM_1}
    E_f(\bar{Q}:M_1)_{\sigma^{1,m}} \leq m\,n_{\bar{Q}}\eta_{2\sqrt{\epsilon}} + H(\eta_{2\sqrt{\epsilon}}).
\end{align}
In words, we see that in the state $\sigma^{1,m}$ systems $\bar{Q}:M_1$ are close to separable, while $\bar{Q}:M_1M_2$ is entangled.

Now we combine our statements so far to show that this can only occur when $R$ is large enough,
\begin{align}
    E_f(\bar{Q}:A)_{(\rho^{1})^{\otimes m}}-m\,n_{\bar{Q}}\eta_\epsilon-H(\eta_\epsilon) &\leq E_f(\bar{Q}:M_1M_2)_{\sigma^{1,m}} \qquad \qquad \,\,\,\,\,\,\,\,\,\text{From eq. \eqref{eq:correct}}\nonumber \\
       &\leq E_f(\bar{Q}:M_1RB)_{\sigma^{1,m}} \qquad \qquad \,\,\,\text{From data processing}\nonumber \\
       &= E_f(\bar{Q}:M_1R)_{\sigma^{1,m}} \nonumber \qquad\qquad \,\,\,\,\,\,\,\,\,\,\text{Because $B$ is tensor product} \\
       &\leq E_f(\bar{Q}:M_1)_{\sigma^{1,m}} + n_R \nonumber \quad\,\,\,\,\,\,\qquad \text{From eq. \eqref{eq:SRcontinuity}}\\
       &\leq m\,n_{\bar{Q}}\eta_{2\sqrt{\epsilon}} + H(\eta_{2\sqrt{\epsilon}}) + n_R, \,\quad \,\,\,\, \text{From eq. \eqref{eq:smallonQM_1}.}\nonumber 
\end{align}
so we have that 
\begin{align}
    n_R  \geq E_f(\bar{Q}:A)_{(\rho^{1})^{\otimes m}} -m(n_{\bar{Q}}\eta_\epsilon + n_{\bar{Q}}\eta_{2\sqrt{\epsilon}}) -H(\eta_\epsilon) - H(\eta_{2\sqrt{\epsilon}})
\end{align}
as claimed. 
\end{proof}

\subsection*{Entropy lower bound for any pure state resource}\label{sec:purestatelowerbound}

Above, we gave a lower bound on the number of qubits of resource system needed in an NLQC implementing a unitary $U_{AB}$ with the controllable entanglement property. 
In this section we translate this into a bound on the entanglement in the resource system, under the assumption that the resource system is pure, which we quantify using the entropy of one side of the resource state. 
We treat this first in the case where the implementation of $U$ is exact, then give a somewhat more complex argument dealing with the approximate case. 

\vspace{0.2cm}
\noindent \textbf{Exact case:} To upgrade our dimension lower bound to a lower bound on entropy, our approach is to make use of Schumacher compression \cite{schumacher1995quantum}, stated in the next theorem.

\begin{theorem}\label{thm:schumacher} \textbf{(Schumacher compression)}
    Suppose we have a quantum source which produces $\ket{\psi}_{LR}^{\otimes m}$. 
    Then, for all $\epsilon, \delta\in(0,1)$, there is a large enough $m$ such that there exists a compression map $\mathcal{E}_{R^m\rightarrow M}$ and decompression map $\mathcal{D}_{M\rightarrow R^m}$ with
    \begin{align}
        \Vert\ketbra{\psi}{\psi}^{\otimes m} - \mathcal{D}_{M\rightarrow R^m}\circ \mathcal{E}_{R^m\rightarrow M}(\ketbra{\psi}{\psi}^{\otimes m}) \Vert_1 \leq \epsilon
    \end{align}
    and where $\log d_M\leq (S(R)+\delta)m$.
\end{theorem}

We use this along with our lower bound on dimension, lemma \ref{lemma:dimensionlowerbound}, to obtain a lower bound on entropy. 
The basic idea is that the resource system $\ket{\psi}_{LR}$ can be compressed to contain $S(R)m$ qubits using Schumacher compression, but from our dimension bound we know the number of qubits must be $E_f(\bar{Q}:A)_{(\rho^1)^{\otimes m}}$, so we must have $S(R)\geq \frac{E_f(\bar{Q}:A)_{(\rho^1)^{\otimes m}}}{m}$. 
Taking $m$ to infinity we obtain the claimed bound. 
We give a more careful proof next. 

\begin{lemma}\label{lemma:exactcaseentropylowerbound}
    Suppose that an NLQC protocol implements $U_{AB}$ exactly, using a pure resource state $\ket{\psi}_{LR}$, and where $U_{AB}$ has $\lambda$-controllable entanglement. Then, 
    \begin{align}
        S(R)\geq \lambda
    \end{align}
\end{lemma}

\begin{proof}
Using Schumacher compression (theorem \ref{thm:schumacher}), for any choice of $\epsilon, \delta>0$ there is an $m$ large enough, choice of compression channel $\mathcal{E}_{R^m\rightarrow M}$, and decompression channel $\mathcal{D}_{M\rightarrow R^m}$ with $n_M\leq (S(R)+\delta)m$, such that if we define
\begin{align}
    \Psi_{L^mM}=\mathcal{E}_{R^m\rightarrow M}(\ketbra{\psi}{\psi}^{\otimes m})
\end{align}
then
\begin{align}
    \left\Vert \ketbra{\psi}{\psi}^{\otimes m}_{LR} - \mathcal{D}_{M\rightarrow R^m}(\Psi_{L^mM})\right\Vert_1 \leq \epsilon.
\end{align}
We define an NLQC protocol to implement $U^{\otimes m}$ as follows. 
The distributed resource state is taken to be $\Psi_{L^mM}$. 
In the first set of operations, on the right, Bob applies $\mathcal{D}_{M\rightarrow R^m}$, leaving Alice and Bob sharing a state $\epsilon$-close to $\ket{\psi}^{\otimes m}$. 
Next, they run $m$ copies of the protocol, using the $m$ copies of $\ket{\psi}$ as resource states. 
By the properties of the diamond norm distance, this will be $\epsilon$-close in diamond norm to an implementation of $U^{\otimes m}$. 

Now we make use of lemma \ref{lemma:dimensionlowerbound}, which tells us that
\begin{align}
    n_M \geq E_f(\bar{Q}:A)_{(\rho^1)^{\otimes m}}-mg_1(\epsilon) - g_2(\epsilon)
\end{align}
But also, at large enough $m$, $(S(R)+\delta)m \geq n_M$, so that
\begin{align}\label{eq:SRlowerbound}
    S(R) \geq \frac{E_f(\bar{Q}:A)_{(\rho^1)^{\otimes m}}}{m}-\delta - g_1(\epsilon)-\frac{g_2(\epsilon)}{m}.
\end{align}
But we can choose $\epsilon$, $\delta$ arbitrarily small while $m$ becomes arbitrarily large, so this simplifies to $S(R)\geq \lim_{m\rightarrow \infty}\frac{E_f(\bar{Q}:A)_{(\rho^1)^{\otimes m}}}{m}=\lambda$, as claimed.   
\end{proof}

\vspace{0.2cm}
\noindent \textbf{Approximate case:} When the NLQC protocol implements $U$ approximately, the asymptotic statement of Schumacher compression doesn't suffice to obtain a lower bound. 
The reason for this can be seen by considering equation \eqref{eq:SRlowerbound}. 
There, the error $\epsilon$ in the implementation of $U^{\otimes m}$ comes from the approximation to the resource state appearing when decompressing from Schumacher's scheme. 
If each $U$ implementation is approximate, there is a contribution to the error from each $U$, so we would replace $\epsilon\rightarrow \epsilon + \gamma\,m$ where $\gamma$ is the error in a single $U$ implementation.
But the lower bound only applies when the (total) error has $\epsilon +\gamma m\in[0,1]$, so we don't obtain a bound in the $m\rightarrow \infty$ limit. 
To remedy this, we will need to consider Schumacher compression for a finite number of copies of the input state. 
This is addressed in \cite{abdelhadi2020second}; we briefly recall one of their results here. 

A compression protocol consists of a compression channel $\mathcal{E}_{A^n\rightarrow M}$ and a decompression channel $\mathcal{D}_{M\rightarrow A^n}$. 
We say the protocol is $\epsilon$-correct if,
\begin{align}
    F_e(\mathcal{D}\circ\mathcal{E}) \geq 1-\epsilon.
\end{align}
We denote the minimal log-dimension of $M$ needed to achieve $\epsilon$-correct compression on $n$ copies of $\rho_A$ by $M(n,\epsilon, \rho)$. 

The value of $M(n,\epsilon, \rho)$ is well understood. 
To state the result, we define the function on density matrices
\begin{align}
    V(A)_\rho = \tr(\rho_A\log^2\rho_A) - \left(\tr(\rho_A \log \rho_A)\right)^2
\end{align}
and the function
\begin{align}
    \Phi^{-1}(x) = \sup \left\{z\in \mathbb{R}:\frac{1}{\sqrt{2\pi}}\int_{-\infty}^{z} e^{-t^2/2} dt\leq x \right\}.
\end{align}
This is known as the \emph{quantile} of the normal distribution; it expresses how far we need to integrate the normal distribution with variance 1 to reach a given value $x$. 
The quantile of the normal distribution is defined on $(0,1)$ and diverges as $x\rightarrow 0,1$. 

Finally, we can state the following theorem, proven in \cite{abdelhadi2020second} as theorem 3. 
\begin{theorem}\label{thm:finite_n_schumacher}\textbf{(Schumacher compression at finite block length)}
The minimal achievable value of $M(n,\epsilon, \rho)$ in performing $\epsilon$-correct compression of the state $\rho_A$ is
\begin{align}
    M(n,\epsilon,\rho)= n S(A)_\rho + \Phi^{-1}(\sqrt{1-\epsilon})\sqrt{n V(A)_\rho} + O(\log n)
\end{align}
\end{theorem}
We use this along with lemma \ref{lemma:dimensionlowerbound} to obtain a lower bound on the entropy. 

\begin{theorem}\label{thm:entropylowerboundapproximate}
    Let $U_{AB}$ be a unitary with $\lambda$-controllable entanglement.
    Suppose that an NLQC protocol implements $U_{AB}$ to within diamond norm distance $\gamma$, using a pure resource state $\ket{\psi}_{LR}$. Then, for small enough $\gamma$,
    \begin{align}
        S(R)_\psi\geq \lambda- P(\gamma).
    \end{align}
    where $P(\gamma)=2\sqrt{2}n_{\bar{Q}}\gamma^{1/8}$. 
\end{theorem}
\begin{proof}
We consider an implementation of $U^{\otimes m}$, where we will choose $m$ later. 
Our implementation uses as a resource state a compressed version of  $\ket{\psi}_{LR}^{\otimes m}$, that is we use 
\begin{align}
    \Psi_{L^mM} = \mathcal{I}\otimes \mathcal{E}_{R^m\rightarrow M}(\ketbra{\psi}{\psi}^{\otimes m})
\end{align}
where $\mathcal{E}_{R^m\rightarrow M}$ is an optimal compression channel. 
The protocol proceeds by first having Bob decompress $M$ into $R^m$, and then running $m$ parallel implementations of $U$ as before. 
We use an $\epsilon$-correct compression protocol where we choose $\epsilon$ later. 
Since the compression protocol is $\epsilon$-correct, and each $U$ implementation is $\gamma$-correct, by the properties of the diamond norm the implementation of $U^{\otimes m}$ using the compressed resource state will be $\epsilon+\gamma\,m$ correct. 

Now we make use of lemma \ref{lemma:dimensionlowerbound}, which tells us that
\begin{align}
    n_M \geq E_f(\bar{Q}:A)_{(\rho^1)^{\otimes m}}-m g_1(\epsilon+\gamma m) - g_2(\epsilon+\gamma m).
\end{align}
Now use theorem \ref{thm:finite_n_schumacher} as a bound on $n_M$, and using that $V(R)_\rho \leq n_R^2$, we have
\begin{align}
    S(R)_\psi &\geq \frac{E_f(\bar{Q}:A)_{(\rho^1)^{\otimes m}}}{m} - \Phi^{-1}(\sqrt{1-\epsilon})\frac{n_R}{\sqrt{m}} - g_1(\epsilon + \gamma m) - \frac{g_2(\epsilon+\gamma m)}{m} - O\left(\frac{\log m}{m}\right)\nonumber
\end{align}
We will use lemma \ref{lemma:ECandfinite_m} to observe that
\begin{align}
    \frac{E_f(\bar{Q}:A)_{(\rho^1)^{\otimes m}}}{m} \geq \lim_{m'\rightarrow \infty}\frac{E_f(\bar{Q}:A)_{(\rho^1)^{\otimes m'}}}{m'} = \lambda
\end{align}
so that we obtain 
\begin{align}
    S(R)_\psi&\geq \lambda - \Phi^{-1}(\sqrt{1-\epsilon})\frac{n_R}{\sqrt{m}} - g_1(\epsilon + \gamma m) - \frac{g_2(\epsilon+\gamma m)}{m} - O\left(\frac{\log m}{m}\right)
\end{align}
For intuition on the above inequality, notice that if we take $\gamma=0$ we can maximize the lower bound by sending $\epsilon\rightarrow0,m\rightarrow \infty$, in which case we recover $S(R)_\psi\geq \lambda$. 
At non-zero $\gamma$ however, sending $m\rightarrow \infty$ would trivialize the lower bound. 
To recover a good lower bound, we need to choose $\epsilon, m$ in a way that depends on $\gamma$ such that the lower bound approaches $\lambda$ as $\gamma\rightarrow 0$.
We will achieve this with a simple choice by taking
\begin{align}
    \epsilon &= \gamma, \qquad m =\frac{1}{\sqrt{\gamma}}.
\end{align}
Inserting this above leads to the lower bound
\begin{align}
    S(R)_\psi \geq \lambda - \Phi^{-1}(\sqrt{1-\gamma})n_R \gamma^{1/4} - g_1(\gamma + \sqrt{\gamma}) - \sqrt{\gamma} g_2(\gamma+\sqrt{\gamma}) - O\left(\sqrt{\gamma}\log \gamma\right).\nonumber 
\end{align}
We can see that as $\gamma\rightarrow 0$ this approaches the lower bound obtained in the exact case, so this bound is equal to that one plus terms that go to zero as $\gamma\rightarrow 0$. 
To obtain the error terms, we expand in a series around $\gamma=0$, obtaining
\begin{align}
    S(R)_\psi \geq \lambda -2\sqrt{2}n_{\bar{Q}}\gamma^{1/8}- \tilde{O}(n_R\gamma^{1/4})- \tilde{O}(n_{\bar{Q}}\gamma^{1/4}) \,\,\,\text{as}\,\,\, \gamma\rightarrow 0
\end{align}
where the $\tilde{O}$ notation hides logarithmic factors.
For $\gamma$ small enough, concretely $\gamma$ such that $\gamma$ is much smaller than $\min\{1/n_R^4,1/n_{\bar{Q}}^{4}\}$ then, we obtain the lower bound
\begin{align}
    S(R)_\psi \geq \lambda- 2\sqrt{2}n_{\bar{Q}}\gamma^{1/8}
\end{align}
 as claimed. 
\end{proof}

\subsection*{Entanglement lower bound}\label{sec:entanglementbound}

We've given lower bounds on the entropy of one end of the resource system for any resource state that allows a unitary $U_{AB}$ with controllable entanglement to be implemented as an NLQC. 
Theorem \ref{thm:mainlowerbound} now follows as a corollary of the pure state entropy lower bound (theorem \ref{thm:entropylowerboundapproximate}) and our general purpose method of upgrading entropy to entanglement lower bounds, theorem \ref{thm:entropytoentanglement}. 

\section{History and further reading}

The controllable correlation and controllable entanglement lower bounds were proven in \cite{cleve2026lower}.
Note that this chapter corrects a number of (repairable) mistakes in earlier versions of \cite{cleve2026lower}.
The lower bound on the simple routing task, which inspires these strategies, was first written down in an appendix of \cite{asadi2025linear}, though the result was previously known as folklore.  

\chapter{Entanglement lower bounds on \texorpdfstring{$f$}{TEXT}-routing}\label{chapter:mostlyclassicallowerbounds}

\minitoc

In the last chapter we made some progress on lower bounding the entanglement needed to implement unitaries as NLQCs. 
We've seen that $f$-routing is a class of special importance to the theory of NLQC; it's a good candidate QPV scheme, plays a special role in the $T$-depth based protocol, and is closely connected to the conditional disclosure of secrets (CDS) primitive studied in information-theoretic classical cryptography. 
All of these applications motivate us to try and understand lower bounds on $f$-routing.

In this chapter we give one lower bound technique for $f$-routing.  
A key property we would like such lower bounds to have is that the cost of implementing $f$-routing should grow with the length of the \emph{classical} strings given as input, call it $n$. 
Our bound has this property. 
Unfortunately, our bound achieves this at the expense of inducing a drawback --- our entanglement lower bound only applies if the $f$-routing protocol is either perfectly correct when $f(x,y)=0$, or perfectly correct when $f(x,y)=1$. 
Obtaining lower bounds on entanglement that grow (faster than logarithmically) with $n$, even when allowing two-sided error, is an important open problem.

Later, in chapter \ref{chapter:QPV}, we discuss an approach which shows that the number of quantum gates needed to implement the closely related $f$-measure protocol grows linearly with $n$. 
In the context of QPV, this means an attacker needs much larger quantum resources than an honest player for the $f$-measure task.
In fact, because of the equivalence between $f$-measure and $f$-route, this also shows a gate lower bound for $f$-route. 
Unlike the result we give in this chapter, the gate lower bound holds even for noisy protocols. 
Here we remain focused on entanglement lower bounds however. 

\section{Preliminaries}

\begin{figure*}
    \centering
    \begin{tikzpicture}[scale=0.5]
    
    \draw[thick] (-5,-5) -- (-5,-3) -- (-3,-3) -- (-3,-5) -- (-5,-5);
    \node at (-4,-4) {$\mathcal{N}^x$};
    
    \draw[thick] (5,-5) -- (5,-3) -- (3,-3) -- (3,-5) -- (5,-5);
    \node at (4,-4) {$\mathcal{M}^y$};
    
    \draw[thick] (5,5) -- (5,3) -- (3,3) -- (3,5) -- (5,5);
    \node at (4,4) {$\mathcal{V}^{x,y}$};
    
    \draw[thick] (-5,5) -- (-5,3) -- (-3,3) -- (-3,5) -- (-5,5);
    \node at (-4,4) {$\mathcal{W}^{x,y}$};
    
    \draw[thick] (-4.5,-3) -- (-4.5,3);
    \node[left] at (-4.5,-2) {$M_0$};
    
    \draw[thick] (4.5,-3) -- (4.5,3);
    \node[right] at (4.5,-2) {$M_1'$};
    
    \draw[thick] (-3.5,-3) to [out=90,in=-90] (3.5,3);
    \node[right] at (-3.25,-2) {$M_0'$};
    
    \draw[thick] (3.5,-3) to [out=90,in=-90] (-3.5,3);
    \node[left] at (3.25,-2) {$M_1$};
    
    \draw[thick] (-3.5,-5) to [out=-90,in=-90] (3.5,-5);
    \draw[black] plot [mark=*, mark size=3] coordinates{(0,-7.05)};

    \draw[thick] (-4.5,-5) to [out=-90,in=-90] (-8.5,-5);
    \node[above] at (-8.5,-5) {$\bar{Q}$};
    
    \draw[thick] (4.5,5) -- (4.5,6);
    \node[above] at (-4.5,6) {$A$};
    \draw[thick] (-4.5,5) -- (-4.5,6);
    \node[above] at (4.5,6) {$B$};
    
    \end{tikzpicture}
    \caption{General NLQC implementing an $f$-routing position verification scheme. The first round operations are quantum channels that depend on the inputs $x\in \{0,1\}^n$ and $y\in\{0,1\}^n$. In the second round Alice and Bob apply channels mapping to qubit systems $A$, $B$. If $f(x,y)=0$ $A$ should be maximally entangled with the reference $\bar{Q}$. If $f(x,y)=1$ then $B$ should be maximally entangled with $\bar{Q}$.}
    \label{fig:froutelabels}
\end{figure*}

A general $f$-routing protocol is shown in figure \ref{fig:froutelabels}, where we also show the system labels we use in this section. 
Note that we label $M_0M_1=M$, $M_0'M_1'=M'$. 
We will often consider taking the $Q$ subsystem of a maximally entangled state $\Psi^+_{\bar{Q}Q}$ as input to the protocol. 
The density matrix $\rho_{\bar{Q}MM'}$ we refer to as the \emph{mid-protocol density matrix}. 

\subsection{Communication complexity}

In looking for lower bounds on $f$-routing, a likely source of inspiration is the field of communication complexity, which also appeared in chapter \ref{chapter:communicationcomplexity}. 
A useful object commonly studied in communication complexity is the \emph{communication matrix} of a function $f:\{0,1\}^n\times\{0,1\}^n\rightarrow \{0,1\}$.
This is defined to be a $2^n\times 2^n$ matrix labelled $M_f$ with rows labelled by $x$, columns labelled by $y$, and entries $f(x,y)$. 
For instance, the equality function 
\begin{align}
    EQ(x,y)=\begin{cases}
        0 \,\,\,\text{if} \,\,\, x\neq y \\
        1 \,\,\,\text{if} \,\,\, x=y
    \end{cases}
\end{align}
has the identity matrix as its communication matrix. 

The rank of the communication matrix $M_f$ is related to the communication complexity of the function $f$. 
For instance, let the minimal number of classical bits of communication needed to compute $f$ (with probability 1) in the two-way communication model be denoted $D(f)$. 
Then, we have that \cite{kushilevitz1996communication}
\begin{align}
    \log( \text{rank}(M_f)) \leq D(f).
\end{align}
We will make use of a second notion of rank, which is called the non-deterministic rank. 
This is defined by
\begin{align}
    \text{nrank}(f) = \min_{G\sim M_f} \rank(G)
\end{align}
where the minimization is over all matrices $G$ with the same zero entries as $M_f$, but arbitrary complex numbers in the non-zero entries. 
The notion of $\text{nrank}$ may seem a bit unwieldy, but we can notice that there are some simple cases where it can be lower bounded easily. 
For instance the equality function has $\text{nrank}(M_{EQ})=2^n$, since $M_{EQ}$ has non-zero entries on the diagonal and zero entries off the diagonal, so any $G$ with the same zeros must also be full rank. 

\subsection{Structural features of \texorpdfstring{$f$}{TEXT}-routing protocols}

To obtain lower bounds, we need to understand how the structure of the function $f$ is reflected in the NLQC protocol, and specifically in the resource state $\Psi_{LR}$. 
We are starting with conditions on the output systems, and we somehow want to wind these conditions backwards and see what they tell us about the resource state. 
Towards this, we want to understand how the correctness requirements on the outputs translate to constraints on the protocol, and specifically on the mid-protocol density matrix. 

Suppose we entangle the quantum input $Q$ with a reference system $\bar{Q}$, then execute the first round operations. 
This produces the mid-protocol density matrix $\rho_{\bar{Q}MM'}(x,y)$. 
Notice that if $f(x,y)=0$ then the reference system $\bar{Q}$ is maximally entangled with the output on the left, so must also be maximally entangled with the systems $M=M_0M_1$ that are sent left. 
Similarly, if $f(x,y)=1$ then $\bar{Q}$ must be maximally entangled with $M'=M_0'M_1'$. 
We state and prove this more carefully in the next lemma. 

\begin{lemma}\label{lemma:f=0tensorproduct}
    Suppose an $f$-routing protocol which uses a pure state resource $\ket{\Psi}_{LR}$ is perfectly correct on zero instances, and $\epsilon_1<\epsilon_1^*$ correct on $1$ instances, where $\epsilon_1^*$ depends on $d_Q$ but is lower bounded by a constant. Then the mid-protocol state it produces, $\rho_{\bar{Q}MM'}$, satisfies $\tr_M\rho_{\bar{Q}MM'}=\rho_{\bar{Q}}\otimes \rho_{M'}$ if and only if $f(x,y)=0$.
\end{lemma}
\begin{proof}\,
    By correctness on $0$ instances, we have that 
    \begin{align}
        \text{when}\,\,\, f(x,y)=0,\,\,\,F(\mathcal{D}^{x,y}_{M\rightarrow Q} \circ\tr_{M'} \circ\mathcal{N}^{x,y}_{Q\rightarrow MM'}(\Psi^+_{\bar{Q}Q}),\Psi^+_{\bar{Q}Q}) =1
    \end{align}
    so that $f(x,y)=0$ implies Alice can produce a Bell state $\ket{\Psi^+}_{\bar{Q}Q}$. 
    By data processing we have
    \begin{align}
        2n_{\bar{Q}}=I(\bar{Q}:Q)_{\Psi^+_{\bar{Q}Q}} \leq I(\bar{Q}:M)_{\rho_{\bar{Q}M}}. 
    \end{align}
    We consider the purification of $\rho_{\bar{Q}MM'}$, by adding a system $E$, and then we have
    \begin{align}
        I(\bar{Q}:ME)_\rho + I(\bar{Q}:M')_\rho \leq 2S(\bar{Q})
    \end{align}
    Since $I(\bar{Q}:ME)_\rho \geq I(\bar{Q}:M)$, we obtain that $I(\bar{Q}:M')=0$, which implies
    \begin{align}
        \rho_{\bar{Q}M'}=\rho_{\bar{Q}}\otimes \rho_{M'}
    \end{align}
    as needed. 
    
    Conversely, if $f(x,y)=1$ we cannot have $\rho_{\bar{Q}M'}=\rho_{\bar{Q}}\otimes \rho_{M'}$. 
    To see why, first recall that $f$-routing which is $\epsilon_1$-correct on $1$ instances has that there exists a family of channels $\{D^{x,y}_{M'\rightarrow Q}\}_{x,y}$ such that
    \begin{align}
        \text{when}\,\,\, f(x,y)=1,\,\,\,F(\mathcal{D}^{x,y}_{M'\rightarrow Q} \circ\tr_{M} \circ\mathcal{N}^{x,y}_{Q\rightarrow MM'}(\Psi^+_{\bar{Q}Q}),\Psi^+_{\bar{Q}Q}) \geq 1-\epsilon_1.
    \end{align}
    Define 
    \begin{align}
        \sigma_{\bar{Q}Q} = \mathcal{D}^{x,y}_{M'\rightarrow Q}(\rho_{\bar{Q}M'})
    \end{align} 
    and use that by data processing, 
    \begin{align}
        I(\bar{Q}:M')_{\rho_{\bar{Q}M'}} \geq I(\bar{Q}:Q)_{\sigma_{\bar{Q}Q}}.
    \end{align}
    But then by the Fuchs--van de Graaf inequalities $\sigma_{\bar{Q}Q}$ is close in trace distance to $\Psi^+_{\bar{Q}Q}$,
    \begin{align}
        \frac{1}{2}||\Psi^+_{\bar{Q}Q} - \sigma_{\bar{Q}Q}||_1\leq \sqrt{\epsilon_1} 
    \end{align}
    and by continuity of the mutual information (lemma \ref{lemma:MIcontinuity}), 
    \begin{align}
        I(\bar{Q}:Q)_{\Psi^+} - I(\bar{Q}:Q)_{\sigma_{\bar{Q}Q}} 
        \leq 2 \sqrt{\epsilon_1} \log d_{\bar{Q}} + (1+\sqrt{\epsilon_1}) h\left( \frac{\sqrt{\epsilon_1}}{1+\sqrt{\epsilon_1}}\right)
    \end{align}
    so that
    \begin{align}
        2\log d_{\bar{Q}} - 2 \sqrt{\epsilon_1} \log d_{\bar{Q}} - (1+\sqrt{\epsilon_1}) h\left( \frac{\sqrt{\epsilon_1}}{1+\sqrt{\epsilon_1}}\right) \leq I(\bar{Q}:Q)_{\sigma}
    \end{align}
    We have that $\sigma$ is not tensor product whenever the left hand side of the equation above is strictly positive. 
    For $d_Q=2$, this occurs for $\epsilon_1 < \epsilon_1^* \approx 0.16$, and $\epsilon_1^*$ approaches $1$ as $d_Q\rightarrow \infty$. 
\end{proof}

This lemma is saying that $f(x,y)$ is already fixed by the NLQC protocol after the \emph{first} round operations, essentially because where the entanglement with $\bar{Q}$ is going must already be decided before the communication or second round operations occur.

The above gives a property that the mid-protocol density matrix must have: it should be product across $\bar{Q}M'$ when $f(x,y)=0$.
Typically, other lower bound strategies also start by finding and exploiting special properties of the mid-protocol density matrix.  
For instance, we can also ask about how instantiations of $\rho_{\bar{Q}MM'}(x,y)$ with differing values of $f(x,y)$ are related to one another. 
One intuition is that since $f(x,y)$ is determined by the first round density matrix, we might expect $f(x,y)=0$ instances of $\rho_{\bar{Q}MM'}(x,y)$ to be distinguishable from $f(x,y)=1$ instances. 
Indeed this is the case, and different formalizations of this idea are used in multiple lower bounds techniques \cite{asadi2025linear, bluhm2022single}. 
For our proof strategy here however the above lemma will suffice. 

\section{Rank lower bound}

We're now ready to develop our entanglement lower bound for $f$-routing. 
In this section we will focus on $f$-routing with $\epsilon_0=0$, $\epsilon_1\geq 0$, so that the protocol is perfectly correct on zero instances of $f(x,y)$, or with $\epsilon_0>0$ and $\epsilon_1=0$, so that the protocol is perfectly correct on $1$ instances. 
We define the entanglement cost of an $f$-routing protocol to be the logarithm of the minimal Schmidt rank of any resource system which can be used to perform the $f$-routing task. 
In notation, we define $\FR_0(f)$ to be entanglement cost for $f$-routing with $\epsilon_0=0$, $\epsilon_1=0.05$, and $\FR_1(f)$ to be the entanglement cost when $\epsilon_0=0.05$, $\epsilon_1=0$.

Note that the Schmidt rank is a somewhat bad measure of the entanglement cost: we can deform a state (in trace distance) slightly yet make a large change to the Schmidt rank. 
This is unfortunately a limitation of our methods so far. 
If we assume the resource system consists of $n$ EPR pairs, then the log Schmidt rank is equal to $n$. 

In lemma \ref{lemma:f=0tensorproduct} we saw that the mid-protocol density matrix in an $f$-routing protocol, $\rho_{\bar{Q}MM'}(x,y)$ has $\rho_{\bar{Q}M'}(x,y)=\rho_{\bar{Q}}\otimes \rho_{M'}(x,y)$ if and only if $f(x,y)=0$. 
To prove our rank lower bound, we will show that the way in which $\rho_{\bar{Q}M'}(x,y)$ depends on $(x,y)$ is constrained by the amount of entanglement in the resource state, and that consequently to have the density matrix be product for only the right values of $(x,y)$, it needs to have some lower bounded amount of entanglement.

To understand how this product structure in the mid-protocol density matrix relates to the entanglement in the resource state, we define what we call a \emph{structure function} for a protocol.

\begin{definition}\label{def:structurefunction}
    Given an $f$-routing protocol with mid-protocol density matrix $\rho_{\bar{Q}MM'}$, define the \textbf{structure function} $g(x,y)$ according to
    \begin{align}
        g(x,y) = \tr\left(\rho_{\bar{Q}M'}-\frac{\mathcal{I}}{d_{\bar{Q}}}\otimes \rho_{M'}\right)^2 .
    \end{align}
\end{definition}

Note that we can also phrase this definition in terms of the Frobenius norm, $\Vert A\Vert_F=\sqrt{\tr(A^\dagger A)}$. 

We claim that $g(x,y)$ captures some aspect of the structure in the function $f$ which must be present in a correct $f$-routing protocol. 
More concretely we have the following. 
\begin{lemma}
    In a perfectly correct $f$-routing protocol, the structure function $g(x,y)$ is zero if and only if $f(x,y)=0$.
\end{lemma}
\begin{proof}\,
    Since $g(x,y)=\Vert \rho_{\bar{Q}M'}-\frac{\mathcal{I}}{d_{\bar{Q}}}\otimes \rho_{M'}\Vert_F^2$, and a norm is zero iff it is evaluated on 0, we see that $g(x,y)=0$ iff $\rho_{\bar{Q}M'}=\frac{\mathcal{I}}{d_{\bar{Q}}}\otimes \rho_{M'}$.
    Then lemma \ref{lemma:f=0tensorproduct} shows we have $\rho_{\bar{Q}M'}=\frac{\mathcal{I}}{d_{\bar{Q}}}\otimes \rho_{M'}$ if and only if $f(x,y)=0$. 
\end{proof}

Our next job is to relate the function $g(x,y)$ to the entanglement available to Alice and Bob. 
We prove the following lemma. 
\begin{lemma}
    An $f$-routing protocol that uses a resource system with Schmidt rank $d_E$ has a structure function of the form
    \begin{align*}
        g(x,y) = \sum_{I=1}^{d_E^4} f_I(x)f_I'(y)\enspace.
    \end{align*}
\end{lemma}
\begin{proof}\,
    From the general form of a non-local quantum computation protocol, the density matrix $\rho_{\bar{Q}M_0'M_1'}(x,y)$ can be expressed as
    \begin{align*}
        \rho_{\bar{Q}M_0'M_1'}=\mathcal{N}^x_{QL\rightarrow M_0'}\otimes \mathcal{M}^y_{R\rightarrow M_1'}(\Psi^+_{\bar{Q}Q}\otimes \ketbra{\Psi}{\Psi}_{LR})\enspace.
    \end{align*}
    We will write $\ket{\Psi}_{LR}$ in the Schmidt basis, 
    \begin{align*}
        \ket{\Psi}_{LR} = \sum_{i=1}^{d_E} \ket{i}_L\ket{i}_R\enspace,
    \end{align*}
    with un-normalized vectors $\ket{i}_L, \ket{i}_R$. 
    Then we get
    \begin{align*}    \rho_{\bar{Q}M_0'M_1'}&=\sum_{i,j=1}^{d_E}\mathcal{N}^x_{QL\rightarrow M_0'}(\Psi^+_{Q\bar{Q}}\otimes \ketbra{i}{j}_L)\otimes \mathcal{M}^y_{R\rightarrow M_1'}(\ketbra{i}{j}_{R}) \nonumber \\
        &= \sum_{i,j=1}^{d_E} A^{x,i,j}_{\bar{Q}M_0'}\otimes B^{y,i,j}_{M_1'}\enspace.
    \end{align*}
    We can also compute the trace over $\bar{Q}$ where we define $A$ and $B$ in the second line of the previous equation and get
    \begin{align*}
        \rho_{M_0'M_1'} &= \sum_{i,j=1}^{d_E} A^{x,i,j}_{M_0'}\otimes B^{y,i,j}_{M_1'}\enspace.
    \end{align*}
    Next, we compute $g(x,y)$. 
    It is convenient to first re-express the function $g(x,y)$ as follows, 
    \begin{align}
        \tr\left(\rho_{AB}-\frac{\mathcal{I}_A}{d}\otimes\rho_B\right)^2
        &= \tr\left(\rho^2_{AB} +\frac{\mathcal{I}_A}{d^2}\otimes\rho^2_B - 2\rho_{AB}\left(\frac{\mathcal{I}_A} {d_A}\otimes \rho_B\right)\right) \\
        &= \tr(\rho^2_{AB})+\frac{1}{d_A^2}\tr(\mathcal{I}_A)\tr(\rho^2_B)-\frac{2}{d}\tr_B(\rho^2_B) \\
        &= \tr(\rho^2_{AB}) - \frac{1}{d_A}\tr(\rho^2_B) \\
        &= \tr(\rho^2_{AB}) - \tr\left(\frac{\mathcal{I}_A}{d_A^2}\otimes \rho^2_B\right) \enspace.
    \end{align}
    so that
    \begin{align*}
        g(x,y) = \tr\left(\rho_{\bar{Q}M'}^2 - \frac{\mathcal{I}}{d_{\bar{Q}}^2}\otimes  \rho_{M'}^2\right)\enspace.
    \end{align*}
    Inserting the forms of $\rho_{\bar{Q}M'}$ and $\rho_{M'}$ into this, we obtain
    \begin{align*}
        g(x,y) &= \sum_{i,j,i',j'=1}^{d_E} \tr\left(\left( A^{x,i,j}_{\bar{Q}M_0'}A^{x,i',j'}_{\bar{Q}M_0'}  - \frac{\mathcal{I}}{d_{\bar{Q}}^2}\otimes A^{x,i,j}_{M_0'}A^{x,i',j'}_{M_0'}\right)\otimes B^{y,i,j}_{M_1'}B^{y,i',j'}_{M_1'}\right) \nonumber \\
        &= \sum_{I=1}^{d_E^4} f_I(x) f'_I(y)\enspace,
    \end{align*}
    as needed. 
\end{proof}

We can view $g(x,y)$ as a matrix, and $f_I(x)$, $f_I'(y)$ as vectors, so that the minimal number of terms appearing in this sum is the rank of the matrix $g(x,y)$. 
Thus we obtain the lower bound
\begin{align}
    \FR_0(f) \geq \frac{1}{4} \log \rank (M_g)
\end{align}
Here $g$ is defined by the construction in the proof given above. 
We know that $g$ has zeros in the same entries as $f$, so the above implies that
\begin{align}\label{eq:rankbound}
    \boxed{\FR_0(f) \geq \frac{1}{4} \log (\text{nrank} (M_f))}
\end{align}
We can also notice that we can reverse the role of $M$ and $M'$, and assume perfect correctness on $1$ instances, leading to a similar bound, 
\begin{align}\label{eq:rankboundnegation}
    \boxed{\FR_1(f) \geq \frac{1}{4} \log (\text{nrank}(M_{\neg f}))}
\end{align}
where $\neg f$ is the negation of $f$. 

We won't describe this in detail here, but we can also rephrase the above bounds in terms of the non-deterministic quantum communication complexity \cite{de2003nondeterministic}, denoted $\QNP^{\cc}(f)$, 
\begin{align}
    \FR_0(f)&=\Omega(\QNP^{\cc}(f)), \nonumber \\
    \FR_1(f)&=\Omega(\coQNP^{\cc}(f)). 
\end{align}
See \cite{asadi2024rank} for details.
This relationship to the $\QNP^{\cc}$ complexity comes about by an apparent coincidence: in \cite{de2003nondeterministic} it was proven that the $\QNP^{\cc}$ complexity is equal to the log of the non-deterministic rank, and we prove a lower bound from the non-deterministic rank here. 
It is an open problem to understand if there is a reduction from the $f$-routing scenario to the $\QNP^{\cc}$ scenario that would explain this. 

One consequence of our lower bound is a new lower bound on randomness complexity in CDS. 
Recall that we had from chapter \ref{chapter:ITCandNLQC}, equation \eqref{eq:FR=CDQS}, 
\begin{align}
    \FR(f)= \Omega(\CDQS(f))
\end{align}
and from equation \eqref{eq:CDSandCDQS} that
\begin{align}
    \CDS(f) \geq \CDQS(f).
\end{align}
Combining these, we obtain
\begin{align}
    \CDS(f)= \Omega(\FR(f)).
\end{align}
In this section we lower bounded $\FR_0(f)$ and $\FR_1(f)$ (perfect correctness in either $0$ or $1$ instances). 
We can check that the above relations go through when we impose either perfect security or perfect correctness on CDS to give a corresponding type of error in $f$-routing. 
Specifically, considering perfectly correct CDS,
\begin{align}
    \pc\CDS(f)=\Omega(\FR_1(f)).
\end{align}
Meanwhile perfectly secure (also called perfectly private) CDS satisfies
\begin{align}
\pp\CDS(f) = \Omega(\FR_0(f)).
\end{align}
Using our rank lower bounds, we obtain lower bounds in terms of the non-deterministic rank, or equivalently in terms of the QNP communication complexity, 
\begin{align}
    \pc\CDS(f) &= \Omega\left(\coQNP^\cc(f) \right), \nonumber \\
    \pp\CDS(f) &= \Omega\left(\QNP^\cc(f) \right).
\end{align}
The perfectly correct lower bound is weaker than one already known classically, where a lower bound from $\NP^\cc$ has been proven \cite{applebaum2021placing}. 
However, the lower bound on perfectly private classical CDS is new. 
In fact, there is no known classical technique for proving this bound.

\subsection*{Evaluating the lower bound}

We can evaluate our lower bound on the Schmidt rank of the resource state explicitly for several simple choices of function. 

Choosing $f(x,y)$ to be the equality function,
\begin{align}
EQ(x,y) = \begin{cases}
		0, & x\neq y\\
            1, & x=y
		 \end{cases}\enspace.
\end{align}
Then $g(x,y)$ is zero except on the diagonal, which forces it to have full rank, so from equation \eqref{eq:rankbound} and equation \eqref{eq:rankboundnegation}
\begin{align}
    \FR_0(EQ) \geq \frac{n}{4}, \qquad \FR_1(\neg EQ) \geq \frac{n}{4}.
\end{align}

Similarly, the `greater than' function, 
\begin{align}
    GT(x,y) = \begin{cases}
			0, & x< y\\
                1, & x\geq y
		 \end{cases}\enspace.
\end{align}
is upper triangular with non-zero elements on the diagonal, so it also has full rank, and we obtain a linear lower bound. 
\begin{align}
    \FR_0(GT) \geq \frac{n}{4}.
\end{align}
Further, because the negation of Greater-Than is also full rank, we can also bound $\FR_1$, 
\begin{align}
    \FR_1(GT)\geq \frac{n}{4}\enspace.
\end{align}
The same bounds hold for the `less than' function.

Set disjointness is upper left triangular.\footnote{To see why, consider that on the diagonal of the truth table running top right to bottom left, we have that $x+y=11...11$, the all $1$'s string. This means $x$ and $y$ must have non-zero values in non-overlapping locations, e.g. for two bits this diagonal consists of $(x=00,y=11)$, $(x=01,y=10)$, $(x=10,y=01)$ and $(x=11,y=00)$. Moving downward from any entry on that diagonal $y$ becomes larger, so must now have an overlapping entry with the $x$ string.}
This implies it is full rank, therefore
\begin{align}
    \FR_0(DISJ) \geq \frac{n}{4}.
\end{align}
Since the negation of set intersection is set disjointness and hence of full rank, we also obtain
\begin{align}
    \FR_1(INT) \geq \frac{n}{4}\enspace.
\end{align}

\section{History and further reading}

In the original garden-hose paper studying $f$-routing \cite{buhrman2013garden} gives logarithmic lower bounds on the size of the communication used in an $f$-routing protocol, in the perfect setting. 
Similar ideas were revived in \cite{bluhm2022single}, who start from similar observations to prove lower bounds. 
Their lower bounds are on the number of qubits in the resource system, and assume a restrictive model of NLQC where all operations are unitary. 
They prove linear (in $n$) lower bounds on this size for typical choices of function $f$. 

The rank lower bound presented in this section was proven in \cite{asadi2024rank}.
Another strategy not discussed here is to consider a model where $f(x,y)$ is not known to the players, but can only be accessed as an oracle. 
In this model an exponential lower bound on the number of oracle calls was proven \cite{unruh2014quantum}.

Recently, the work \cite{bogner2026robust} shows a logarithmic lower bound on the entanglement of formation needed to implement $f$-routing as an NLQC, when $f$ is the inner product function. 

Using an extension of the rank lower bound technique, the work \cite{hasegawa2026new} proves a lower bound on Schmidt rank in $f$-routing based on the sign-rank of $f$ (a notion similar to the non-deterministic rank). 
This gives, for instance, a linear lower bound in the case where $f$ is the inner product function.  

\chapter{Lower bounds for measurement NLQCs from monogamy games}\label{chapter:monogamygames}

\minitoc

In this chapter we expand on the technique given in section \ref{sec:measuringintro}, where we argued entanglement is necessary to complete the measure NLQC task by giving a reduction to \emph{monogamy of entanglement} (MoE) games \cite{tomamichel2013monogamy}. 
We will discuss more general MoE games here, and consider the parallel repetition of MoE games. 
Using these ideas, we show that as the measure task is repeated in parallel, the success probability of the game when using zero entanglement goes to zero. 
Upper bounds on success probabilities in the zero entanglement case can then be translated to lower bounds on the entanglement needed to complete the task with high probability.  

\section{Monogamy games}

Monogamy of entanglement games are played by three players, call them player 1, player 2 and the referee.
A monogamy game is defined by a set of measurements $\{\mathcal{M}^\theta\}_{\theta}$ on a $d$-dimensional system, along with a winning condition, which we define below.
We focus on the $d=2$ case here. 
Each measurement $\mathcal{M}^\theta$ consists of a complete set of projectors, $\mathcal{M}^\theta=\{\Pi^\theta_1,...,\Pi^\theta_k\}$.
For example, we could consider a set of measurements consisting of both the computational and Hadamard basis measurements, so that
\begin{align}
    \{\mathcal{M}^\theta\}_{\theta} = \{\mathcal{M}^0, \mathcal{M}^1\}
\end{align}
with 
\begin{align}\label{eq:BB84measurements}
    \mathcal{M}^0 &=\{\ketbra{0}{0}, \ketbra{1}{1}\}, \nonumber \\
    \mathcal{M}^1 &=\{\ketbra{+}{+}, \ketbra{-}{-}\}.
\end{align}
To carry out the game, player 1, player 2, and the referee implement the following steps. 

\vspace{0.2cm}
\noindent \textbf{Monogamy game:}
\begin{itemize}
    \item \textbf{Preparation phase}: Players 1 and 2 prepare a quantum state $\rho_{S_1S_2R}$. They then send system $R$ to the referee, where $R$ consists of a single qubit. Player 1 holds $S_1$ and player 2 holds $S_2$. Once this is done, the players are separated and no longer communicate.
    \item \textbf{Question phase}: The referee chooses a random bit $\theta$. She then measures $R$ using measurement $\mathcal{M}^{\theta}$, obtaining outcome $x$. The referee then announces $\theta$ to player 1 and player 2.
    \item \textbf{Answer phase}: Player 1 and 2 each act on $S_1$ and $S_2$ respectively to form independent guesses, call them $x',x''$, of the referee's measurement outcomes.
\end{itemize}
We define the players to have won the game if $x=x'=x''$. 

We will be interested in the \emph{parallel repetition} of MoE games. 
To understand the setting, consider an MoE game $G$ with measurements $\mathcal{M}^\theta$. 
Then the $n$-fold parallel repetition of $G$, denoted $G^{n}$, involves repeating $G$ $n$ times in parallel: player 1 and player 2 get a string $\Theta$ consisting of $n$ measurement settings, and should produce outcome strings that match the outcomes of $n$ measurements in corresponding bases made by the referee. 
We can also consider a relaxation of this where player 1 and player 2 are only required to correctly guess a fraction $1-\delta$ of the referee's $n$ measurement outcomes, in which case we denote the game by $G^{n,\delta}$. 

The most commonly studied monogamy game is the BB84 game, which we introduced in chapter \ref{chapter:NLQC}.
We will denote this game by $G_{BB84}$; recall that in that case the bases $\theta=0$ and $\theta=1$ are the computational and Hadamard bases respectively.  
Recall that lemma \ref{lemma:BB84MoE} showed an upper bound on the success probability of this game of $\cos^2(\pi/8)$. 

We will be interested in the game $G_{BB84}^{n,\delta}$, where the $n$ superscript means we are considering the $n$-fold parallel repetition of $G_{BB84}$, and the $\delta$ means we declare player 1 and player 2 to have won if they succeed on a fraction $1-\delta$ of the outputs.
The success probability is bounded in the next lemma. 
\begin{theorem}\label{thm:BB84MoEupperbound}
The success probability of the $G_{BB84}^{n,\delta}$ task is upper bounded according to
\begin{align}
    p_{suc}(G_{BB84}^{n,\delta}) \leq \left(2^{h(\delta)}\cos^2\left( \frac{\pi}{8}\right) \right)^n
\end{align}
where $h(x)$ is the binary entropy function $h(x) = -x\log x - (1-x)\log(1-x)$.
\end{theorem}
We won't prove this theorem here, but see \cite{tomamichel2013monogamy}. 
Our goal below is only to explain the relevance of monogamy games and this theorem in particular to NLQC. 

\section{Reduction from measurement NLQC to monogamy games}

A measurement NLQC is specified by a collection of 2 or more bases, which we label with $\theta$. 
Thus $\{\ket{\psi^\theta_i}\}_i$ is a basis for each choice of $\theta$. 
The task is then defined as follows. 

\vspace{0.2cm}
\noindent \textbf{Measurement NLQC:}
\begin{itemize*}
    \item \textbf{Input:} On the left, a quantum system $Q$ in state $\ket{\psi^\theta_i}_Q$. On the right, a classical input $\theta$ which labels the choice of basis.
    \item \textbf{Output:} On both sides, the variable $i$. 
\end{itemize*}
We will also consider the $n$ fold parallel repetition of measurement NLQCs, which we denote by $N^{n}$. 
When we relax the winning condition to declare the NLQC succeeds when a fraction $1-\delta$ of the measurement outcomes are correct we denote the NLQC by $N^{n,\delta}$. 

We claim that a strategy to complete the measurement NLQC task with probability $p$ \emph{without using any entanglement} implies the existence of a strategy to complete an associated monogamy game with the same probability. 
To see why, let's start with an (entanglement free) NLQC protocol for a measurement NLQC.
We assume that Alice and Bob share a separable state $\rho_{AB}$ as their resource. 
Our figure of merit that we are interested in is the probability that both Alice and Bob output $i$. 
This measurement probability is a function of the input state $\rho$, so we write $p_{suc}=p_{suc}(\rho)$. 
We can notice that
\begin{align}
    p_{suc}\left(\sum_j p_j \rho_A^j\otimes \rho_B^j \right) \leq  \sum_j p_j p_{suc}(\rho_A^j\otimes \rho_B^j).
\end{align}
This is because the left hand side denotes the success probability when given $\rho_A^j\otimes \rho_B^j$ with probability $p_j$ and not told $j$, while the right hand side denotes the success probability in the same setting but when you are told $j$, averaged over $p_j$. 
Since our goal is to upper bound the success probability $p_{suc}=p_{suc}(\rho)$ with $\rho$ a separable state in terms of the MoE game success probability, the above shows that it suffices to focus on product states. 

Now we can focus on NLQC protocols using only product states as a resource. 
In this case, given input $Q$ on the left, Alice's most general strategy is to apply a channel $\mathcal{N}_{Q\rightarrow AB}$, and then to keep $A$ and send $B$ to Bob.
Meanwhile Bob holds only classical input, and a fixed quantum state. 
His most general strategy is to prepare a quantum state $\sigma_{A'B'}^\theta$ and send $A'$ to Alice and keep $B'$.

In fact, we can simplify Bob's strategy: we have Alice prepare all of the states $\{\sigma^{\theta'}\}_{\theta'}$ and distribute them. 
Bob need only send $\theta$ to both sides, then Alice can trace out the states with $\theta'\neq\theta$, reproducing the effect of the earlier protocol. 
This state preparation channel can be absorbed into the definition of $\mathcal{N}_{Q\rightarrow AB}$.

We see that protocols consisting of Alice applying a channel $\mathcal{N}_{Q\rightarrow AB}$ and Bob forwarding $\theta$ to both sides fully capture all possible protocols. 
A final transformation will show that these protocols also define MoE game strategies. 
Consider having the referee prepare a maximally entangled state on $\bar{Q}Q$, and then prepare her input on $Q$ by measuring the $\bar{Q}$ system in the $\theta$ basis. 
This leaves $Q$ in one of the states $\ket{\psi^\theta_i}$ with $i$ determined by the measurement outcome, so this produces the correct output. 
But now consider having the referee hand $Q$ to Alice before making her measurement, and then later measure to determine $\theta$. 
Since actions on $\bar{Q}$ and $Q$ commute, this cannot change the success probability of the game. 
We will consider the game from this perspective. 

Now consider Alice's operation. 
She is given one end of a maximally entangled state $\ket{\Psi^+}_{\bar{Q}Q}$, and can act on $Q$ to prepare a state on $\bar{Q}AB$. 
Then, $A$ and $B$ get split up and then at that point measured in a way that can depend on $\theta$, and $\bar{Q}$ is always measured in the $\theta$ basis. 
This is exactly the scenario in an MoE game, where the measurements in the MoE game are chosen to correspond to the set of bases labelled by $\theta$ in the measurement NLQC. 
This means the success probability in the measurement NLQC is upper bounded by the MoE game probability, as claimed. 

We can summarize our result by saying that
\begin{align}\label{eq:pbound}
    p_{suc}(N^{n,\delta}, \sigma) \leq p_{suc}(G^{n,\delta})
\end{align}
where $p_{suc}(N^{n,\delta}, \sigma)$ denotes the success probability of the NLQC $N$, repeated $n$ times and allowing a fraction $\delta$ of outcomes to be in error, and where $p_{suc}(G^{n,\delta})$ denotes the success probability in the monogamy game with the same choice of measurement bases. 
The state $\sigma$ is the resource state used in the NLQC, and can be any separable state. 
The MoE game and measurement NLQC should consider the same set of bases.
Considering the BB84 monogamy game in particular, we obtain the following. 
\begin{lemma}\label{lemma:separableNLQCbound}
    The success probability of completing the BB84 measurement NLQC using any separable state $\sigma$ satisfies
    \begin{align}
        p_{suc}(N_{BB84}^{n,\delta},\sigma) \leq \left(2^{h(\delta)}\cos^2\left( \frac{\pi}{8}\right) \right)^n
    \end{align}
\end{lemma}
This follows from equation \eqref{eq:pbound} and lemma \ref{thm:BB84MoEupperbound}.

\section{Entanglement lower bounds from probability upper bounds}\label{sec:robustness}

The upper bound on success probability in the measurement NLQC in the zero entanglement case means that to achieve a probability near 1, we need some entanglement. 
To make quantitative how much entanglement is needed, there are various approaches one can follow. 
Maybe the most elementary, used for example in \cite{may2020holographic,may2022connected}, is to observe that a resource state that works well for the NLQC can be easily distinguished from one that works poorly (by using the success in the NLQC as a test to distinguish states). 
A second technique makes use of the robustness of entanglement.
We treat each approach below. 

\vspace{0.2cm}
\noindent \textbf{Distinguishing game technique:} We give a lower bound using the idea that we can treat the NLQC protocol as a method of distinguishing quantum states.
The lower bound is on the relative entropy of entanglement, defined by
\begin{align}
    E_R(A:B)_{\rho}=\min_{\sigma\in \mathcal{S}} D(\rho_{AB}||\sigma_{AB})
\end{align}
where the minimization is over separable states. 

In this technique, we will study lower bounds for completing the $N_{BB84}^{n,\delta}$ measurement NLQC with an exponentially small error probability. 
This is motivated as follows. 
If we assume each round of the NLQC can be implemented with probability of failure $\epsilon<\delta$, then the probability of having a fraction of rounds larger than $\delta$ fail will be exponentially small. 
In fact, by Hoeffding's inequality this probability can be seen to be as small as $e^{-2n(\epsilon-\delta)^2}$. 
Thus we state our results in terms of an overall failure probability of $e^{-2n(\epsilon-\delta)^2}$.
We also define $\beta=\cos^2(\pi/8)$.

\begin{lemma}(Relative entropy of entanglement lower bound)     
Completing the $N_{BB84}^{n,\delta}$ measurement NLQC with probability $1-e^{- 2 n(\delta - \epsilon)^2}$ requires relative entropy of entanglement
\begin{align}
    E_R(L:R)_\rho \geq - n \log_2(2^{h(\delta)} \beta) - 1 + O\left( (2^{h(\delta)} \beta)^n, e^{-2 n (\delta - \epsilon)^2} (2^{h(\delta)} \beta)^{-n} \right).
\end{align}
\end{lemma}
\begin{proof}
    The one-norm distance between two states $\rho$ and $\sigma$ is defined by
    \begin{equation}
        \lVert \rho - \sigma \rVert_{1} = \tr(|\rho - \sigma|).
    \end{equation}
    If you are given either state $\rho$ or $\sigma$ with probability $1/2,$ and are tasked with guessing which state you have been given, then your maximum probability of success optimized over all strategies is \cite{helstrom1969quantum,wilde2013quantum}
    \begin{equation} \label{eq:trace-distance-formula}
        p_{\text{dist}}(\rho, \sigma)
            = \frac{1}{2} + \frac{1}{4} \lVert \rho - \sigma \rVert_{1}.
    \end{equation}
    Given a pair of quantum tasks $\mathbf{T}_{\rho}$ and $\mathbf{T}_{\sigma},$ with the only difference being that $\mathbf{T}_{\rho}$ has resource state $\rho$ while $\mathbf{T}_{\sigma}$ has resource state $\sigma,$ we can devise a strategy for distinguishing $\rho$ and $\sigma$ as follows.
    We pick an optimal strategy for completing the task $\mathbf{T}_{\rho},$ and perform that strategy.
    If we succeed at our task, we guess that we were given the state $\rho$.
    If we fail, we guess that we were given the state $\sigma.$
    The probability of successfully distinguishing the states using this strategy is
    \begin{equation} \label{eq:task-distinguishing-probability}
        \frac{1}{2} \text{Prob}(\text{success} | \rho) + \frac{1}{2} (1 - \text{Prob}(\text{success} | \sigma)) \leq p_{\text{dist}}(\rho, \sigma).
    \end{equation}
    Since the strategy we choose for completing the task is optimal for $\mathbf{T}_{\rho}$ but potentially suboptimal for $\mathbf{T}_{\sigma},$ we have
    \begin{align}
        \text{Prob}(\text{success} | \rho)
            & = p_{\text{suc}}(\mathbf{T}_{\rho}), \\
        \text{Prob}(\text{success} | \sigma)
            & \leq p_{\text{suc}}(\mathbf{T}_{\sigma}).
    \end{align}
    Combining these statements with inequality \eqref{eq:task-distinguishing-probability} and equation \eqref{eq:trace-distance-formula} gives
    \begin{equation}
         p_{\text{suc}}(\mathbf{T}_{\rho}) - p_{\text{suc}}(\mathbf{T}_{\sigma}) \leq \frac{1}{2} \lVert \rho - \sigma \rVert_{1}.
    \end{equation}
    Making the substitution ``$\rho \leftrightarrow \sigma$'' everywhere in the above discussion gives an analogous inequality that, combined with this one, becomes
    \begin{equation} \label{eq:pdistandTD}
         |p_{\text{suc}}(\mathbf{T}_{\rho}) - p_{\text{suc}}(\mathbf{T}_{\sigma})| \leq \frac{1}{2} \lVert \rho - \sigma \rVert_{1}.
    \end{equation}

    Trace distance can be related to the relative entropy as follows. 
    First, recall that the trace distance is related to the fidelity by the Fuchs--van de Graaf inequality (equation \ref{eq:FVdG}), which leads to
    \begin{equation}
        \lVert \rho - \sigma \rVert_{1} \leq 2 \sqrt{1 - F(\rho, \sigma)}.
    \end{equation}
    Fidelity is related to relative entropy by
    \begin{equation}
        - \log_2 F(\rho, \sigma) \leq D(\rho \Vert \sigma),
    \end{equation}
    which can be seen from the fact that both sides of this equation are sandwiched R\'{e}nyi relative entropies \cite{muller2013quantum, wilde2014strong} --- the left is $\alpha=1/2$ and the right is $\alpha=1$ --- and these are monotonically increasing in $\alpha$.
    Combining these inequalities we have
    \begin{equation}
        D(\rho||\sigma) \geq - \log_{2} (1 - \lVert \rho - \sigma \rVert_{1}^2 /4).
    \end{equation}
    Using equation \eqref{eq:pdistandTD} to express this in terms of success probabilities, we have
    \begin{equation} \label{eq:Ilowerboundunfinished}
        D(\rho||\sigma) \geq - \log_{2} (1 - |p_{\text{suc}}(\mathbf{T}_{\rho}) - p_{\text{suc}}(\mathbf{T}_{\sigma})|^2).
    \end{equation}
    
    Now, recall our assumption
    \begin{align}
        p_{\text{suc}}(\mathbf{T}_{\rho_{LR}}) \geq 1 - e^{-2 n (\delta - \epsilon)^2},
    \end{align}
    and that from lemma \ref{lemma:separableNLQCbound} we have, for any separable state $\sigma$,
    \begin{align}
        p_{\text{suc}}(\mathbf{T}_{\sigma}) \leq (2^{h(\delta)}\beta)^n.
    \end{align}
    Inserting these into \eqref{eq:Ilowerboundunfinished} and expanding at large $n$ using the assumption $2^{h(\delta)} \beta > e^{-2 (\delta - \epsilon)^2}$, we obtain 
    \begin{align}
        D(\rho||\sigma) \geq - n \log_2(2^{h(\delta)} \beta) - 1 + O\left( (2^{h(\delta)} \beta)^n, e^{-2 n (\delta - \epsilon)^2} (2^{h(\delta)} \beta)^{-n} \right).
    \end{align}
    for $\sigma$ any separable state, and $\rho$ any state that gives a success probability of at least $1-e^{-2n(\delta-\epsilon)^2}$. 
    Taking the minimum over all choices of separable state $\sigma$ on both sides of this expression returns the desired bound on  $E_R$. 
\end{proof}

\vspace{0.2cm}
\noindent \textbf{Robustness of entanglement approach:} A more immediate method to translate success probabilities into entanglement lower bounds uses the \emph{robustness of entanglement} \cite{vidal1999robustness}, defined as follows. 
\begin{align}
    \mathcal{R}(\rho)=\min_{\sigma\in \mathcal{S}} \left\{ s\geq0: \frac{1}{1+s}\rho + \frac{s}{1+s}\sigma \in \mathcal{S}\right\}
\end{align}
Here $\mathcal{S}$ denotes the separable states. 
The robustness of entanglement asks how much we need to mix $\rho$ with a separable state to produce another separable state. 
For $n$ EPR pairs, the robustness of entanglement is $2^n-1$. 
This motivates defining the log-robustness as
\begin{align}
    r(\rho)=\log\left(\mathcal{R}(\rho)+1\right)
\end{align}
so that $r(\rho)=n$ when evaluated on $n$ EPR pairs.

Now we can prove the following lower bound. 

\begin{theorem}
    Completing the BB84 measurement NLQC $N^{n,\delta}_{BB84}$ with probability $1-\gamma$\footnote{Note that this means the probability that more than $n\delta$ rounds of the NLQC fail is $\gamma$. If we have some constant error rate $\epsilon$ for each round considered separately $\gamma$ will be exponentially small in $n$, but we don't need this in the technique considered here.} requires a resource state $\rho_{LR}$ with log-robustness of entanglement $r(\rho)$ lower bounded according to
    \begin{align}
        r(\rho)\geq n \left(-\log\left( 2^{h(\delta)}\beta\right)\right) + \log(1-\gamma)
    \end{align}
    where $\beta=\cos^2(\pi/8)$.
\end{theorem}
\begin{proof}
    Let $s$ be the robustness of entanglement, so that $s$ is the smallest positive real number such that
    \begin{align}
        \kappa = \frac{1}{1+s}\rho + \frac{s}{1+s}\sigma
    \end{align}
    is separable, where $\sigma$ is any choice of separable state. 
    Then consider that
    \begin{align}\label{eq:ppbound}
        p_{suc}(N_{BB84}^{n,\delta},\kappa) \geq \frac{1}{1+s} p_{suc}(N_{BB84}^{n,\delta},\rho)
    \end{align}
    This holds because one strategy for implementing the NLQC given input $\kappa$ is to run the strategy for $\rho$, which succeeds with probability $p_{suc}(N_{BB84}^{n,\delta},\rho)$ if given input $\rho$. 
    Since $\kappa$ can be seen as an ensemble where you are given the state $\rho$ with probability $1/(1+s)$, this strategy succeeds with at least probability $\frac{1}{1+s} p_{suc}(N_{BB84}^{n,\delta},\rho)$ when given $\kappa$. 

    Since $\kappa$ is separable, the bound in lemma \ref{lemma:separableNLQCbound} gives an upper bound on $p_{suc}(N_{BB84}^{n,\delta},\kappa)$. 
    Further, by assumption we have that $p_{suc}(N_{BB84}^{n,\delta},\rho)\geq 1-\gamma$. 
    Combining both these statements with equation \eqref{eq:ppbound}, we get
    \begin{align}
        \left(2^{h(\delta)}\beta \right)^n \geq \frac{1}{1+s} (1-\gamma)
    \end{align}
    Rearranging to bound $1+s$ and recalling that $r(\rho)=\log(s+1)$ leads to the claimed lower bound. 
\end{proof}

\section{History and further reading}

The original reduction to what amounts to a MoE game (it was not given that name at the time) appears in \cite{buhrman2014position}.
The bound they gave was based on entropic arguments, and leads to a somewhat weaker bound. 
The MoE game was defined formally and given the bound we quote in \cite{tomamichel2013monogamy}. 

Notice that the MoE game setting is somewhat more general than is needed to lower bound measurement based NLQC tasks. 
A more minimal variant of this task, more tightly related to NLQC, was studied in \cite{george2025orthogonality}.

\chapter{Application: Entanglement and spacetime in quantum gravity}\label{chapter:gravity}

\minitoc

One of the most surprising applications of NLQC has been to the understanding of quantum gravity, and in particular the understanding of the role of entanglement in the emergence of spacetime from quantum mechanical degrees of freedom. 
We review these developments here. 

\section{Gravity and holography}

\subsection{Black holes and the holographic principle}\label{sec:blackhole}

A black hole is a region in spacetime out of which it is impossible to escape, at least without travelling faster than the speed of light. 
A cartoon of a black hole horizon, the surface that defines the boundary of a black hole, is shown in figure \ref{fig:blackhole}. 
There, we see the future light cone of a point tilt as it approaches the horizon. 
Once inside the horizon, the light cone has fully moved onto its side: signals can only go \emph{into} the black hole. 

\begin{figure}
    \centering
    \begin{tikzpicture}

        \node[above] at (0,5) {horizon};
        \draw[dashed] (0,0) -- (0,5);

        \node[above] at (5,5) {singularity};
        \draw[thick] (5,0) -- (5,5);

        \begin{scope}[xshift=-4cm, yshift=2cm,rotate=0]
            \draw[->] (0,0) -- (0.6,0.6);
            \draw[->] (0,0) -- (-0.6,0.6);
            \node[below] at (0,0) {$p_1$};
            \draw plot [mark=*, mark size=1.5] coordinates{(0,0)};
        \end{scope}

        \begin{scope}[xshift=-2cm, yshift=2cm,rotate=-20]
            \draw[->] (0,0) -- (0.6,0.6);
            \draw[->] (0,0) -- (-0.6,0.6);
            \node[below] at (0,0) {$p_2$};
            \draw plot [mark=*, mark size=1.5] coordinates{(0,0)};
        \end{scope}

        \begin{scope}[xshift=0cm, yshift=2cm,rotate=-45]
            \draw[->] (0,0) -- (0.6,0.6);
            \draw[->] (0,0) -- (-0.6,0.6);
            \node[left] at (0,0) {$p_3$};
            \draw plot [mark=*, mark size=1.5] coordinates{(0,0)};
        \end{scope}

        \begin{scope}[xshift=2cm, yshift=2cm,rotate=-65]
            \draw[->] (0,0) -- (0.6,0.6);
            \draw[->] (0,0) -- (-0.6,0.6);
            \node[below] at (0,0) {$p_4$};
            \draw plot [mark=*, mark size=1.5] coordinates{(0,0)};
        \end{scope}
        
    \end{tikzpicture}
    \caption{A cartoon of a black hole horizon, shown as the dashed line. The trajectory of light rays starting from points $p_1$, $p_2$, $p_3$, and $p_4$ are shown. Far from the horizon, light rays can travel away from the black hole. As the point approaches the horizon, the light rays tilt toward the hole. Inside the horizon, both light rays point inward, towards the black hole singularity.}
    \label{fig:blackhole}
\end{figure}

Astrophysically, black holes form via the collapse of stars. 
Within general relativity however, our modern theory of classical gravity, we can have black hole solutions without any matter present at all --- they are solutions purely within the confines of general relativity. 
An interesting aspect of black hole solutions in general relativity is that static black holes\footnote{Roughly, this means black holes that are not changing in time --- we can consider any black hole and then wait a long time until it settles down, and it will become static in this sense.} are completely characterized by just three numbers: their charge, angular momentum, and mass. 

This situation is very different than ordinary matter. 
Consider for example a box filled with gas. 
We can specify the mass $M$, angular momentum $J$, and charge of the gas $Q$, but by doing so we certainly haven't fully described the gas. 
Instead, the particles in the gas can be in many different possible microstates, all of which correspond to the same value of $M$, $J$, and $Q$. 
In thermodynamics, we describe this by saying the $M$, $J$, $Q$ numbers are thermodynamic quantities that describe the macroscopic state of the system, and then we have many different possible microstates consistent with a given macrostate. 
We count the number of available microstates using the entropy $S$, which is defined as the logarithm of the number of available microstates, 
\begin{align}
    S = \log \Omega(M, J, Q).
\end{align}
Because our black hole solution only has these macroscopic parameters present, it is suggestive that general relativity is only giving a macroscopic description of the black hole. 
Presumably a microscopic description would be provided by a theory of quantum gravity. 

In the absence of a theory of quantum gravity to provide this microscopic description, we can try and understand how the entropy of the black hole is related to the other parameters, and in particular to the mass. 
We can hope to extract the entropy from the classical theory of gravity just like we can define and study the entropy in thermodynamics, even before we know anything about statistical mechanics and the microscopic description of the microstates.
Hopefully, at least knowing how many microstates the quantum gravity theory has will provide some hint towards developing such a theory. 

An important hint as to how the classical theory describes entropy is given by a theorem due to Hawking, who proved that the area of black holes always increases. 
For example, if we have two black holes and we let them merge into each other, the remaining single black hole will have an area larger than the sum of the original two. 
Or less dramatically, if we take some matter and throw it into our black hole the area will increase. 
This monotonically increasing area might remind you of the second law of thermodynamics, which says that in any closed system, entropy increases. 
This raises the area of the black hole as a possible candidate for the entropy. 

The idea that the black hole area counts the entropy received a beautiful validation when Hawking showed that black holes radiate, and do so at a temperature related to the entropy by the usual Clausius statement from thermodynamics, $TdS=dE$, where $dE$ is the change in the energy of the black hole and $dS$ the change in the entropy. 
This calculation also lets us fix the constants, and we find
\begin{align}
    S_{bh}=\frac{\text{Area}}{4G_N}.
\end{align}
Here $G_N$ is Newton's constant, the same constant that appears in the famous gravitational force law $F=\frac{G_Nm_1m_2}{r^2}$, and we've set all other physical constants (Planck's constant, the speed of light, and Boltzmann's constant) to be 1. 

The black hole entropy being proportional to the \emph{area} of the black hole, rather than a volume, is really strange. 
For ordinary matter we always find an entropy related to the volume, which happens for a very simple reason. 
To see this, consider a box of gas with entropy $S_T$. 
We can divide this box into small unit cells of volume $\Delta V$. 
Let the number of microstates in each cell be $\Omega$. 
Then the total number of microstates of the gas, assuming the small cells are independent, is just
\begin{align}
    \Omega_T = \Omega^{V/\Delta V}
\end{align}
and the total entropy is
\begin{align}
    S_T = \frac{V}{\Delta V} \log \Omega
\end{align}
which is proportional to the volume. 
One can see that this will happen whenever the microstates of the small cells are independent, so that the total number of states is $\Omega^{V/\Delta V}$. 

Apparently, ordinary matter and black holes work very differently. 
But in a world with gravity, we can always turn ordinary matter into black holes: we need only add matter (and hence add entropy) and our ordinary system will collapse to a black hole. 
For large systems, where the volume is much larger than the area, if this is to avoid decreasing entropy (and violating the second law) it must have been the case that the ordinary matter too had only an area's worth of entropy. 

This is a radical conclusion: all matter should have at most an area $A/4G_N$ worth of entropy, and hence $\Omega$, the number of microstates, is proportional to $2^{A/4G_N}$.
This is a hint from black hole physics that quantum gravity is \emph{holographic}: the fundamental theory should have only an area's worth of degrees of freedom. 
This idea was proposed by 't Hooft \cite{hooft1993dimensional} and then promoted and developed by Susskind \cite{susskind1995world}. 

While the holographic principle was argued for on the basis of the physics of black holes, we can also find that something similar is suggested by the existence of NLQC protocols for all quantum operations, and the consequent insecurity of QPV.
To appreciate this, let's imagine we have a $2+1$ dimensional spacetime, and we draw a cylinder in that spacetime. 
We can imagine an external observer trying to probe the cylinder from outside. 
They are sending in signals from far away and looking for certain responses, trying to check if there is something happening inside the cylinder or if their observations can be explained by processes happening on the surface of the cylinder. 
The insecurity of quantum position verification, which comes from the existence of NLQC protocols, says that in principle it is possible to have a process occurring on the cylinder that simulates, to the far away observer, any physics occurring inside.
The outside observer can't tell the difference between the $2+1$ dimensional physics occurring inside the cylinder and the $1+1$ dimensional physics on its boundary. 
This suggests there is some kind of equivalence possible between the higher and lower dimensional descriptions.\footnote{Something unnatural about this setup is that the $1+1$ dimensional physics either needs to be highly non-local, or needs to send signals through the $2+1$ dimensional space. 
This is just like in the set-up of figure \ref{fig:1dQPV} where the communication goes through the interior region. 
We will see in the next section though that this issue is resolved when we consider actual models of quantum gravity.} 

\subsection{The AdS/CFT correspondence}\label{sec:AdSCFT}

We argued at the end of the last section for an intuitive connection between NLQC and the holographic principle in quantum gravity. 
We can make this more precise in concrete models of quantum gravity. 
Specifically, we will consider the AdS/CFT correspondence, which is perhaps the best studied model. 

\begin{figure}
    \centering
    \subfloat[\label{subfig:B84bulk}]{
    \tdplotsetmaincoords{10}{0}
    \begin{tikzpicture}[scale=1.0,tdplot_main_coords]
    \tdplotsetrotatedcoords{0}{20}{0}
    \draw (-2,0,0) -- (-2,4,0);
    \draw (2,0,0) -- (2,4,0);
    
    \begin{scope}[tdplot_rotated_coords]
    \begin{scope}[canvas is xz plane at y=0]
    \draw (0,0) circle [radius=2];
    \draw[->,domain=0:70] plot ({2.5*cos(\x)},{2.5*sin(\x)});
    \node[above,left] at ({3*cos(75)},{3*sin(75)}) {$\theta$};
    \end{scope}
    
    \begin{scope}[canvas is xz plane at y=4]
    \draw (0,0) circle [radius=2];
    \end{scope}

    \draw[->] (-3,0,0) -- (-3,0.75,0);
    \node[above] at (-3,0.75) {$t$};

    \draw[->] (0,0,0) -- (1,0,0);
    \node[right] at (1,0,0) {$r$};
    
    \end{scope}
    \end{tikzpicture} 
    }
    \hfill
    \subfloat[\label{subfig:B84boundary}]{
    \begin{tikzpicture}[scale=1.25]
    
    \draw (-2.5,0) -- (2.5,0) -- (2.5,2.5) -- (-2.5,2.5) -- (-2.5,0);

    \draw[->] (-3,0) -- (-3,0.75);
    \node[above] at (-3,0.75) {$t$};

    \draw[->] (-2.5,-0.5) -- (-1.75,-0.5);
    \node[right] at (-1.75,-0.5) {$\theta$};
    
    \end{tikzpicture}
    }
    \caption{a) The cylinder representation of $2+1$ dimensional AdS space. Time runs upwards in the picture, while the radial and angular coordinates are in the plane. b) The boundary of $2+1$ dimensional AdS space.}
    \label{fig:AdScylinder}
\end{figure}

The AdS/CFT correspondence asserts that gravity in a special kind of spacetime called \emph{anti de Sitter space} (AdS) is equivalent to a lower dimensional theory that lives at the boundary of that spacetime. 
We will unpack what is meant by `equivalent' here in a moment, but first let's understand a bit about the geometry of AdS space. 
We can visualize AdS space as the cylinder drawn in figure \ref{fig:AdScylinder}.
In the picture the time direction goes upwards, and the two directions in the plane are spatial directions. 
In AdS/CFT we can consider any number of dimensions, but here we will focus on $2+1$ dimensional AdS space, corresponding to $2$ spatial and $1$ time dimension. 
We will unpack slowly what this cylinder picture means. 

\begin{figure}
    \centering
    \includegraphics[scale=0.25]{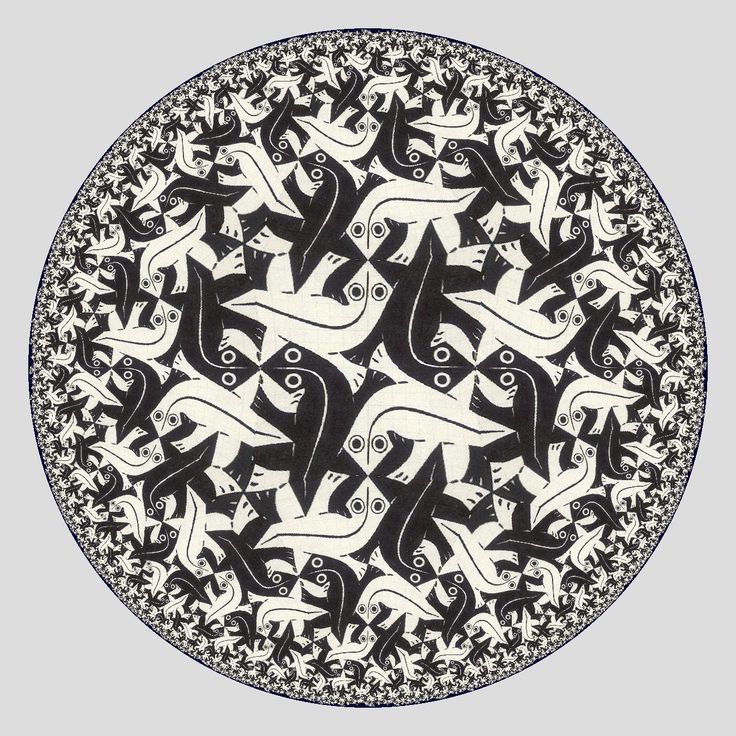}
    \caption{A drawing by M. C. Escher representing the geometry of a constant time slice of AdS space. Each lizard is of unit physical size, but the disk representation of AdS distorts their sizes. To travel from the center of the disk to the boundary involves crossing an infinite number of lizards, corresponding to an infinite physical distance.}
    \label{fig:lizards}
\end{figure}

Let's first focus on a constant time slice of AdS space. 
This constant time slice has constant negative curvature. 
To understand what this means, consider that if you draw a triangle on a sheet of paper and add up the angles in the triangle you get $180^\circ$. 
This corresponds to the sheet of paper being flat. 
If you draw a triangle on a sphere you get more than $180^\circ$, so we call the sphere positively curved, and if you draw a triangle on a horse-saddle you get less than $180^\circ$ and we call the saddle negatively curved. 
Aside from being negatively curved, the constant time slice also has infinite spatial extent --- if you're standing in AdS space, space extends outward from you in all directions forever. 
However, we can draw our constant time, negatively curved, space on our finite page by choosing our coordinates in a way that compresses distances more and more as we continue outwards. 
This geometry is represented in figure \ref{fig:lizards}, in a famous drawing by the artist M. C. Escher. 
In the picture physical distances are represented by lizards: if you live in the spacetime and measure distances with a ruler, every lizard is of equal size. 
The disk picture, which we are using to represent an infinite space in a finite picture, necessarily distorts how it represents some of the lizards. 
In particular as you approach the boundary of the disk the lizards become very small, so that an infinite distance as measured by a ruler (an infinite number of lizards) can be fit into the picture. 

The Escher drawing makes apparent a notion of a boundary for the AdS space. 
In the picture this is the circle where the disk ends. 
Mathematically, you can describe the boundary in terms of the directions you can travel to infinity in. 
In fact, we can think of this space of directions as a manifold in its own right, equipped with its own geometry and living ``at infinity'' from the perspective of AdS. 
In AdS/CFT, the quantum theory that describes gravity in AdS space lives on this boundary geometry. 

A further convenient fact about AdS space is shown in figure \ref{fig:lightrays}, where we see that a light ray that travels through the bulk AdS space crosses the space in the same amount of time as a light ray that travels through the boundary. 
For more general spaces that are asymptotically AdS (where we have some matter sitting in the bulk, so the metric looks like the AdS space above only as you go out near the boundary) bulk light rays are never faster than light rays around the boundary. 
This is an important geometrical fact: if light rays travelled faster through the bulk, the boundary theory would have to include non-local interactions so that it can reproduce the effect of these interactions. 

\begin{figure}
    \centering
    \subfloat[\label{subfig:fastbulkray}]{
    \tdplotsetmaincoords{10}{0}
    \begin{tikzpicture}[scale=1.0,tdplot_main_coords]
    \tdplotsetrotatedcoords{0}{-25}{0}
    \draw (-2,0,0) -- (-2,4,0);
    \draw (2,0,0) -- (2,4,0);
    
    \begin{scope}[tdplot_rotated_coords]
    \begin{scope}[canvas is xz plane at y=0]
    \draw (0,0) circle [radius=2];
    \end{scope}
    
    \begin{scope}[canvas is xz plane at y=4]
    \draw (0,0) circle [radius=2];
    \end{scope}

    \draw[red,mid arrow,->] (-2,0,0) -- (2,4,0);

    \draw[red,domain=0:180,mid arrow,->] plot ({-2*cos(-\x)},\x/45, {2*sin(-\x)});
    
    \end{scope}
    \end{tikzpicture} 
    }
    \hfill
    \subfloat[\label{subfig:slowbulkray}]{
    \tdplotsetmaincoords{10}{0}
    \begin{tikzpicture}[scale=1.0,tdplot_main_coords]
    \tdplotsetrotatedcoords{0}{-25}{0}
    \draw (-2,0,0) -- (-2,4,0);
    \draw (2,0,0) -- (2,4,0);
    
    \begin{scope}[tdplot_rotated_coords]
    \begin{scope}[canvas is xz plane at y=0]
    \draw (0,0) circle [radius=2];
    \end{scope}
    
    \begin{scope}[canvas is xz plane at y=4]
    \draw (0,0) circle [radius=2];
    \end{scope}

    \draw[red,mid arrow,->] (-2,0,0) -- (2,4,0);

    \draw[red,domain=0:180,mid arrow,->] plot ({-2*cos(-\x)},\x/50, {2*sin(-\x)});
    
    \end{scope}
    \end{tikzpicture}
    }
    \caption{a) Light rays through an AdS space with no matter present. Light rays through the center of the bulk and around the boundary take the same amount of time. b) Light rays through an AdS space with matter present. A light ray extending through the center of the bulk is slower than one travelling through the boundary.}
    \label{fig:lightrays}
\end{figure}

At this point, we can start to understand why holography in AdS space is particularly easy to make sense of. 
In our flat space cylinder mentioned at the end of the last section, the boundary could only simulate the bulk if we let signals pass through the bulk, so that we weren't really getting a boundary-only description. 
Now, this issue is resolved, and the door is open to a local boundary theory that fully describes the bulk physics. 

So far we haven't said much about the quantum mechanical theory that lives in the boundary. 
In fact, we will get away with not needing to describe this in any detail. 
It'll suffice to note that the theory is \emph{only} quantum mechanical, meaning the spacetime it lives on is fixed, and that it is local, so it has a good notion of causality and light cones. 
In a bit more detail the boundary theory is a conformal field theory (CFT), which is somewhat like the usual field theories that describe particle physics, but which has additional (conformal) symmetry. 

Let's now understand a bit more what it means to say the bulk AdS and boundary CFT theories are equivalent. 
We can describe the two theories as a Hilbert space along with a Hamiltonian, 
\begin{align}
    \mathcal{H}_{CFT}, H_{CFT} \,\,\,\,\,\,\,\text{and}\,\,\,\,\,\,\, \mathcal{H}_{AdS}, H_{AdS}.
\end{align}
One way to formally state the equivalence of these theories is to specify that there exists a unitary mapping
\begin{align}
    V: \mathcal{H}_{CFT}\rightarrow \mathcal{H}_{AdS}
\end{align}
which preserves time evolution, 
\begin{align}
    V^\dagger H_{AdS}V = H_{CFT}.
\end{align}
We believe such a mapping exists between the CFT and AdS Hilbert spaces, though we don't know how to write down the full map exactly. 
Instead, we have various objects that we know how to equate on either side of the duality, and some settings where we can define the map implicitly in terms of an equality of partition functions. 

\subsection{Entanglement and spacetime}\label{sec:entanglementandspacetime}

We discussed in section \ref{sec:blackhole} the understanding of the black hole area as giving the (thermodynamic) entropy of the hole. 
We can also take the area to be the von Neumann entropy of the hole, since the thermodynamic and von Neumann entropy agree for thermal states, and the black hole is in a thermal state. 
The black hole entropy formula naturally raises a deep question: what are the degrees of freedom which are in the thermal state? 
As we discussed, general relativity doesn't describe these degrees of freedom, but instead only captures certain coarse quantities. 
From the perspective of AdS/CFT, we expect that the black hole entropy is the entropy of the boundary degrees of freedom in the CFT. 
In fact, AdS/CFT tells us exactly that: the thermal state of the CFT corresponds to a bulk state with a black hole. 
Thus AdS/CFT provides the microscopic description of this thermal system that was missing from general relativity. 

Inspired by this, it is natural to ask about the von Neumann entropy of other CFT states, and ask if they too have an entropy which is described geometrically in terms of bulk areas. 
Again, AdS/CFT is able to point the way here.
The Ryu-Takayanagi formula is a broad generalization of the black hole entropy formula. 
It states that the von Neumann entropy of a boundary subregion $A$ is given by
\begin{align}
    S(A) = \min_{\gamma_A \in \text{Hom(A)}}  \frac{\text{Area}[\gamma_A]}{4G_N}.
\end{align}
Understanding this formula requires some unpacking. 
$A$ denotes a subregion of the CFT, whose entropy we are interested in calculating. 
To calculate it using bulk data, the formula instructs us to look at surfaces which are \emph{homologous} to $A$, and pick the one with minimal area. 
Surface $\gamma_A$ is said to be homologous to a boundary region $A$ if $A \cup \gamma_A$ forms the boundary of a region, which will be called the entanglement wedge of $A$, $\mathcal{E}_W(A)$. 

In $2+1$ dimensional space, minimal surfaces are geodesics. 
Let's consider what these look like in AdS space.
Consider the lizard representation by Escher of a constant time slice of AdS space. 
We want to draw a minimal surface enclosing the region $A$ shown. 
To do this we should take our surface in from the boundary as quickly as possible, since the lizards there are tightly packed and hence count for a large physical distance. 
As we go in we can start bending the curve over, until eventually heading back out to the boundary and meeting the other side of $A$. 
Solving for the minimal surface here precisely reveals that the minimal surface is a semi-circle. 
Thus the area of this semi-circle computes the entropy of the region $A$.

The RT formula is actually a generalization of the black hole entropy formula. 
If we take the CFT to be in a thermal state and choose $A$ to be the entire CFT, the homology condition is satisfied by taking the surface to be the black hole horizon. 
Indeed this turns out to be the minimal homologous surface, so the RT formula correctly computes the entropy. 

\begin{figure}
    \centering
    \begin{subfigure}{0.45\textwidth}
    \begin{tikzpicture}[scale=0.8]
    
    \draw[thick] (0,0) circle (3);
    
    \draw [domain=60:120,fill=lightgray,opacity=0.8] plot ({3*cos(\x)}, {3*sin(\x)}) -- (-1.5, 2.60) to [out=-60,in=-120] (1.5, 2.60);
     \draw [domain=-60:-120,fill=lightgray,opacity=0.8] plot ({3*cos(\x)}, {3*sin(\x)}) -- (-1.5, -2.60) to [out=60,in=120] (1.5, -2.60);
    
    \draw [green,ultra thick,domain=60:120] plot ({3*cos(\x)}, {3*sin(\x)});
    \draw [green,ultra thick,domain=-60:-120] plot ({3*cos(\x)}, {3*sin(\x)});
    
    \draw[blue, thick] (1.5, 2.60) to [out=-120,in=-60] (-1.5, 2.60);
    \draw[blue, thick] (1.5, -2.60) to [out=120,in=60] (-1.5, -2.60);
    
    \draw[<->,gray,domain=59:-59] plot ({3.3*cos(\x)},{3.3*sin(\x)});
    \node[right,gray] at (3.3,0) {$2\alpha$};
    
    \draw[<->,gray,domain=61:119] plot ({3.3*cos(\x)},{3.3*sin(\x)});
    \node[above,gray] at (0,3.3) {$x$};
    
    \end{tikzpicture}
    \caption{}
    \label{fig:disconnectedsurfaces}
    \end{subfigure}
    \hfill
    \begin{subfigure}{0.45\textwidth}
    \begin{tikzpicture}[scale=0.8]
    
    \draw [lightgray,domain=80:20,fill=lightgray,opacity=0.8] plot ({3*cos(\x)}, {3*sin(\x)}) -- (2.82,1.03) to [out=-160,in=+90] (2.25,0) -- (0.205,0) to [out=90,in=-100] (0.52,2.96);
    \draw [lightgray,domain=-80:-20,fill=lightgray,opacity=0.8] plot ({3*cos(\x)}, {3*sin(\x)}) -- (2.82,-1.03) to [out=160,in=-90] (2.25,-0.012) -- (0.205,-0.012) to [out=-90,in=100] (0.52,-2.96);   
    
    \draw[thick] (0,0) circle (3);
    \draw [green,ultra thick,domain=20:80] plot ({3*cos(\x)}, {3*sin(\x)});
    \draw [green,ultra thick,domain=-20:-80] plot ({3*cos(\x)}, {3*sin(\x)});
    
    \draw[blue,thick] (2.82,1.03) to [out=-160,in=+160] (2.82,-1.03);
    \draw[blue,thick] (0.52,2.96) to [out=-100,in=100] (0.52,-2.96);
    
    \draw[<->,gray,domain=19:-19] plot ({3.3*cos(\x)},{3.3*sin(\x)});
    \node[right,gray] at (3.3,0) {$2\alpha$};
    
    \draw[<->,gray,domain=21:79] plot ({3.3*cos(\x)},{3.3*sin(\x)});
    \node[above right,gray] at (2.12,2.52) {$x$};
    
    \node[above,gray] at (0,3.3) {$ $};
    
    \end{tikzpicture}
    \caption{}
    \label{fig:connectedsurfaces}
    \end{subfigure}
    \caption{Extremal surfaces (shown in blue) for two intervals $R_1$ and $R_2$ (shown in green) of equal size $x$ sitting on a constant time slice of AdS$_{2+1}$. The intervals are separated by an angle $2\alpha$. The entanglement wedge $\mathcal{E}(R_1R_2)$ (shown in grey) is the region whose boundary is the union of the regions $R_1$ and $R_2$ and their minimal surfaces. For large $\alpha$ and small $x$ the entanglement wedge of the region $R_1\cup R_2$ is disconnected, while for small $\alpha$ or large  enough $x$, the entanglement wedge becomes connected, as shown at right. The entanglement wedge being connected indicates the mutual information is $O(1/G_N)$, while a disconnected entanglement wedge indicates the mutual information is $O(1)$.}
    \label{fig:min_surfaces}
\end{figure}

You might be worried that since the distance to the boundary is infinite, the entropy as computed by the RT formula is often infinite. 
In fact that's correct: both sides of the RT formula are divergent when we take $A$ to be a subregion of the CFT. 
To extract a meaningful quantity in this case, one thing we can do is look at a mutual information of two separated regions $A$ and $B$. 
Recall that
\begin{align}
    I(A:B)=S(A)+S(B)-S(AB).
\end{align}
An important feature of this formula is that, if A and B are separated, the divergences in $S(A)+S(B)$ will always cancel with those in $S(AB)$. 
The reason is that the divergences come from the endpoints of the intervals $A$ and $B$ (where the minimal surfaces run to infinity) and the two terms have contributions from the same endpoints. 

In fact the mutual information has an interesting behaviour that will feature in our later discussion. 
In figure \ref{fig:min_surfaces} we consider computing the entropy $S(AB)$ using the RT formula.
There are two locally minimal surfaces homologous to $A\cup B$, which are candidates to be the global minimum.  
First, we can have $\gamma_{A\cup B}' = \gamma_{A}\cup \gamma_{B}$.
This is a candidate minimal surface composed of two pieces, one homologous to $A$ and the other homologous to $B$. 
This is shown in figure \ref{fig:disconnectedsurfaces}.
We call this the disconnected configuration, because the entanglement wedge of $A\cup B$ has two disconnected pieces in this case.
A second candidate minimal surface is of the type shown in figure \ref{fig:connectedsurfaces}, which we will call $\gamma_{A\cup B}$. 
This surface is also composed of two components, but each component is connected across $A$ and $B$.
The entanglement wedge in this case is connected, so we call this the connected configuration.
The RT formula says that we should find the minimal area surface, and use its area as the entropy. 
Thus we have
\begin{align}
    I(A:B)&=S(A)+S(B)-S(AB), \nonumber \\
    &= \frac{1}{4G_N}\left(\text{area}(\gamma_A)+\text{area}(\gamma_B)-\min\{\text{area}(\gamma_{A})+\text{area}(\gamma_B),\text{area}(\gamma_{A\cup B})\} \right) .\nonumber 
\end{align}
Notice that if we are in the regime where the entanglement wedge of $A\cup B$ is disconnected, then we find $I(A:B)=0$. 
Meanwhile if we have that the entanglement wedge of $A\cup B$ is connected, then $I(A:B)=\Theta(1/G_N)$.

\section{The connected wedge theorem}

So far we've reviewed some of the background on AdS/CFT and we've seen that NLQC is, at least at a conceptual level, related to the holographic principle and the AdS/CFT correspondence. 
We're ready now to make the connection between AdS/CFT and NLQC precise and understand the consequences of this connection. 

To do so, we consider the set-up shown in figure \ref{subfig:cylinder}. 
We consider a $2+1$ dimensional AdS space, and we pick four points, which we've labelled $c_0, c_1, r_0,r_1$. 
We've picked these points in a careful way, so that they satisfy certain constraints. 
To describe these, consider two points in spacetime $p$ and $q$. 
We will write $p\rightarrow q$ if it is possible to travel from $p$ to $q$ without ever moving faster than the speed of light. 
We then define
\begin{align}
    J^+(p)=\{q:p\rightarrow q\}, \nonumber \\
    J^-(p)=\{q:q\rightarrow p\}.
\end{align}
$J^+(p)$ is called the \emph{causal future} of $p$, and $J^-(p)$ is called the causal past. 
In the context of AdS/CFT, we can consider two different notions of the causal future and past. 
When we mean all those points in the bulk geometry in past or future of $p$, we write $J^\pm(p)$; when we mean all those points in the boundary geometry in the past or future of $p$ we write $\hat{J}^\pm(p)$.

\begin{figure*}
\begin{center}
\begin{subfigure}[b]{0.45\textwidth}
\centering
\tdplotsetmaincoords{15}{0}
\begin{tikzpicture}[scale=1.05,tdplot_main_coords]
\tdplotsetrotatedcoords{0}{35}{0}
\draw[gray] (-2,0.5,0) -- (-2,6.25,0);
\draw[gray] (2,0.5,0) -- (2,6.25,0);
    
    \begin{scope}[tdplot_rotated_coords]
    
    \draw[domain=0:45,variable=\x,smooth, fill=black!60!,opacity=0.8] plot ({-2*sin(\x)}, {1+\x/45}, {2*cos(\x)}) -- plot ({-2*sin((45-\x))}, {3-(45-\x)/45}, {2*cos(45-\x)}) --  plot ({2*sin(\x)}, {3-\x/45}, {2*cos(\x)}) -- plot ({2*sin(45-\x)}, {1+(45-\x)/45}, {2*cos(45-\x)});
    
    \draw[domain=0:45,variable=\x,smooth,thick] plot ({-2*sin(\x)}, {1+\x/45}, {2*cos(\x)});
    \draw[domain=0:45,variable=\x,smooth,thick] plot ({2*sin(\x)}, {1+\x/45}, {2*cos(\x)});
    \draw[domain=0:45,variable=\x,smooth,thick] plot ({-2*sin(\x)}, {3-\x/45}, {2*cos(\x)});
    \draw[domain=0:45,variable=\x,smooth,thick] plot ({2*sin(\x)}, {3-\x/45}, {2*cos(\x)});
    
    \begin{scope}[canvas is xz plane at y=0.5]
    \draw[gray] (0,0) circle[radius=2] ;
    \end{scope}
    
    \begin{scope}[canvas is xz plane at y=6.25]
    \draw[gray] (0,0) circle[radius=2] ;
    \end{scope}
    
    \draw[red] (0,1,-2) -- (0,2,-1);
    \draw[red] (0,1,2) -- (0,2,1);
    
    \begin{scope}[canvas is xz plane at y=2]
    
    \draw[gray] (0,0) circle (2);
    
    \draw [domain=-45:45,fill=lightgray,opacity=0.8] plot ({2*cos(\x+90)}, {2*sin(\x+90)}) -- (-1.41,1.41) to [out=-45,in=45] (-1.41,-1.41) -- plot ({2*cos(\x-90)}, {2*sin(\x-90)}) -- (1.41,-1.41) to [out=135,in=-135] (1.41,1.41);
    
    \draw [green,ultra thick,domain=-45:45] plot ({2*cos(\x+90)}, {2*sin(\x+90)});
    
    \draw[blue,thick] (1.41,1.41) to [out=-135,in=+135] (1.41,-1.41);
    \draw[blue,thick] (-1.41,1.41) to [out=-45,in=45] (-1.41,-1.41);
    
    \end{scope}
    
    \draw[domain=0:90,variable=\x,smooth,dashed] plot ({2*sin(\x+180)}, {3+\x/45}, {2*cos(\x+180)});
    \draw[domain=0:90,variable=\x,smooth,dashed] plot ({2*sin(\x+180)}, {3+\x/45}, {-2*cos(\x+180)});
    \draw[domain=0:90,variable=\x,smooth,dashed] plot ({-2*sin(\x+180)}, {3+\x/45}, {2*cos(\x+180)});
    \draw[domain=0:90,variable=\x,smooth,dashed] plot ({-2*sin(\x+180)}, {3+\x/45}, {-2*cos(\x+180)});
    
    \draw[thick, red,-triangle 45] (0,3,0) -- (2,5,0);
    \draw[thick, red,-triangle 45] (0,3,0) -- (-2,5,0);
    
    \draw[thick,red,-triangle 45] (0,2,-1) -- (0,3,0);
    \draw[thick,red,-triangle 45] (0,2,1) -- (0,3,0);
    
    \draw plot [mark=*, mark size=1.5] coordinates{(2,5,0)};
    \node[above] at (2,5,0) {$r_0$};
    \draw plot [mark=*, mark size=1.5] coordinates{(-2,5,0)};
    \node[above] at (-2,5,0) {$r_1$};
    
    \draw[domain=0:45,variable=\x,smooth, fill=black!50!,opacity=0.8] plot ({-2*sin(\x+180)}, {1+\x/45}, {2*cos(\x+180)}) -- plot ({-2*sin((45-\x)+180)}, {3-(45-\x)/45}, {2*cos(45-\x+180)}) --  plot ({2*sin(\x+180)}, {3-\x/45}, {2*cos(\x+180)}) -- plot ({2*sin(45-\x+180)}, {1+(45-\x)/45}, {2*cos(45-\x+180)});
    
    \begin{scope}[canvas is xz plane at y=2]
    \draw [green,ultra thick,domain=-45:45] plot ({2*cos(\x-90)}, {2*sin(\x-90)});
    \end{scope}
    
    \draw[domain=0:45,variable=\x,smooth,thick] plot ({-2*sin(\x+180)}, {1+\x/45}, {2*cos(\x+180)});
    \draw[domain=0:45,variable=\x,smooth,thick] plot ({2*sin(\x+180)}, {1+\x/45}, {2*cos(\x+180)});
    \draw[domain=0:45,variable=\x,smooth,thick] plot ({-2*sin(\x+180)}, {3-\x/45}, {2*cos(\x+180)});
    \draw[domain=0:45,variable=\x,smooth,thick] plot ({2*sin(\x+180)}, {3-\x/45}, {2*cos(\x+180)});
    
    \draw plot [mark=*, mark size=1.5] coordinates{(0,1,-2)};
    \node[left] at (0,1,-2) {$c_0$};
    \draw plot [mark=*, mark size=1.5] coordinates{(0,1,2)};
    \node[below] at (0,1,2) {$c_1$};
    \draw plot [mark=*, mark size=1.5] coordinates{(0,3,0)};
    \node[left] at (0,3,0) {$J_{01\rightarrow 01}$};
    
    \end{scope}
    
    \begin{scope}[canvas is xz plane at y=2]
    \draw[lightgray,domain=10:45] plot ({2*cos(\x-90)}, {2*sin(\x-90)});;
    \end{scope}
    \end{tikzpicture}
\caption{}
\label{subfig:cylinder}
\end{subfigure}
\hfill
\begin{subfigure}[b]{.45\textwidth}
\begin{center}
\begin{tikzpicture}[scale=1.5]

 \draw (-2,0) -- (2,0) -- (2,2) -- (-2,2) -- (-2,0);
    
    \draw[gray,fill=gray,opacity=0.25] (-2,2) -- (0,0) -- (2,2);
    
    \draw[gray,fill=gray,opacity=0.25] (-2,2) -- (-2,0) -- (0,2) -- (-2,2);
    
    \draw[gray,fill=gray,opacity=0.25] (2,2) -- (2,0) -- (0,2) -- (2,2);
    
    \draw[blue,fill=blue,opacity=0.25] (-1,2) -- (-2,1) -- (-2,0) -- (1,0) -- (-1,2);
    
    \draw[blue,fill=blue,opacity=0.25] (1,2) -- (2,1) -- (2,0) -- (-1,0) -- (1,2);
    
    \draw[blue,fill=blue,opacity=0.25] (2,1) -- (2,0) -- (1,0) -- (2,1);
    \draw[blue,fill=blue,opacity=0.25] (-2,1) -- (-2,0) -- (-1,0) -- (-2,1);
    
    \draw[black] plot [mark=*, mark size=2] coordinates{(-2,0)};
    \node[below left] at (-2,0) {$c_1$};
    \node at (1.8,0.5) {$V_1$};
    \node at (-1.8,0.5) {$V_1$};
    
    \draw[black] plot [mark=*, mark size=2] coordinates{(2,0)};
    \node[below right] at (2,0) {$c_1$};
    
    \draw[black] plot [mark=*, mark size=2] coordinates{(0,0)};
    \node[below left] at (0,0) {$c_0$};
    \node at (0,0.5) {$V_0$};
    
    \draw[blue] plot [mark=*, mark size=2] coordinates{(-1,2)};
    \node[above right] at (-1,2) {$r_0$};
    \node[below] at (-1,1.75) {$W_0$};

    \draw[blue] plot [mark=*, mark size=2] coordinates{(1,2)};
    \node[above right] at (1,2) {$r_1$};
    \node[below] at (1,1.75) {$W_1$};
    
    \node at (0,-1) {$ $};

\end{tikzpicture}
\end{center}
\caption{}
\label{subfig:boundary}
\end{subfigure}
\begin{subfigure}[b]{.45\textwidth}
\begin{center}
\begin{tikzpicture}[scale=0.55]
    
    \draw[postaction={on each segment={mid arrow}}] (-4,0) -- (0,4);
    \draw[postaction={on each segment={mid arrow}}] (4,0) -- (0,4);
    \draw[postaction={on each segment={mid arrow}}] (0,4) -- (4,8);
    \draw[postaction={on each segment={mid arrow}}] (0,4) -- (-4,8);
    
    \node[below left] at (-4,0) {$c_0$};
    \draw[fill=black] (-4,0) circle (0.15);

    \node[below right] at (4,0) {$c_1$};
    \draw[fill=black] (4,0) circle (0.15);

    \node[above right] at (4,8) {$r_1$};
    \draw[fill=blue] (4,8) circle (0.15);

    \node[above left] at (-4,8) {$r_0$};
    \draw[fill=blue] (-4,8) circle (0.15);
    
    \draw[fill=yellow] (0,4) circle (0.30);
    \node[right] at (0.2,4) {$J_{01\rightarrow01}$};
    
    \node[below] at (0,-0.53) {$ $};
    
    \node[above left] at (-3,1) {$A_0$};
    \node[above right] at (3,1) {$A_1$};
    
    \node[below left] at (-3,7) {$B_0$};
    \node[below right] at (3,7) {$B_1$};
    
    \end{tikzpicture}
\end{center}
\caption{}
\label{subfig:localschematicgravity}
\end{subfigure}
\hfill
\begin{subfigure}[b]{.45\textwidth}
\begin{center}
\begin{tikzpicture}[scale=0.55]

    \node[below left] at (-4,0) {$c_0$};
    \draw[fill=black] (-4,0) circle (0.15);

    \node[below right] at (4,0) {$c_1$};
    \draw[fill=black] (4,0) circle (0.15);

    \node[above right] at (4,8) {$r_1$};
    \draw[fill=blue] (4,8) circle (0.15);

    \node[above left] at (-4,8) {$r_0$};
    \draw[fill=blue] (-4,8) circle (0.15);
    
    \draw[postaction={on each segment={mid arrow}}] (-4,0) -- (-2,2) -- (-2,6) -- (-4,8);
    \draw[postaction={on each segment={mid arrow}}] (4,0) -- (2,2) -- (2,6) -- (4,8);
    \draw[postaction={on each segment={mid arrow}}] (-2,2) -- (0,4) -- (2,6);
    \draw[postaction={on each segment={mid arrow}}] (2,2) -- (0,4) -- (-2,6);
    
    \draw[dashed] (2,2) -- (0,0) -- (-2,2);
    \node[below] at (0,0) {$\Psi_{LR}$};
    
    \draw[fill=yellow] (-2,2) circle (0.3);
    \node[right] at (-1.8,2) {$V_0$};
    
    \draw[fill=yellow] (2,2) circle (0.3);
    \node[left] at (1.8,2) {$V_1$};
    
    \draw[fill=yellow] (-2,6) circle (0.3);
    \node[right] at (-1.8,6) {$W_0$};
    
    \draw[fill=yellow] (2,6) circle (0.3);
    \node[left] at (1.8,6) {$W_1$};
    
    \node[above left] at (-3,1) {$A_0$};
    \node[above right] at (3,1) {$A_1$};
    
    \node[below left] at (-3,7) {$B_0$};
    \node[below right] at (3,7) {$B_1$};
    
\end{tikzpicture}
\end{center}
\caption{}
\label{subfig:nonlocalschematicgravity}
\end{subfigure}

\caption{(a) A view of the bulk of AdS. Quantum systems $A_0,A_1$ travel inward from $c_0$ and $c_1$ to the region $J_{01\rightarrow 01}$, where they interact under the channel $\mathcal{N}_{A_0A_1\rightarrow B_0B_1}$. Systems $B_0$, $B_1$ travel outward to $r_0$ and $r_1$. (b) The same set-up viewed in the boundary, showing the regions $V_0$ and $V_1$. Note that there is no location where $A_0$ and $A_1$ can meet while still being in the past of $r_0,r_1$. (c) A computation on systems $A_0$, $A_1$ happening locally. The yellow circle represents a channel acting on input systems $A_0$ and $A_1$, and producing output systems $B_0$ and $B_1$. Causally, this has the same form as the interaction happening in the bulk of the AdS space, with the central yellow dot playing the role of the scattering region $J_{01\rightarrow01}$. (d) A non-local quantum computation. The causal connections in this form of computation match those in the boundary view above; the four yellow dots play the role of the four regions $V_0,V_1,W_0,W_1$ shown in the figure above.}
\label{fig:cylinder1}
\end{center}
\end{figure*}

Returning to our four points $c_0,c_1,r_0,r_1$, we require that
\begin{align}
    J_{0,1\rightarrow 0,1}:=J^+(c_0)\cap J^+(c_1) \cap J^-(r_0) \cap J^-(r_1) \neq \emptyset.
\end{align}
This expression defines $J_{0,1\rightarrow 0,1}$, which we name the \emph{bulk scattering region}. 
The fact that the bulk scattering region is non-empty means that signals can travel in from $c_0$ and $c_1$, meet somewhere, and then travel out again to $r_0$ and $r_1$.
A process like this is shown in figure \ref{fig:cylinder1}; we call this scattering. 
We will place a second requirement however, which is that
\begin{align}
    \hat{J}_{0,1\rightarrow 0,1}:=\hat{J}^+(c_0)\cap \hat{J}^+(c_1) \cap \hat{J}^-(r_0) \cap \hat{J}^-(r_1) = \emptyset.
\end{align}
This is saying that we can't have a scattering process in the boundary. 
The set up shown in figure \ref{subfig:boundary} has no scattering in the boundary; you can convince yourself that the four lightcones from the four points don't overlap anywhere in this boundary picture. 

After constructing geometrical settings like this with bulk-only scattering, we can ask ourselves the following. 
Suppose some quantum systems fall into the bulk from $c_0$ and $c_1$, interact inside of the scattering region, and then re-emerge at $r_0$ and $r_1$. 
The equivalence of the bulk and boundary pictures tells us that this process must be simulated in the boundary: we need the same inputs to transform to the same outputs. 
However, there's no location for this interaction to happen, since the boundary scattering region is empty. 

This is puzzling, but the resolution of course is to recall our results about NLQC: we know that the local interaction that happens in the bulk could also occur as an NLQC. 
In an NLQC, the inputs never need to be brought together, so we don't necessarily need a scattering region. 
Instead, we only need communication from each input to each output, like the pattern shown in figure \ref{subfig:nonlocalschematicgravity}. 
Indeed, we do have this more limited set of causal connections in the boundary. 
This always occurs, because a consequence of the bulk scattering region being non-empty is that it must be the case that $c_i\rightarrow r_j$ through the bulk. 
We commented earlier (figure \ref{fig:lightrays}) that if two points are causally connected through the bulk, then they must also be through the boundary. 
This guarantees that whenever we have a bulk interaction, we always have the right set of causal connections in the boundary to support that interaction as an NLQC. 

Just as we suggested earlier, we are seeing that the higher dimensional bulk physics is reproduced in one less dimension via NLQC. 
This makes holography possible, but it also places constraints on the boundary CFT. 
In the boundary picture, the role of the entangled state $\Psi_{LR}$ used in the NLQC is played by the state of the CFT in the two regions
\begin{align}
    V_0=\hat{J}^+(c_0)\cap \hat{J}^-(r_0)\cap \hat{J}^-(r_1),\nonumber \\
    V_1=\hat{J}^+(c_1)\cap \hat{J}^-(r_0)\cap \hat{J}^-(r_1).
\end{align}
These are shown in figure \ref{subfig:boundary}. 
These are the regions in the future of one input, and the past of the two outputs, so causally they play the same role as Alice and Bob's first round operations in an NLQC.
For the boundary to support the interactions occurring in the bulk then, we need the state $\Psi_{V_0V_1}$ to be entangled. 
More specifically, we need there to be enough entanglement in $\Psi_{V_0V_1}$ to support whatever interactions occur in the bulk scattering region. 

Recall that in the last section we studied the mutual information in the boundary CFT. 
This mutual information had a sharp transition, from being small to being $O(1/G_N)$, which occurred suddenly as the regions came closer together or grew larger. 
Our reasoning above indicates that whenever there is a non-empty bulk scattering region, we should have $I(V_0:V_1)=\Omega(1/G_N)$, so that there is large correlation in the boundary subregions supporting the bulk interaction. 
To make this more precise, suppose we throw $n$ qubits into the scattering region from $c_0$ in states $H^{q_i}\ket{b_i}$ with $q_i,b_i\in \{0,1\}$ and randomly selected, and throw the bits $b_i$ in from $c_1$. 
We arrange for them to interact in the bulk so as to complete the measuring task, sending $\{b_i\}_i$ to both $r_0$ and $r_1$.\footnote{Certainly this is causally allowed, but we're assuming we can arrange for the correct interaction to occur.}
Then, according to our entanglement lower bounds proven in chapter \ref{chapter:monogamygames} on the parallel repeated measuring task, we know that
\begin{align}
    \frac{1}{2}I(V_0:V_1)\geq \alpha n
\end{align}
for constant $\alpha$. 
If we take $n$ too large then we expect that the qubits we throw in will deform the geometry we're trying to investigate, but if we keep $n=o(1/G_N)$ we don't have to worry about this effect.\footnote{This is a claim about gravity I won't get into here, but see \cite{may2020holographic} for a discussion.} 
But now since we can use the above to produce any lower bound which is $o(1/G_N)$, we know the mutual information must be at least $\Omega(1/G_N)$. 

To summarize our reasoning: if there is a non-empty bulk scattering region but empty boundary scattering region, then the boundary must reproduce bulk interactions using NLQC. 
To do this, there must be entanglement in $V_0:V_1$, and in particular the mutual information must be $\Omega(1/G_N)$. 
We can also re-frame the conclusion as the statement that the entanglement wedge of $V_0\cup V_1$ should be in the connected configuration, since the Ryu Takayanagi formula tells us this is equivalent to the mutual information being $\Theta(1/G_N)$.

We can summarize this as the following theorem, which we so far are only conjecturing. 
\begin{theorem}[Connected wedge theorem] \label{thm:main}
    Let $\{c_0, c_1, r_0, r_1\}$ be a bulk-only scattering configuration on the boundary of an asymptotically AdS spacetime with a holographic dual. Let $V_0$ and $V_1$ be boundary regions defined by
    \begin{align}
        V_0 &= \hat{J}^+(c_0)\cap \hat{J}^-(r_0)\cap \hat{J}^-(r_1) \nonumber \\
        V_1 &= \hat{J}^+(c_1)\cap \hat{J}^-(r_0)\cap \hat{J}^-(r_1)
    \end{align}
    Then the entanglement wedge of $V_0\cup V_1$ is connected.
\end{theorem}

Usefully, the final statement here is purely a geometrical claim about AdS spacetimes. 
Our reasoning about NLQC relates bulk light-cones to boundary entanglement, and then the RT formula relates boundary entanglement back to bulk geometry, now in the form of statements about minimal surfaces. 
This makes the claim checkable in a straightforward way: we just pick a geometry and look at whether we have bulk only scattering, and if so whether the entanglement wedge is connected. 
In fact, the work \cite{may2020holographic, may2022connected} provided a proof in general relativity of the connected wedge theorem, which was conjectured earlier \cite{may2019quantum} on the basis of the argument we've given about NLQC. 
That the connected wedge theorem is in fact true provides a validation of our arguments that bulk interactions can be understood as supported by NLQC in the boundary. 

\section{Gravity and complexity}

We learned in the last section that computations happening inside of the scattering region are supported as NLQCs in the boundary picture, and we argued that the relevant boundary entanglement is in the state on $\Psi_{V_0V_1}$. 
From this we can put requirements on boundary entanglement if we assume non-trivial operations can happen inside the scattering region, and this is borne out by the connected wedge theorem, which guarantees that there is $O(1/G_N)$ mutual information in $V_0\!:\!V_1$ whenever there is a non-empty scattering region. 

We can also view this setting from a somewhat different perspective. 
To illustrate this, take the simple case of implementing $n$ instances of the measuring task in parallel.
Then, our lower bounds on entanglement cost in NLQC dictate that
\begin{align}
    \frac{1}{2}I(V_0:V_1)\geq \alpha n
\end{align}
which means that we can't implement more than $\frac{1}{2}I(V_0:V_1)$ parallel instances of the measuring task inside the scattering region, since these tasks are completed as NLQCs in the boundary, which has only $\frac{1}{2}I(V_0:V_1)$ correlation available.
From our perspective so far this is a fairly obvious claim, but we should step back and appreciate that this is a non-trivial statement about gravity: apparently something prevents us from carrying out certain computations in the presence of gravity, at least within particular regions. 

We can arrive at these constraints on computation from the boundary, NLQC, perspective, but it must be true that there is some bulk physics that enforces these constraints. 
To understand how the upper bound on implementations of the measuring task is enforced in the bulk, let's consider the geometry of the scattering region in a bit more detail. 
In figure \ref{fig:scatteringregion} we show a typical scattering region. 
The region is formed by the intersection of four causal future or past regions, so it has four faces corresponding to the light cones of the four points. 
We are especially interested in the lower edge of the region, which we call the \emph{ridge}. 
The gravitational proof of the connected wedge theorem tells us a bit more than that the mutual information is $O(1/G_N)$, it actually tells us the mutual information is lower bounded by the ridge area, 
\begin{align}
    \frac{1}{2}I(V_0:V_1)\geq \frac{\text{area}(r)}{4G_N}.
\end{align}
This is good, because now we've related one part of our constraint to some kind of bulk object. 
If we could say that for some reason bulk physics requires
\begin{align}\label{eq:CEBconsequence}
    \frac{\text{area}(r)}{4G_N} \geq n
\end{align}
that is if we knew that bulk physics required that we can't fit more than an area's worth of qubits into the region, then we would have our bulk explanation of these constraints. 

\begin{figure}
\centering
\tdplotsetmaincoords{15}{0}
\begin{tikzpicture}[scale=1.05,tdplot_main_coords]
\tdplotsetrotatedcoords{0}{35}{0}
    
    \begin{scope}[tdplot_rotated_coords]
    
    \draw[thick,dashed] (-3,0,0) -- (0,4,3);
    
    \draw[opacity=0.25,fill=red] (0,4,3) -- (-3,0,0) -- (3,0,0) -- (0,4,3);
    \draw[opacity=0.25,fill=blue] (-3,0,0) -- (0,4,3) -- (0,4,-3) -- (-3,0,0);
    \draw[opacity=0.6,fill=blue] (3,0,0) -- (0,4,3) -- (0,4,-3) -- (3,0,0);
    \draw[opacity=0.7,fill=red] (0,4,-3) -- (-3,0,0) -- (3,0,0) -- (0,4,-3);
    
    \draw[thick] (-3,0,0) -- (3,0,0);
    \draw[thick] (0,4,-3) -- (0,4,3);
    
    \draw[thick] (-3,0,0) -- (0,4,-3);
    
    \draw[thick] (3,0,0) -- (0,4,-3);
    \draw[thick] (3,0,0) -- (0,4,3);
    
    \node[below] at (0,-0.25,0) {$r$};
    \draw[->] (0.25,-0.5,0) to [out=0,in=-90] (0.5,0,0);
    
    \end{scope}
    \end{tikzpicture}
    \caption{A scattering region in AdS$_{2+1}$. The lower edge is the ridge, $r$.}
    \label{fig:scatteringregion}
\end{figure}

In fact, exactly the needed statement about gravity has been conjectured and argued for long ago, under the name the \emph{covariant entropy bound} (CEB) \cite{bousso1999covariant}. 
Very roughly, the CEB says that we can't pack more qubits into a spacetime region than the area of that region. 
More carefully, the statement only applies to regions defined in a certain way, but happily our scattering region is of the needed form, so we can indeed apply the bound here and recover statement \eqref{eq:CEBconsequence}. 
Note that the covariant entropy bound has not been proven in full generality, so it's already non-trivial that we obtain a new argument for the CEB from the NLQC perspective. 
That said, the CEB can be convincingly argued for from several other perspectives, so we haven't found anything too radical so far. 

The best lower bounds we've been able to prove so far for NLQC are linear. 
Any linear or sub-linear lower bound on NLQC can be enforced in the bulk by the CEB. 
Suppose for a second though that we were to obtain a super-linear lower bound, so for implementing a family of unitaries $\{U_n\}_n$, where $U_n$ is an $n$ qubit unitary. 
Then, we would obtain constraints like
\begin{align}
    \frac{1}{2}I(V_0:V_1) \geq f(n)
\end{align}
where $n$ is the size of the unitary we can implement in the bulk and $f(n)$ is super-linear.
For any super-linear $f$, $n$ is forced to grow more slowly than $1/G_N$ as $G_N\rightarrow 0$, so this constraint cannot be enforced by the CEB.
This would tell us that there is a non-trivial constraint on the computations that can happen in the scattering region, as opposed to just a constraint on the size of the inputs we can fit into the region. 

There is a long history of authors speculating on the limits of computation in the presence of gravity, see e.g. \cite{lloyd2000ultimate}. 
For instance, does quantum gravity make more computations efficient than quantum mechanics alone? Or does it place basic constraints?
AdS/CFT provides a concrete model of quantum gravity where we can begin to explore this.
The NLQC picture lets us relate computation in the presence of gravity to NLQC without gravity, so gives us a concrete and well defined setting where we can address this question. 

Stepping back, we can notice that super-linear lower bounds on NLQC would have many other impressive implications: we clearly would get better QPV security guarantees, and from the $T$-depth upper bound we also would get non-trivial $T$-depth lower bounds. Another consequence which we haven't covered are good Hamiltonian simulation bounds \cite{apel2024security}. 
Thus super-linear NLQC lower bounds provide a single problem that captures something hard, fundamental, and admittedly mysterious that crosses through many applications. 

It is also intriguing to notice that we might expect to be able to do reasonably complex operations within the scattering region. 
For instance, suppose that we could implement a circuit of complexity polynomial in the area of the scattering region. 
Then, since the area is of the same order (in $1/G_N$) as the entanglement available in the boundary, we should also be able to implement circuits as non-local quantum computations using polynomial entanglement. 
This is a surprising claim: the best protocols we have currently are the $T$-depth protocol of chapter \ref{chapter:Tdepth}, which scales exponentially in the $T$-depth, or various techniques based on the $T$-count described in chapter \ref{chapter:Tcount}, which scale exponentially in the $T$-count. 
As well, even in the more limited context of $f$-routing, we only know how to do polynomial size formulas efficiently, and not polynomial size circuits. 

What could be going on here? This observation highlights a basic tension in the connection between NLQC and AdS/CFT. 
If we assume `reasonable' (polynomial size circuits) are possible to implement in the bulk, there must be much more efficient NLQC protocols out there waiting to be discovered. 
This would be very surprising, but at the same time there are no lower bounds that contradict this possibility. 
Alternatively, it could be the case that something unexpected is happening in AdS/CFT: perhaps bulk physics is somehow such that computations are severely constrained, in a way they are not in the real world. 

\section{History and further reading}

The connection between NLQC and the AdS/CFT correspondence was first suggested in \cite{may2019quantum}, where the connected wedge theorem was also conjectured. 
The connected wedge theorem was then proven in \cite{may2020holographic}, which also made improvements to the quantum information perspective. 
Later in \cite{may2022connected} the connected wedge theorem was generalized to a richer set of scattering scenarios, and the theorem was proven to a higher level of rigour. 
The work \cite{may2020holographic} already considered the connection between strong NLQC lower bounds and constraints on gravity, but this was elaborated on further in \cite{may2022complexity}.

\chapter{Application: Quantum position verification}\label{chapter:QPV}

\minitoc

We introduced the idea of quantum position verification in chapter \ref{chapter:introduction}, noting that it is one of the main motivating problems behind the study of entanglement cost in NLQC.
In this chapter we return to study QPV in more detail. 
We begin by considering a number of challenges that appear in the context of experimental settings, including noise, finite computation time, and other issues. 

Our starting point and candidate protocol for practical QPV will be the $f$-measure task. 
This is a generalization of the measure task we defined in chapter \ref{chapter:NLQC}. 
We give a definition next. 
\begin{definition}\label{def:qubitfbb84-initial}
    An $f$-measure task is defined by a choice of Boolean function $f:\{0,1\}^{n}\times \{ 0,1\}^{n}\rightarrow \{0,1\}$, and a $2$ dimensional Hilbert space $\mathcal{H}_Q$.
    Inputs $x\in \{0,1\}^{n}$ and system $Q$ are given to Alice, and input $y\in \{0,1\}^{n}$ is given to Bob.
    The system $Q$ is in the state $\rho^{f(x,y)}_z=H^{f(x,y)}\ketbra{z}{z}H^{f(x,y)}$.  
    Alice and Bob exchange one round of communication, with the combined systems received or kept by Alice labelled $M$ and the systems received or kept by Bob labelled $M'$.
    Alice and Bob can output values $a,b$ respectively with $a,b\in\{0,1\}$.
    We say the $f$-measure task is completed $C$-correctly if, on average over $(x,y)$, $C=\Pr[a=b=z]$.
\end{definition}

The $f$-measure task seems favourable for an experimental implementation, at least before we start worrying about issues around imperfect implementations: the honest player performs a classical computation on $\Theta(n)$ size inputs, does $\Theta(1)$ quantum gates, and then responds with classical messages. 
As an advantage compared to similar task of $f$-routing, the quantum communication is only in one direction, from verifier to prover, potentially simplifying the implementation. 
Meanwhile, we saw in chapter \ref{chapter:mostlyclassicallowerbounds} that one similar setting, namely $f$-routing, requires $\Omega(n)$ entanglement to attack, at least for suitable choices of function $f$, and without considering noise.
In fact, the theory of reductions among NLQC instances (see chapter \ref{chapter:reductions}\footnote{Understanding chapter \ref{chapter:reductions} isn't essential to this chapter; we will only reference it to motivate our approach here.}) shows that this same entanglement lower bound applies to $f$-measure, so, again without worrying about noise, an attacker needs $\Omega(n)$ entanglement.
While we don't yet have a statement about entanglement cost for $f$-measure in the noisy setting, this is at least suggestive that an attack on $f$-measure in realistic settings will require $\Omega(n)$ quantum resources.
We could then base the security of QPV on the difference in the quantum resources needed by the honest prover ($O(1)$) and quantum resources needed by an attacker (hopefully, $\Omega(n)$).

The natural approach is to try to show $\Omega(n)$ lower bounds on the entanglement needed to attack $f$-measure even in the presence of noise.
Unfortunately this problem is still open: the best bounds known to date are $\Omega(\log(n))$. 
Further, there is good reason to think resolving this will be hard: $f$-measure entanglement lower bounds imply $f$-routing entanglement lower bounds (chapter \ref{chapter:reductions}), which in turn imply CDS lower bounds (chapter \ref{chapter:ITCandNLQC}). 
Meanwhile, proving super-logarithmic lower bounds on CDS is open, has been for awhile, and has been identified as a stepping stone towards proving $\mathsf{AM}^{\cc}$ lower bounds \cite{applebaum2021placing}, a major open problem in communication complexity.
Fortunately though, there are ways we can obtain good security guarantees for $f$-measure even without resolving this apparently difficult problem. 
In particular, we shift attention from entanglement to the computational power of the attacker, showing $\Omega(n)$ lower bounds on the number of quantum gates the attacker needs, even while the honest player needs $O(1)$ gates.
This avoids the CDS barrier and may be a better approach, at least in the short term, towards practical QPV.
This is the approach we take in this chapter: our goal is to develop $\Omega(n)$ lower bounds on quantum gates for $f$-measure, even while accounting for noise and other implementation challenges. 

\section{Implementation challenges}\label{sec:QPVchallenges}

In figure \ref{fig:1dQPV} we illustrated the basic idea of a QPV protocol: the honest player completes a computation locally, while the dishonest player tries to simulate the same computation as an NLQC. 
If we can find computations that are harder to implement locally than non-locally we have a basis on which to set the security of QPV. 

Real experiments are much messier than the simplified picture we started with though, where we made a number of idealizations. 
The idealizations we will try to relax in this chapter are:
\begin{itemize*}
    \item \textbf{Computational delay:} The intended operation will not be implemented instantaneously, as our spacetime diagram suggests, but with a delay.
    \item \textbf{Noise:} The intended operation will not be implemented perfectly, but noisily. 
    \item \textbf{Loss:} The quantum communication link from verifier to prover will lose many of the qubits sent over it. This is a form of noise, but sometimes loss can be handled separately from more general forms of noise.
    \item \textbf{Imperfect challenges:} The states sent to the prover may not be exactly the BB84 states, but instead noisy variants of these.
    \item \textbf{Slow quantum information:} In many implementations, quantum information is sent over optical fibre, where it travels slower than the speed of light. An attacker might have access to a faster communication link. 
\end{itemize*}
Potentially, these complications make an attacker's job easier: rather than simulate the actions of a perfect prover, they just need to simulate the delayed, noisy, and lossy prover. 
Further, they may be able to exploit the imperfections in the input states to make their job easier. 
We will consider each of the above implementation challenges in turn below, and understand how each of them could be exploited by an attacker. 
In the next section we develop bounds on the quantum resources needed by an attacker that continue to apply in the context of these implementation challenges. 

\textbf{Computational delay:} Let's first consider delay coming from the time the prover needs to perform the computation. 
The situation is shown in figure \ref{fig:timingconstraint}.
The honest player requires time $T$ to execute the local computation involved in the QPV protocol, for instance in $f$-measure this would be the time needed to compute $f(x,y)$, make the conditional measurement, and then copy and broadcast the outcome. 
The spacetime location of the outputs (points $r_1$ and $r_2$) needs to be delayed to allow the honest prover to respond at the correct time. 
For a given size of region being verified however, if the delay is too large the attacker gains a simple attack: the attacker performs the needed computation locally, to one side of the verified region. 
If the attacker can compute the challenge fast enough, and the region being verified is small enough, they can successfully cheat. 
Specifically, careful consideration of the spacetime diagram reveals that they can cheat when their computation time $T'$, the honest players computation time $T$, and the radius $L$ of the verified region satisfy $L<\frac{c(T-T')}{2}$. 
This means security requires the prover be able to run the computation in time
\begin{align}
    T<2L/c + T'.
\end{align}
For a practical scheme, we want to be able to take $L$ as small as possible.\footnote{If we want to check a message came from a particular location, like a bank, the verified region needs to be smaller than the bank.}
We also want to make as relaxed of assumptions as possible about the attacker, so we don't want to assert that $T'$ is large. 
This means we want be able to take $T$ as small as possible. 

\begin{figure}
\centering
    \begin{tikzpicture}[scale=0.65]
    \node[below left]  at (-4,0) {$c_1$};
    \draw[fill=black] (-4,0) circle (0.15);
    \node[below right] at (4,0) {$c_2$};
    \draw[fill=black] (4,0) circle (0.15);
    
    \node[above right] at (4,11) {$r_2$};
    \draw[fill=blue] (4,11) circle (0.15);
    \node[above left]  at (-4,11) {$r_1$};
    \draw[fill=blue] (-4,11) circle (0.15);

    \draw[fill=gray,opacity=0.5] (-1,1) rectangle (1,10);

    \draw[thick, mid arrow] (-4,0) -- (0,3.7);
    \node[below right] at (-2.8,1.1) {$Q,x$};
    \draw[thick, mid arrow] (4,0) -- (0,3.7);
    \node[above right] at (2.6,1.1) {$y$};

    \draw[thick, mid arrow] (0,7.3) -- (-4,11);
    \node[below left] at (-2.6,9.9) {$a$};
    \draw[thick, mid arrow] (0,7.3) -- (4,11);
    \node[below right] at (2.6,9.9) {$b$};

    \draw[thick] (0,3.7) -- (0,7.3);

    \draw[red, thick, mid arrow] (4,-0.2) -- (-1.3,4.7);
    \draw[red, thick, mid arrow] (-4,0) -- (-1.3,4.7);
    \draw[red, thick, mid arrow] (-1.3,4.7) -- (-1.3,6.4);
    \draw[red, thick, mid arrow] (-1.3,6.4) -- (3.9,11.15);
    \draw[red, thick, mid arrow] (-1.3,6.4) -- (-4,11);

    \node[right] at (0,5.5) {$T$};
    \node[left] at (-1.3,5.5) {$T'$};
    
\end{tikzpicture}
\caption{An honest prover executes a QPV protocol, but their actions take time $T$. A dishonest player, executing the same operations but running the computation in time $T'$ is able to spoof the scheme if they are within distance $c(T-T')/2$ of the honest position, where $c$ is the speed of light. Thus the region considered the honest region, shown in grey, must be made larger; the precision of the scheme is limited to a distance $c(T-T')/2$.}
\label{fig:timingconstraint}
\end{figure}

In the specific setting of $f$-measure, the security of the scheme relies on taking $n$ large, since (we aim to have) the quantum resources used by the attacker be $\Omega(n)$. 
This is in tension with the idea that we need to keep $T$ small, since to keep $T$ small we need to make $f$ as quick to compute as possible, and $f$ takes in inputs of size $n$. 
To get as favourable of a setting as possible then, we want to make the time to compute $f$ to grow with $n$ as slowly as possible. 
Specifically, $f$ should be computable with a low-depth circuit, since the depth of the circuit will control the time needed to compute it. 
A good choice we will return to is the inner-product function, 
\begin{align}
    f(x,y)=\sum_{i=1}^n x_i y_i \,\, \text{mod}\,\, 2. 
\end{align}
This function can be computed by a log-depth circuit, as we show in figure \ref{fig:innerproductcircuit}. 
This is the minimal depth of a circuit where the output depends on all of the inputs, so we don't expect to find (useful) functions with a lower depth than this.

Our take-away for considerations of delay in an implementation of $f$-measure then is to \emph{use the inner product function}, or another function with a log-depth circuit. 

\begin{figure}
\centering
\begin{tikzpicture}[
    gate/.style  = {circle, draw, thick, minimum size=8mm, inner sep=0pt},
    wire/.style  = {->, >=stealth, thick},
    layer/.style = {font=\small\itshape, anchor=west, gray}
  ]

  \foreach \i in {1,...,4} {
    \pgfmathsetmacro{\px}{2.4*(\i-1)}
    \node (x\i) at (\px-0.5, 0) {$x_{\i}$};
    \node (y\i) at (\px+0.5, 0) {$y_{\i}$};
    \node[gate] (a\i) at (\px, 1.5) {$\wedge$};
    \draw[wire] (x\i) -- (a\i);
    \draw[wire] (y\i) -- (a\i);
  }

  \node[gate] (b1) at (1.2, 3.0) {$\oplus$};
  \node[gate] (b2) at (6.0, 3.0) {$\oplus$};
  \draw[wire] (a1) -- (b1);
  \draw[wire] (a2) -- (b1);
  \draw[wire] (a3) -- (b2);
  \draw[wire] (a4) -- (b2);

  \node[gate] (c1) at (3.6, 4.5) {$\oplus$};
  \draw[wire] (b1) -- (c1);
  \draw[wire] (b2) -- (c1);

  \node (out) at (3.6, 5.9) {$\displaystyle\sum_{i=1}^{4} x_i y_i \bmod 2$};
  \draw[wire] (c1) -- (out);

  \node[layer] at (8.3, 1.5) {depth 1};
  \node[layer] at (8.3, 3.0) {depth 2};
  \node[layer] at (8.3, 4.5) {depth 3};
  \node[layer] at (8.3, 5.9) {total: $1+\log_2 n$};
\end{tikzpicture}
\caption{Log-depth circuit computing the inner-product function, shown for input size $n=4$.}
\label{fig:innerproductcircuit}
\end{figure}

\textbf{Noise:} To quantify how noisy an implementation of $f$-measure is, we consider the probability with which both outputs match the verifiers input $z$, 
\begin{align}
    C=\Pr[z=a=b].
\end{align}
We also write $C=1-\epsilon$, so that small $\epsilon$ means low noise.
We could also consider taking a more refined look at the outputs: for instance in an honest implementation, we should essentially always have $a=b$, since $b$ is just copied classically from $a$, and this can be done with very small error. 
Here though, we will just consider the value $\epsilon$ as our noise quantification. 

How large of a noise parameter can we expect to tolerate in our implementation? 
We can observe that if the noise is too large, an attacker can always simulate the output using just $O(1)$ quantum resources. 
In particular an attacker can, upon receiving $Q$, measure always in the Breidbart basis (see \eqref{eq:breidbartbasis}), and send the outcome to both sides.
As we saw in chapter \ref{chapter:monogamygames}, this succeeds with probability $\cos^2(\pi/8)$, so achieves $\epsilon=\sin^2(\pi/8)\approx 0.146$. 
We can't hope to prove security of the scheme then if the noise is larger than this value. 

\textbf{Loss:} Losing the input qubit is, of course, a form of noise. 
By loss however we mean that the qubit gets lost, and the prover knows it is lost and can report that. 
This motivates updating the definition of the scheme we want to prove security of, to allow the players to respond with a third output, call it $\perp$, that signals ``loss''.
In doing so, we need to be careful now about what we mean by the parameter we already introduced, $\epsilon$. 
We define
\begin{align}
    1-\epsilon := \Pr[z=a=b\,|\, a\neq \perp \, \text{or}\,\, b\neq \perp]. 
\end{align}
Thus $\epsilon$ is the probability of error, conditioned on at least one player actually responding with something other than ``loss''. 

How much loss can the $f$-measure QPV scheme tolerate? 
A simple attack strategy that exploits the declaration of loss is as follows. 
The attacker, upon receiving $Q$, randomly selects a basis $\theta\in\{0,1\}$ (labelling the computational and Hadamard bases) and measures $Q$ in that basis, obtaining outcome $a$.
They copy $a$ and forward it to both sides.
As well, they copy the classical inputs and forward these to both sides.
In the second round, they compute $f(x,y)$ on both sides, and then output $a$ if they happened to have guessed $\theta=f(x,y)$. 
This occurs with probability $1/2$; when it does they will have $z=a=b$ and hence be correct. 
When they guess the basis incorrectly, they both output $\perp$, declaring a loss. 
Thus the attacker can also, using a strategy with $O(1)$ quantum resources, achieve $L=1/2$, $\epsilon=0$.  

We can also put the Breidbart measurement attack and the guess-the-basis attack together to achieve other values of $L,\epsilon$. 
Consider first flipping a coin with two outcomes, which occur with probabilities $(p,1-p)$. 
If we get the first outcome, we execute the Breidbart basis attack. 
If we get the second outcome, we execute the guess-the-basis attack. 
Then, the error parameter is
\begin{align}
    \epsilon &= \Pr[\,\text{error}\,|\,\text{not declaring loss}] \nonumber \\
    &= \frac{\Pr[\,\text{error}\,,\,\text{not declaring loss}]}{\Pr[\text{not declaring a loss}]}.
\end{align}
Note that $\Pr[\text{not declaring a loss}]=1-L$, so we just need to work out the numerator of this expression. 
This is 
\begin{align}
\Pr[\,\text{error}\,,\,\text{not declaring loss}] &= p \Pr[\,\text{error}\,,\,\text{not declaring loss}\,|\,\text{Breidbart basis attack}] \nonumber \\
&\quad + (1-p)\Pr[\,\text{error}\,,\,\text{not declaring loss}\,|\,\text{guess-the-basis attack}] \nonumber \\
&= p\sin^2(\pi/8)+ (1-p) \cdot 0.
\end{align}
Putting this together, we find that
\begin{align}
    \epsilon &= \frac{p \sin^2(\pi/8)+(1-p)\cdot 0}{1-L}, \nonumber \\
    &=\frac{p \sin^2(\pi/8)}{1-L}
\end{align}
Meanwhile, the loss parameter is
\begin{align}
    L &= p\cdot 0+ (1-p)\cdot 1/2, \nonumber \\
    &= \frac{1-p}{2}.
\end{align}
Eliminating $p$ between these two equations, we find the curve in the $(\epsilon,L)$ parameter space achieved by these strategies is fixed by
\begin{align}
    \frac{\epsilon(1-L)}{\sin^2(\pi/8)}+ 2L = 1.
\end{align}
We can also, of course, purposefully worsen our strategy by sometimes outputting random outputs to achieve higher error, or randomly declaring a loss to increase the loss probability, so that we can actually achieve any $\epsilon, L$ parameters with
\begin{align}
    \frac{\epsilon(1-L)}{\sin^2(\pi/8)}+ 2L \geq 1
\end{align}
Thus the region where the attacker can succeed with just $O(1)$ quantum resources at least includes the above, and could be larger.

Notice that the loss parameter $L$ can be as large as $1/2$, which is larger than the maximal error parameter $\epsilon$. 
Thus it is favourable to be able to identify losses and declare them as such, instead of grouping all types of error together into the noise parameter $\epsilon$. 

\textbf{Imperfect challenges:} Another difficulty comes from noise in the verifiers operations. 
Their intention is to prepare BB84 states $H^{f(x,y)}\ket{z}$, but they won't be able to do this perfectly. 
To characterize how well they prepare these states, a simple approach is to define a parameter $\eta$ such that the actual prepared states $\rho^{f(x,y)}_z$ satisfy
\begin{align}
    \eta \geq \frac{1}{2}\max_{\theta, z}\Vert \rho^{\theta}_z-H^{\theta}\ketbra{z}{z}H^{\theta}\Vert_1. 
\end{align}
Then, we expect that as $\eta$ gets large, the cheater is able to exploit the poorly prepared input states to cheat. 
For instance, a large enough $\eta$ would allow for both the $\theta=0$ and $\theta=1$ states to be in the same basis, so that a single measurement setting would allow the cheater to always determine $z$. 
Many other cheating possibilities occur, depending on the exact form of the density matrices $\rho^{\theta}_z$. 
We won't investigate these systematically, though for any given experimental set-up it would be interesting to characterize these density matrices and understand possible cheating protocols. 
We will find more generically that for small enough $\eta$ however, the protocol remains secure.

\textbf{Slow quantum information:} Another challenge is delay coming from the transmission time.
In a typical experiment, the quantum information is sent over fibre, which means it will travel slower than the speed of light. 
If all of the signals, classical and quantum, faced this delay the effect would be similar to what is caused by the computational delay we discussed above, and the delay would impose a limit on how small the verified region can be taken. 
However, we can hope to send classical signals much nearer to the speed of light, for instance by sending them over free space. 
The idea then will be to send the quantum information early, and rely on the classical signals to fix the region precisely. 
The situation is shown in figure \ref{fig:slowquantumspacetime}. 

\begin{figure}
\centering
\begin{tikzpicture}[scale=0.65]
    \node[below left]  at (-4,0) {$c_1$};
    \draw[fill=black] (-4,0) circle (0.15);
    \node[below right] at (4,0) {$c_2$};
    \draw[fill=black] (4,0) circle (0.15);

    \draw[dashed] (-4,-4) -- (-4,0);
    \draw[fill=black] (-4,-4) circle (0.15);

    \node[above right] at (4,9) {$r_2$};
    \draw[fill=blue] (4,9) circle (0.15);
    \node[above left]  at (-4,9) {$r_1$};
    \draw[fill=blue] (-4,9) circle (0.15);

    \draw[fill=gray,opacity=0.5] (-1,1) rectangle (1,8);

    \draw[thick, mid arrow] (-4,-4) -- (0,3.7);
    \node[right] at (-3.1,-2.3) {$Q$};
    \draw[thick, mid arrow] (-4,0) -- (0,3.7);
    \node[above left] at (-2.2,1.7) {$x$};
    \draw[thick, mid arrow] (4,0) -- (0,3.7);
    \node[above right] at (2.6,1.1) {$y$};

    \draw[thick, mid arrow] (0,5.3) -- (-4,9);
    \node[below left] at (-2.6,7.9) {$a$};
    \draw[thick, mid arrow] (0,5.3) -- (4,9);
    \node[below right] at (2.6,7.9) {$b$};

    \draw[thick] (0,3.7) -- (0,5.3);
    \node[right] at (0,4.5) {$T$};
\end{tikzpicture}
\caption{A quantum position verification protocol where the quantum input $Q$ travels slower than the classical signals $x,y$. Potentially, an attacker who can move quantum information more quickly can exploit this to attack the scheme. The appropriate attack model, allowing arbitrarily slow quantum information, is shown in figure \ref{fig:attack_earlyquantum}.}
\label{fig:slowquantumspacetime}
\end{figure}

The honest prover in this set-up receives the classical and quantum data at the same time, and acts just as before. 
The attacker however, who might be able to send quantum information quickly, has a potential advantage. 
In the worst case, where they receive the quantum signal far in the past, they can act on the input $H^{f(x,y)}\ket{z}_Q$ in an arbitrary way, producing a state $\psi_{AB}^{f(x,y),z}=\mathcal{N}_{Q\rightarrow AB}(H^{f(x,y)}\ketbra{z}{z}_QH^{f(x,y)})$ with $A$ held on the left and $B$ held on the right. 
To show security of the scheme then, we should show that even given this advantage, any attack still requires $\Omega(n)$ quantum resources. 
The new attack model is shown in figure \ref{fig:attack_earlyquantum}. 

\section{Gate lower bounds in practical settings}

\subsection{Setting and overview}

We're now ready to collect all of the challenges discussed above that appear in real implementations, and define a new task that captures these complications.
The task is defined as follows. 

\begin{definition}\label{def:qubitfbb84}
    An $\eta$-\textbf{ideal, lossy $f$-measure} task is defined by a choice of Boolean function $f:\{0,1\}^{n}\times \{ 0,1\}^{n}\rightarrow \{0,1\}$, and a $2$ dimensional Hilbert space $\mathcal{H}_Q$.
    Inputs $x\in \{0,1\}^{n}$ and system $Q$ are given to Alice, and input $y\in \{0,1\}^{n}$ is given to Bob.
    The system $Q$ is in the state $\rho^{f(x,y)}_z$, with $z$ chosen uniformly at random, and $\frac{1}{2}\Vert \rho^{\theta}_z-V^{\theta}_z\Vert_1\leq \eta$, for all $\theta,z$.  
    Alice and Bob can output values $a,b$ respectively with $a,b\in\{0,1,\perp\}$.
    We say the lossy $f$-measure task is completed $C$-correctly with loss $L$ on input $(x,y)$ if $C=\Pr[a=b=z|x,y]$ and $L=\Pr[a=b=\perp|x,y]$
\end{definition}

We would like to show a lower bound on the quantum resources needed to implement this task in the attack model shown in figure \ref{fig:attack_earlyquantum}.
Note that compared to an NLQC, in this model the quantum input can be acted on arbitrarily and distributed between the left and right, and then an NLQC circuit is implemented once the classical inputs are received.

To prove a lower bound on this setting, recall that for the (basic) measuring task, we used a reduction to a monogamy-of-entanglement game (theorem \ref{thm:measuretoBB84MoE}). 
Here as well we will make use of MoE games, but this time we will need a more general definition that allows the players of the MoE game to declare loss.
The other ingredient we will need is the simultaneous message passing model in communication complexity, illustrated in figure \ref{fig:SMP*}. 
Recall that the $\Rsim^*$ model also made an appearance in chapter \ref{chapter:communicationcomplexity}.

\begin{figure*}
    \centering
    \begin{subfigure}{0.45\textwidth}
    \centering
    \begin{tikzpicture}[scale=0.45]
    
    \draw[thick] (-5,-5) -- (-5,-3) -- (-3,-3) -- (-3,-5) -- (-5,-5);
    \node at (-4,-4) {$\mathcal{N}^x$};
    
    \draw[thick] (5,-5) -- (5,-3) -- (3,-3) -- (3,-5) -- (5,-5);
    \node at (4,-4) {$\mathcal{M}^y$};
    
    \draw[thick] (5.5,5) -- (5.5,3) -- (2.5,3) -- (2.5,5) -- (5.5,5);
    
    \draw[thick] (-5.5,5) -- (-5.5,3) -- (-2.5,3) -- (-2.5,5) -- (-5.5,5);
    
    \draw[thick, mid arrow] (-4.5,-3) -- (-4.5,3);
    \node[left] at (-4.5,-2) {$M_0$};
    
    \draw[thick, mid arrow] (4.5,-3) -- (4.5,3);
    \node[right] at (4.5,-2) {$M_1'$};
    
    \draw[thick, late arrow] (-3.5,-3) to [out=90,in=-90] (3.5,3);
    \node[right] at (-3.25,-2) {$M_0'$};
    
    \draw[thick, late arrow] (3.5,-3) to [out=90,in=-90] (-3.5,3);
    \node[left] at (3.25,-2) {$M_1$};

    \draw[thick] (-1,-9) -- (1,-9) -- (1,-7) -- (-1,-7) -- (-1,-9);
    \draw[thick, mid arrow] (-0.5,-7) to [out=90,in=-90] (-3.5,-5);
    \draw[thick, mid arrow] (0.5,-7) to [out=90,in=-90] (3.5,-5);

    \draw[thick, mid arrow] (-2.5,-8) -- (-1,-8);
    \node[left] at (-2.5,-8) {$Q$};
    
    \draw[thick, mid arrow] (-4.5,-6.5) -- (-4.5,-5);
    \node[below] at (-4.5,-6.5) {$x$};
    \draw[thick, mid arrow] (4.5,-6.5) -- (4.5,-5);
    \node[below] at (4.5,-6.5) {$y$};
    
    \draw[thick, late arrow] (4.5,5) -- (4.5,6.5);
    \node[above] at (-4.5,6.5) {$a$};
    \draw[thick, late arrow] (-4.5,5) -- (-4.5,6.5);
    \node[above] at (4.5,6.5) {$b$};
    
    \end{tikzpicture}
    \caption{}\label{fig:attack_earlyquantum}
    \end{subfigure}
    \hfill
    \begin{subfigure}{0.45\textwidth}
    \centering
    \begin{tikzpicture}[scale=0.4]
    
    \draw[thick] (-5,-5) -- (-5,-3) -- (-3,-3) -- (-3,-5) -- (-5,-5);
    
    \draw[thick] (5,-5) -- (5,-3) -- (3,-3) -- (3,-5) -- (5,-5);
    
    \draw[thick] (5,5) -- (5,3) -- (3,3) -- (3,5) -- (5,5);
    
    \draw[thick, mid arrow, double] (4,-3) -- (4.5,3);
    
    \draw[thick, mid arrow, double] (-4,-3) to [out=90,in=-90] (3.5,3);=
    
    \draw[thick] (-3.5,-5) to [out=-90,in=-90] (3.5,-5);
    \draw[black] plot [mark=*, mark size=3] coordinates{(0,-7.05)};
    
    \draw[thick] (-4.5,-6) -- (-4.5,-5);
    \node[below] at (-4.5,-6) {$x$};
    
    \draw[thick] (4.5,-6) -- (4.5,-5);
    \node[below] at (4.5,-6) {$y$};
    
    \draw[thick] (4,5) -- (4,6);
    \node[above] at (4,6) {$f(x,y)$};
    
    \end{tikzpicture}
    \caption{}
    \label{fig:SMP*}
    \end{subfigure}
    \caption{a) An $f$-measure protocol in the attack model where the adversary receives the quantum input $Q$ early. The adversary is able to act on $Q$ with an arbitrary channel, distribute the state, and then execute an NLQC once give the classical inputs. We define $M=M_0M_1$, $M'=M_0'M_1'$. b) An SMP$^*$ protocol. Alice receives $x$ on the left, Bob receives $y$ on the right. Alice and Bob share entanglement but must send only classical messages to the referee. We give a reduction from any $f$-measure protocol to a corresponding $\Rsim^*$ protocol. The number of bits of communication in the resulting $\Rsim^*$ protocol is set by the number of quantum gates needed to implement $\mathcal{N}^x$ and $\mathcal{N}^y$.} 
    \label{fig:reductionfig}
\end{figure*}

The general strategy of the proof is as follows. 
After the first round operations in the strategy, Alice and Bob have prepared a state $\rho_{\bar{Q}MM'}(x,y)$.\footnote{We take the viewpoint that $Q$ is prepared in the maximally entangled state with $\bar{Q}$, and $\bar{Q}$ is later measured in the basis $\theta$. 
Recall that in chapter \ref{chapter:NLQC} we saw this was equivalent to preparing one of the BB84 states at random.}
After the communication round, system $M$ will be held by Alice, and system $M'$ will be held by Bob.
To try and match the referee's measurement outcome, who measures in the $\theta=f(x,y)$ basis, Alice and Bob use a three outcome POVM, with one outcome corresponding to outputting $\perp$ and hence declaring a loss. 
The lossy monogamy game bound expresses that for a given state $\rho_{\bar{Q}MM'}(x,y)$, Alice and Bob can only succeed when the referee measures in one of the two bases. 
This means the density matrix $\rho_{\bar{Q}MM'}(x,y)$ should be better for matching measurements in the $\theta=f(x,y)$ basis than in the $\theta=\neg f(x,y)$ basis. 

Since the density matrix $\rho_{\bar{Q}MM'}(x,y)$ predicts $f(x,y)$, we can (at least heuristically) build a communication protocol for $f(x,y)$ out of the $f$-measure protocol in the following way.
In the $\Rsim^*$ protocol, Alice sends a description of her operation $\mathcal{N}^x$, and Bob sends a description of his operation $\mathcal{M}^y$. 
The referee then works out a classical description of $\rho_{\bar{Q}MM'}(x,y)$, determines if this density matrix is better for guessing measurement outcomes in the $\theta=0$ or $\theta=1$ basis, and then outputs their guess of $f(x,y)$ to be the corresponding $\theta$ value.
The above is only the rough idea however, in that the above leaves out a number of complications in the proof. 
In particular, the above would have a description of all of Alice and Bob operations to the referee, so the $\Rsim^*$ cost would be based on their full computations. 
In fact, we only want Alice and Bob to send a description of their quantum gates, and classical processing should be free. 
We handle these complications when we discuss the full proof below. 

\subsection{Technical tools}

\subsubsection{Lossy monogamy games}

We first define the notion of a lossy monogamy-of-entanglement game \cite{escola2023single, may2026quantum}, which will be one of our main tools used in lower bounding attacks on the lossy $f$-measure task. 
Recall that we first introduced the BB84 MoE game in chapter \ref{chapter:NLQC}, and then studied more general MoE games and their relationship to NLQC in chapter \ref{chapter:monogamygames}. 

\begin{definition}\label{def:lossyMoE}
    A lossy BB84 monogamy-of-entanglement game consists of two stages, a state preparation phase and then measurement phase. 
    \begin{itemize}
        \item \textbf{Preparation phase:} Alice and Bob prepare an arbitrary state $\Psi_{\bar{Q}AB}$ with $\bar{Q}$ a single qubit, and give $\bar{Q}$ to the referee. Alice takes system $A$, Bob system $B$, and then the players separate and can no longer communicate. 
        \item \textbf{Measurement phase:} The referee chooses $\theta\in\{0,1\}$ uniformly at random, then measures $\bar{Q}$ in the computational basis if $\theta=0$, or the Hadamard basis if $\theta=1$. She labels her measurement outcome $z$. The referee then reveals $\theta$ to both Alice and Bob. Alice and Bob then respond with values $a,b$ from the set $\{0,1,\perp\}$, where $\perp$ denotes `loss'.
    \end{itemize} 
    We define $L:=\Pr[a=b=\perp]$, $C:=\Pr[a=b=z]$.
\end{definition}

In the case of loss, a general strategy involves Alice and Bob performing three-outcome POVMs with possible outcomes $\mathcal{O} = \{0,1,\perp\}$. 
We define the joint-correctness and joint-abort probabilities operators and their respective expectations:
\begin{align}
    \tilde C_\theta = \sum_{x=0}^{1} (V_x^\theta \otimes A_x^\theta \otimes B_x^\theta), \qquad C_\theta &= \tr[ \rho_{QAB} \tilde{C}_\theta ], \\
    \tilde L_\theta = (\mathcal{I}_Q \otimes A^\theta_\perp \otimes B^\theta_\perp), \qquad L_\theta &= \tr[ \rho_{QAB} \tilde{L}_\theta].
\end{align}
Averaged over the basis we define:
\begin{align}
    &\tilde C = \frac{\tilde C_0 + \tilde C_1}{2}, &&C = \frac{C_0 + C_1}{2}, &&\tilde L = \frac{\tilde L_0 + \tilde L_1}{2}, &L = \frac{L_0 + L_1}{2}.
\end{align}
Note that we do not assume $a=b$ always, and that $C$, $L$ are unconditional probabilities. 
This should be contrasted with $\epsilon$, which is defined as the probability of error conditioned on at least one player not declaring a loss.

\begin{theorem}[Optimal lossy BB84 bound] \label{thm:optimal_lossy_bb84_bound}
Every strategy for completing the lossy BB84 monogamy-of-entanglement game state satisfies
\begin{align}
    C+\frac{L}{\sqrt 2}\leq \cos^2\left(\frac{\pi}{8}\right).
\end{align}
\end{theorem}

\begin{proof} \,
We relax the operators $\tilde C_\theta, \tilde L_\theta$ again, like in the proof for the BB84 MoE bound (lemma \ref{lemma:BB84MoE}), and introduce `weakened loss operators':
    \begin{align}
        \tilde C_0 &\preceq \sum_{x=0}^{1} V_x^0 \otimes A_x^0 \otimes \mathcal{I}_B, \\
        \tilde C_1 &\preceq \sum_{x=0}^{1} V_x^1 \otimes \mathcal{I}_A \otimes B_x^1, \\
        \tilde L_0 &\preceq \mathcal{I}_Q \otimes A_\perp^0 \otimes \mathcal{I}_B, \\
        \tilde L_1 &\preceq \mathcal{I}_Q \otimes \mathcal{I}_A \otimes B_\perp^1.
    \end{align}
    Next, we define:
    \begin{align}
        \tilde T := \frac{1}{2} \left( \sum_{x=0}^{1} V_x^0 \otimes A_x^0 \otimes \mathcal{I}_B + \sum_{x=0}^{1} V_x^1 \otimes \mathcal{I}_A \otimes B_x^1 + \frac{\mathcal{I}_Q}{\sqrt{2}} \otimes A_\perp^0 \otimes \mathcal{I}_B +  \frac{\mathcal{I}_Q}{\sqrt{2}} \otimes \mathcal{I}_A \otimes B_\perp^1 \right),
    \end{align}
    such that
    \begin{align}
        \tilde C + \frac{\tilde L}{\sqrt{2}} &= \frac{1}{2} \left( \tilde C_0 + \tilde C_1 + \frac{\tilde L_0 + \tilde L_1}{\sqrt{2}} \right) \preceq \tilde T.
    \end{align}
    Now, we expand all the operators in $\tilde T$ using $\sum_{a \in \{0,1,\perp\}} A^0_a = \mathcal{I}_A, \sum_{b \in \{0,1,\perp\}} B^1_b = \mathcal{I}_B$, for instance the first term in $\tilde{T}$ is expressed as
    \begin{align}
        \sum_{a=0}^{1} V_a^0 \otimes A_a^0 \otimes \mathcal{I}_B = \sum_{a=0}^{1} \sum_{b \in \{0,1,\perp\}} V_a^0 \otimes A_a^0 \otimes B_b^1 = \sum_{a,b \in \{0,1,\perp\}} \delta_{a\neq\perp} V_a^0 \otimes A_a^0 \otimes B_b^1,
    \end{align}
    where we inserted $\delta_{a\neq\perp}:=\delta_{a,0}+\delta_{a,1}$ to get $a,b$ both summed over $\{0,1,\perp\}$. 
    Then, continuing in this way for the remaining terms, we get 
    \begin{align}
        \tilde T = \sum_{a,b \in \{0,1,\perp\}} K_{ab} \otimes A_a^0 \otimes B_b^1, 
    \end{align}
    where we have defined the operator on $Q$:
    \begin{align}
        K_{ab} = \frac{1}{2} \left( \delta_{\{a\neq \perp\}} V_a^0 + \delta_{\{b \neq \perp\}} V_b^1 + \frac{1}{\sqrt{2}}(\delta_{\{a,\perp\}} + \delta_{\{b,\perp\}}) \mathcal{I}_Q \right).
    \end{align}
    We now claim that 
    \begin{align}
        K_{ab} \preceq \cos^2(\pi/8) \mathcal{I}_Q, \qquad \text{for every } a,b \in \{0,1,\perp\}. 
    \end{align}
    There are 3 cases: both $a,b \in \{0,1\}$, exactly one output is $\perp$, or $a=b=\perp$.

    Suppose both $a,b \in \{0,1\}$ then:
    \begin{align}
        K_{ab} = \frac{V_a^0 + V_b^1}{2},
    \end{align}
    which has maximal eigenvalue $\cos^2(\pi/8)$, thus $K_{ab} \preceq \cos^2(\pi/8) \mathcal{I}_Q$. 

    If exactly one output is $\perp$ then either:
    \begin{align}
        K_{\perp b} = \frac{1}{2} \left( \frac{\mathcal{I}_Q}{\sqrt{2}} + V_b^1 \right) \quad \text{or} \quad K_{a \perp} = \frac{1}{2} \left( \frac{\mathcal{I}_Q}{\sqrt{2}} + V_a^0 \right),
    \end{align}
    which both have the same largest eigenvalue $\frac{1}{2}\left( \frac{1}{\sqrt{2}} + 1\right) = \cos^2(\pi/8)$, thus both $K_{\perp b}\preceq \cos^2(\pi/8) \mathcal{I}_Q$ and $K_{a \perp} \preceq \cos^2(\pi/8) \mathcal{I}_Q$.

    If both outputs are $\perp$ then:
    \begin{align}
        K_{\perp \perp} := \frac{\mathcal{I}_Q}{\sqrt{2}} \preceq \cos^2(\pi/8) \mathcal{I}_Q,
    \end{align}
    since $1/\sqrt{2} \leq \cos^2(\pi/8)$. This proves the claim.

    Now taking the inequality on $\tilde T$:
    \begin{align}
        \tilde T &= \sum_{a,b \in \{0,1,\perp\}} K_{ab} \otimes A_a^0 \otimes B_b^1 \preceq \cos^2(\pi/8) \sum_{a,b \in \{0,1,\perp\}} \mathcal{I}_Q \otimes A_a^0 \otimes B_b^1 \\
        &= \cos^2(\pi/8) \mathcal{I}_Q \otimes \left( \sum_{a \in \{0,1,\perp\}} A_a^0 \right) \otimes \left( \sum_{b \in \{0,1,\perp\}} B_b^1 \right) = \cos^2(\pi/8) \mathcal{I}_{QAB}.
    \end{align}
    Thus we have
    \begin{align}
        C + \frac{ L}{\sqrt{2}} = \tr[\left( \tilde C + \frac{\tilde L}{\sqrt{2}} \right) \rho_{QAB} ] \leq \tr[ \tilde T \rho_{QAB} ] \leq \tr[\cos^2(\pi/8) \mathcal{I}_{QAB} \ \rho_{QAB}] = \cos^2(\pi/8),
    \end{align}
    as claimed.
\end{proof}

The bound $C+L/\sqrt{2}\leq \cos^2(\pi/8)$ is somewhat counter-intuitive: it says that the fewer rounds we actually require Alice and Bob to respond, the more constrained $C$ is, since $C\leq \cos^2(\pi/8)-L/\sqrt{2}$. 
This behaviour comes from $C$ being an unconditional probability, and the behaviour is more intuitive is we discuss the probability the players are correct, conditioned on them responding. 
Define this to be $C_{R}$. 
Then $C_R=\Pr[\text{correct}\,|\,a\neq \perp \, \text{or}\,\, b\neq \perp]=\frac{C}{1-L}$.
This leads to
\begin{align}
    C_R=\frac{C}{1-L} \leq \frac{1}{1-L}\left(\cos^2(\pi/8)-\frac{L}{\sqrt{2}} \right). 
\end{align}
Now, we see that the bound on $C_R$ actually relaxes as $L$ increases, which says that when the loss can be large, Alice and Bob can succeed with high probability on the remaining cases where they choose not to declare loss. 
In fact, the upper bound reaches 1 and hence becomes trivial at $L=1/2$. 
This must occur, because when $L=1/2$ Alice and Bob can guess a basis and measure before separating, and then both return the measurement outcome if it turns out they guessed the basis right. 

\subsubsection{Communication complexity}

We will make use of a reduction from $f$-routing and $f$-measure to communication complexity scenarios. 
Specifically, we will be interested in the \emph{simultaneous message passing with shared entanglement} scenario, denoted $\Rsim^*$.
We introduced the simultaneous message passing model in chapter \ref{chapter:communicationcomplexity}, where we showed it can be separated from $\Rtwo$. 
We didn't give a full definition there, so lets do that now. 

\begin{definition}[$\Rsim$ complexity]\label{def:SMP}
Let $f : \{0,1\}^n\times \{0,1\}^n\rightarrow \{0,1\}$ be a function, and $\epsilon\in [0,1]$. A $\Rsim$ protocol $P$ for $f$ consists of three algorithms Alice, Bob, and a referee. Alice receives $x\in \{0,1\}^n$ as input and outputs $m_A \in \{0,1\}^*$, Bob receives $y\in \{0,1\}^n$ as input and outputs $m_B \in \{0,1\}^*$, and the referee receives $m_A, m_B$ and outputs a bit $c=P(x,y)$. 
A protocol $P$ is $\epsilon$-correct on the input distribution $\mu$ if
\begin{align*}
    \Pr_{\mu,R}[P(x,y)=f(x,y)] \geq 1-\epsilon \enspace.
\end{align*}
The $\Rsim_{\epsilon,\mu}$ complexity of $f$ is defined as
\begin{equation*}
    \Rsim_{\mu,\epsilon}(f) = \min_{P: P \text{ is $\epsilon$-correct}}(|m_A| + |m_B|) \enspace.
\end{equation*}
Similarly, we can define $\Rsim^*_{\mu,\epsilon}(f)$ for the case where Alice and Bob share entanglement.
\end{definition}

A standard function studied in communication complexity is the inner product, 
\begin{align*}
    \text{IP}_n(x,y) = \sum_{i=1}^n x_i y_i \,\, \text{mod}\,\, 2 \enspace.
\end{align*}
Intuitively, this is a difficult function to compute in communication complexity scenarios because the output depends sensitively on every bit of the input. 
More concretely, we will make use of the following lower bound, which follows from simple arguments (see \cite{may2026quantum}) starting with a bound from \cite{nayak2002communication}, and taking $\mu=u$ to be the uniform distribution, 
\begin{align}
    \boxed{\Rsim^*_{u,\epsilon}(\text{IP}_n)\geq n + 2\log(1-2\epsilon).}
\end{align}
More generally, several techniques are available for lower bounding quantum communication complexity costs, see e.g.  \cite{buhrman2009non}. 

\subsubsection{Continuity of the loss and error parameters}

The reduction we sketched above, and will prove below, takes an $f$-measure protocol with perfect BB84 input states and shows it defines a $\Rsim^*$ protocol with communication cost related to the quantum gate complexity of the initial $f$-measure protocol. 
We then will use communication lower bounds on the $\Rsim^*$ model to obtain a gate lower bound on the $f$-measure protocol. 
We will need to relax this, to obtain gate lower bounds on $f$-measure protocols that take in noisy inputs. 
To do this we will use a continuity argument, to show that an $f$-measure protocol that succeeds with certain error and loss parameters on noisy inputs can't have very different error and loss parameters when given perfect inputs. 
This is shown in the next lemma. 

\begin{lemma}\label{lemma:imperfectpreparation}
    Consider an $f$-measure protocol which is $C_\eta$-correct and $L_\eta$-lossy for some set of $\eta$ ideal inputs $\{\rho^{\theta}_x\}$. 
    Then, when given the ideal inputs $V^\theta_x=H^\theta\ketbra{x}{x}H^\theta$, the protocol's new correctness and loss parameters $C,L$ satisfy
    \begin{align}
        \left|\left(E_\eta+2sL_\eta\right)-\left(E+2sL\right) \right|\leq \eta
    \end{align}
    where $s=\sin^2(\pi/8)$, and where $E_\eta=1-C_\eta-L_\eta$ and $E=1-C-L$ is the error in the $\eta$-ideal and perfect input cases, respectively. 
\end{lemma}
\begin{proof}\,
    The proof proceeds by using the behaviour of the protocol as a method of distinguishing the possible inputs $\rho^\theta_z$ and $V^\theta_z$. 
    Since the states are close, they must be hard to distinguish, so the behaviour of the protocol cannot be too different. 

    In more detail, we design a distinguishing game as follows. 
    Fix values of $(x,y)$ and $z$.
    The referee computes $\theta=f(x,y)$, then prepares either $\rho^\theta_z$ or $V^\theta_z$, each with probability $1/2$. 
    The referee then gives the resulting state to the player. 
    The player runs the $f$-measure protocol for the chosen $(x,y)$, then guesses the state according to the following rules:
    \begin{itemize*}
        \item If the game is correct, guess the state is $V^\theta_z$ with probability $0$. 
        \item If the game is incorrect, guess the state is $V^{\theta}_z$ with probability $1$. 
        \item If the game declares a loss, guess the state is $V^{\theta}_z$ with probability $2s=2\sin^2(\pi/8)$. 
    \end{itemize*}
    Then the probability the player correctly determines the state is
    \begin{align}
        p_{\text{dist}}(V^{\theta}_z, \rho^\theta_z)=&\Pr[V]\Pr[\text{correct}\,|\,V]\Pr[\text{guess}\,V\,|\,\text{correct}] +\nonumber \\
        &\Pr[V]\Pr[\text{incorrect}\,|\,V]\Pr[\text{guess}\,V\,|\,\text{incorrect}]+\nonumber \\
    &\Pr[V]\Pr[\text{loss}\,|\,V]\Pr[\text{guess}\,V\,|\,\text{loss}]+\nonumber\\
    &\Pr[\rho]\Pr[\text{correct}\,|\,\rho]\Pr[\text{guess}\,\rho\,|\,\text{correct}]+\nonumber \\
    &\Pr[\rho]\Pr[\text{incorrect}\,|\,\rho]\Pr[\text{guess}\,\rho\,|\,\text{incorrect}]+\nonumber \\
    &\Pr[\rho]\Pr[\text{loss}\,|\,\rho]\Pr[\text{guess}\,\rho\,|\,\text{loss}].
    \end{align}
    Evaluating this using our probability assignments above, we find that
    \begin{align}
        p_{\text{dist}}(V^{\theta}_z, \rho^\theta_z) = \frac{1}{2}+\frac{1}{2}\left(\left( E+2sL\right)-\left(E_\eta+2sL_\eta\right)\right).
    \end{align}
    From the Holevo-Helstrom theorem, we know that the optimal distinguishing probability is at most $1/2+\Vert V_z^\theta-\rho_z^\theta\Vert_1/4$. 
    Since our strategy cannot be better than the optimal one, we obtain
    \begin{align}
        \left( E+2sL\right)-\left( E_\eta+2sL_\eta\right)\leq \eta.
    \end{align}
    We can then repeat the argument, reversing the role of $V$ and $\rho$ in the guessing strategy above to derive
    \begin{align}
        \left( E_\eta+2sL_\eta\right)-\left( E+2sL\right)\leq \eta.
    \end{align}
    Combining the last two claims completes the proof. 
\end{proof}

\subsection{Reduction from \texorpdfstring{$f$}{TEXT}-measure to \texorpdfstring{$\Rsim^*$}{TEXT}}

We prove the following reduction from $f$-measure protocols with ideal BB84 state inputs to an $\Rsim^*$ protocol. 
Note that by a $f$-measure protocol, we mean a protocol in the attack model shown in figure \ref{fig:attack_earlyquantum}.

\begin{theorem}\label{thm:reduction}
    Suppose $P$ is an $f$-measure protocol that is $C$-correct and $L$-lossy, where the inputs are distributed according to distribution $\mu$, and where the inputs to the game are exactly the BB84 states.
    Further, suppose the $f$-measure protocol uses, in the first round operations, $C_G(f)$ gates drawn from a gate set of size $4$, $C_M(f)$ single-qubit computational basis measurements, and circuits acting on $q$ qubits.
    Then this implies the existence of a $\Rsim^*$ protocol with communication cost
    \begin{align}\label{eq:fRgatelowerbound}
        (\log(q)+1)(2C_G(f) + C_M(f)) \geq \Rsim^*_{\mu,\epsilon'}(f),
    \end{align}
    where $\Rsim^{*}_{\mu,\epsilon'}(f)$ denotes the minimal message size needed to compute $f(x,y)$ in the $\Rsim^*$ model with correctness $\epsilon'$ satisfying 
    \begin{align}
        \epsilon' < \frac{1}{2s}\left(E + 2sL\right)
    \end{align}
    where $E=1-C-L$.
\end{theorem}
\begin{proof}\,
We define an $\Rsim^*$ protocol as follows. 
We let the referee hold a classical description of the initial resource state shared by Alice and Bob in the $f$-measure protocol. 
As well, Alice and Bob share a copy of the resource system from the $f$-measure protocol. 
Alice and Bob's strategy will be to send the referee a description of their local operations. 
We consider a decomposition of Alice and Bob's operations into gates and measurements. 
Alice and Bob apply their operations to their shared resource state and the input system. 
As they do so, they keep a record of the gates they apply (which may be computed using mid-circuit measurement outcomes) and their measurement outcomes $m$, then send this to the referee. 
The referee will then compute a classical description of the state $\rho_{\bar{Q}MM'}^{x,y}(m)$, which would result if one end of the maximally entangled state $\Psi^+_{\bar{Q}Q}$ were input to the $f$-measure protocol.
For each gate, they specify the gate choice, requiring $2$ bits, and the location of the gate, which requires $2\log q$ bits for a contribution of $(2\log q + 2) C_G(f)$ bits. 
Further, to specify each measurement requires $\log q$ bits to specify where the measurement occurs plus $1$ bit to specify the measurement outcome, for a contribution of $(\log q+1)C_M(f)$. 
The total message size sent by Alice and Bob then is the left hand side of \cref{eq:fRgatelowerbound}. 

We claim that from the description of $\rho_{\bar{Q}MM'}^{x,y}(m)$, the referee can make a guess of $f(x,y)$ which is biased towards the correct value, with a bias that depends on the $(C,L)$ parameters of the $f$-measure protocol. 
To understand this, define
\begin{align}
    E^i(\sigma) &= \text{Pr}[\text{at least one player answers, incorrect}\,|\,\text{referee measures in basis} \,\,i] \nonumber \\
    C^i(\sigma) &= \text{Pr}[\text{at least one player answers, correct}\,|\,\text{referee measures in basis} \,\,i] \nonumber \\
    L^i(\sigma) &= \text{Pr}[\text{neither player answers}\,|\,\text{referee measures in basis} \,\,i]
\end{align} 
Note that $C^i+E^i+L^i=1$. 
Note also that these are the probabilities that occur in the $f$-measure protocol, if we were to insert the density matrix $\sigma$ into the second round operations.  
Define the averaged (over the choice of basis) quantities,
\begin{align}
    {E}(\sigma) &= \frac{1}{2}\left(E^0+E^1\right),\nonumber \\
    {C}(\sigma) &= \frac{1}{2}\left(C^0+C^1\right),\nonumber \\
    {L}(\sigma) &= \frac{1}{2}\left(L^0+L^1\right).
\end{align}
Note that to complete the $f$-measure protocol, Alice and Bob must be able to match the referee's measurement outcomes using the mid-protocol density matrix $\rho_{\bar{Q}MM'}^{x,y}$, with the referee holding $\bar{Q}$. 
This is exactly the lossy MoE game, so \cref{thm:optimal_lossy_bb84_bound} gives
\begin{align}\label{eq:MoEapplied}
    C+\frac{L}{\sqrt{2}} \leq \cos^2(\pi/8).
\end{align}
Substituting $C=1-E-L$, defining $s=\sin^2(\pi/8)$, and simplifying, we find
\begin{align}
    2s \leq (E^0+2sL^0) + (E^1+2sL^1). 
\end{align}
Define $Z^i(\sigma)\equiv E^i(\sigma)+2sL^i(\sigma)$, so that
\begin{align}
    \boxed{2s \leq Z^0+Z^1}
\end{align}
is the MoE bound. 

The variable $Z^i$ is small for $f(x,y)=i$ when the protocol is highly correct, in the sense that both the error and loss are low. 
The value of $Z^i$ depends on the density matrix $\rho$, and is computed from the probabilities resulting from running Alice and Bob's strategies using the density matrix $\rho$.
We can also consider $Z^i$ for the post-selected density matrix $\rho_{\bar{Q}MM'}^{x,y}(m)$, which we abbreviate as $Z_{m,x,y}^i\equiv Z^i(\rho^{x,y}(m))$. 
This is computed from the probabilities that occur using the measurement strategy defined by the $f$-measure protocol, but now replacing $\rho^{x,y}$ with $\rho^{x,y}(m)$. 
We also define
\begin{align}
    Z_{\text{opt}}^i(\sigma) = \min_{\text{strategies}}Z^i(\sigma)
\end{align}
where the minimization is over all allowed measurement strategies, which act separately on $M$ and $M'$. 
Notice that while the referee in the $\Rsim^*$ protocol may not be able to compute $Z^i(\rho^{x,y}(m))$ (since it depends on Alice and Bob's choice of measurements, which depends on $x,y$) they can compute $Z_{\text{opt}}^i(\rho^{x,y}(m))$, since this only depends on the density matrix $\rho^{x,y}(m)$, which they know. 
We can also observe that
\begin{align}
    Z_{\text{opt}}^0(\rho^{x,y}(m))+ Z_{\text{opt}}^1(\rho^{x,y}(m)) \geq 2s
\end{align}
by the MoE bound, which applies for all strategies, and that
\begin{align}
    Z^{i}_{\text{opt}}(\rho^{x,y}(m))\leq Z_{m,x,y}^i
\end{align}
since the actual strategy used in the $f$-measure protocol may be worse than the optimal one. 

We take the referee's strategy in the SMP protocol to be as follows. 
Given a description of $\rho^{x,y}(m)$, they compute $Z^{0}_{\text{opt}}(\rho^{x,y}(m))$ and $Z^{1}_{\text{opt}}(\rho^{x,y}(m))$. 
Then, if $Z^{i}_{\text{opt}}(\rho^{x,y}(m))$ is the smaller of the two, they bias their response towards $i$. 
This is sensible, as this is saying that $\rho^{x,y}(m)$ is better for guessing the outcome in basis $i$ than $\neg i$, so it seems likely to be sampled from a $\rho^{x,y}$ that works well for basis $i$, which occurs when $f(x,y)=i$. 
More specifically, they compute $D(m,x,y)=Z^{1}_{\text{opt}}(\rho^{x,y}(m))-Z^{0}_{\text{opt}}(\rho^{x,y}(m))$ and output $0$ with probability
\begin{align}\label{eq:p0}
    p_0(m,x,y)=\frac{1}{2}+\frac{D(m,x,y)}{4s}.
\end{align}
Note that $|D|\leq 2s$, because $Z^{j}_{\text{opt}}(\rho^{x,y}(m))\leq 2s$ (using the strategy that always outputs $\perp$, so has $E=0,L=1$), so $0\leq p_0 \leq 1$ is a probability.

Now say that the actual basis is $0$. 
Then $Z^{0}_{\text{opt}}(\rho^{x,y}(m))\leq Z_{m,x,y}^0$, so 
\begin{align}
    D(m,x,y) &= Z^{1}_{\text{opt}}(\rho^{x,y}(m))-Z^{0}_{\text{opt}}(\rho^{x,y}(m)) \nonumber \\
    &\geq 2s - 2Z^{0}_{\text{opt}}(\rho^{x,y}(m)) \nonumber \\
    &\geq 2s - 2Z_{m,x,y}^0
\end{align}
where we used $Z^0_{\text{opt}}+Z^1_{\text{opt}}\geq 2s$ (MoE) in the first inequality, and $Z^{0}_{\text{opt}}(\rho^{x,y}(m))\leq Z_{m,x,y}^0$ in the second inequality. 
But then on average over $m$, this is
\begin{align}
    \langle D(m,x,y) \rangle \geq 2s-2\langle Z_{m,x,y}^0\rangle  
\end{align}
so we need to bound $\langle Z_{m,x,y}^0\rangle$. 
This is
\begin{align}
    \langle Z_{m,x,y}^0 \rangle = \langle E_{m,x,y}^0\rangle  + 2s\langle L_{m,x,y}^0 \rangle
\end{align}
where $E_{m,x,y}^0$ is the probability at least one of the players respond and the response is incorrect, and $L_{m,x,y}^0$ is the probability both players don't respond. 
We have that
\begin{align}
    \langle E_{m,x,y}^0 \rangle &= E^{0}_{x,y} \nonumber \\
    \langle L^0_{m,x,y}\rangle &= L^0_{x,y}
\end{align}
so that
\begin{align}
    \langle Z_{m,x,y}^0 \rangle = E^0_{x,y}+2sL^0_{x,y}.
\end{align}
This gives a lower bound on $\langle D(m,x,y)\rangle $ of
\begin{align}
    \langle D(m,x,y)\rangle \geq 2s-2(E^0_{x,y}+2sL^0_{x,y}).
\end{align}
Returning to \cref{eq:p0}, we have that $p_0(x,y)=\langle p_0(m,x,y)\rangle$, when $(x,y)\in f^{-1}(0)$, is at least
\begin{align}
    p_{0}(x,y) \geq \frac{1}{2} + \frac{1}{2}\left(1 - \frac{1}{s}\left({E^0_{x,y}} +2sL_{x,y}^0\right)\right).
\end{align}
Similarly, we find that when $(x,y)\in f^{-1}(1)$, 
\begin{align}
    p_{1}(x,y) \geq \frac{1}{2} + \frac{1}{2}\left(1 - \frac{1}{s}\left({E^1_{x,y}} +2sL_{x,y}^1\right)\right).
\end{align}
So then the winning probability, averaged over inputs $(x,y)$, is
\begin{align}
    p_{win}&=\sum_{x,y}\mu(x,y) \,p_{f(x,y)}(x,y),  \nonumber \\
    &\geq \frac{1}{2} + \frac{1}{2}\left(1 - \frac{1}{s}\left(E +2sL\right)\right).
\end{align}
Then using $p_{win}=1-\epsilon'$, we obtain the claimed bound on $\epsilon'$.
\end{proof} 

\subsection{Gate lower bounds}

Next, we combine the above reduction with \cref{lemma:imperfectpreparation} to bound an $f$-measure protocol with imperfect inputs. 

\begin{figure}
    \centering
    \includegraphics[width=0.5\linewidth]{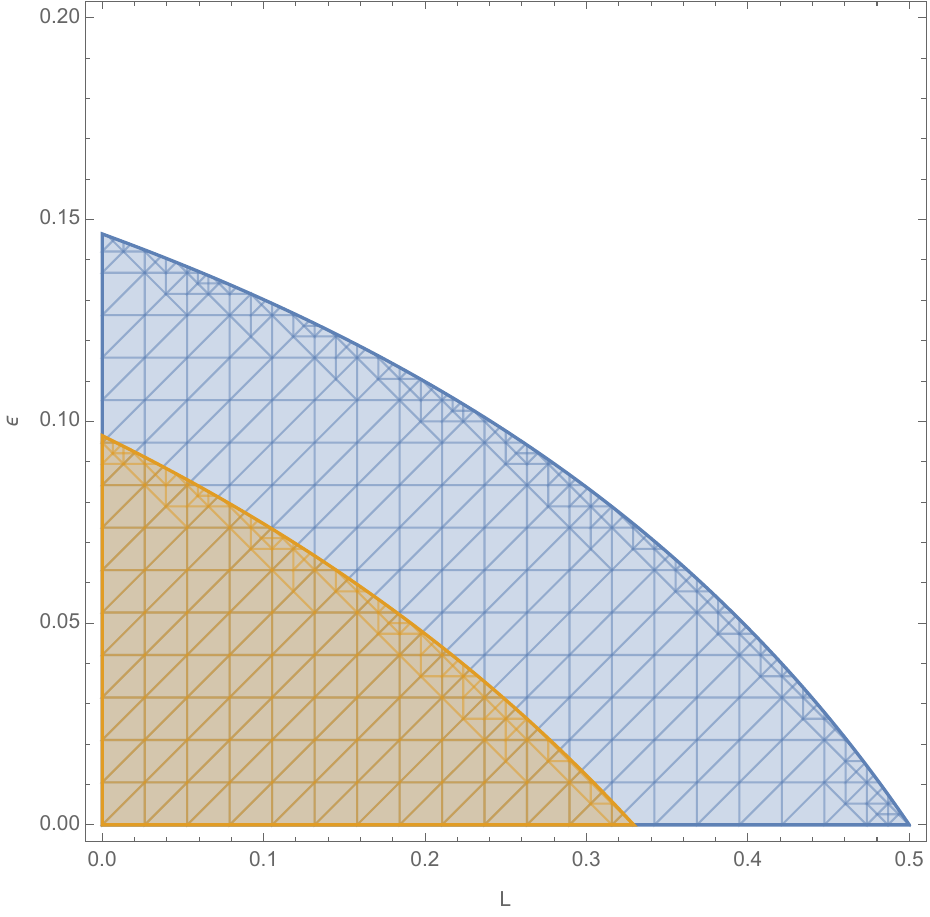}
    \caption{The region in $(\epsilon, L)$ parameter space where we show security of the $f$-measure protocol. The larger region corresponds to setting $\eta=0$, so that the inputs to the protocol are prepared perfectly. The smaller region corresponds to setting $\eta=0.05$. Analytically the security region is given by $\frac{\epsilon(1-L)+\eta}{\sin^2(\pi/8)}+2L<1$.}
    \label{fig:securityregion}
\end{figure}

\begin{theorem}\label{thm:reduction_imperfect}
    Suppose $P$ is an $f$-measure protocol that is $C$-correct and $L$-lossy, where the inputs are distributed according to distribution $\mu$, and where the inputs to the game are $\eta$-close in trace distance to the ideal states.
    Further, suppose the $f$-measure protocol uses, in the first round operations, $C_G(f)$ gates drawn from a gate set of size $4$, $C_M(f)$ single-qubit computational basis measurements, and circuits acting on $q$ qubits.
    Then this implies the existence of a $\Rsim^*$ protocol with communication cost
    \begin{align}
        (\log(q)+1)(2C_G(f) + C_M(f)) \geq \Rsim^*_{\mu,\epsilon'}(f),
    \end{align}
    where $\Rsim^{*}_{\mu,\epsilon'}(f)$ denotes the minimal message size needed to compute $f(x,y)$ in the $\Rsim^*$ model with correctness $\epsilon'$ satisfying 
    \begin{align}
        \epsilon' < \frac{1}{2}\left(\frac{\epsilon(1-L)}{s} + 2L+\frac{\eta}{s}\right).
    \end{align}
    where $\epsilon=\Pr[\text{incorrect response}\,|\,\text{at least one player responds}\,]$.
\end{theorem}
\begin{proof}\,
    By \cref{lemma:imperfectpreparation}, we have that if we take the given $f$-measure protocol (with correctness $C$ and loss $L$) and feed it the ideal input states, we have correctness $C'$ and loss $L'$ satisfying
    \begin{align}
        \left| (E+2sL)-(E'+2sL')\right|\leq \eta
    \end{align}
    Applying \cref{thm:reduction} then, we obtain an SMP protocol with
    \begin{align}
        \epsilon' < \frac{1}{2s}\left(E'+2sL'\right)
    \end{align}
    so then, in terms of the parameters $C,L,E=1-C-L$ describing the protocol given non-ideal inputs,
    \begin{align}
        \epsilon' < \frac{1}{2s}\left(E'+2sL'\right) \leq \frac{1}{2s}\left(E+2sL+\eta \right).
    \end{align}
    Finally we can use that
    \begin{align}
        E&=\Pr[\text{incorrect response}] \nonumber \\
        &=\Pr[\text{incorrect response}\,|\,\text{at least one player responds}]\Pr[\text{at least one player responds}] \nonumber \\
        &=\epsilon(1-L)
    \end{align}
    which yields the final expression.
\end{proof}

Finally we are ready to apply this to the $f$-measure protocol with $f$ chosen to be the inner product function. 

\begin{corollary}\label{corrolary:IPapplication}
    Consider any $f$-measure protocol with $f$ chosen to be the inner product function, 
    \begin{align}
        \text{IPfunc}(x,y) = \sum_{i=1}^n x_i y_i \,\, \text{mod}\,\, 2\,.
    \end{align}
    Assume the protocol has
    \begin{itemize*}
        \item $L:=\Pr[\text{both players declare a loss}]$
        \item $1-\epsilon:=\Pr[\text{both players respond correctly}\,|\, \text{at least one player responds}]$
    \end{itemize*}
    Further, suppose the inputs are distributed according to the uniform distribution, that the inputs are $\eta$-ideal, and that the first round operations involve $C_G$ two qubit gates drawn from a set of size $4$, $C_M$ single-qubit measurements, and act on at most $q$ qubits (shared between the left and right). 
    Then, the first round operations in any such $f$-measure protocol must satisfy
    \begin{align}
        (\log(q)+1)(2C_G + C_M) \geq n + 2\log(1-2\epsilon')
    \end{align}
    where $\epsilon'=\frac{1}{2}\left(\frac{\epsilon(1-L)}{s} + 2L+\frac{\eta}{s}\right)$, whenever $\left(\frac{\epsilon(1-L)}{s} + 2L+\frac{\eta}{s}\right)<1$, corresponding to the security region shown in \cref{fig:securityregion}.
\end{corollary}

\section{Towards long-range QPV}

In the last section we showed security of a QPV protocol even in the presence of a loss of up to 50\%. 
We also saw that this limitation is inherent to the $f$-measure protocol as we've defined it: if the loss is 50\% or higher, the honest protocol can be simulated using a $O(1)$ quantum resource attack. 
In practice, the 50\% loss barrier limits the implementation of QPV to be over only a few kilometres, due to absorption of photons in fibre optic wires. 
To perform long-range QPV then, new ideas are required. 
In this section we present two ways of modifying the $f$-measure protocol to achieve long-range QPV. 
Both variants involve the use of a quantum memory, and consequently are more challenging to implement in practice.

\subsection{Two round protocol}

Our first method to obtain long-range QPV involves exploiting the fact that the $f$-measure protocol remains secure when the quantum input is sent early. 
The basic setup is shown in figure \ref{fig:tworoundprotocol}. 
We call this the \emph{two round protocol}, because it involves sending the quantum information and getting a signal back (the first round), then sending the classical information and getting the outputs of the $f$-measure protocol back (the second round). 
In the first round, the verifier sends the quantum input, and the prover attempts to load this into a quantum memory. 
We assume the prover has a \emph{heralded} memory, which means that they know (with good probability) whether they successfully moved the qubit they received into their memory. 
When they have successfully done this, they send a signal indicating they are ready to begin the QPV protocol. 
Upon receiving the ``ready'' signal, the verifiers send the classical inputs $x,y$ to the prover, who should respond with the correct measurement outcome. 

\begin{figure}
    \begin{subfigure}{0.45\textwidth}
    \centering
    \begin{tikzpicture}[scale=0.6]

    \node at (2,-1) {Verifier 1};
    \draw[fill=black] (2,0) circle (0.15);

    \node at (5.5,-1) {Prover};
    \draw[fill=black] (5.5,0) circle (0.15);
    
    \node at (9,-1) {Verifier 2};
    \draw[fill=black] (9,0) circle (0.15);

    \node[draw, thick, fill=white, inner sep=3pt] (load) at (5.5,3.5) {Load};
    \node[draw, thick, fill=white, inner sep=3pt] (measure) at (5.5,10.5) {measure};

    \draw[snake it, blue,thick ] (2,0) -- (load);

    \draw[thick, mid arrow] (load) -- (2,7);
    \node[rotate=-45] at (3.5,4.5) {``Ready''};
    
    \draw[thick, mid arrow] (load) -- (9,7);
    \node[rotate=45] at (7.5,4.5) {``Ready''};

    \draw[snake it, blue,thick ] (load) -- (measure);

    \draw[thick, mid arrow] (2,7) -- (measure);
    \node at (3.4,9.1) {$x$};
    
    \draw[thick, mid arrow] (9,7) -- (measure);
    \node at (7.6,9.1) {$y$};

    \draw[thick, mid arrow] (measure) -- (9,14);
    \draw[thick, mid arrow] (measure) -- (2,14);

    \draw[thick, ->] (0,0) -- (0,10);
    \node[above] at (0,10) {Time};
    \draw[thick, ->] (0,0) -- (10,0);
    \node[right] at (10,0) {Position};
    
    \end{tikzpicture}
    \caption{}
    \end{subfigure}
    \hfill
    \begin{subfigure}{0.45\textwidth}
    \centering
    \begin{tikzpicture}[scale=0.45]
    
    \draw[thick] (-5,-5) -- (-5,-3) -- (-3,-3) -- (-3,-5) -- (-5,-5);
    \node at (-4,-4) {$\mathcal{N}^x$};
    
    \draw[thick] (5,-5) -- (5,-3) -- (3,-3) -- (3,-5) -- (5,-5);
    \node at (4,-4) {$\mathcal{M}^y$};
    
    \draw[thick] (5.5,5) -- (5.5,3) -- (2.5,3) -- (2.5,5) -- (5.5,5);
    
    \draw[thick] (-5.5,5) -- (-5.5,3) -- (-2.5,3) -- (-2.5,5) -- (-5.5,5);
    
    \draw[thick, mid arrow] (-4.5,-3) -- (-4.5,3);
    \node[left] at (-4.5,-2) {$M_0$};
    
    \draw[thick, mid arrow] (4.5,-3) -- (4.5,3);
    \node[right] at (4.5,-2) {$M_1'$};
    
    \draw[thick, late arrow] (-3.5,-3) to [out=90,in=-90] (3.5,3);
    \node[right] at (-3.25,-2) {$M_0'$};
    
    \draw[thick, late arrow] (3.5,-3) to [out=90,in=-90] (-3.5,3);
    \node[left] at (3.25,-2) {$M_1$};

    \draw[thick] (-1,-9) -- (1,-9) -- (1,-7) -- (-1,-7) -- (-1,-9);
    \draw[thick, mid arrow] (-0.5,-7) to [out=90,in=-90] (-3.5,-5);
    \draw[thick, mid arrow] (0.5,-7) to [out=90,in=-90] (3.5,-5);

    \draw[thick, mid arrow] (-2.5,-8) -- (-1,-8);
    \node[left] at (-2.5,-8) {$Q$};
    
    \draw[thick, mid arrow] (-4.5,-6.5) -- (-4.5,-5);
    \node[below] at (-4.5,-6.5) {$x$};
    \draw[thick, mid arrow] (4.5,-6.5) -- (4.5,-5);
    \node[below] at (4.5,-6.5) {$y$};
    
    \draw[thick, late arrow] (4.5,5) -- (4.5,6.5);
    \node[above] at (-4.5,6.5) {$a$};
    \draw[thick, late arrow] (-4.5,5) -- (-4.5,6.5);
    \node[above] at (4.5,6.5) {$b$};

    \draw[thick] (0,-7) -- (0,-6);
    \draw[thick] (-0.75,-6) -- (0.75,-6) -- (0,-4.75) -- (-0.75,-6);
    \node at (0,-5.5) {$0$};

    \node at (0,-12) {$ $};
    
    \end{tikzpicture}
    \caption{}
    \label{fig:tworoundattack}
    \end{subfigure}
\caption{a) The two round variant of the $f$-measure QPV protocol. One verifier sends the quantum signal (blue curvy line) to the prover, who moves the quantum data into a heralded quantum memory. If the transfer is successful, they send a ``Ready'' message to both verifiers. The verifiers then respond with the classical inputs $x,y\in\{0,1\}^n$. The prover receives $x,y$, computes $f(x,y)$, measures in the corresponding basis, then responds to both verifiers with the measurement outcomes. b) The attack model appropriate to the two-round $f$-measure QPV protocol. The model is similar to the one shown in figure \ref{fig:attack_earlyquantum}, but now the attacker can apply post-selected operations before dividing the input into systems sent left and right. This post-selection is implemented in practice by the attacker choosing to send the ``ready'' signal only when they obtain the desired measurement outcome.}
\label{fig:tworoundprotocol}
\end{figure}

We claim that $f$-measure is secure even in the two round protocol.
The basic intuition comes from the security of $f$-measure even with slow quantum information; the two round protocol gives the attacker early access to the quantum input, but this shouldn't matter. 
However, this doesn't amount to a full proof --- in particular the attacker gets extra power in this situation, coming from their ability to control which rounds they actually play. 
The attack model appropriate to this setting is shown in figure \ref{fig:tworoundattack}; the attacker now receives the quantum state early as in figure \ref{fig:attack_earlyquantum}, but rather than apply an arbitrary channel early, they can now apply an arbitrary channel and post-selection.
The attack implements the post-selection by running the channel, measuring, and then declaring the qubit was lost whenever the measurement outcome is unfavourable.

To establish security of the two round protocol, we point out that the gate lower bound proof we gave earlier remains valid even in the context of this stronger attack model.
The reason for this is that the proof uses only that the mid-protocol density matrix satisfies the MoE bound, but this is true regardless of whether post-selection was used in preparing it or not.
In this setting, we can still apply the lossy $f$-measure bound, but only loss occurring after the post-selection is relevant to the parameter $\eta$. 
In practice this means $\eta$ counts the loss in the quantum memory used by the prover, and does not count the loss in the channel. 

\subsubsection{Prover generated quantum challenge}

Another variation, similar to the above, has the prover generate the quantum challenge rather than the verifier. 
The protocol is shown in figure \ref{fig:proverstateprepprotocol}, and attack model in figure \ref{fig:proverstateprepattack}.
If the qubit is lost when travelling from the prover to the verifier, the verifier aborts that round and the protocol is repeated.
This removes channel loss as an obstruction to security: the verifiers abort ability post-selects onto cases where a qubit is received. 

It is immediate that the gate lower bound proof continues to hold: this follows by exploiting a strength of the monogamy game bound that we previously didn't make full use of. 
In particular, recall that in the monogamy game setting, Alice and Bob can prepare an \emph{arbitrary} state $\rho_{\bar{Q}AB}$ and send $\bar{Q}$ to the referee. 
In the protocols where the verifier prepares the entangled state $\Psi^+_{\bar{Q}Q}$ and sends $Q$ to the prover, the attack is limited to states of the form $\rho_{\bar{Q}AB}=\mathcal{N}_{Q\rightarrow AB}(\Psi^+_{\bar{Q}Q})$, but we didn't need this to apply the MoE bound. 
Re-applying the earlier proof then, we find that in the security region an attacker must use large quantum resources to attack the protocol. 
As in the two round protocol, the relevant loss parameter is now the loss in the quantum memory, between photon generation and the measurement. 

\begin{figure}
    \begin{subfigure}{0.45\textwidth}
    \centering
    \begin{tikzpicture}[scale=0.6]

    \node at (2,-1) {Verifier 1};
    \draw[fill=black] (2,0) circle (0.15);

    \node at (5.5,-1) {Prover};
    \draw[fill=black] (5.5,0) circle (0.15);

    \node at (9,-1) {Verifier 2};
    \draw[fill=black] (9,0) circle (0.15);

    \node[draw, thick, fill=white, inner sep=3pt] (generate) at (5.5,1) {Generate};
    \node[draw, thick, fill=white, inner sep=3pt] (vmeasure) at (2,4.5) {measure};
    \node[draw, thick, fill=white, inner sep=3pt] (measure)  at (5.5,9.5) {measure};

    \draw[snake it, blue, thick] (generate) -- (vmeasure);
    \draw[snake it, blue, thick] (generate) -- (measure);

    \draw[thick, mid arrow] (2,6) -- (measure);
    \node at (3.4,8.1) {$x$};

    \draw[thick, mid arrow] (9,6) -- (measure);
    \node at (7.6,8.1) {$y$};

    \draw[thick, mid arrow] (measure) -- (9,13);
    \draw[thick, mid arrow] (measure) -- (2,13);

    \draw[thick, ->] (0,0) -- (0,10);
    \node[above] at (0,10) {Time};
    \draw[thick, ->] (0,0) -- (10,0);
    \node[right] at (10,0) {Position};

\end{tikzpicture}
\caption{}
\label{fig:proverstateprepprotocol}
    \end{subfigure}
    \hfill
    \begin{subfigure}{0.45\textwidth}
    \centering
    \begin{tikzpicture}[scale=0.45]
    
    \draw[thick] (-5,-5) -- (-5,-3) -- (-3,-3) -- (-3,-5) -- (-5,-5);
    \node at (-4,-4) {$\mathcal{N}^x$};
    
    \draw[thick] (5,-5) -- (5,-3) -- (3,-3) -- (3,-5) -- (5,-5);
    \node at (4,-4) {$\mathcal{M}^y$};
    
    \draw[thick] (5.5,5) -- (5.5,3) -- (2.5,3) -- (2.5,5) -- (5.5,5);
    
    \draw[thick] (-5.5,5) -- (-5.5,3) -- (-2.5,3) -- (-2.5,5) -- (-5.5,5);
    
    \draw[thick, mid arrow] (-4.5,-3) -- (-4.5,3);
    \node[left] at (-4.5,-2) {$M_0$};
    
    \draw[thick, mid arrow] (4.5,-3) -- (4.5,3);
    \node[right] at (4.5,-2) {$M_1'$};
    
    \draw[thick, late arrow] (-3.5,-3) to [out=90,in=-90] (3.5,3);
    \node[right] at (-3.25,-2) {$M_0'$};
    
    \draw[thick, late arrow] (3.5,-3) to [out=90,in=-90] (-3.5,3);
    \node[left] at (3.25,-2) {$M_1$};

    \draw[thick] (-1,-9) -- (1,-9) -- (1,-7) -- (-1,-7) -- (-1,-9);
    \draw[thick, mid arrow] (-0.5,-7) to [out=90,in=-90] (-3.5,-5);
    \draw[thick, mid arrow] (0.5,-7) to [out=90,in=-90] (3.5,-5);

    \draw[thick, mid arrow]  (-1,-8) -- (-2.5,-8);
    \node[left] at (-2.5,-8) {$\bar{Q}$};
    
    \draw[thick, mid arrow] (-4.5,-6.5) -- (-4.5,-5);
    \node[below] at (-4.5,-6.5) {$x$};
    \draw[thick, mid arrow] (4.5,-6.5) -- (4.5,-5);
    \node[below] at (4.5,-6.5) {$y$};
    
    \draw[thick, late arrow] (4.5,5) -- (4.5,6.5);
    \node[above] at (-4.5,6.5) {$a$};
    \draw[thick, late arrow] (-4.5,5) -- (-4.5,6.5);
    \node[above] at (4.5,6.5) {$b$};

    \node at (0,-12) {$ $};
    
    \end{tikzpicture}
    \caption{}
    \label{fig:proverstateprepattack}
    \end{subfigure}
\caption{a) A $f$-measure style protocol where the prover generates the quantum message. An honest prover generates $\Psi^+_{\bar{Q}Q}$ with $\bar{Q}$ sent to the referee and $Q$ stored locally. b) The new attack model. Note that system $\bar{Q}$ is now prepared by the attacker and sent to the verifier. }
\label{fig:proverstate}
\end{figure}

This protocol has similar implementation requirements as the two round protocol in the last section. 
In particular, channel loss tolerance is achieved at the cost of introducing a quantum memory. 
The implementation of this protocol is somewhat different however, and may be favourable in some cases. 

\subsection{Round-and-a-half protocols}\label{sec:roundandhalfprotocols}

Looking at both the variants of the two round protocols, it is natural to ask if we really needed the prover to store the qubit for two light travel times, or if less time would suffice. 
Indeed we can, though we will find that this complicates the security proofs considerably, and we will not give complete proofs here. 

\begin{figure}
    \begin{subfigure}{0.45\textwidth}
    \centering
    \begin{tikzpicture}[scale=0.6]

    \node at (2,-1) {Verifier 1};
    \draw[fill=black] (2,0) circle (0.15);

    \node at (5.5,-1) {Prover};
    \draw[fill=black] (5.5,0) circle (0.15);
    
    \node at (9,-1) {Verifier 2};
    \draw[fill=black] (9,0) circle (0.15);

    \node[draw, thick, fill=white, inner sep=3pt] (load) at (5.5,3.5) {Load};
    \node[draw, thick, fill=white, inner sep=3pt] (measure) at (5.5,7) {measure};

    \draw[snake it, blue,thick ] (2,0) -- (load);
    \draw[snake it, blue,thick ] (load) -- (measure);

    \draw[thick, mid arrow] (2,3.5) -- (measure);
    \node at (2.6,4.7) {$x$};
    
    \draw[thick, mid arrow] (9,3.5) -- (measure);
    \node at (8.4,4.7) {$y$};

    \draw[thick, late arrow] (load) -- (2,7);
    \node[rotate=-45] at (2.9,6.6) {``Ready''};
    
    \draw[thick, late arrow] (load) -- (9,7);
    \node[rotate=45] at (8.1,6.6) {``Ready''};

    \draw[thick, mid arrow] (measure) -- (9,10.5);
    \draw[thick, mid arrow] (measure) -- (2,10.5);

    \draw[thick, ->] (0,0) -- (0,10.5);
    \node[above] at (0,10.5) {Time};
    \draw[thick, ->] (0,0) -- (10,0);
    \node[right] at (10,0) {Position};
\end{tikzpicture}
    \caption{}
    \label{fig:commitmentprotocol}
    \end{subfigure}
    \hfill
    \begin{subfigure}{0.45\textwidth}
    \centering
    \begin{tikzpicture}[scale=0.45]
    
    \draw[thick] (-5.5,-5) -- (-5.5,-3) -- (-2.5,-3) -- (-2.5,-5) -- (-5.5,-5);
    \node at (-4,-4) {$\mathcal{N}^x$};
    
    \draw[thick] (5.5,-5) -- (5.5,-3) -- (2.5,-3) -- (2.5,-5) -- (5.5,-5);
    \node at (4,-4) {$\mathcal{M}^y$};
    
    \draw[thick] (5.5,5) -- (5.5,3) -- (2.5,3) -- (2.5,5) -- (5.5,5);
    
    \draw[thick] (-5.5,5) -- (-5.5,3) -- (-2.5,3) -- (-2.5,5) -- (-5.5,5);
    
    \draw[thick, mid arrow] (-5,-3) -- (-5,3);
    \node[left] at (-5,-2) {$M_0$};
    
    \draw[thick, mid arrow] (5,-3) -- (5,3);
    \node[right] at (5,-2) {$M_1'$};
    
    \draw[thick, late arrow] (-3,-3) to [out=90,in=-90] (3,3);
    \node[right] at (-2.75,-2) {$M_0'$};
    
    \draw[thick, late arrow] (3,-3) to [out=90,in=-90] (-3,3);
    \node[left] at (2.75,-2) {$M_1$};

    \draw[thick] (-4,-3) -- (-4,-2);
    \draw[thick] (-4.75,-2) -- (-3.25,-2) -- (-4,-0.75) -- (-4.75,-2);
    \node at (-4,-1.5) {$0$};

    \draw[thick] (4,-3) -- (4,-2);
    \draw[thick] (3.25,-2) -- (4.75,-2) -- (4,-0.75) -- (3.25,-2);
    \node at (4,-1.5) {$0$};

    \draw[thick] (-1,-9) -- (1,-9) -- (1,-7) -- (-1,-7) -- (-1,-9);
    \draw[thick, mid arrow] (-0.5,-7) to [out=90,in=-90] (-3.5,-5);
    \draw[thick, mid arrow] (0.5,-7) to [out=90,in=-90] (3.5,-5);

    \draw[thick, mid arrow] (-2.5,-8) -- (-1,-8);
    \node[left] at (-2.5,-8) {$Q$};
    
    \draw[thick, mid arrow] (-4.5,-6.5) -- (-4.5,-5);
    \node[below] at (-4.5,-6.5) {$x$};
    \draw[thick, mid arrow] (4.5,-6.5) -- (4.5,-5);
    \node[below] at (4.5,-6.5) {$y$};
    
    \draw[thick, late arrow] (4,5) -- (4,6.5);
    \node[above] at (-4,6.5) {$a$};
    \draw[thick, late arrow] (-4,5) -- (-4,6.5);
    \node[above] at (4,6.5) {$b$};

    \draw[thick] (0,-7) -- (0,-6);
    \draw[thick] (-0.75,-6) -- (0.75,-6) -- (0,-4.75) -- (-0.75,-6);
    \node at (0,-5.5) {$0$};
    
\end{tikzpicture}
    \caption{}
    \end{subfigure}
\caption{a) The honest protocol when the verifier generates entanglement, and the prover should commit to playing the protocol before receiving the classical inputs. b) Attack model appropriate to this setting. By declaring a loss, the prover can post-select on measurement outcomes in their first round operations.}
\label{fig:commitmentprotocol_attack}
\end{figure}

In the case where the verifier generates the entanglement, the protocol is shown in figure \ref{fig:commitmentprotocol}. 
The corresponding attack model is shown in figure \ref{fig:commitmentprotocol_attack}.
Our attack model allows for slow quantum information, which means the attacker receives $Q$ early. 
Since they could act on this and choose to declare a loss based on a measurement outcome, they again have the ability to post-select in their early operations. 
Additionally however, because the ``Ready'' signal is delayed, they are choosing to play or not play later, after having received the inputs $x,y$. 
This allows them to post-select also in the first round operations, rather than just in the early processing of the quantum input. 

In fact however, this model is too generous, and effectively Alice and Bob are restricted to post-selection operations occurring in the state preparation phase. 
To see why, notice that if the prover was being honest, the prover in the middle decides once to declare a loss or not, and sends their decision in both directions. 
Thus the classical signal to both sides will always match. 
In the attack meanwhile, Alice and Bob can independently decide to return the ``Ready'' signal or declare a loss. 
If they do not always declare the same output, this will be detected and the verifier can reject. 
It turns out that this means Alice and Bob cannot really make use of post-selection in the first round --- in effect, the ``Ready'' decision had to happen early, so that it is coordinated between both players. 

As a simple example, suppose we try to execute the attack corresponding to Bob post-selecting on teleporting $\ket{y}$ to Alice with no corrections. 
Bob makes the needed measurements, and with small probability gets the measurement outcome corresponding to no corrections. 
If he does, he returns ``Ready'', and otherwise he declares a loss. 
But Alice meanwhile doesn't know if Bob's measurements succeeded, so will go ahead and declare ``Ready'' even while Bob declares a loss, which reveals that they are cheating. 
This effect returns us to the attack model of figure \ref{fig:tworoundattack}. 
We won't give a more formal statement of this claim, but see \cite{allerstorfer2025making}. 

\subsubsection{Prover generated quantum challenge}

As in the two round protocol, we can also modify the scheme to have the prover generate the quantum state, so they now send system $\bar{Q}$ to the verifier. 
The protocol is shown in figure \ref{fig:commitmentprotocol_proverquantum}, and the corresponding attack model in figure \ref{fig:commitmentprotocol_proverquantum_attackmodel}. 
As in the last section, we can also use the fact that ``Ready'' signals from the left and right must be coordinated to reduce the attack model here, which allows projections in the first round, to one which only uses a projection in the state preparation stage. 
At that point we can apply our reduction to communication complexity and obtain a linear gate lower bound. 

\begin{figure}
    \begin{subfigure}{0.45\textwidth}
    \centering
    \begin{tikzpicture}[scale=0.6]

    \node at (2,-1) {Verifier 1};
    \draw[fill=black] (2,0) circle (0.15);

    \node at (5.5,-1) {Prover};
    \draw[fill=black] (5.5,0) circle (0.15);

    \node at (9,-1) {Verifier 2};
    \draw[fill=black] (9,0) circle (0.15);

    \node[draw, thick, fill=white, inner sep=3pt] (generate) at (5.5,1) {Generate};
    \node[draw, thick, fill=white, inner sep=3pt] (vmeasure) at (2,4.5) {measure};
    \node[draw, thick, fill=white, inner sep=3pt] (measure)  at (5.5,5) {measure};

    \draw[snake it, blue, thick] (generate) -- (vmeasure);
    \draw[snake it, blue, thick] (generate) -- (measure);

    \draw[thick, mid arrow] (2,1.5) -- (measure);
    \node at (2.0,2.2) {$x$};

    \draw[thick, mid arrow] (9,1.5) -- (measure);
    \node at (8.8,2.2) {$y$};

    \draw[thick, mid arrow] (measure) -- (9,8.5);
    \draw[thick, mid arrow] (measure) -- (2,8.5);

    \draw[thick, ->] (0,0) -- (0,10);
    \node[right] at (10,0) {Position};
    \node[above] at (0,10) {Time};
    \draw[thick, ->] (0,0) -- (10,0);

\end{tikzpicture}
    \caption{}
    \label{fig:commitmentprotocol_proverquantum}
    \end{subfigure}
    \hfill
    \begin{subfigure}{0.45\textwidth}
    \centering
    \begin{tikzpicture}[scale=0.45]
    
    \draw[thick] (-5.5,-5) -- (-5.5,-3) -- (-2.5,-3) -- (-2.5,-5) -- (-5.5,-5);
    \node at (-4,-4) {$\mathcal{N}^x$};
    
    \draw[thick] (5.5,-5) -- (5.5,-3) -- (2.5,-3) -- (2.5,-5) -- (5.5,-5);
    \node at (4,-4) {$\mathcal{M}^y$};
    
    \draw[thick] (5.5,5) -- (5.5,3) -- (2.5,3) -- (2.5,5) -- (5.5,5);
    
    \draw[thick] (-5.5,5) -- (-5.5,3) -- (-2.5,3) -- (-2.5,5) -- (-5.5,5);
    
    \draw[thick, mid arrow] (-5,-3) -- (-5,3);
    \node[left] at (-5,-2) {$M_0$};
    
    \draw[thick, mid arrow] (5,-3) -- (5,3);
    \node[right] at (5,-2) {$M_1'$};
    
    \draw[thick, late arrow] (-3,-3) to [out=90,in=-90] (3,3);
    \node[right] at (-2.75,-2) {$M_0'$};
    
    \draw[thick, late arrow] (3,-3) to [out=90,in=-90] (-3,3);
    \node[left] at (2.75,-2) {$M_1$};

    \draw[thick] (-4,-3) -- (-4,-2);
    \draw[thick] (-4.75,-2) -- (-3.25,-2) -- (-4,-0.75) -- (-4.75,-2);
    \node at (-4,-1.5) {$0$};

    \draw[thick] (4,-3) -- (4,-2);
    \draw[thick] (3.25,-2) -- (4.75,-2) -- (4,-0.75) -- (3.25,-2);
    \node at (4,-1.5) {$0$};

    \draw[thick] (-1,-9) -- (1,-9) -- (1,-7) -- (-1,-7) -- (-1,-9);
    \draw[thick, mid arrow] (-0.5,-7) to [out=90,in=-90] (-3.5,-5);
    \draw[thick, mid arrow] (0.5,-7) to [out=90,in=-90] (3.5,-5);

    \draw[thick, mid arrow] (-1,-8) -- (-2.5,-8);
    \node[left] at (-2.5,-8) {$\bar{Q}$};
    
    \draw[thick, mid arrow] (-4.5,-6.5) -- (-4.5,-5);
    \node[below] at (-4.5,-6.5) {$x$};
    \draw[thick, mid arrow] (4.5,-6.5) -- (4.5,-5);
    \node[below] at (4.5,-6.5) {$y$};
    
    \draw[thick, late arrow] (4,5) -- (4,6.5);
    \node[above] at (-4,6.5) {$a$};
    \draw[thick, late arrow] (-4,5) -- (-4,6.5);
    \node[above] at (4,6.5) {$b$};

    \draw[thick] (0,-7) -- (0,-6);
    \draw[thick] (-0.75,-6) -- (0.75,-6) -- (0,-4.75) -- (-0.75,-6);
    \node at (0,-5.5) {$0$};
    
\end{tikzpicture}
    \caption{}
    \label{fig:commitmentprotocol_proverquantum_attackmodel}
    \end{subfigure}
\caption{a) Protocol where the prover generates the quantum state, and the classical input is timed to arrive shortly after the state is generated. b) Attack model appropriate for this scheme. By choosing not to send the state, the attacker can force the protocol to abort. This gives them the ability to post-select on measurement outcomes.}
\end{figure}

\section{History and further reading}

Serious interest in implementing QPV in the lab in large part began with the work \cite{bluhm2022single}; this paper argued $f$-measure and $f$-route needed $\Omega(n)$ quantum resources to attack, even while the honest player needed only $O(1)$ quantum resources. 
The argument there begins with observations dating back to \cite{buhrman2013garden}, in particular the observation that, in $f$-routing, the routing direction is fixed by the first round operations. 
A key drawback of the argument in \cite{buhrman2013garden} though was that the lower bound required a perfect implementation of the $f$-routing protocol; \cite{bluhm2022single} extends the lower bound to the case with errors allowed. 
Both works rely on counting arguments and lower bound the quantum resources needed to implement a random function $f$. 

One limitation of this initial approach is that in the attack model discussed in \cite{bluhm2022single}, all of Alice and Bob's operations are unitary, with the choice of unitary fixed by the inputs $x,y$. 
Lower bounding quantum resources in this model need not correspond to a lower bound on the quantum resources used by a physical attacker, who may (for instance) perform mid-circuit measurements and adaptively choose quantum gates.
Dilating a process like this to a unitary view, the classical workspace is counted as quantum, so the claimed separation between quantum resources used by an attacker vs an honest player may not be maintained. 
Inspired by \cite{bluhm2022single} however, a number of works sought to make the scheme more practical, by addressing this issue in the attack model or by addressing one or more of the challenges mentioned in section \ref{sec:QPVchallenges}.

The separate consideration of a ``loss'' output and study of the loss-tolerance of $f$-measure was studied in \cite{escola2023single}, who considered loss tolerance in the random function setting.
To allow arbitrary loss, and hence longer distance QPV, \cite{allerstorfer2025making} proposed modifying the standard $f$-measure protocol to include a commitment step, which corresponds to our round-and-a-half protocol from section \ref{sec:roundandhalfprotocols}.  
Modifying the protocol to have the prover generate the quantum challenge, which can also be used to establish arbitrary loss tolerance, was proposed in \cite{kon2026losstolerant}.

One aspect of security of QPV schemes we have not addressed here is the security under parallel repetition of the $f$-measure or similar schemes. 
This may be desirable for instance if we want to increase the quantum resources needed by an attacker, but we don't want to degrade precision by increasing $n$ too far (recall that this incurs a time delay). 
This has been studied in the context of random functions and purified strategies in \cite{escola2025quantum}. 
It is open to prove security under parallel repetition for the gate bounds discussed here. 

Another approach to the security of QPV is to consider the quantum random oracle model (QROM). 
In this approach, we consider $f$ to be a random function, unknown to the players, that can be queried on any choice of inputs. 
This is called the random oracle. 
Since we are in the quantum setting, the oracle can be queried on superpositions of inputs.
We let both Alice and Bob, in the attack model, query the oracle. 
Unruh \cite{unruh2014quantum} shows that for a random oracle, an exponential (in the classical input size) number of queries are needed in the attack to implement $f$-measure. 
Meanwhile, the honest player may bring the inputs together, query the oracle once on the full input, and then measure in the correct basis, so they only need a single oracle call. 
Unruh modifies the standard $f$-measure scheme which uses Boolean functions to use a function $f:\{0,1\}^n\rightarrow \{0,1\}^\ell$, and corresponding uses a $\ell$ qubit input, with the $i$th bit of the input measured in the basis specified by the $i$th bit of the output. 

Given a lower bound on oracle calls, the standard argument in cryptography is to replace the oracle with a concrete function, usually called a \emph{hash function}, which is a concrete function that has certain nice properties that make it look similar to a random function in some ways. 
We then hope that the random oracle is a good model of the hash function, in that the security of the protocol remains even when using the hash rather than the random function. 
The intuition is that the behaviour of the function is random enough, there will not be enough structure for the attack to exploit and do significantly better than attacking the random function. 

Unfortunately, many of the standard functions which are assumed to be good hash functions are actually easy to implement in the $f$-measure protocol, so this intuition does not seem to hold in the context of QPV. 
For instance, we could choose the concrete function to be pseudo-random, and hope the pseudo-randomness property is strong enough to lead to security. 
Recall that pseudo-random functions are those for which no poly-time algorithm, querying the function, can distinguish the given function from a random function. 
These also have nice properties that might be thought to help, like being difficult to invert, and unlikely to have outputs collide. 
However, pseudo-random functions are believed to be constructable in the class $\mathsf{TC}^0$, which is inside of $\LogSpace$ and hence has an efficient $f$-measure protocol. 
Thus pseudo-randomness does not suffice to obtain $f$-measure with exponential security. 

More generally, reasoning in the random oracle model explicitly gives up on looking at any description of $f$ and using the structure of $f$ to attack the protocol. 
The lower bound in this viewpoint is saying that without exploiting the structure of $f$, we can't have an efficient attack. 
At the same time, the known attacks (e.g. code-routing) explicitly do use the structure of $f$ in a detailed way, for instance using a formula decomposition of $f$. 
Thus efficient attacks do not use $f$ in a black-box way. 
Thus a proof that without going beyond querying the function as a black box we won't get an efficient attack is good to know, but also expected, and doesn't make clear how to identify good candidate explicit functions $f$ for security in $f$-measure. 
In particular the usual notion of ``unstructured enough'', which means unstructured looking when you sample the function, explicitly doesn't suffice here. 
Thus the key problem to obtaining super-linear security of $f$-routing or $f$-measure is left open, which is to identify what the right properties of an explicit function $f$ are that make it hard in the context of $f$-routing or $f$-measure.

\part{Relational approach to NLQC}\label{part:structure}

\chapter{Reductions among NLQC families}\label{chapter:reductions}

\minitoc

In parts II and III of this book, we studied upper and lower bounds on NLQC. 
We looked for bounds expressed in terms of the specification of the NLQC being considered. 
For instance in $f$-routing, the NLQC is specified by the choice of Boolean function $f$, and we looked for bounds in terms of properties of $f$.  
In part IV, we take instead a \emph{relational} approach to entanglement cost in NLQC: we try to understand when one example of NLQC is harder or easier to implement than another. 
This lets us address the question of what makes an NLQC hard without ever proving explicit bounds. 

We are motivated to take this approach for a number of reasons. 
First, we found our lower bound techniques are limited, with all proven lower bounds being at best linear. 
Thus we are not able to compare the hardness of any NLQCs that require super-linear entanglement by finding entanglement bounds directly, but we can hope to circumvent this and directly relate the hardness of examples. 

Another motivation is from upper bounds: for certain NLQCs, for instance $f$-routing, upper bounds are well studied and we have various approaches, including the sub-exponential upper bound for all functions, the garden-hose technique, and span-program based upper bounds. 
The theory of relations among NLQCs allows us to re-apply these upper bound strategies across a wide range of NLQC examples. 
More generally, relating NLQCs gives an economy of effort --- we don't need to reprove the same results across many examples, but will find that new examples of NLQC inherit properties of better studied ones via these relations. 

Perhaps most broadly, the study of relations among NLQC examples is inspired by the role of \emph{reductions} in the study of computational complexity theory. 
In that context, a computational problem A can be reduced to B if (a small number of) calls to a machine that solves B can be used to solve A, perhaps along with the use of small additional resources. 
Reductions are central to complexity theory in that they reveal the relative hardness of computational problems, and allow the organization of problems into complexity classes. 
Because of the many close relations between entanglement cost in NLQC and computational complexity, it is natural to adopt this strategy in NLQC as well. 
Our treatment follows \cite{bluhm2025complexity, bluhm2026reductions}.

\section{NLQC reductions}

Recalling our definition of a $2\rightarrow 2$ quantum task from chapter \ref{chapter:NLQC}, we will also be interested in \emph{families} of $2\rightarrow 2$ quantum tasks, which are collections of $2\rightarrow 2$ tasks parameterized by a natural number $n$.
The parameter $n$ will correspond to an input size with the exact relation specified in the definition of each family of tasks.  
We will label families of tasks with capital letters $F$, $G$, etc., where these denote sets of tasks, so that $F=\{F_n\}_n$, $G=\{G_n\}_n$, etc. 
As an example, $f$-routing with a choice of Boolean function family $\{f_n\}_n$ defines a family of $2\rightarrow 2$ tasks, with each member of the family labelled by an element of $\{f_n\}_n$.  

We are interested in understanding the relative difficulty of implementing different $2\rightarrow 2$ tasks as NLQCs.
To do this, we will define a notion of reduction between $2\rightarrow 2$ tasks. 
Heuristically, our notion of reduction says that the task $G$ reduces to $F$ when (a few copies of) the resources to implement $F$ as an NLQC can be used to implement $G$ as an NLQC. 

To formalize our notion of a reduction among $2\rightarrow 2$ tasks, it is helpful to recall the notion of reduction among computational problems. 
There we say a function family $A=\{a_n\}_n$ is polynomial-time Turing reducible to function family $B=\{b_n\}$ if a poly$(n)$ time machine with oracle access to $B$ can solve $A$. 
More generally, we can replace `poly-time' with any other complexity class, call it $\mathcal{X}$; the notion of reduction is most meaningful when $\mathcal{X}$ is itself too weak to implement $A$ or $B$.
Inspired by this definition, we give the following definition of reduction among NLQC classes. 
\begin{definition}\label{def:reduction}
    Let $F=\{F_n\}_n$ and $G=\{G_n\}_n$ be families of $2\rightarrow 2$ tasks, and $\alpha$, $\beta$, $\delta$ all be functions of $(n,\lambda,\epsilon)$, with $\delta\rightarrow 0$ as $\lambda\rightarrow \infty, \epsilon\rightarrow 0$.
    Then we say there is a $(\alpha, \beta,\delta)$-\textbf{reduction} from $G$ to $F$ if, for any resource state $\Psi^{n,\epsilon}_{LR}$ which can be used to implement $F_n$ at least $\epsilon$-correctly as an NLQC, it is possible to implement $G_n$ $\delta$-correctly using $\alpha$ copies of $\Psi^{n,\epsilon}_{LR}$ along with $\beta$ additional qubits of shared resource state.
\end{definition}
When there is a reduction from family $F$ to family $G$, we will also say there is an \textbf{implication} from $G$ to $F$, and write $G \Rightarrow F$.
When $\alpha, \beta$ are both $O(g(n))$, we say we have a \emph{$O(g(n))$ reduction}. 
In practice, we will construct $O(1)$ reductions, by which we mean that $\alpha,\beta$ have no $n$ dependence. 
Note that $\alpha,\beta$ may still depend on $\lambda,\epsilon$. 
The parameter $\lambda$ should be interpreted as parameterizing a family of protocols, which implement $F$ well as we increase $\lambda$. 
When it is necessary to distinguish this notion of a reduction from other similar notions, we refer to it as a \emph{resource state reduction}.  

One further notion of reduction that we make use of requires that the protocol for $G_n$ be given by using the protocol for $F_n$ used as an oracle, meaning it has access only to copies of the implementation of $F_n$ but not access directly to the resource state $\Psi^{n,\epsilon}_{LR}$. 
We give a definition of this notion of reduction next. 
\begin{definition}\label{def:oraclereduction}
    Let $F=\{F_n\}_n$ and $G=\{G_n\}_n$ be families of $2\rightarrow 2$ tasks, and $\alpha$, $\beta$, $\delta$ all be functions of $(n,\lambda,\epsilon)$, with $\delta\rightarrow 0$ as $\lambda\rightarrow \infty, \epsilon\rightarrow 0$.
    Then we say that there is a $(\alpha,\beta,\delta)$ \textbf{oracle reduction} from $G$ to $F$ if $G$ can be implemented $\delta$ correctly by using $\alpha$ parallel implementations of $F$ along with $\beta$ additional qubits of resource system and communication. 
\end{definition}
The structure of an implementation of $G$ using $F$ as an oracle is shown in figure \ref{fig:oraclereduction}. 

We should also comment on the requirement that $\delta\rightarrow 0$ as $\lambda\rightarrow \infty, \epsilon\rightarrow 0$, which appears in both notions of reduction. 
This requirement ensures that our notion of reduction is non-trivial. 
For instance, consider $f$-routing: without use of any entanglement, it is always possible to bring the input quantum system to the correct side with probability $1/2$, and hence trivially achieve some $\epsilon=\epsilon_0$ correctness parameter. 
Thus, without the $\delta\rightarrow 0$ requirement, any NLQC implies $f$-routing under a $(\alpha,\beta,\delta=\epsilon_0)$ reduction. 
Our definition excludes trivial constructions like this from being considered as reductions. 
Concretely, requiring $\delta\rightarrow 0$ in the double limit $\lambda\rightarrow \infty$,$\epsilon\rightarrow 0$ imposes that as the implementations of $F$ become perfect, and we are allowed to increase how many of them we use $(\alpha$ can grow with $\lambda$) or how much additional resources we use ($\beta$ can grow with $\lambda$), the implementation of $G$ becomes perfect. 

\begin{figure}
    \centering
    \begin{tikzpicture}[scale=0.5]

    \draw[thick, fill=gray,opacity=0.5] (-3,-1.5) -- (3,-1.5) -- (3,1.5) -- (-3,1.5) -- (-3,-1.5);
    
    \draw[thick] (-5,-5) -- (-5,-3) -- (-3,-3) -- (-3,-5) -- (-5,-5);
    
    \draw[thick] (5,-5) -- (5,-3) -- (3,-3) -- (3,-5) -- (5,-5);
    
    \draw[thick] (5,5) -- (5,3) -- (3,3) -- (3,5) -- (5,5);
    
    \draw[thick] (-5,5) -- (-5,3) -- (-3,3) -- (-3,5) -- (-5,5);
    
    \draw[thick,mid arrow,dashed] (-4.5,-3) -- (-4.5,3);
    
    \draw[thick,mid arrow,dashed] (4.5,-3) -- (4.5,3);
    
    \draw[thick,mid arrow, dashed] (-4,-3) -- (-4,-1.5) to [out=90,in=-90] (4,3);
    
    \draw[thick,mid arrow, dashed] (4,-3) -- (4,-1.5) to [out=90,in=-90] (-4,3);

    \draw[thick,gray] (-3.5,-3) -- (-2,-1.5);
    \draw[thick,gray] (3.5,-3) -- (2,-1.5);
    \draw[thick,gray] (3.5,3) -- (2,1.5);
    \draw[thick,gray] (-3.5,3) -- (-2,1.5);
    
    \draw[thick] (-3.5,-6) -- (3.5,-6) -- (0,-8) -- (-3.5,-6);
    \draw[thick] (-3.25,-6) -- (-3.25,-5);
    \draw[thick] (3.25,-6) -- (3.25,-5);
    \node at (0,-7) {$\Phi$};

    \node at (0,-0.25) {$F_n^{\otimes \alpha(n)}$};
    
    \draw[thick] (-4.5,-6) -- (-4.5,-5);
    \draw[thick] (4.5,-6) -- (4.5,-5);
    
    \draw[thick] (4.5,5) -- (4.5,6);
    \draw[thick] (-4.5,5) -- (-4.5,6);
    
    \end{tikzpicture}
    \caption{An oracle reduction from an NLQC $G$ to $F$. The protocol for $G_n$ uses $\alpha(n)$ instances of $F_n$, plus at most $\beta(n)$ qubit resource system $\Phi$ and at most $\beta(n)$ qubits of communication.}
    \label{fig:oraclereduction}
\end{figure}

Compared to the notion of resource state reduction in definition \ref{def:reduction}, an oracle reduction gives a tighter relationship between task families.
This is because in a resource state reduction, we've shown that a resource state that works for $F$ also (up to certain overheads) works for $G$, while an oracle reduction implies not only that, but that additionally the local operations used in $F$ can also be applied to complete $G$. 
Resource state reductions allow greater flexibility in what operations are performed locally to exploit the non-local resources, and reductions under this definition focus on the power of non-local resources for completing NLQCs.

A useful fact is that oracle reductions have convenient error-propagation properties. 
\begin{remark}\label{remark:blackboxerror}
    Suppose that there is a protocol which uses parallel implementations of channels $\mathcal{N}_1, \mathcal{N}_2,...$ to execute some target channel $\mathcal{N}$, so that
    \begin{align}
        \mathcal{N} = \mathcal{F}\circ \left(\bigotimes_{i=1}^m \mathcal{N}_i \right) \circ \mathcal{P}.
    \end{align}
    Then suppose we replace the exact implementations of the $\mathcal{N}_i$ with approximate implementations $\overline{\mathcal{N}}_i$ which satisfy $\Vert \mathcal{N}_i-\overline{\mathcal{N}}_i\Vert_\diamond \leq \epsilon_i$. 
    Then
    \begin{align}
        \overline{\mathcal{N}} = \mathcal{F}\circ \left(\bigotimes_{i=1}^m \overline{\mathcal{N}}_i \right) \circ \mathcal{P}
    \end{align}
    is $\sum_i \epsilon_i$ close in diamond norm to $\mathcal{N}$. 
\end{remark}
This remark follows from the properties of the diamond norm. 

To apply this remark to our oracle reductions, note that any realization of $G_n$ using oracle instances of $F_n$ must be exactly correct when $F_n$ is exactly correct. 
This is because we require that $\delta \rightarrow 0$ as $\epsilon\rightarrow 0$. 
When we instead use $\epsilon$ correct realizations of $F_n$, we get from the remark above that the diamond norm distance to the perfectly correct channel is at most $\alpha \cdot \epsilon$, where $\alpha $ is the number of instances of $F_n$ used. 

\section{\texorpdfstring{$f$}{TEXT}-routing is \texorpdfstring{$f$}{TEXT}-measure}

After $f$-routing, the best studied $2\rightarrow 2$ task is $f$-measure. 
This modifies the standard measure task, considered in chapter \ref{chapter:NLQC}, by hiding the basis information in the output of a Boolean function $f:X\times Y\rightarrow \{0,1\}$, with $x\in X=\{0,1\}^n$ given to Alice and $y\in Y=\{0,1\}^n$ given to Bob. 
This task is of interest especially in the context of position verification. 
In that context, it has the same advantage as $f$-routing in that the quantum operations of the honest player are $O(1)$ size (they do not grow with $n$), but all known NLQC protocols require entanglement that grows with $n$. 
$f$-measure has the additional advantage that quantum communication is only required from one verifier to the prover, and not back to both verifiers as in $f$-routing. 
In fact, recent experimental implementations of position verification protocols implement the $f$-measure or very similar protocols \cite{fan2026relativistic,kavuri2025device}. 

We develop a new definition of $f$-measure next. 
This definition is somewhat different than the one we saw earlier as definition \ref{def:qubitfbb84-initial}. 
In particular in definition \ref{def:qubitfbb84-initial} we take there to be a promise on the inputs, which is that the inputs are drawn from the BB84 states. 
Here, we need a definition that works for all possible input states. 
Define the measurement channels
\begin{align}\label{eq:measurechannels}
    \mathcal{M}^{0}_{Q\rightarrow ZZ'}(\rho_Q) &= \sum_{b} \bra{b}\rho\ket{b}_Q \, \ketbra{b}{b}_Z\otimes \ketbra{b}{b}_{Z'},\nonumber \\
    \mathcal{M}^{1}_{Q\rightarrow ZZ'}(\rho_Q) &= \sum_{b} \bra{b}H\rho H\ket{b}_Q \, \ketbra{b}{b}_Z\otimes \ketbra{b}{b}_{Z'}.
\end{align}
We can then define the $f$-measure task as follows. 

\begin{definition}\label{def:qubitfbb84-channel}
    A \textbf{qubit $f$-measure} task is defined by a choice of Boolean function $f:\{ 0,1\}^{2n}\rightarrow \{0,1\}$, and a $2$ dimensional Hilbert space $\mathcal{H}_Q$.
    Inputs $x\in \{0,1\}^{n}$ and system $Q$ are given to Alice, and input $y\in \{0,1\}^{n}$ is given to Bob.
    The $f$-measure task is completed $\epsilon$-correctly on input $(x,y)$ if channel $\mathcal{N}^{x,y}$ executed on input $x,y$ satisfies
    \begin{align}
        \Vert \mathcal{M}^{f(x,y)} - \mathcal{N}^{x,y} \Vert_\diamond \leq \epsilon,
    \end{align}
    where $\mathcal{M}^0$, $\mathcal{M}^1$ are as defined in equation \eqref{eq:measurechannels}. We say the task is implemented $\epsilon$-correctly if the above holds for all inputs $(x,y)$. 
\end{definition}
We are interested in implementing the $f$-measure task as an NLQC. 

A natural question is: is $f$-routing or $f$-measure harder as an NLQC?
In some ways these tasks are similar --- they both involve small quantum operations controlled off of large classical data --- but apparently the core difficulty of what makes them non-trivial differs. 
For $f$-routing, no cloning prevents us from sending the quantum state to both sides. 
For $f$-measure, the incompatibility of non-commuting measurements means we can't measure simultaneously in the two possible bases. 
As well, algorithmically, NLQC protocols for the two cases seem quite different. 
For instance, we can apply the connection to CDS to obtain sub-exponential protocols for every function $f$ when $f$-routing, but it's not clear how to use similar protocols for $f$-measure. 

In fact, there is a hidden relationship between $f$-routing and $f$-measure: the two settings are equivalent under $O(1)$ reductions. 
From this fact we inherit new protocols and lower bounds for $f$-measure: all of the upper bounds from chapter \ref{chapter:f-routing} and all of the lower bounds from chapter \ref{chapter:mostlyclassicallowerbounds} apply to $f$-measure. 
Perhaps more fundamentally, this result indicates that the routing task and unknown basis measuring task are fundamentally the same from the perspective of NLQC. 

To show the equivalence of $f$-route and $f$-measure, we go through a long sequence of steps through intermediate tasks:
\begin{itemize*}
    \item Section \ref{sec:routetoBB84}: $f$-route implies $f$-measure 
    \item Section \ref{sec:BB84toBell}: $f$-measure implies $f$-Bell
    \item Section \ref{sec:BelltoClifford}: $f$-Bell implies $f$-Clifford
    \item Section \ref{sec:SWAPtoroute}: $f$-SWAP (a special case of $f$-Clifford) implies $f$-route
\end{itemize*}
The equivalence of many of these other tasks to $f$-measure and $f$-route is also of independent interest.

\subsection{\texorpdfstring{$f$}{TEXT}-route implies \texorpdfstring{$f$}{TEXT}-measure}\label{sec:routetoBB84}

Our first implication is from $f$-route to $f$-measure. 

\begin{lemma} \label{lem:frouteTOfmeasure}
    \textbf{$f$-route $\Rightarrow$ $f$-measure:} $f$-measure can be implemented using one instance of an $f$-route oracle and one instance of a $\neg f$-route oracle. 
\end{lemma}
\begin{proof} We begin by describing the $f$-measure protocol and then check its correctness below. 

\vspace{0.2cm}
\noindent \emph{Protocol:} In the $f$-measure protocol, we are given input $x\in \{0,1\}^n$, $\rho_A$ on Alice's side\footnote{Note that for convenience we've relabelled the $Q$ input as system $A$.}, $y\in\{0,1\}^n$ on Bob's side. 
Alice's protocol begins by copying system $A$ in the computational basis into a register $C$, then copying $A$ in the Hadamard basis into $B$, and $C$ in the Hadamard basis into $D$. 
We consider the effect of these operations on the four states $\ket{0}, \ket{1}, \ket{+}, \ket{-}$.
Thus for example given input $\ket{0}_A$ this procedure gives
\begin{align}
    \ket{0}_A\overset{\text{copy}}{\rightarrow} \ket{00}_{AC} &=(\ket{+}_A+\ket{-}_A)(\ket{+}_C+\ket{-}_C) \nonumber \\
    &=\ket{++}_{AC}+ \ket{+-}_{AC}+ \ket{-+}_{AC} + \ket{--}_{AC} \nonumber \\
    &\overset{\text{copy}\, H}{\rightarrow} \ket{++++}_{ABCD}+ \ket{++--}_{ABCD}+ \ket{--++}_{ABCD}+ \ket{----}_{ABCD} \nonumber\\
    &=\ket{0000}_{ABCD}+ \ket{0011}_{ABCD}+ \ket{1100}_{ABCD}+\ket{1111}_{ABCD}, \nonumber 
\end{align}
where we haven't kept track of normalization. 
Repeating this for all four input states, we find
\begin{align}
    \ket{\Psi_{f=0,b=0}}_{ABCD} &= \frac{1}{2}\left(\ket{0000}+ \ket{0011}+ \ket{1100}+\ket{1111} \right), \nonumber \\
    \ket{\Psi_{f=0,b=1}}_{ABCD} &= \frac{1}{2}\left(\ket{0101}+ \ket{0110} + \ket{1001} + \ket{1010} \right), \nonumber \\
    \ket{\Psi_{f=1,b=0}}_{ABCD} &= \frac{1}{\sqrt{2}}\left(\ket{++++}+\ket{----} \right), \nonumber \\
    \ket{\Psi_{f=1,b=1}}_{ABCD} &=\frac{1}{\sqrt{2}}\left(\ket{++--}+\ket{--++} \right).
\end{align}
Alice and Bob perform $f$-route on the $B$ system, and $\neg f$-route on the $C$ system.
Alice always sends $D$ to Bob. 
Afterwards, Alice labels the two systems she holds (which may be $AB$ if $f=0$ or $AC$ if $f=1$) as $AB$, and Bob labels the two systems he holds ($CD$ if $f=0$ or $BD$ if $f=1$) as $CD$.
This produces the four states
\begin{align}\label{eq:postSWAP}
    \ket{\Psi_{0}}_{ABCD} &= \frac{1}{2}\left(\ket{0000}+ \ket{0011}+ \ket{1100}+\ket{1111} \right), \nonumber \\
    \ket{\Psi_{1}}_{ABCD} &= \frac{1}{2}\left(\ket{0101}+ \ket{0110} + \ket{1001} + \ket{1010} \right), \nonumber \\
    \ket{\Psi_{+}}_{ABCD} &= \frac{1}{\sqrt{2}}\left(\ket{++++}+\ket{----} \right), \nonumber \\
    \ket{\Psi_{-}}_{ABCD} &=\frac{1}{\sqrt{2}}\left(\ket{+-+-}+\ket{-+-+} \right).
\end{align}
In the second round, Alice and Bob know $x,y$ and hence know $f(x,y)$.
Then, Alice measures $AB$ and Bob measures $CD$; they measure each qubit in the computational basis if $f(x,y)=0$ and in the Hadamard basis if $f(x,y)=1$. 
Each of Alice and Bob then output the parity of their two measurement outcomes. 

\vspace{0.2cm}
\noindent \emph{Correctness:} To see that this procedure implements the $f$-measure task correctly, consider each of the $f(x,y)=0$ and $f(x,y)=1$ cases separately. 
We will show perfect correctness in both cases for pure state inputs. 
But we know that if two channels agree on all pure state inputs, they are in fact equal as channels and in particular the diamond norm between them is zero, so this shows perfect correctness according to our definition. 

Beginning with $f(x,y)=0$, and considering the action of the protocol described above for inputs $\ket{0}, \ket{1}$ by linearity, we have that just before the measurement is made
 \begin{align}
     \ket{\psi}=\alpha \ket{0}+\beta\ket{1}\rightarrow \alpha \ket{\Psi_0}+\beta\ket{\Psi_1}.
 \end{align}
Label the overall action of the protocol, including the measurement, by the channel $\mathcal{N}^{x,y}_{A\rightarrow ZZ'}$. 
Then because $\Psi_0$ returns even parity and $\Psi_1$ returns odd parity, the channel acts according to
\begin{align}
     \mathcal{N}^{x,y}(\psi_A) &= |\alpha|^2 \ketbra{0}{0}\otimes \ketbra{0}{0}+|\beta|^2\ketbra{1}{1}\otimes \ketbra{1}{1} \nonumber \\
     &= \sum_b \bra{b}\psi\ket{b}_A \ketbra{b}{b}_Z \otimes \ketbra{b}{b}_{Z'}
\end{align}
This is exactly the channel $\mathcal{M}^0$, so the protocol is correct on $(x,y)\in f^{-1}(0)$ instances. 

Now we consider $(x,y)\in f^{-1}(1)$ instances. 
In that case, we write the input $\psi$ in the Hadamard basis, 
\begin{align}
    \ket{\psi}=a\ket{+}+b\ket{-}
\end{align}
Then the protocol described above, just before the measurement, produces the map
\begin{align}
    a\ket{+}+b\ket{-}\rightarrow a\ket{\Psi_+} + b\ket{\Psi_-}
\end{align}
After the measurement is made, we obtain
\begin{align}
    \mathcal{N}^{x,y}(\psi_A) &= |a|^2\ketbra{0}{0}_Z\otimes \ketbra{0}{0}_{Z'} +|b|^2\ketbra{1}{1}_Z\otimes \ketbra{1}{1}_{Z'}  \nonumber \\
    &= \sum_b\bra{b}H \psi_A H\ket{b}\, \ketbra{b}{b}_Z\otimes \ketbra{b}{b}_{Z'}
\end{align}
This is exactly the action of the $\mathcal{M}^1$ channel, so the protocol is perfectly correct in $(x,y)\in f^{-1}(1)$ instances.
\end{proof}

Note that $f$-route and $\neg f$-route can be performed using the same resource system. 
This is because Alice and Bob can run the first round of the $f$-route protocol, then keep the system they would have sent, and send the system they would have kept. 
This effectively performs a SWAP operation on the outputs, so that it implements the $\neg f$-route protocol. 
This means the above also shows a resource state implication from $f$-route to $f$-measure where two copies of the $f$-route resource state are used. 

\subsection{\texorpdfstring{$f$}{TEXT}-measure to \texorpdfstring{$f$}{TEXT}-Bell}\label{sec:BB84toBell}

The next step in our path back to $f$-route is a task we call $f$-Bell. 
The $f$-Bell task is defined using the following measurement channels. 
\begin{align}
    \mathcal{B}^0(\rho_Q)&=\sum_{a,b} \bra{a,b}\rho_Q\ket{a,b}\, \ketbra{a,b}{a,b}\otimes \ketbra{a,b}{a,b}\nonumber \\
    \mathcal{B}^1(\rho_Q)&=\sum_{a,b} \bra{\Psi_{a,b}}\rho_Q\ket{\Psi_{a,b}}\, \ketbra{a,b}{a,b}\otimes \ketbra{a,b}{a,b}
\end{align}

\begin{definition}\label{def:fBell}
    A \textbf{$f$-Bell} task is defined by a choice of Boolean function $f:\{ 0,1\}^{2n}\rightarrow \{0,1\}$, and a two qubit Hilbert space $\mathcal{H}_Q$.
    Inputs $x\in \{0,1\}^{n}$ and system $Q$ are given to Alice, and input $y\in \{0,1\}^{n}$ is given to Bob.
    The $f$-Bell task is completed $\epsilon$-correctly on input $(x,y)$ if channel $\mathcal{N}^{x,y}$ executed on input $x,y$ satisfies
    \begin{align}
        \Vert \mathcal{B}^{f(x,y)} - \mathcal{N}^{x,y} \Vert_\diamond \leq \epsilon \enspace.
    \end{align}
    We say the task is implemented $\epsilon$-correctly if the above holds for all inputs $(x,y)$. 
\end{definition}

We now proceed to show $f$-Bell is implied by $f$-measure. 

\begin{lemma} \label{lem:fBB84TOfBell}
    \textbf{$f$-measure $\Rightarrow$ $f$-Bell:} $f$-Bell can be implemented using one $f$-measure oracle. 
\end{lemma}

\begin{proof} 
We begin with a description of the $f$-Bell protocol and then check correctness below. 
\vspace{0.2cm}
\noindent \textbf{Protocol:}
Alice and Bob's protocol is as follows. 
We will track the action of the protocol for the computational basis input and Bell basis inputs, beginning with the computational basis states.  
First, Alice applies a $CNOT$ to her input state. 
In the computational basis, this permutes the input states:
\begin{align}\label{eq:CNOTpermutation}
    \{ \ket{00}, \ket{01}, \ket{10}, \ket{11} \} \xrightarrow{CNOT} \{ \ket{00}, \ket{01}, \ket{11}, \ket{10} \}.
\end{align}
If the input is in the Bell basis, we obtain the following states:
\begin{align}\label{eq:Bellpermuation}
    \{ \ket{\Psi_{00}}, \ket{\Psi_{01}}, \ket{\Psi_{10}}, \ket{\Psi_{11}} \} \xrightarrow{CNOT} \{\ket{+} \ket{0}, \ket{-} \ket{0}, \ket{+} \ket{1}, \ket{-} \ket{1} \}
    \end{align}
Next, we apply the $f$-measure protocol on the \textit{first} qubit and we measure the second qubit in the computational basis. 
In the second round, Alice and Bob learn $f(x,y)$ and obtain the two measurement outcomes from the $f$-measure protocol and the computational basis measurement. 
If $f(x,y)=0$, the $f$-measure protocol measures the first qubit in the computational basis. 
Alice and Bob undo the permutation of equation \eqref{eq:CNOTpermutation} and output the corresponding computational basis state label. 
If $f(x,y)=1$, the $f$-measure protocol measures the first qubit in the Hadamard basis. 
Alice and Bob interpret this outcome and the outcome from the computational basis measurement according to the mapping in equation \eqref{eq:Bellpermuation} and output the label of the corresponding Bell basis state. 

\vspace{0.2cm}
\noindent \textbf{Correctness:} To check that this protocol is correct, we consider $(x,y)\in f^{-1}(0)$ and $(x,y)\in f^{-1}(1)$ instances separately. 
Begin with the $(x,y)\in f^{-1}(0)$ instances, in which case the protocol should implement the $\mathcal{B}^0$ channel. 
Write the input state in the computational basis, and then observe that before the measurement the protocol produces the state
\begin{align}
    \ket{\psi}_Q=\alpha_{00} \ket{00} + \alpha_{01}\ket{01}+\alpha_{10}\ket{10} + \alpha_{11}\ket{11} \rightarrow \alpha_{00} \ket{00} + \alpha_{01}\ket{01}+\alpha_{11}\ket{11} + \alpha_{10}\ket{10}
\end{align}
Alice and Bob then both learn measurement outcomes from measuring both qubits in the computational basis. 
Their output is then described by,
\begin{align}
    \mathcal{N}^{x,y}_{Q\rightarrow ZZ'} (\psi_Q) = \sum_{a,b}|\alpha_{ab}|^2 \ketbra{a,b}{a,b}\otimes \ketbra{a,b}{a,b}
\end{align}
But since $|\alpha_{a,b}|^2=\bra{a,b} \psi_Q \ket{a,b}$, this is exactly the channel $\mathcal{B}^0$, as needed. 

A similar argument, beginning with the input written in the Bell basis, shows that the protocol is perfectly correct on $1$ instances. 

Finally, we observe that since the protocol is perfectly correct for all pure state inputs, it is also perfectly correct in diamond norm, as needed. 
\end{proof}

\subsection{\texorpdfstring{$f$}{TEXT}-Bell to \texorpdfstring{$f$}{TEXT}-Clifford}\label{sec:BelltoClifford}

For $C$ any choice of Clifford, we define the measurement channel
\begin{align}
    \mathcal{M}^{C}_{A\rightarrow ZZ'}(\rho_A)=\sum_{b} \bra{b}C^\dagger\rho C \ket{b}\,\ketbra{b}{b}_Z\otimes \ketbra{b}{b}_{Z'}.
\end{align}
Here $b\in\{0,1\}^{n_A}$ ranges over the computational basis of $\mathcal{H}_A$.
In words, this channel measures in the basis $\{C\ket{b}\}$, and then copies the measurement outcome into the two registers $Z$, $Z'$. 

Our next implication shows that $f$-Bell implies the NLQC which implements the above channel with the choice of Clifford controlled by classical inputs. 
We call this the $f$-Clifford-measure task. 

\begin{definition}\label{def:fCmeasure}
    A \textbf{$f$-Clifford-measure} task is defined by a choice of Boolean function $f:\{ 0,1\}^{2n}\rightarrow \{0,1\}$, and two choices of Clifford unitary $C_A^0, C_A^1$, acting on a Hilbert space $A$.
    Inputs $x\in \{0,1\}^{n}$ and system $A$ are given to Alice, and input $y\in \{0,1\}^{n}$ is given to Bob.
    The $f$-Clifford-measure task is completed $\epsilon$-correctly on input $(x,y)$ if channel $\mathcal{N}^{x,y}$ executed on input $x,y$ satisfies
    \begin{align}
        \Vert \mathcal{M}^{C^{f(x,y)}} - \mathcal{N}^{x,y} \Vert_\diamond \leq \epsilon \enspace.
    \end{align}
    We say the task is implemented $\epsilon$-correctly if the above holds for all inputs $(x,y)$. 
\end{definition}

\begin{lemma}
    \textbf{$f$-Bell $\Rightarrow$ $f$-Clifford-measure:} Given $n_A$ copies of a perfect $f$-Bell oracle, $f$-Clifford-measure can be implemented perfectly, where the Clifford acts on $n_A$ qubit inputs.
\end{lemma}

\begin{proof} We first give the protocol then comment on correctness.
Because the correctness argument is similar to the last two lemmas, we omit details. 

\vspace{0.2cm}
\noindent \textbf{Protocol:} Alice prepares the maximally entangled state on $\mathcal{H}_{B}\otimes \mathcal{H}_C$ with $d_A=d_B=d_C$, then applies Clifford $C_A^0\otimes (C_C^0)^\dagger C_{C}^1$. 
Alice measures the $C$ register in the computational basis and broadcasts the measurement outcome.
Then, Alice and Bob take $\mathcal{H}_A\otimes \mathcal{H}_B$ as the input to $f$-Bell oracles (one oracle use per qubit of $A$ is needed). 

In the second round, Alice and Bob learn $x,y$ and hence $f(x,y)$. 
If $f(x,y)=0$, Alice and Bob record the measurement outcomes from the $f$-Bell oracle in the $Z$ and $Z'$ registers as their outputs. 
If $f(x,y)=1$, observe that $AB$ has been measured in the Bell basis, so that $A$ has been teleported into the $C$ Hilbert space. 
The measurement outcomes from the $f$-Bell protocol then determine Pauli corrections on $C$. 
Alice and Bob determine these Pauli corrections and use them, along with the computational basis measurement outcome from Alice measuring $C$ in the first round, to determine the output $b$. 
Alice and Bob record $b$ into the $Z$ and $Z'$ registers as their outputs. 

\vspace{0.2cm}
\noindent \textbf{Correctness:} Follows by a similar argument as lemmas \ref{lem:frouteTOfmeasure} and \ref{lem:fBB84TOfBell}. 
We analyze the $(x,y)\in f^{-1}(0)$ case by writing the input state in the computational basis and checking the action of the protocol is the same as measuring in the $\{{C}^0\ket{b}\}$ basis; similarly we analyze the $(x,y)\in f^{-1}(1)$ case by writing the input in the basis $\{C^1\ket{b}\}$ and checking the action of the protocol is the same as measuring in the $\{{C}^1\ket{b}\}$ basis. 
\end{proof}

Next, we move from Clifford measurements to the controlled application of Cliffords. 
Specifically, we wish to implement the channel
\begin{align}
    \mathcal{C}^s_{AB\rightarrow AB}(\rho_{AB}) = C^{s}\rho_{AB}(C^s)^\dagger
\end{align}
In the NLQC context, we want $A$ and $B$ to start and end on Alice and Bob's side respectively. 

\begin{definition}\label{def:fClifford}
A \textbf{$f$-Clifford} task is defined by a choice of Boolean function $f:\{ 0,1\}^{2n}\rightarrow \{0,1\}$, and a pair of Clifford unitaries $C_{AB}^0,C_{AB}^1$, acting on a Hilbert space $\mathcal{H}_A\otimes \mathcal{H}_B$.
Inputs $x\in \{0,1\}^{n}$ and system $A$ are given to Alice, and input $y\in \{0,1\}^{n}$ is given to Bob.
The $f$-Clifford task is completed $\epsilon$-correctly on input $(x,y)$ if channel $\mathcal{N}^{x,y}$ executed on input $x,y$ satisfies
\begin{align}
    \Vert \mathcal{C}^{f(x,y)} - \mathcal{N}^{x,y} \Vert_\diamond \leq \epsilon \enspace.
\end{align}
We say the task is implemented $\epsilon$-correctly if the above holds for all inputs $(x,y)$. 
\end{definition}

Next we define a particular choice of pair of Cliffords $C^0,C^1$ such that $f$-Clifford-measure for that pair implies $f$-Clifford. 
Define
\begin{align}
    C^0_{ABC}=B_{AB}\otimes \mathcal{I}_C\nonumber \\
    C^1_{ABC}=B_{AC}\otimes \mathcal{I}_B
\end{align}
where $B$ is the (Clifford) unitary mapping from the computational to the Bell basis. 
We find that an $f$-Clifford-measure oracle for this pair implies $f$-Clifford (for any choice of Clifford). 

\begin{lemma}
    \textbf{$f$-Clifford-measure $\Rightarrow$ $f$-Clifford:} One oracle implementation of $f$-Clifford-measure protocol for function $f$ and Clifford pairs $B_{AB}\otimes \mathcal{I}_C, B_{AC}\otimes \mathcal{I}_B$ gives a perfect implementation of an $f$-Clifford protocol for the same function and any pair of Cliffords $C^0_{AB},C^1_{AB}$.
\end{lemma}

\begin{proof}
    We only give the protocol. Correctness can be checked as in the previous lemmas. 

    Alice is given input $A$ and prepares two maximally entangled states $\ket{\Psi^+}_{A_0'B'_0 A''_0B''_0}$, $\ket{\Psi}^+_{A_1'B'_1 A''_1B''_1}$, where each of the $A_i, B_i$ are the same size as $A, B$, and these states are maximally entangled across $A_i'B_i':A_i''B_i''$. 
    Alice applies $C_{A''_0B''_0}^0\otimes C_{A''_1B''_1}^1$, then sends $B_0''B_1''$ to Bob.
    Alice and Bob then apply $f$-Clifford-measure to $ABC$, with $C^0_{ABC}=B_{AB}\otimes \mathcal{I}_C$, $C^1_{ABC}=B_{AC}\otimes \mathcal{I}_B$ as the choice of Clifford pair for the measure protocol.

    In the second round Alice and Bob learn $f(x,y)$. 
    The $f$-Clifford-measure protocol teleports$^*$ the $AB$ systems into either the primed or doubly primed systems. 
    If $f(x,y)=0$, they take the singly primed systems as their outputs. 
    If $f(x,y)=1$, they take the doubly primed systems as their outputs. 
    In either case, the measurement outcomes from the $f$-Clifford-measure determine a set of Pauli corrections that appear on the output systems. 
    Alice and Bob undo these and return the appropriate output systems. 
\end{proof}

\subsection{\texorpdfstring{$f$}{TEXT}-SWAP to \texorpdfstring{$f$}{TEXT}-route}\label{sec:SWAPtoroute}

To complete our chain of implications, we finally note that $f$-route can be implemented using an $f$-Clifford with the choice of unitaries $I_{AB}, SWAP_{AB}$.
We refer to this particular $f$-Clifford as $f$-SWAP.

\begin{lemma}
    \textbf{$f$-SWAP $\Rightarrow$ $f$-route:} One oracle implementation of $f$-SWAP for function $f$ can be used to implement an $f$-route protocol for the same function.
\end{lemma}
\begin{proof}
    Let the input system to the $f$-route protocol be labelled $A$. 
    Alice prepares an ancilla register $B$ in the $\ket{0}_B$ state. Alice and Bob insert $AB$ into an $f$-SWAP oracle. 
    This keeps $A$ on the left if $f=0$ and moves $A$ to the right if $f=1$, so that correctness is clear in the perfect case. 
\end{proof}

\section{Further reductions}

The equivalence of $f$-route and $f$-measure, and the equivalences we found along the way, all dealt with classically controlled Clifford operations. 
To go further, we should ask about NLQC examples involving non-Clifford operations. 
For instance, if we do a classically controlled $T$-gate, could this be much harder to do than a classically controlled Clifford? 
Another way of phrasing this is to ask if our upper bound strategies for $f$-route, like the formula size upper bound or the sub-exponential generic upper bound, relied on Clifford structure, or only on the fact that a small quantum operation is classically controlled. 

To a large extent, the answer is no: many classically controlled non-Clifford operations can be reduced to the Clifford case. 
We won't review these developments in depth here. 
Indeed, these reductions are only beginning to be explored and the status of what is known may change rapidly. 
However, we briefly mention what is known so far. 

One natural class of NLQCs involves measuring a single qubit in either the computational basis, or a basis rotated by angle $\theta$ from the computational basis, $\{R_X(\theta)\ket{0}, R_X(\theta)\ket{1}\}$, where recall
\begin{align}
    R_X(\theta)=\begin{pmatrix}
        \cos(\theta/2) & -i\sin(\theta/2) \\
        -i\sin(\theta/2) & \cos(\theta/2)
    \end{pmatrix}
\end{align}
Except for special angles, the rotation $R_X(\theta)$ is non-Clifford, so in general this is a single qubit non-Clifford operation. 
The task is to measure in this basis if $f(x,y)=1$, or measure in the computational basis if $f(x,y)=0$, and then produce the measurement outcome on both Alice and Bob's side.
We call this the $f$-measure$(I,R_X(\theta))$ task. 
When necessary to emphasize the distinction, the usual notion of $f$-measure we refer to as the $f$-measure$(I,H)$ task.

In \cite{bluhm2026reductions}, it is shown how to use an $O(1)$ number of copies of $f$-measure$(I,H)$ to implement $f$-measure$(I,R_X(\theta))$ for any fixed angle $\theta$. 
Conversely, for any angle $\theta\neq 0$ a constant number of $f$-measure$(I,R_X(\theta))$ oracles can be used to implement $f$-measure$(I,H)$.
Once the ability to measure in non-Clifford bases is granted, gate-teleportation like tricks can be used to implement non-Clifford gates. 
In fact, one can also show that $f$-controlled unitaries of the form $U=C_2 D C_1$ can be reduced to $O(1)$ copies of $f$-route, where $D$ is an arbitrary diagonal unitary. 
It is not known if we can extend this to arbitrary classically controlled unitaries.  

The initial observations made in this direction suggest that perhaps all NLQCs involving classically controlled quantum operations, where the quantum operations involve $O(1)$ qubits, can be reduced to $f$-route. 
This would be a striking result about the relationship of different NLQC examples. 
It would also indicate that there is a basic limitation to the security of QPV schemes that rely on classical controls to enhance the entanglement required and hence security of the scheme. 
This is because such a result would show that all such protocols can be implemented with sub-exponential entanglement using the generic protocol inherited from $f$-routing. 

\section{History and further reading}

The notion of reduction among NLQC classes was first formally defined in \cite{bluhm2025complexity}. 
However, the results in \cite{allerstorfer2024relating} relating CDS to $f$-routing, and PSM to coherent function evaluation, already hint at similar relationships, and indeed inspired the more formal notion of reduction. 
The treatment given here follows \cite{bluhm2026reductions}, who first showed an oracle relationship between $f$-routing and $f$-measure, and established the reductions we show here as well as several others. 

A key open problem in this direction is to understand if all $f$-controlled $O(1)$ size unitaries can be reduced to $O(1)$ copies of $f$-route. 
So far, the result of \cite{bluhm2026reductions} showing this can be done for unitaries of the form $C_2DC_1$ with $D$ diagonal, $C_i$ Clifford is the furthest result in this direction that has been shown. 

\chapter{Addendum: Open problems}

Before concluding, we recall some open problems that have appeared in this book. 
These open problems are highlighted here both because they are important, and because there are no known obstructions to their solutions (for instance, they do not imply complexity theory breakthroughs).
Thus they seem to be natural points of focus for future study in NLQC. 

\begin{itemize}
    \item \textbf{Existence of an exponentially costly channel:} In chapter \ref{chapter:allchannels} we gave an upper bound strategy based on port-teleportation that shows every channel can be implemented as an NLQC if we allow exponential entanglement. For the related setting of a universal processor, we highlighted that exponential lower bounds are known. It is natural then to ask: \emph{Can we similarly prove an exponential entanglement lower bound on some, perhaps non-explicit, choice of channel?}
    \item \textbf{Polynomial, robust, lower bounds on entanglement for $f$-routing:} In chapter \ref{chapter:ITCandNLQC} we explained that entanglement lower bounds in $f$-routing imply randomness lower bounds on conditional disclosure of secrets. Proving robust, polynomial CDS lower bounds is an important open problem in classical information theoretic cryptography and communication complexity. We also discussed that polynomial $f$-routing lower bounds, especially when considering noisy protocols, are key to proving the security of experimentally feasible quantum position verification schemes. Finally, we saw in chapter \ref{chapter:reductions} that $f$-routing is equivalent to many other NLQC families, including most other QPV candidate schemes, so good $f$-routing lower bounds lead to good lower bounds on many candidate QPV schemes. To resolve these problems then, it suffices to understand: \emph{Can we prove a polynomial entanglement lower bound on $f$-routing?}
    \item \textbf{Are all classically controlled NLQCs reducible to $f$-routing?} In chapter \ref{chapter:reductions}, we gave the first results in exploring the relationships among NLQC families. We found that many NLQC families involving $O(1)$ size quantum operations controlled off of a large classical computation can be reduced to $f$-routing. To better understand the landscape of relationships among NLQCs, we should ask: \emph{Are all classically controlled quantum operations acting on $O(1)$ qubits reducible to $f$-routing?} A negative answer would imply that the security of some QPV schemes is independent of the security of $f$-routing. An answer in either direction would clarify the structure relating different NLQC examples. 
    \item \textbf{A faithful property signalling entanglement cost:} In chapter \ref{chapter:quantumlowerbound}, we studied the controllable correlation, which is defined for unitaries. 
    We have proven that a unitary being controllably correlated implies it has a non-zero entanglement cost to implement as an NLQC. Numerical evidence is consistent with a unitary having a non-zero cost implying that it is controllably correlated. \emph{Is the controllable correlation faithful, in the sense that it is non-zero if and only if there is a non-zero entanglement cost? Or can we find a faithful signal for non-zero entanglement cost in NLQC?} As well, so far this discussion is limited to unitary channels - can we find a faithful test for entanglement cost even for general quantum channels?
\end{itemize}

\appendix

\chapter{Notation and basics}

\section{Quantum states and distances}\label{section:statesanddistances}

We label the dimension of a Hilbert space $\mathcal{H}_A$ by $d_A$, and the (base 2) log dimension by $n_A=\log d_A$.
Throughout this work $\log$ denotes the base 2 logarithm, while $\ln$ denotes the natural logarithm.
When considering entanglement across bipartitions of a quantum state $\ket{\psi}_{AB}$, we refer to entanglement across $A:B$ where the colon indicates the partitioning of the systems. 
We use the notation 
\begin{align}
    \ket{\Psi^+}_{AB}=\frac{1}{\sqrt{2}}\left(\ket{00}_{AB}+\ket{11}_{AB} \right)
\end{align}
for this particular maximally entangled state of two qubits, and the notation
\begin{align}
    \rho_{cc} = \frac{1}{2}\left(\ketbra{00}{00}+\ketbra{11}{11} \right)
\end{align}
for the maximally classically correlated state of two qubits. 

We quantify the distance between quantum states with the one-norm distance, 
\begin{align}
    \Vert\rho-\sigma\Vert_1 = \tr|\rho-\sigma|.
\end{align}
Note that $\Vert\rho-\sigma\Vert_1/2$ is known as the trace distance.
We also use the fidelity, which we define as
\begin{align}
    F(\rho,\sigma)=\Vert\sqrt{\rho}\sqrt{\sigma} \Vert_1^2.
\end{align}
This is related to the trace distance by the Fuchs--van de Graaf inequalities, 
\begin{align}\label{eq:FVdG}
    1-\sqrt{F(\rho,\sigma)}\leq \frac{1}{2}\Vert \rho-\sigma \Vert_1 \leq \sqrt{1-F(\rho,\sigma)}.
\end{align}
Uhlmann's theorem states that 
\begin{align}
    F(\rho_A,\sigma_A)=\max_{\ket{\psi_\rho}_{AB}} |\braket{\psi_\rho}{\psi_\sigma}|^2
\end{align}
where $\ket{\psi_\sigma}$ is any purification of $\sigma$, and the maximization is over purifications of $\rho$. 
We can always take $d_B=d_A$.

We record the following useful consequence of Uhlmann's theorem combined with the Fuchs--van de Graaf inequalities. 
\begin{lemma}\label{lemma:traceUhlmann}
    Suppose that $\Vert \sigma_A-\rho_A\Vert \leq \epsilon$, and consider any extension of $\sigma_A$ to the AB Hilbert space, call it $\sigma_{AB}$. Then there exists an extension of $\rho_A$ to the $AB$ Hilbert space, call it $\rho_{AB}$, such that
    \begin{align}
        \Vert \sigma_{AB}-\rho_{AB}\Vert_1\leq 2\sqrt{\epsilon}
    \end{align}
\end{lemma}
\begin{proof}
    Starting with $\Vert \sigma_A-\rho_A\Vert \leq \epsilon$, use Fuchs--van de Graaf to bound the fidelity from below, 
    \begin{align}
        F(\sigma_A,\rho_A) \geq 1-\epsilon.
    \end{align}
    Now consider any purification of $\sigma_{AB}$ into the $ABX$ Hilbert space, call it $\ket{\psi_\sigma}_{ABX}$. 
    Then by Uhlmann's theorem we have that there exists a state $\ket{\psi_\rho}_{ABX}$ such that
    \begin{align}
        F(\sigma_A,\rho_A)=|\braket{\psi_\sigma}{\psi_\rho}|^2.
    \end{align}
    But then we also have that the fidelity increases under the partial trace, so that
    \begin{align}
        F(\sigma_{AB},\rho_{AB})\geq |\braket{\psi_\sigma}{\psi_\rho}|^2 \geq 1-\epsilon.
    \end{align}
    Here $\rho_{AB}$ is defined by tracing out $X$ from $\ket{\psi_\rho}_{ABX}$. 
    Now we use Fuchs--van de Graaf again to bound the trace distance between $\sigma_{AB}$ and $\rho_{AB}$, giving
    \begin{align}
        \Vert \sigma_{AB}-\rho_{AB}\Vert_1 \leq 2\sqrt{\epsilon}
    \end{align}
    as needed. 
\end{proof}

We quantify the distance between quantum channels using the diamond norm distance. 
\begin{definition} Let $\mathcal{N}_{B\rightarrow C}, \mathcal{M}_{B\rightarrow C}$ be quantum channels. 
The \textbf{diamond norm distance} is defined by 
\begin{align}
    \Vert\mathcal{N}_{B\rightarrow C}-\mathcal{M}_{B\rightarrow C}\Vert_\diamond = \sup_{d} \max_{\Psi_{A_dB}}\Vert\mathcal{N}_{B\rightarrow C}(\Psi_{A_dB}) - \mathcal{M}_{B\rightarrow C}(\Psi_{A_dB})\Vert_1
\end{align}
where $\mathcal{H}_{A_d}$ is a $d$ dimensional Hilbert space. 
\end{definition}

The average fidelity of a channel $F_{\text{avg}}(\mathcal{N})$ is defined by
\begin{align}
    F_{\text{avg}}(\mathcal{N}) = \int d\psi \, F(\psi, \mathcal{N}(\psi)).
\end{align}
The entanglement fidelity is defined by
\begin{align}
    F_e(\mathcal{N}_A) = F(\Psi^+_{RA},\mathcal{I}_R\otimes \mathcal{N}_A(\Psi^+_{RA})). 
\end{align}
The average fidelity and entanglement fidelity are related by \cite{horodecki1999general, nielsen2002simple}
\begin{align}\label{eq:avgandent}
    F_{\text{avg}}(\mathcal{N})=\frac{d F_e(\mathcal{N})+1}{d+1}
\end{align}
The average fidelity is related to the diamond norm distance by \cite{wallman2014randomized}
\begin{align}
    \frac{d+1}{d}(1-F_{\text{avg}}(\mathcal{N}))\leq \frac{1}{2}\Vert \mathcal{N}-\mathcal{I} \Vert_\diamond \leq \sqrt{d(d+1)(1-F_{\text{avg}}(\mathcal{N}))}
\end{align}
In terms of the entanglement fidelity, using \ref{eq:avgandent} this is
\begin{align}\label{eq:diamondandFe}
    1-F_e(\mathcal{N}) \leq \frac{1}{2}\Vert \mathcal{N}-\mathcal{I} \Vert_\diamond\leq d \sqrt{1-F_e(\mathcal{N})}
\end{align}

\section{List of quantum gates}\label{sec:gates}

In this appendix we give matrix expressions for the two qubit unitaries appearing in table~\ref{table:lowerbounds}. 

\begin{align}
CNOT &= \begin{pmatrix}
        1 & 0 & 0 & 0\\
        0 & 1 & 0 & 0\\
        0 & 0 & 0 & 1\\
        0 & 0 & 1 & 0
    \end{pmatrix} 
\end{align}
\begin{align}
\text{DCNOT} &=\begin{pmatrix}
        1 & 0 & 0 & 0\\
        0 & 0 & 1 & 0\\
        0 & 0 & 0 & 1\\
        0 & 1 & 0 & 0
    \end{pmatrix}
\end{align}
\begin{align}
\text{B} &= \begin{pmatrix}
        \cos(\pi/8) & 0 & 0 & i\sin(\pi/8)\\
        0 & \cos(3\pi/8) & i \sin(3\pi/8) & 0 \\
        0 & i\sin(3\pi/8) & \cos(3\pi/8) & 0 \\
        i\sin(\pi/8) & 0 & 0 & \cos(\pi/8)
    \end{pmatrix}
\end{align}
\begin{align}
\text{RXX}(\pi/2) &=\exp\left(-i\frac{\pi}{4}X\otimes X\right) = \begin{pmatrix}
        \cos(\pi/4) & 0 & 0 & -i\sin(\pi/4)\\
        0 & \cos(\pi/4) & -i\sin(\pi/4) & 0 \\
        0 & -i\sin(\pi/4) & \cos(\pi/4) & 0 \\
        -i\sin(\pi/4) & 0 & 0 & \cos(\pi/4)
    \end{pmatrix}
\end{align}
\begin{align}
i\SWAP&=\begin{pmatrix}
        1 & 0 & 0 & 0\\
        0 & 0 & i & 0\\
        0 & i & 0 & 0\\
        0 & 0 & 0 & 1
    \end{pmatrix}
\end{align}
\begin{align}
\sqrt{\SWAP}&=\begin{pmatrix}
        1 & 0 & 0 & 0\\
        0 & \frac{1}{2}(1+i) & \frac{1}{2}(1-i) & 0\\
        0 & \frac{1}{2}(1-i) & \frac{1}{2}(1+i) & 0\\
        0 & 0 & 0 & 1
    \end{pmatrix}
\end{align}
\begin{align}
\text{Sycamore}&=\begin{pmatrix}
        1 & 0 & 0 & 0\\
        0 & 0 &-i & 0\\
        0 & -i & 0 & 0\\
        0 & 0 & 0 & e^{-i\pi/6}
    \end{pmatrix}
\end{align}
\begin{align}
\text{Magic}&=\frac{1}{\sqrt{2}} \begin{pmatrix}
        1 & i & 0 & 0\\
        0 & 0 & i & 1\\
        0 & 0 & i & -1\\
        1 & -i & 0 & 0
    \end{pmatrix}
\end{align}
\begin{align}
\text{Dagwood Bumstead}&= \begin{pmatrix}
        1 & 0 & 0 & 0\\
        0 & \cos(3\pi/8) & -i\sin(3\pi/8) & 0 \\
        0 & -i\sin(3\pi/8) & \cos(3\pi/8) & 0 \\
        0 & 0 & 0 & 1
    \end{pmatrix}
\end{align}
\begin{align}
\text{CS}&= \begin{pmatrix}
        1 & 0 & 0 & 0\\
        0 & 1 & 0 & 0 \\
        0 & 0 & 1 & 0 \\
        0 & 0 & 0 & i
    \end{pmatrix}
\end{align}
\begin{align}
\text{CT}&= \begin{pmatrix}
        1 & 0 & 0 & 0\\
        0 & 1 & 0 & 0 \\
        0 & 0 & 1 & 0 \\
        0 & 0 & 0 & e^{i\pi/4}
    \end{pmatrix}
\end{align}
\begin{align}
\text{Echoed cross resonance}&=\frac{1}{\sqrt{2}} \begin{pmatrix}
        0 & 0 & 1 & i\\
        0 & 0 & i & 1 \\
        1 & -i & 0 & 0 \\
        -i & 1 & 0 & 0
    \end{pmatrix}
\end{align}
\begin{align}
    CSX = \begin{pmatrix}
        1 & 0 & 0 & 0\\
        0 & 1 & 0 & 0 \\
        0 & 0 & e^{i\pi/4}/\sqrt{2} & e^{-i\pi/4}/\sqrt{2} \\
        0 & 0 & e^{-i\pi/4}/\sqrt{2} & e^{i\pi/4}/\sqrt{2}
    \end{pmatrix}
\end{align}

\chapter{Fidelity of Pauli adapted port-teleportation}\label{sec:pauliadaptedPT}

\noindent \textbf{AI disclosure:} \emph{The proof in this appendix was developed with AI assistance. In particular Claude Opus 4.0 was asked to develop a lower bound on $\tr(X^{1/2})$ using known values of $\tr(X)$ and $\tr(X^2)$, which then returned the approach given by lemmas \ref{lemma:rootGlowerbound}, \ref{lemma:algebraicinequal}, and \ref{lemma:rootAlowerbound}.}\\

\noindent In this appendix, we give the calculation establishing equation \ref{eq:mixedteleportation}.

\vspace{0.2cm}
\noindent \textbf{Set-up:} Recall that we defined the states
\begin{align}
    g^{i,s} = \frac{1}{d_{\bar{A}_i}} \Psi^s_{A_0A_i}\otimes \mathcal{I}_{\bar{A}_i}
\end{align}
where $\Psi^s_{A_0A_i}$ is the maximally entangled state acted on by Pauli $P[s]$.
We will allow $s\in S$, for $S$ a subset of the full Pauli group. 
We also defined the operators
\begin{align}
    G &=\sum_{i,s} g^{i,s} \nonumber \\
    \Sigma^{i,s} &= G^{-1/2} g^{i,s}G^{-1/2} \nonumber \\
    \Delta &= \frac{1}{m}\left(I - \sum_{i,s} \Sigma^{i,s}\right)
\end{align}
Here $m=N\cdot |S|$, i.e. $m$ is the total number of distinct states $g^{i,s}$. 
Then we define the POVM elements 
\begin{align}
    \Pi^{i,s}=\Sigma^{i,s} + \Delta.
\end{align}
We are interested in the entanglement fidelity of the teleportation channel, which recall is
\begin{align}
    \Lambda^S(\rho_{A_0}) = \sum_{i,s} \mathcal{P}^s\left(\tr_{A_0A}\tr_{\bar{B}_{i}}\left(\sqrt{\Pi^{i,s}_{A_0A}}\left(\rho_{A_0} \otimes \Psi_{AB} \right) \sqrt{\Pi^{i,s}_{A_0A}}\right)\right)_{B_i\rightarrow A_0}.
\end{align}

\vspace{0.2cm}
\noindent \textbf{Calculation:} To begin, first notice that we can simplify this expression by taking the trace over $\bar{B}_i$ explicitly, and collecting the $\sqrt{\Pi}$'s into one factor,
\begin{align}
    \Lambda^S(\rho_{A_0}) = \sum_{i,s} \mathcal{P}^s\left(\tr_{A_0A}\left(\Pi^{i,s}_{A_0A}\left(\rho_{A_0} \otimes \Psi_{A_iB_i} \otimes \pi_{\bar{A}_i}\right) \right)\right)_{B_i\rightarrow A_0}.
\end{align}
Then, we can also apply $\mathcal{P}^s$ to get
\begin{align}
    \Lambda^S(\rho_{A_0}) = \sum_{i,s}\left(\tr_{A_0A}\left(\Pi^{i,s}_{A_0A}\left(\rho_{A_0} \otimes \Psi_{A_iB_i}^s\otimes \pi_{\bar{A}_i} \right) \right)\right)_{B_i\rightarrow A_0}.
\end{align}
We will quantify how well $\Lambda^S$ works as a teleportation by computing its entanglement fidelity, 
\begin{align}
    F_e(\Lambda^S) = F(\Psi_{RA_0}, \mathcal{I}_R\otimes \Lambda^S(\Psi_{RA_0})).
\end{align}
That is, we are checking how close this channel is to the identity when acting on one end of a maximally entangled state. 
Using that $\Psi_{RA_0}$ is pure, we can rewrite the entanglement fidelity as
\begin{align}
    F_e(\Lambda^S) &= \tr\left(\Psi_{RA_0} \Lambda^S_{A_0}(\Psi_{RA_0})\right) \nonumber \\
    &= \sum_{i,s}\tr\left(\Psi_{RA_0} \Pi^{i,s}_{A'_0A}(\Psi_{RA_0'}\otimes \Psi_{A_iA_0}^s\otimes \pi_{\bar{A}_i})\right).
\end{align}
Then we can use that
\begin{align}
    \tr_{R}(\Psi_{RA_0}\Psi_{RA_0'}) = \frac{1}{d^2}SWAP_{A_0A_0'}
\end{align}
where $d=d_{A_0}=d_R=d_{A_i}$, to get this to
\begin{align}
    F_e(\Lambda^S) &= \frac{1}{d^2}\sum_{i,s}\tr\left(\Pi^{i,s}_{A_0A}(\Psi_{A_iA_0}^s\otimes \pi_{\bar{A}_i})\right)
\end{align}
and then finally,
\begin{align}\label{eq:Fe}
    \boxed{F_e(\Lambda^S)=\frac{1}{d^2}\sum_{i,s}\tr\left(\Pi^{i,s}_{A_0A} g^{i,s}_{A_0A}\right).}
\end{align}
Now we would like to bound the entanglement fidelity from below in terms of $N$ and $|S|$. 

To do this, we can first use that
\begin{align}
    \tr(g^{i,s}\Delta)=0.
\end{align}
To see this, 
\begin{align}
    \tr(g^{i,s}\Delta) \propto &\, \tr(g^{i,s}) - \sum_{j,t} \tr(g^{i,s} \Sigma^{j,t}), \nonumber \\
    =&\, 1 - \sum_{j,t} \tr(g^{i,s} G^{-1/2} g^{j,t} G^{-1/2}), \nonumber \\
    =&\, 1 - \tr(g^{i,s} I_{span(\{g^{i,s}\})}), \nonumber \\
    =&\, 0,
\end{align}
as claimed. 
Thus we can replace $\Pi^{i,s}$ with $\Sigma^{i,s}$ in expression \ref{eq:Fe}, 
\begin{align}
    F_e(\Lambda^S)&=\frac{1}{d^2}\sum_{i,s}\tr\left(\Sigma^{i,s}_{A_0A} g^{i,s}_{A_0A}\right), \nonumber \\
    &=\frac{1}{d^2}\sum_{i,s}\tr\left(G^{-1/2}g^{i,s}_{A_0A} G^{-1/2} g^{i,s}_{A_0A}\right).
\end{align}
Our goal is to bound this from below as tightly as possible. 

Towards this, let's first define the projectors
\begin{align}
    Q^{i,s} = d^{N-1} g^{i,s}.
\end{align}
Further, we define $\tilde{G}=\sum_{i,s}Q^{i,s} = d^{N-1}G$. 
Then the fidelity becomes
\begin{align}
    F_e(\Lambda^S)&=\frac{1}{d^{N+1}}\sum_{i,s}\tr\left(\tilde{G}^{-1/2}Q^{i,s}_{A_0A} \tilde{G}^{-1/2} Q^{i,s}_{A_0A}\right).
\end{align}
Now we prove the following. 

\begin{lemma}\label{lemma:rootGlowerbound}
    \begin{align}
        F_e(\Lambda^S) \geq \frac{1}{md^{2N}} \left(\tr(\tilde{G}^{1/2})\right)^2
    \end{align}
\end{lemma}
\begin{proof}
Let's first rewrite the trace in terms of the $2$-norm,
    \begin{align}
        \tr\left(\tilde{G}^{-1/2}Q^{i,s}_{A_0A} \tilde{G}^{-1/2} Q^{i,s}_{A_0A}\right) &= \tr\left( \left(Q^{i,s}_{A_0A}\tilde{G}^{-1/2}Q^{i,s}_{A_0A}\right)\left(Q^{i,s}_{A_0A}\tilde{G}^{-1/2}Q^{i,s}_{A_0A}\right) \right) \nonumber \\
        &= \left\Vert \left(Q^{i,s}_{A_0A}\tilde{G}^{-1/2}Q^{i,s}_{A_0A}\right)\right\Vert_2^2
    \end{align}
    Define $M^{i,s}:=\left(Q^{i,s}_{A_0A}\tilde{G}^{-1/2}Q^{i,s}_{A_0A}\right)$, so that
    \begin{align}
        F_e(\Lambda^S)=\frac{1}{d^{N+1}}\sum_{i,s} \Vert M^{i,s}\Vert^2_2
    \end{align}
    Since $M^{i,s}$ is Hermitian, we have that for $\{\lambda_k\}$ the eigenvalues, 
    \begin{align}
        \Vert M^{i,s}\Vert^2_2 = \sum_{k} \lambda_k^2 = \Vert \vec{\lambda}\Vert_2^2
    \end{align}
    where the norm on the right is the vector $2$ norm. Recall that $\Vert \vec{x}\Vert_2 \geq \Vert \vec{x}\Vert_1/\sqrt{r}$ for $r$ the dimension of the space. Here, $M$ lives on a $d^{N-1}$ dimensional space, so $r=d^{N-1}$. 
    Then note that here we have $\tr (M^{i,s})=\Vert \vec{\lambda} \Vert_1$, we get that
    \begin{align}
        \Vert M^{i,s}\Vert^2_2 \geq \frac{1}{d^{N-1}}[\tr(M^{i,s})]^2
    \end{align}
    The fidelity then is lower bounded according to
    \begin{align}
        F_e(\Lambda^S)\geq\frac{1}{d^{2N}} \sum_{i,s}[\tr( M^{i,s})]^2
    \end{align}
    Now we use $\tr(M^{i,s})=\tr(\tilde{G}^{-1/2}Q^{i,s})$, and that $\Vert \vec{x}\Vert_2\geq \frac{1}{\sqrt{r}}\Vert \vec{x}\Vert_1$ again, to get
    \begin{align}
        F_e(\Lambda^S)\geq \frac{1}{md^{2N}} \left(\sum_{i,s}\tr( M^{i,s})\right)^2=\frac{1}{md^{2N}} \left(\sum_{i,s}\tr( \tilde{G}^{-1/2}Q^{i,s})\right)^2 = \frac{1}{md^{2N}}\left(\tr(\tilde{G}^{1/2})\right)^2
    \end{align}
    as claimed. 
\end{proof}

Now, we use that we can lower bound $\tr(\tilde{G}^{1/2})$ in terms of $\tr \tilde{G}$ and $\tr(\tilde{G}^2)$, as follows. 

\begin{lemma}\label{lemma:algebraicinequal}
For $x\geq 0$, $\beta>0$, 
\begin{align}
    \sqrt{x} \;\geq\; \sqrt\beta + \frac{x-\beta}{2\sqrt\beta} - \frac{(x-\beta)^2}{2\beta^{3/2}}.
\end{align}
\end{lemma}
\begin{proof}
    Divide through by $\sqrt{\beta}$, and set $v=\sqrt{x/\beta}$, to obtain
    \begin{align}
        v \geq 1 + \frac{v^2-1}{2} - \frac{(v^2-1)^2}{2}.
    \end{align}
    Rearranging, 
    \begin{align}
        0 &\geq 1-v - \frac{1-v^2}{2} - \frac{(1-v^2)^2}{2} \nonumber \\
        &=(1-v)\left(1 - \frac{1+v}{2} - \frac{(1-v^2)(1+v)}{2}\right) \nonumber \\
        &=(1-v)\left(\frac{1-v}{2} - \frac{(1-v^2)(1+v)}{2}\right) \nonumber \\
        &=\frac{(1-v)^2}{2}\left(1 - (1+v)^2\right) \nonumber \\
        &= -v\frac{(1-v)^2}{2}\left(2+v\right)
    \end{align}
    Since $v\geq 0$, this obviously holds, so our initial statement does as well. 
\end{proof}

\begin{lemma}\label{lemma:rootAlowerbound}
Let $A$ be a positive semi-definite Hermitian operator, which acts on a space of dimension $D$, and set $\beta = \tr A/D$. Then
\begin{align}
    \tr \sqrt{A} \geq D\sqrt{\beta} - \frac{1}{2\beta^{3/2}}\left(\tr A^2- 2\beta \tr A + \beta^2 D \right)
\end{align}
\end{lemma}
\begin{proof}
    Consider an eigenvalue $\lambda_k$ of $A$, and apply lemma \ref{lemma:algebraicinequal} with $x=\lambda_k$, obtaining
    \begin{align}
        \sqrt{\lambda_k}\geq \sqrt{\beta}+\frac{\lambda_k-\beta}{2\sqrt{\beta}} - \frac{(\lambda_k-\beta)^2}{2\beta^{3/2}}
    \end{align}
    Now sum this over $k$, obtaining
    \begin{align}
        \tr \sqrt{A}\geq \sqrt{\beta}D + \frac{\tr(A)-D\beta}{2\sqrt{\beta}} - \frac{1}{2\beta^{3/2}} \tr(A-\beta \mathcal{I})^2
    \end{align}
    The second term drops out because of our choice of $\beta$. Removing this and expanding the second term, we have
    \begin{align}
        \tr \sqrt{A} \geq D\sqrt{\beta} - \frac{1}{2\beta^{3/2}}\left(\tr A^2- 2\beta \tr A + \beta^2 D \right)
    \end{align}
    as claimed. 
\end{proof}

Now we have a bound on $\tr (\tilde{G}^{1/2})$ in terms of $\tr (\tilde{G})$ and $\tr(\tilde{G}^2)$.
We can actually compute $\tr (\tilde{G})$ and $\tr(\tilde{G}^2)$ straightforwardly. 
We start with $\tr(\tilde{G})$, 
\begin{align}
    \tr (\tilde{G}) &= \sum_{i,s}\tr Q^{i,s} \nonumber \\
    &= d^{N-1}m
\end{align}
For $\tilde{G}^2$ we need a bit more work, 
\begin{align}
    \tr(\tilde{G}^2) &= \sum_{i,j,s,t}\tr\left(Q^{i,s}Q^{j,t}\right) \nonumber \\
    &= \sum_{i,j,s,t}\tr\left((\Psi^{s}_{A_0A_i}\otimes \mathcal{I}_{\bar{A}_i})(\Psi^{t}_{A_0A_j}\otimes \mathcal{I}_{\bar{A}_j})\right) \nonumber \\
    &= \sum_{i=j}\sum_{s,t}\tr\left((\Psi^{s}_{A_0A_i}\otimes \mathcal{I}_{\bar{A}_i})(\Psi^{t}_{A_0A_i}\otimes \mathcal{I}_{\bar{A}_i})\right) + \sum_{i\neq j}\sum_{s,t}\tr\left((\Psi^{s}_{A_0A_i}\otimes \mathcal{I}_{\bar{A}_i})(\Psi^{t}_{A_0A_j}\otimes \mathcal{I}_{\bar{A}_j})\right) \nonumber \\
    &= \sum_{i=j}\sum_{s,t}\delta_{s,t} d^{N-1} + \sum_{i\neq j}\sum_{s,t} \frac{1}{d^3} d^N \nonumber \\
    &= md^{N-1} + |S|^2(N^2-N)d^{N-3}.
\end{align}

Now we can collect everything to obtain our lower bound on the fidelity. 
From lemma \ref{lemma:rootGlowerbound}, we have
\begin{align}
    F_e(\Lambda^S) \geq \frac{1}{md^{2N}}\left[\tr(\tilde{G}^{1/2})\right]^2
\end{align}
and then from lemma \ref{lemma:rootAlowerbound} we have
\begin{align}
    F_e(\Lambda^S) \geq \frac{1}{md^{2N}}\left( D\sqrt{\beta} - \frac{1}{2\beta^{3/2}}\left(\tr \tilde{G}^2- 2\beta \tr \tilde{G} + \beta^2 D \right)\right)^2
\end{align}
and inserting the values of $\tr \tilde{G}$, $\tr \tilde{G}^2$, $D=d^{N+1}$, we obtain
\begin{align}
    F_e(\Lambda^S) \geq \left(1-\frac{d^2-|S|}{2N |S|} \right)^2
\end{align}
as claimed. 

\bibliographystyle{unsrt}
\bibliography{biblio.bib}

\end{document}